\documentclass[english,11pt,b5paper,twoside,openright]{extbook}
\usepackage[T1]{fontenc}
\usepackage[latin9]{inputenc}
\usepackage[b5paper]{geometry}
\usepackage{refstyle}
\usepackage{textcomp}
\usepackage{eqnarray,amsmath}
\usepackage{amssymb}
\usepackage{graphicx}
\usepackage{multirow}
\usepackage{tabularx}
\usepackage{pdfpages}
\usepackage{natbib}
\usepackage{times}
\usepackage{appendix}
\newcommand{\bmath}[1]{\mbox{ \boldmath $\!#1\!$ \unboldmath}}
\let\newcommand=\providecommand

\newcommand{\chaptertoc}[1]{\chapter*{#1}
\addcontentsline{toc}{chapter}{#1}
\markboth{\slshape\MakeUppercase{#1}}{\slshape\MakeUppercase{#1}}}

\author{Vincent Pelgrims}
\title{\textsc{Cosmic Anisotropies from Quasars
\\ \normalsize{\large{From polarization to structural-axis alignments}}}}

\begin{document}
\frontmatter
\thispagestyle{empty}

\begin{figure}
\centering
\begin{minipage}{\textwidth}
\centering
\includegraphics[width=2.6cm]{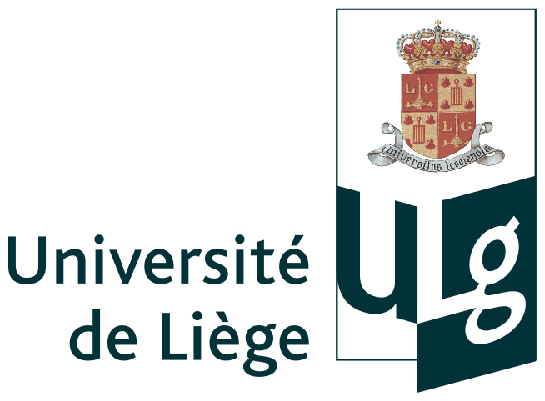}
\end{minipage}
\end{figure}

\vspace{64.5mm}
\begin{minipage}{\textwidth}
\centering
{\textbf{{FACULT{\'E} DES SCIENCES}
\\ \normalsize{{Institut d'Astrophysique et G{\'e}ophysique de Li{\`e}ge}}}}
\end{minipage}

\vspace{50mm}
\begin{minipage}{\textwidth}
\centering
\huge{\textbf{\textsc{Cosmic Anisotropies from Quasars
\\ \normalsize{\LARGE{From polarization to structural-axis alignments}}}}}
\end{minipage}
\vspace{20mm}
\begin{center}
\begin{minipage}{.5\textwidth}
Reading Committee:
\begin{flushright}
Prof. Jean Ren{\'e} Cudell (promoter)\\
Dr. Damien Hutsem{\'e}kers (co-promoter)\\
Prof. Pierre Magain (president)\\
Dr. Dominique Sluse (secretary)\\
Dr. Maret Einasto\\
Dr. Olaf Wucknitz
\end{flushright}
\end{minipage}
\end{center}
\vfill
\begin{minipage}[b]{\textwidth}
\begin{center}
\begin{minipage}{.4\textwidth}
Ann{\'e}e Acad{\'e}mique 2015--2016
\end{minipage}\hfill
\begin{minipage}[t]{.4\textwidth}
\begin{flushright}
Dissertation pr{\'e}sent{\'e}e par\\
Vincent PELGRIMS\\
en vue de l'obtention du grade\\
de Docteur en Sciences
\end{flushright}
\end{minipage}
\end{center}
\end{minipage}

\cleardoublepage
\thispagestyle{empty}
\vspace*{\stretch{1}}
\begin{flushright}
\itshape
Je sais que, curieusement, ceci a un sens et qu'on peut\\
s'amuser {\`a} voir jusqu'o{\`u} on comprend,\\
comme par-dessus les cols.\\
\medskip
\textsc{Ito Naga}
\end{flushright}
\vspace{\stretch{3}}

\tableofcontents

\chapter*{Acknowledgments}
\addcontentsline{toc}{chapter}{Acknowledgments}
I would like to warmly thank my promoter Prof. Dr. Jean Ren{\'e} Cudell, Associate Professor at the University of Li{\`e}ge, for his valuable cooperation and wisdom all along this exciting doctoral thesis.
I am also particularly indebted to my co-promoter Dr. Damien Hutsem{\'e}kers, Senior Researcher Associate at F.R.S.-FNRS, for his assistance and cooperation.
It was a real pleasure to collaborate with them and with Dominique Sluse and Lorraine Braibant, whose own research in the field provided an invaluable resource. My thanks to them for their extended awareness and responsiveness.
\newline
\newline
I extend my sincere thanks to the members of my reading committee: Dr. Maret Einasto, Senior Research Fellow at Tartu Observatory,
Prof. Dr. Pierre Magain, Professor at the University of Li{\`e}ge, Dr. Dominique Sluse, Post-doctoral Researcher at the University of Li{\`e}ge and Dr. Olaf Wucknitz, Scientific Staff at the Max-Planck-Institut for Radioastronomy in Bonn, for their careful reading of the text and the constructive criticism they will provide me with.
\newline
\newline
I am grateful to the scientific community of the Department of Astrophysics, Geophysics and Oceanography and especially the members of the Fundamental Interactions in Physics and Astrophysics group for inspiring discussions about physics and life during coffee breaks.
\newline
\newline
I further express my warmest gratitude to all the members of my (extended) family and my close friends for their enduring support.
My final thanks go to my dear fianc{\'e}e, Lou, for her immeasurable support, encouragement and comprehension.

\chapter*{R{\'e}sum{\'e}}
\addcontentsline{toc}{chapter}{R{\'e}sum{\'e}}
La comparaison des orientations des vecteurs de polarisation de
la lumi{\`e}re visible provenant de quasars s{\'e}par{\'e}s par des milliards d'ann{\'e}es-lumi{\`e}re
a men{\'e} {\`a} une d{\'e}couverte surprenante. Ces vecteurs sont align{\'e}s au lieu d'{\^e}tre
orient{\'e}s al{\'e}atoirement comme on pouvait s'y attendre.
La d{\'e}couverte de ces corr{\'e}lations qui impliquent des {\'e}chelles de l'ordre du gigaparsec
a {\'e}t{\'e} confirm{\'e}e et renforc{\'e}e par de nouvelles observations et de nouvelles analyses.
{\`A} ce jour, cependant, aucun sc{\'e}nario n'a pu rendre compte de ces corr{\'e}lations de mani{\`e}re satisfaisante.

Nous avons dedi{\'e} cette th{\`e}se de doctorat {\`a} l'{\'e}tude minutieuse de ces observations qui se
r{\'e}v{\`e}lent {\^e}tre potentiellement en d{\'e}saccord avec le mod{\`e}le cosmologique actuel.
Il nous a sembl{\'e} important de confirmer de fa\c{c}on ind{}pendante ces alignments {\`a} grande {\'e}chelle qui n'avaient {\'e}t{\'e} caract{\'e}ris{\'e}s qu'{\`a} l'aide des deux m{\^e}mes m{\'e}thodes statistiques.
{\`A} cette fin, nous {\'e}laborons une nouvelle m{\'e}thode statistique dans le Chapitre~1. Celle-ci est
d{\'e}di{\'e}e {\`a} l'analyse et {\`a} la carat{\'e}risation de la distribution de vecteurs axiaux perpendiculaires aux lignes de vis{\'e}e d'un {\'e}chantillon de sources dispers{\'e}es sur la sph{\`e}re c{\'e}leste.
Cette nouvelle m{\'e}thode statistique nous permet, dans le Chapitre~2, de confirmer ind{\'e}pendamment la pr{\'e}sence des alignments {\`a} grandes {\'e}chelles des vecteurs de polarisation optique des quasars, mais aussi de red{\'e}finir objectivement les r{\'e}gions d'alignement.
Nous d{\'e}dions le Chapitre~3 {\`a} une analyse d{\'e}taill{\'e}e d'un {\'e}chantillon de donn{\'e}es polarim{\'e}triques obtenues en longueur d'onde radio. Celle-ci r{\'e}v{\`e}le le m{\^e}me genre d'alignement des vecteurs de polarisations. Les r{\'e}gions du ciel dans lesquelles se concentrent les alignements radio se trouvent au voisinage de celles d{\'e}finies dans le visible. Ceci sugg{\`e}re que les axes des quasars eux-m{\^e}mes pourraient {\^e}tre align{\'e}s.
Afin d'explorer cette possibilit{\'e}, nous analysons dans le Chapitre~4 de nouvelles donn{\'e}es de polarisation obtenues pour des quasars se trouvant dans deux grands amas de quasars. En tenant compte du lien entre l'orientation des vecteurs de polarisation optique et l'orientation des quasars par rapport {\`a} la ligne de vis{\'e}e, nous concluons que les axes de rotation des trous noirs supermassifs situ{\'e}s au centre des quasars sont align{\'e}s avec la structure de l'amas auquel ils appartiennent. Dans le Chapiter~5, nous confirmons notre d{\'e}couverte en utilisant un {\'e}chantillon d'amas de quasars et des mesures de polarisation radio. Nous observons {\'e}galement que l'orientation pr{\'e}f{\'e}rentielle des axes de rotation des trous noirs supermassifs d{\'e}pend de la richesse de l'amas de quasars dans lequel ils sont contenus.

L'ensemble de ces r{\'e}sultats sugg{\`e}re que les alignements {\`a} tr{\`e}s grandes {\'e}chelles des vecteurs de polarisation des quasars soient li{\'e}s aux corr{\'e}lations au sein des amas de quasars. Ces corr{\'e}lations qui seraient elles-m{\^e}mes dues aux alignements des axes de rotation des trous noirs dans la toile cosmique.

\chapter*{Abstract}
\addcontentsline{toc}{chapter}{Abstract}
The comparison of the orientations of the optical-polarization
vectors of quasars that are separated by billions of light-years has led to the striking discovery
that they are aligned instead of pointing in random directions as expected.
This discovery has been confirmed and the significance of the correlations enhanced
but no satisfactory scenario has been provided so far to account properly for the specificities of
these gigaparsec-scale correlations.

We devoted this doctoral thesis to an in-depth analysis of these observations that may constitute an
anomaly to the current cosmological paradigm.
As the large-scale polarization alignments had always been
characterized through the two same statistical methods, we found it important to
independently confirm them.
Therefore, in Chapter~1, we develop a new and independent statistical method which is
dedicated to the study and the characterization of the distribution of the orientations of vectorial
quantities that are perpendicular to the lines of sight of a set of sources spread
on the celestial sphere. This allows us, in Chapter~2, to confirm independently the large-scale
optical-polarization-vector alignments and, further, to refine the limits of the aligned regions through
an unbiased characterization of the signal.
In Chapter~3, we provide a detailed analysis of a large sample of polarization measurements made
at radio wavelengths. We report on similar polarization-vector alignments. The regions of alignments
of the quasar-radio-polarization vectors are found to be close to the regions of optical alignments.
This suggests that quasar axes themselves could be aligned.
Thus, in Chapter~4, based on new observations, we analyse the optical-polarization vectors of quasars
that belong to two large groups. Taking into account the link between the optical-polarization
vectors and the morphologies of the quasars, we find that the spin axes of the supermassive black
holes located at the centres of quasars align with the axis of the large-quasar group to which they
belong. We use radio-polarization data to reinforce our findings in Chapter~5 where we consider
a sample of quasar groups drawn from the Sloan Digital Sky Survey. We additionally find that
the preferred orientations of the spin axes of the supermassive black holes depend on the richness
of their host large-quasar groups.

These results suggest that the very-large-scale alignments of quasar-polarization vectors and the
correlations with the large-quasar groups could be due to the alignments of the supermassive black
hole spin axes within the cosmic web.

\mainmatter

\chaptertoc{Introduction}
Cosmology is the part of science that aims at the understanding of the
working of the Universe as a whole. The main pillar of the current standard
cosmological model, the cosmological principle, states that the Universe has to
be isotropic and homogeneous when it is viewed at sufficiently large scales.
The philosophical reason for this assumption is the generalized Copernican
principle which states that there is no privileged observer in the Universe.
This implies that the part of it which we can observe and study is a
representative sample of its entirety and that its components, their properties
and the physical laws at play have to apply throughout, whoever is observing
from wherever. Fundamentally, the cosmological principle requires that the
physics is the same in every reference frame. Naturally, general relativity has
proven to be the favoured framework within which scientists can build a theory
that fairly describes the Universe in its entirety. Provided with the cosmological
principle, the Friedman-Lema{\^i}tre-Robertson-Walker model stemmed from
this theory and was shown to be the most suitable model to account for our
world. Following this description, we live in a three-plus-one dimensional space
filled with particles and radiation that interact with each other.
Throughout the years, astronomers and astrophysicists have noticed that our
Universe is expanding, emerging from a singularity called the Big-Bang, that it
is mainly filled by unseen matter, the still elusive dark matter, and that its
expansion is accelerated by a  positive cosmological constant covering an
unknown component of negative pressure acting as a vacuum energy, the
dark energy.

Consequently, cosmologists and astrophysicists ended up with the standard
cosmological model often referred to as the $\Lambda$CDM model, where
the $\Lambda$ stands for the cosmological constant and the CDM for the
cold dark matter. Recent studies of the cosmic microwave background (CMB),
the relic radiation of the hot and dense phase of the early Universe, have shown
that this model works fairly well at explaining these data. However, given this
representation of the cosmos, more than ninety five per cents of its energy content
remain unknown. In a sceptical sense, the $\Lambda$CDM cosmological model
turns out to be a very convincing parametrization of our ignorance and most of
Nature is still part of our incomprehension.

In agreement with the theory and the cosmological principle, the relic CMB
radiation is found to be very isotropic and amazingly homogeneous, but also
seeded by sparse temperature fluctuations. Those mark small under- and
over-densities of the early maelstrom. Under the pull of gravity, these
inhomogeneities have grown to form the voids and the clusters and
superclusters of galaxies that we observe in our neighbouring Universe.
The galaxies in the Universe are indeed not evenly distributed through space
but rather form a cosmic web composed of filaments and clumps around huge
voids where galaxies are scarce. This wonderful arrangement of the matter
distribution forms what is called the large-scales structures of the Universe.
However, when viewed on sufficiently large scales, i.e. when a sufficiently
large fraction of the cosmos is considered, the cosmological principle tells
us that the overall distribution has to be homogeneous and isotropic.
The study of the (observable) matter distribution, which is highly inhomogeneous
and anisotropic at small scales ($\lesssim 100\, \rm{Mpc}$), also helps refine our
understanding of the whole picture. However, as the samples of galaxies and other
objects have become larger and deeper in the distant Universe, large-scale
structures of a size comparable to that of the studied samples have always been
found. These inhomogeneities are sometimes able to challenge the
cosmological principle. These studies thus stress the importance of getting more
complete and deeper data sets in order to examine robustly the distribution of matter
and, perhaps, to refine our knowledge. However, the further away the galaxies,
the fainter they appear, which makes the large-scale structures of the
young --or distant-- Universe difficult to study. In this regard, galaxies
harbouring very bright active galactic nuclei, and especially quasars,
are a promising tool as they can be observed at the far reaches of the
cosmos and thus they could be used to trace the evolution of the matter
distribution over very large scales.
What makes the nuclei of certain galaxies active, i.e. very luminous, is
the accretion of the inner material of these galaxies into the massive
black hole located at their centre. Quasars are the most energetic members
of this class of extragalactic objects. It is remarkable, though,
that even in large samples of such bright sources that become available
nowadays, incredibly large systems have been observed and their nature,
while still controversial, could constitute a serious anomaly for the
cosmological paradigm.

Beside the large-scale anisotropies in the space distribution of these
galaxies harbouring active nuclei, other intriguing observations are
available and some could potentially compete with our current view of
the Universe. One of these concerns the polarization of the light emitted
from quasars.
Indeed, more than seventeen years ago, \citeauthor{Hutsemekers1998}
reported on large-scale anisotropies of the orientations of the polarization
vectors of quasars. These correlations involve sources separated by
amazingly huge distances. Until now and despite the variety of models that
have been proposed, none have been able to provide a satisfactory scenario
to account for these observations, which might consequently challenge the
cosmological principle. We can present these unexplained observations as
follow.

\section*{The anomaly}
When studying the polarization of an electromagnetic wave emitted by a
source, a polarization vector is defined in the plane orthogonal to the
propagation direction of the wave, the line of sight. When the directions
of the oscillating electric fields composing the light beam are not
isotropically distributed in that plane, the radiation is said to be partially
or fully polarized and the polarization vector has a non-vanishing norm
which reflects the degree of linear polarization of the waves.
The direction of the polarization vector with respect to an arbitrary
direction defines the polarization position angle. In astronomy, the arbitrary
direction is often chosen as being the North-South axis of the equatorial
coordinate system and the position angles are counted positively in the
East-of-North convention.
The study of the polarization of the light from astronomical source allows
us to infer the geometrical arrangement of their components and, in
general, to probe their orientation even when these objects are too far
to be resolved by telescopes. The analysis of the polarization is thus a
powerful tool in astronomy.

Regarding polarization studies of extragalactic sources, there have been
puzzling observations that could potentially have cosmological importance.
In 1998, \citeauthor{Hutsemekers1998} reported on the striking observation
that the polarization vectors of the visible light from luminous active galactic
nuclei\footnote{Throughout this work and unless specified, we will referred to
active galactic nuclei as quasars, regardless of the observational specificities
that make the latter class more restrictive.} separated by cosmological
distances are coherently oriented in certain regions of space. This analysis,
originally presented in (\citealt{Hutsemekers1998}), was based on a sample of
170 quasars compiled from the literature and carefully selected in order to
eliminate at best the contamination by our hosting Milky Way.
Since the discovery of the regions of polarization alignments, dedicated
observational campaigns have been conducted and the present sample is
composed of 355 quasars with high-quality polarization measurements at
optical wavelengths (\citealt{Lamy-Hutsemekers2000};
\citealt{Hutsemekers-Lamy2001}; \citealt{Sluse-et-al2005} and
\citealt{Hutsemekers-et-al2005}). With the latest sample, statistical tests
that study the coherence of the polarization position angles in groups of
neighbouring sources show that the observed correlations have probabilities
between $3 \times 10^{-5}$ and $2 \times 10^{-3}$ to be due to chance
(\citealt{Hutsemekers-et-al2005}). The probabilities depend on the chosen test
as they are not equally sensitive to the same alignment patterns.
We shall summarize below the main characteristics of the anisotropies in the
orientations of the optical polarization vectors from quasars.

\subsection*{Alignment regions}
The anisotropies of the orientations of the polarization vectors involve
quasars that are grouped in three distinct regions of the three-dimensional
comoving space. These comoving regions, two located in the North Galactic
Cap and one in the South Galactic Cap, are found towards two regions of the
sky that are roughly antipodal. While the borders of these regions are loosely
determined due to the (observational) inhomogeneous character of the data
set, they are defined in redshift and equatorial coordinates as
\begin{itemize}
\item $168^\circ \leq \alpha \leq 218^\circ$; $\delta \leq 50^\circ$ and $0.0 \leq z \leq 1.0$
\item $168^\circ \leq \alpha \leq 218^\circ$; $\delta \leq 50^\circ$ and $1.0 \leq z \leq 2.3$ (A1)
\item $320^\circ \leq \alpha \leq 360^\circ$; $\delta \leq 50^\circ$ and $0.7 \leq z \leq 1.5$ (A3)
\end{itemize}
where $\alpha$, $\delta$ and $z$ are the right ascension, the declination and
the cosmological redshift, respectively.
\begin{figure}[h]
\begin{center}
\begin{minipage}{0.8\textwidth}
\begin{center}
\begin{tabular}{@{}cc}
\includegraphics[width=0.5\columnwidth]{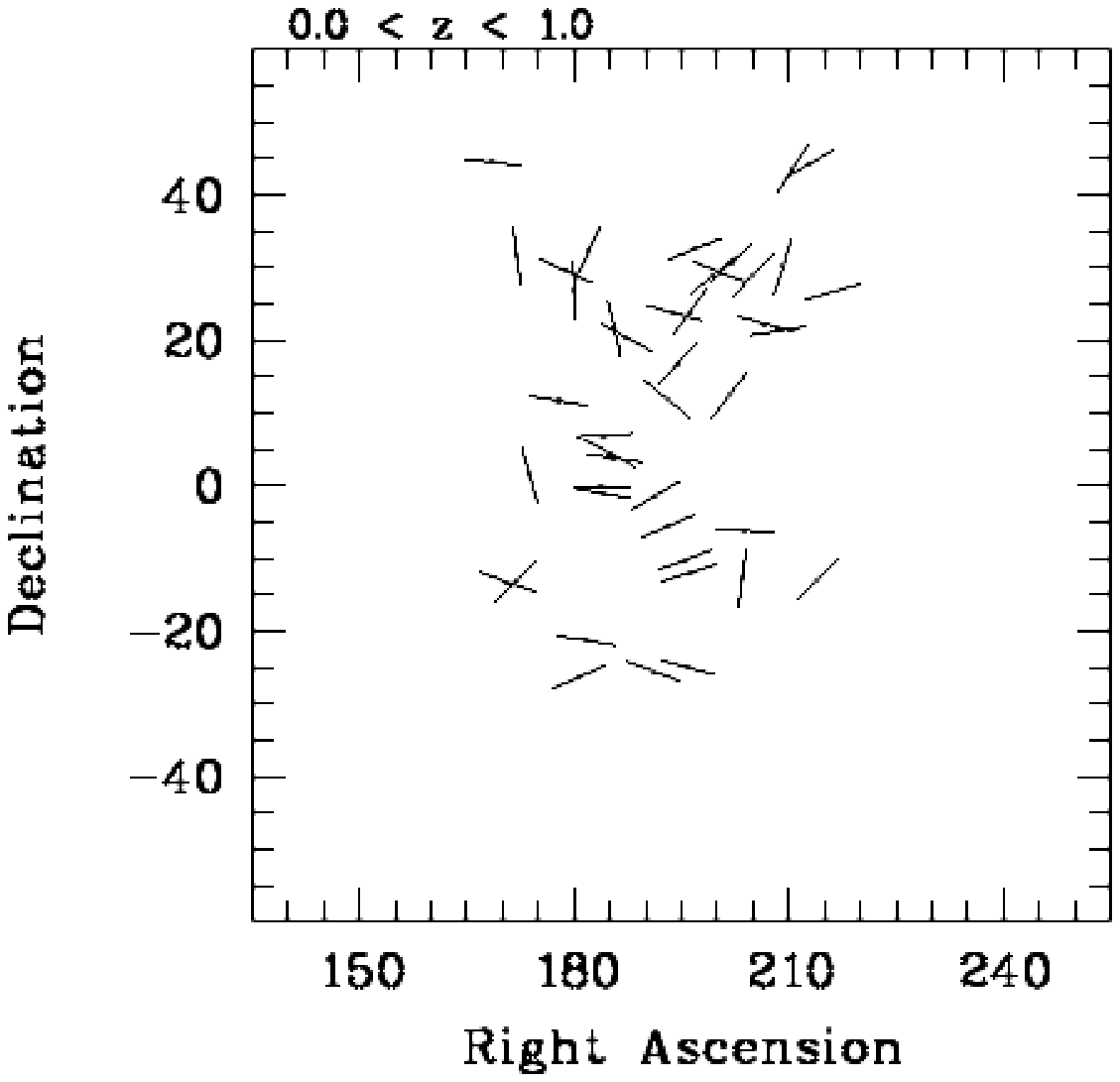} &
\includegraphics[width=0.5\columnwidth]{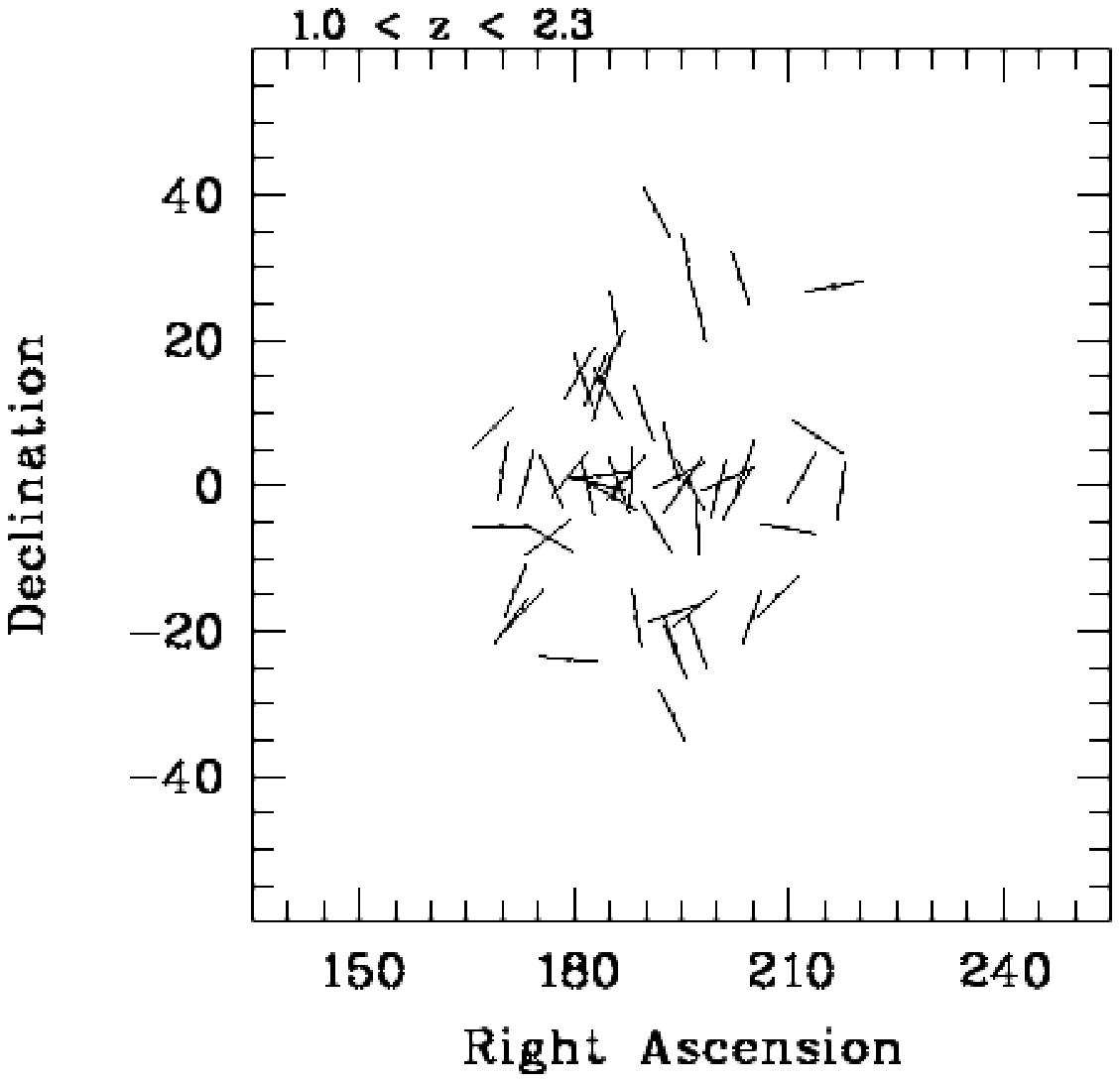} \\
[1.5ex]
\multicolumn{2}{c}{
\includegraphics[width=0.5\columnwidth]{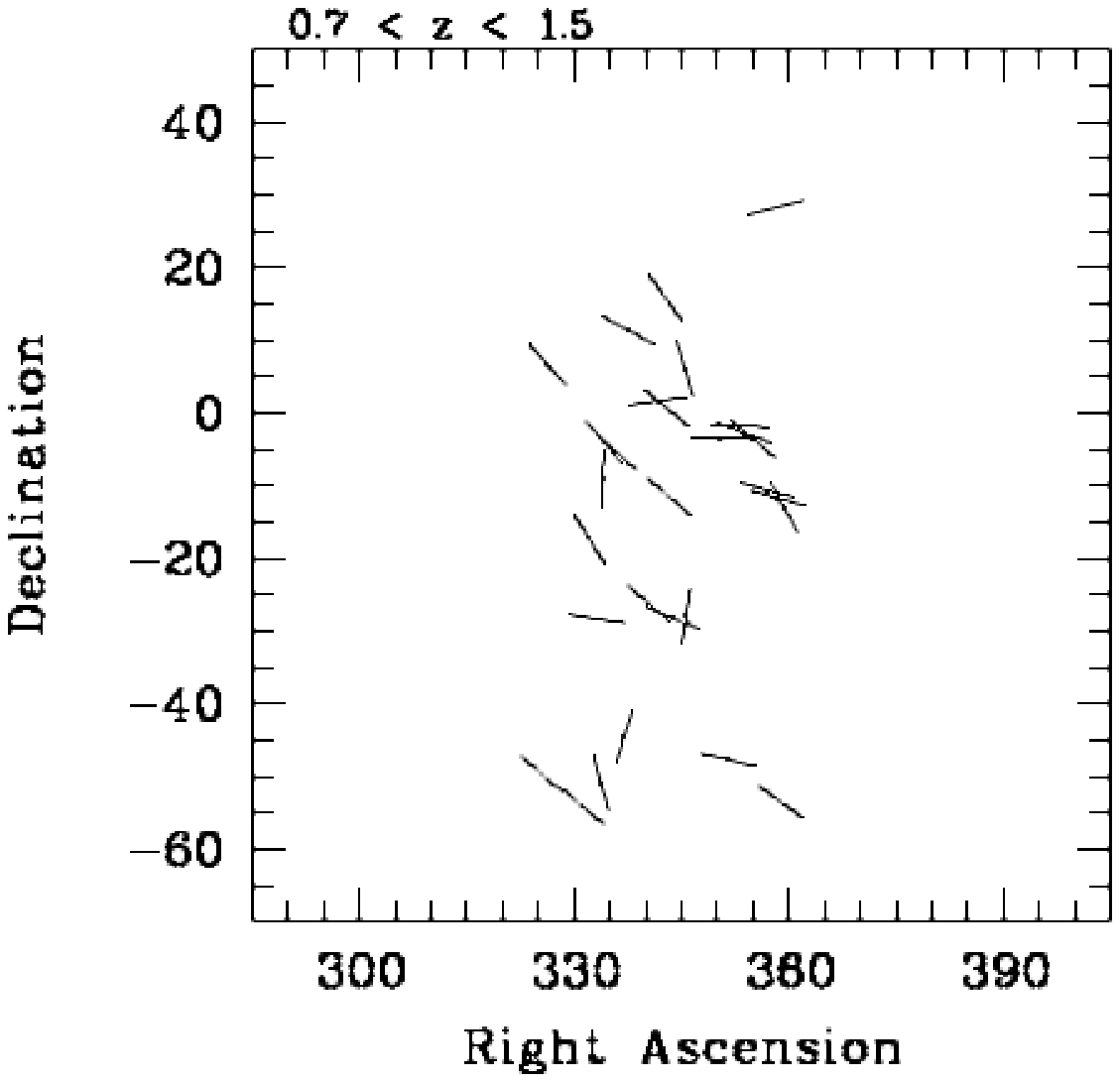}} \\
\end{tabular}
\end{center}
\caption{\small{Maps of quasar polarization vectors in the North Galactic Cap
(\textit{top}) and in the South Galactic Cap (\textit{bottom}).
The dots represent the quasars
and the segments the normalized polarization vectors. Coherent orientations
are visually detectable in each of these regions of the comoving space.
The regions illustrated are the A1 region (\textit{top right}), its low-redshift counterpart
(\textit{top left}) and the A3 region (\textit{bottom}).
Figures taken from \citet{Hutsemekers-et-al2005}.}}
\label{fig:alignedRegions_H05}
\end{minipage}
\end{center}
\end{figure}

These regions are illustrated in Fig.~\ref{fig:alignedRegions_H05} where the
dots represent the quasars and the segments the normalized polarization
vectors. For historical reasons, the region of the South Galactic Cap is referred
to as the A3 region while the region called A1 is the high-redshift region of the
North Galactic Cap, the ``A'' standing for alignment.
Note that in the original paper of 1998, there was an A2 region defined as
$150^\circ \leq \alpha \leq 250^\circ$; $\delta \leq 50^\circ$ and
$0.0 \leq z \leq 0.5$ which is somehow included in the low-redshift part of the
data set towards the same region of the sky as A1. The A2 region has been
dropped because of the adopted observational strategies to increase the
sample size from 1998 to 2005. As we will see in
Section~\ref{subsecPC-1:bestOptRegion}, though, the alignment in this region
is still there.

\begin{figure}[h]
\centering
\begin{minipage}{0.8\textwidth}
\centering
\includegraphics[width=\columnwidth]{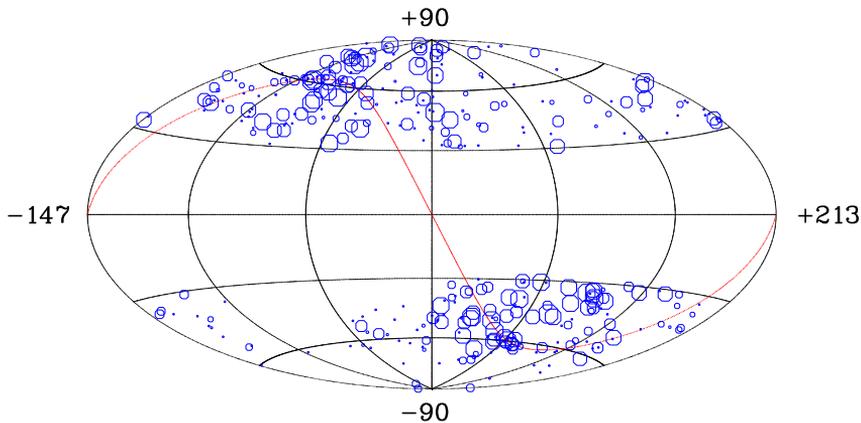}
\caption{\small{Hammer-Aitoff projection of the quasar positions on the sky,
in Galactic coordinates. The 355 objects are plotted. The radius of the circles
is give by $\rho_i \propto \exp s_i -0.9$, where $s_i$ refers the statistics
defined in Section~\ref{secPH-1:IdentifRegions} from Eq.~\ref{eq:Z_i} for the
Z test with parallel transport correction and $n_v = 40$
(see Section~\ref{stat_SandZ}). The larger the circle, the more significant the
alignment at that point. The superimposed red line gives the location of the
celestial equator. Figure taken from \citet{Hutsemekers-et-al2005}.}}
\label{fig:locStat_H05}
\end{minipage}
\end{figure}

Based on the very fact that the A1 and A3 regions are roughly opposite on the sky,
\cite{Hutsemekers-et-al2005} defined the A1--A3 axis as the combination of the
two sky windows pointing towards these regions.
Statistical tests have shown that most of the alignment signal
comes from the quasars that belong to this so-called A1--A3 axis.
This fact is illustrated by Fig.~\ref{fig:locStat_H05}
where the size of the circles reflect the contribution to the overall statistics of the
local alignments of the neighbouring groups. We clearly see that the most ``aligned''
groups (larger circles) concentrate in a dipole pattern.

It is worth noting that the direction defined by the A1--A3 axis points
roughly towards the Virgo cluster. Other large-scale anisotropies have
been reported towards the same direction. They include
the alignment of the directions towards which the amplitudes of the
low-multipoles of the cosmic microwave background are significantly
suppressed (e.g., \citealt{Tegmark-et-al2003}; \citealt{PlanckXXIII_2013}),
the direction of the strongest anisotropy in the offset angles of radio galaxy
symmetry axes relative to their average polarization angles
(\citealt{Jain-Ralston1999}) and the directions of maximum accelerating
expansion rate of the Universe inferred from the Union2 Supernova type
Ia data set (\citealt{Antoniou-Perivolaropoulos2010}).
The coincidental alignment of these axes of anisotropies drawn from
different probes, including the quasar-optical-polarization alignments, has
been questioned by several authors (e.g., \citealt{Ralston-Jain2004};
\citealt{Land-Magueijo2007} and \citealt{Antoniou-Perivolaropoulos2010}).

\subsection*{Redshift dependence}
\begin{figure}[h]
\centering
\begin{minipage}{0.8\textwidth}
\centering
\includegraphics[width=\columnwidth]{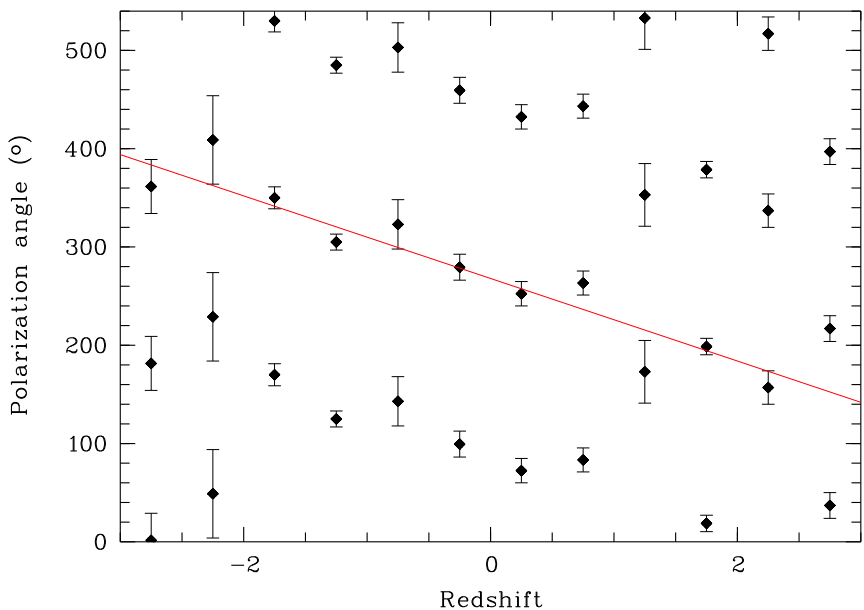}
\caption{\small{The quasars polarization angles, averaged over redshift bins $\Delta z = 0.5$, as a function of the redshift. Redshifts are counted positively for object located in the North Galactic Cap and negatively for those on in the South Galactic Cap. Only the 183 quasars belonging to the A1--A3 axis are considered. Error bars represent $68\,\%$ angular confidence intervals for the circular mean (\citealt{Fisher1993}); they must be seen with caution when the number of quasars per redshift bin is small, i.e. at large $z$. Data points are replicated at $(z,\,\bar{\psi})$, at $(z,\,\bar{\psi} + 180^\circ)$ and at $(z,\,\bar{\psi} + 360^\circ)$. The fitted red line is given by $\bar{\psi} = 268^\circ - 42^\circ z$ (see \citealt{Hutsemekers-et-al2005} for details). Figure taken from \citet{Hutsemekers-et-al2005}.}}
\label{fig:zRot_H05}
\end{minipage}
\end{figure}

The preferred orientations with respect to which the polarization vectors
concentrate differ for the three regions of alignments.
The mean polarization position angles are $79^\circ$ for the low-$z$
counterpart of the A1 region, $8^\circ$ for the A1 region and $128^\circ$
for the A3 region.
The redshift dependence of the anisotropies, which can be observed by
comparing the low- and high-redshift parts of the window towards the A1
region in Fig.~\ref{fig:alignedRegions_H05}, was further tested by \citet{Hutsemekers-et-al2005}.
Concentrating on the 183 quasars that belong to the A1--A3 axis, the authors
potentially unveiled an overall rotation of the mean polarization PA with respect
to the redshift of the sources. This significant trend, which is illustrated in
Fig.~\ref{fig:zRot_H05}, might suggest that the mechanism responsible for the
anisotropic patterns is of cosmological nature.

\subsection*{Surviving possible biases}
The redshift dependence of the importance of the anisotropies and of the
preferential directions constitutes the most stringent argument against
potential biases of the polarization data. Indeed, as was extensively
discussed in (\citealt{Hutsemekers1998}; \citealt{Hutsemekers-Lamy2001};
\citealt{Sluse-et-al2005}; \citealt{Hutsemekers-et-al2005}),
Galactic dust contamination through
dichroism is unable to account for all the anisotropies.
Specifically, if one tries to cancel the alignment in one of the regions by
correcting the polarization measurements from quasars to account for an
hypothetical (strong) Galactic dust (de)polarization, then, inevitably, one
would produce new (or stronger) alignments in the same field but at other
redshift(s). Furthermore, it is worth mentioning that the cut in the degree of
linear polarization ($p_{\rm{lin}}\geq 0.6\,\%$) has been introduced while
defining the sample especially to ensure limited contamination from the
Galactic interstellar medium.
The cut in Galactic latitude ($|b_{\rm{gal}}|\geq 30^\circ$), which can be
readily observed in Fig.~\ref{fig:locStat_H05}, was introduced for the same reason.
Beside, an instrumental bias is also unlikely as approximately half of the
sample comes from different surveys and instruments compiled in
\citet{Hutsemekers1998}. Nevertheless, even if that would be the case,
instrumental bias would produce preferred polarization position angles at
the particular values of $0^\circ$ and $90^\circ$, inconsistent with the
reported ones. The reason is that, at optical wavelengths, the detectors and
the optics are usually oriented following the North-South axis, whatever the
pointing direction.

\subsection*{Search for the underlying mechanism}
Several scenarios were proposed to explain the large-scale anisotropies
of the optical-polar-ization-vector orientations from quasars.
One can consider two main classes of models.
Either the quasar polarization orientation is modified along the line of sight,
or the structural axes of the sources involved in the observed anisotropies
are coherently oriented among themselves over cosmological scales, as the
position angles of the optical polarization vectors and the morphological axes
of quasars are known to be correlated.

\begin{itemize}
\item[] \textbf{Photon path effects} \hspace{0.15cm}
There are two different types of physical phenomena that may describe
the propagation effects that could produce the observed polarization vector
alignments.
\begin{itemize}
\item[] \textbf{Polarization state modulation} \hspace{0.15cm}
The observations might be explained through models that affect the
polarization state of the light during its travel towards Earth.
The effects that belong to this category are those that lead to an
effective degree of linear polarization towards a specific direction
through, e.g., anisotropic absorption or emission of the electromagnetic
waves. A mechanism that leads to an anisotropic redistribution of the
directions of the incoming electric vector fields would also be part of
this category.
\item[] \textbf{Polarization vector rotation} \hspace{0.15cm}
One might think that the intrinsic polarization state of the sources are
not altered along the path from the source to the observer but, instead,
that the polarization vectors are rotated by an underlying mechanism
that produces the correlations. In order to produce an alignment, though,
it is required that the rotation happens asymptotically towards a preferential
direction. As already mentioned in (\citealt{Hutsemekers-et-al2005}), a
Universe in anisotropic expansion could induce polarization vectors to be
coherently oriented through the Brans mechanism (\citealt{Brans1975}).
\end{itemize}
\item[] \textbf{Quasar structural axes alignments} \hspace{0.15cm}
There is observational evidence for correlations between the orientations
of the optical polarization vectors and the morphological axes of quasars
(e.g., \citealt{Rusk1990}; \citealt{Impey-Lawrence-Tapia1991};
\citealt{Smith-et-al2004}). An immediate alternative to mechanisms acting
along the line of sight is therefore the assumption that the polarization
alignments actually reflect alignments of the structural axes of quasars
over cosmological scales ($\sim 1 \, \rm{Gpc}$).
\end{itemize}

\subsection*{Status in 2012}
It is clear that the possible scenarios discussed above lead to
different characteristics of the anisotropic polarization patterns.
For instance, the detection of a wavelength dependence of the
alignment features would clearly help to differentiate models.
It is in this regard that searches for the same kind of anisotropies
in the distribution of the orientations of the polarizations from
extragalactic sources have been made at radio wavelengths by
\citet{Joshi-et-al2007}. Based on a large sample of flat-spectrum
radio sources, these authors did not find any evidence for coherently
oriented polarization vectors, even towards the regions of alignments
at optical wavelengths.

Some models can be discarded if the alignment mechanism is wavelength
dependent.
The assumption that a cosmic anisotropic expansion underlies the observed
polarization anisotropies is unlikely. This is true for every other scenarios
involving wavelength-indepen-dent rotation of the polarization vectors.
The scenario in which the quasar morphological axes are aligned is also
disfavoured because of the non-detection of polarization alignments at radio
wavelengths. Indeed, given that there are (observational) relations between
the orientations of the quasar morphological axes and the polarization vectors, not
only at optical wavelengths but also at radio wavelengths
(e.g., \citealt{Rusk-Seaquist1985}), one would expect to observe alignments
of the polarization vectors in both data sets. This is true even if the correlations
are expected to be present at slightly different levels due to the specificity
of the relation between morphological axes and polarization vectors
for each spectral band (\citealt{Rusk-Seaquist1985}).

As a consequence, the models implying a wavelength-dependent
modulation of the polarization states of the light during its travel to
Earth gained some credits.
Among such models is the hypothetical and elusive axion-like particles
that have been considered as the most serious candidate for a time
(e.g., \citealt{Das-et-al2005}; \citealt{Agarwal-Kamal-Jain2011};
\citealt{Payez-Cudell-Hutsemekers2008}). However, this scenario
has been discarded from the possible explanations for the large-scale
polarization alignments at optical wavelengths
(\citealt*{Payez-Cudell-Hutsemekers2011}). The principal reason is that
this model predicts non-negligible circular polarization which is not
observed (\citealt{Hutsemekers-et-al2010}).

\bigskip

Convinced that the study of this anomaly could help at
a better understanding of the cosmology, I dedicated my doctoral
thesis to a detailed analysis of the large-scale polarization alignments
from quasar light that did not find a satisfactory explanation.
Rather than proposing various scenarios and testing them against
the observations, this work is driven by the data.
We adopted a phenomenological approach in the sense that we
attempted to extract from the observations as much information as
possible in order to draw the most general picture of these anisotropies.
However, as we shall see throughout this thesis, our investigations are
often limited by the size of the available data sets.

\medskip

This thesis is structured as follow.
The first chapter of this work gives a review of the statistical tests
that have been extensively used throughout the years to study the
polarization correlations. We also present there a new independent
tool that we have developed ourselves.
In the second chapter we will use our new method to confirm the
presence of correlations between the orientations of the polarizations
of quasars when observed at optical wavelengths.
The third chapter consists in the search for such alignments of polarization
vectors in a large sample of radio-wavelength observations and the
presentation of our positive results. We also show that a combination
of the polarization alignments from both spectral bands is not feasible
due to the poor overlap of the data sets.
In chapter four and five, we will then report on a new type of
correlations between the polarization orientations of light from quasars
and the spatial distribution of the sources,
at optical wavelengths and at radio wavelengths.
Our conclusion summarize what we have learned from the polarization
alignments of quasars during this four years of research and contains
an outlook for further investigations.

\chapter{Dedicated statistical methods}
\label{Ch-Statistics}
Comparison of polarization orientations from sources that are scarce
and non-uniformly distributed on the sky requires dedicated statistical
methods. We find important to dedicate a full chapter to introduce the
main tools that we will use throughout this work.
Hence, we review two statistical tests that have been widely used
in the literature to characterize the polarization orientations and further
introduce a new method that we have developed.
At the end of this chapter, we also discuss an additional method that is
widely used in the framework of the studies of the alignments of the
morphological axes of galaxies embedded in clusters.

\section{The S and Z tests}
\label{stat_SandZ}
In his original paper, \citet{Hutsemekers1998} developed his own test
and further used a more sophisticated one originally developed by
Andrews and Wasserman and reviewed by \citet{Bietenholz1986}.
They are referred to as the S and Z tests, respectively. In brief, they
are elaborate integrated 2-point correlation functions adapted to
axial data. We introduce these two methods in the next sub-sections
and provide technical details.

\subsection{The S test}
\label{stat_Stest}
The S test was originally developed by \citet{Hutsemekers1998} in
order to detect and statistically characterize the alignment features
of polarization-vector orientations which were first visually detected.
Later, \citet*{Jain-Narain-Sarala2004} built a variant of this test which
actually, and as we shall see, is entirely equivalent of the original
one but is much faster on computers. This is convenient
when a detailed study is to be considered.

\subsubsection{Hutsem{\'e}kers version}
The S test is based on dispersion measures of the polarization position
angles (PA) for groups of $n_v$ neighbouring sources among the sample.
For each object $i$, the quantity 
\begin{equation}
d_i (\psi)=90-\frac{1}{n_v}\sum_{k=1}^{n_v} |90-|\psi_k - \psi||\, ,
\label{eq:d_theta}
\end{equation}
is computed, where the $\psi_k$'s are the polarization PAs of the
objects of the group of $n_v$ neighbours, including the central one,
in degree and defined in the range $0^\circ - 180^\circ$.
This positively defined function accounts for the axial nature of the
polarization vectors through the use of the absolute values
(\citealt{Fisher1993}).
For the object $i$, the mean dispersion of the PAs of the $n_v$ objects
is computed to be the minimum value of $d_i (\psi)$ and is denoted $S_i$.
This value will be small for coherently oriented polarization vectors.
If $N$ is the size of the whole sample under consideration, the degree of
alignment inside groups among the whole sample is given by a statistics
with the free parameter $n_v$, defined as
\begin{equation}
S_D = \frac{1}{N} \sum_{i=1}^{N} S_i \, .
\label{eq:S_D}
\end{equation}
$S_D$ measures the concentration of angles for groups of $n_v$ objects
close to each other in space (in 2 or 3 dimensions, as we shall see later).
If the polarization vectors are on average locally aligned, the value of $S_D$
will be smaller than in the case where the PAs are distributed following a
uniform distribution on the objects.
The significance level (SL) with which one may assign the observed
alignment patterns to chance has to be evaluated through simulated
data sets because of the mutual dependence between groups.
The percentage of simulations with a value of $S_D$ lower than that
of the data gives the SL, i.e. the probability that the dispersion of
position angles is due to chance.
The generation of simulated data sets is explained in
Section~\ref{substat_RsampleGen}.

\subsubsection{Jain et al. version}
The variant of the S test introduced by \citet{Jain-Narain-Sarala2004} is
very similar to the original one except that, instead of using the dispersion
measure in Eq.~\ref{eq:d_theta}, they use
\begin{equation}
d_i^J(\psi) = \frac{1}{n_v} \sum_{k=1}^{n_v} \cos \left(2 \left(\psi - \psi_k \right) \right)
\label{eq:d_theta_J}
\end{equation}
where the sum is, here also, over the $n_v$ neighbouring sources of the
object $i$, including the latter, and where the factor two takes into account
the axial nature of the polarization PAs. The position angle $\bar{\psi_i}$ with
respect to which the dispersion of the $n_v$ PAs of the group around object $i$
is the least maximizes the function $d_i^J(\psi)$. The measure of the dispersion
of the PAs is given by this maximal value, as it was given by the minimal value of 
Eq.~\ref{eq:d_theta}. The larger the value of $d_i^J(\bar{\psi_i})$, the more
concentrated are the PAs. The statistics of a given sample of $N$ PAs is then
introduced as
\begin{equation}
S_D^J = \frac{1}{N} \sum_{i=1}^{N} d_i^J(\bar{\psi_i})
\end{equation}
where the sum is over the entire data sample. A large value of $S_D^J$
implies a strong alignment between the polarization vectors. $S_D^J=1$
is the largest value that is possible and would imply that all the polarization
vectors of the sample are aligned with each other in groups of $n_v$
neighbours.
It is worth to note that the value $\bar{\psi}$ that maximizes Eq.~\ref{eq:d_theta_J}
minimizes in the same time Eq.~\ref{eq:d_theta}, for the same group of sources.
This shows that the statistics $S_D$ and $S_D^J$ are fully equivalent.

However, the interesting property in the version of \citet{Jain-Narain-Sarala2004}
is that the maximization of the $d_i^J(\psi)$, as well as the determination of the
$\bar{\psi_i}$'s can be performed analytically instead of having to minimize
Eq.~\ref{eq:d_theta} via time-consuming trials.

Indeed, searching for the value $\bar{\psi_i}$ that maximized
Eq.~\ref{eq:d_theta_J} by the usual procedure, one gets
\begin{equation}
\tan 2\bar{\psi_i} = \frac{\sum_k \sin 2 \psi_k}{\sum_k \cos 2 \psi_k}\;.
\label{eq:MeanPA}
\end{equation}
Injecting this value in Eq.~\ref{eq:d_theta_J} squared, one finally found after
trivial calculations that
\begin{equation}
d_i^J(\bar{\psi_i}) = \frac{1}{n_v} \left\lbrace
							\left( \sum_{k=1}^{n_v} \cos 2\psi_k \right)^2
							+ \left( \sum_{k=1}^{n_v} \sin 2\psi_k \right)^2
							\right\rbrace^{1/2}
\label{eq:d_iMAX}
\end{equation}
which is convenient to implement in a computer.

The significance level to which one assigns the observed PA correlations
is also obtained through the generation of Monte Carlo simulations.

\subsection{The Z test}
\label{stat_Ztest}
The Z test is a non-parametric test originally introduced by
Andrews \& Wasserman (\citealt{Bietenholz1986}) to quantify the correlation
in groups of objects between the PAs and the position of sources on the sky.
The results returned by this test are henceforth not expected to be in full
agreement with those from the S test since they are not probing exactly the
same kind of correlations.

The basic idea of the Z test is to compute for each object $j$ the mean direction
$\bar{\psi}_j$ of its $n_v$ neighbours, excluding this time the central object $j$,
and to compare this local average to the actual polarization PA of the object $j$,
namely, $\psi_j$.
Specifically, the PAs of the $n_v$ nearest neighbours around each object $j$
but excluding the latter, are used to compute the mean resultant vector
\begin{equation}
\bmath{Y}_j = \frac{1}{n_v} \left(\sum_{k=1}^{n_v} \cos 2\psi_k ,\, \sum_{k=1}^{n_v} \sin 2\psi_k \right)\, ,
\label{eq:Y_i}
\end{equation}
where the factor two, again, accounts for the axial nature of the polarization.
The mean direction $\bar{\psi}_j$ is given by the normalized mean vector
$\bmath{\bar{Y}}_j$ through
\begin{equation}
\bmath{\bar{Y}}_j = \left(\cos 2\bar{\psi}_j, \, \sin 2\bar{\psi}_j \right) \, .
\label{eq:mean_Y_i}
\end{equation}
The inner product $D_{i,j}=\bmath{y}_i \cdot \bmath{\bar{Y}}_j$, where
$\bmath{y}_i = \left( \cos 2\psi_i,\, \sin 2\psi_i \right)$, gives a measure
of the closeness of $\psi_i$ to $\bar{\psi}_j$.
If the PAs are correlated to the source positions, then, on average,
$\psi_i$ will be closer to $\bar{\psi}_{j=i}$ than to $\bar{\psi}_{j\neq i}$
which, in turn, implies $D_{i,j}$ to be larger for $i=j$ than for $i \neq j$.

The statistics $Z_c$ is then introduced as follow (cf. \citealt{Bietenholz1986})
\begin{equation}
Z_c=\frac{1}{N} \sum_{i=1}^{N} Z_i 
\label{eq:Z_c}
\end{equation}
where
\begin{equation}
Z_i=\frac{r_i-\left(N+1\right)/2}{\sqrt{N/12}}
\label{eq:Z_i}
\end{equation}
and where $r_i$ is the rank of $D_{i,j=i}$, when the $D_{i,j=1,n}$'s are
sorted in increasing order (i.e. the position of $D_{i,j=i}$ in the ordered
list of the $D_{i,j=1,N}$'s) and $N$ is the size of the studied sample.
If the position angles are independent of the sky positions of the objects,
then $r_i$ is expected to follow a discrete uniform distribution in the
interval $\left[1,\, N \right]$.
The expected value of $r_i$ is thus $(N+1)/2$ with a standard
deviation of $\sqrt{(N^2 -1) / 12}$. \citet{Bietenholz1986} states that,
for $N>8$, the value $N/ \sqrt{12}$ is within $1\%$ of the true standard
deviation and used it to build the statistics $Z_c$. Given the central limit
theorem, the statistics as defined in Eq.~\ref{eq:Z_c} should be normally
distributed. However, for $i \in \left[1, \, N \right]$, the $Z_i$
(from Eq.~\ref{eq:Z_i}) are not independent because of the overlaps
between the groups of nearest neighbours.

The statistics $Z_c$, which would give the average number of sigma with
which the observations differ from pure chance if the $r_i$ were independent,
is therefore approximately normally distributed. This implies that the use
of random samples is required in order to evaluate thoroughly the
significance level of the dependence of the PAs on the source locations.
The SL will therefore be defined as the percentage of simulations that
show higher value of $Z_c$ than that of the observed data.

\medskip

A modification of the Andrews \& Wasserman test was proposed by
\citet{Hutsemekers1998}.
Instead of considering the inner product with $\bmath{\bar{Y}}_j$, he
proposed to compute $D_{i,j}=\bmath{y}_i \cdot \bmath{Y}_j$.
This definition is expected to give more weight to the groups of sources
having similar PA values. Indeed, aligned polarization vectors imply a
large norm of $\bmath{Y}_i$ which provides a natural measure of the
dispersion of the position angles and leads to a large $D_{i,j=i}$ in the
case of coherent orientations between $\bmath{y}_i$ and the
$\bmath{Y}_i$ of its neighbourhood.

\begin{figure}[h]
\centering
\begin{minipage}{0.8\textwidth}
\centering
\includegraphics[width=\columnwidth]{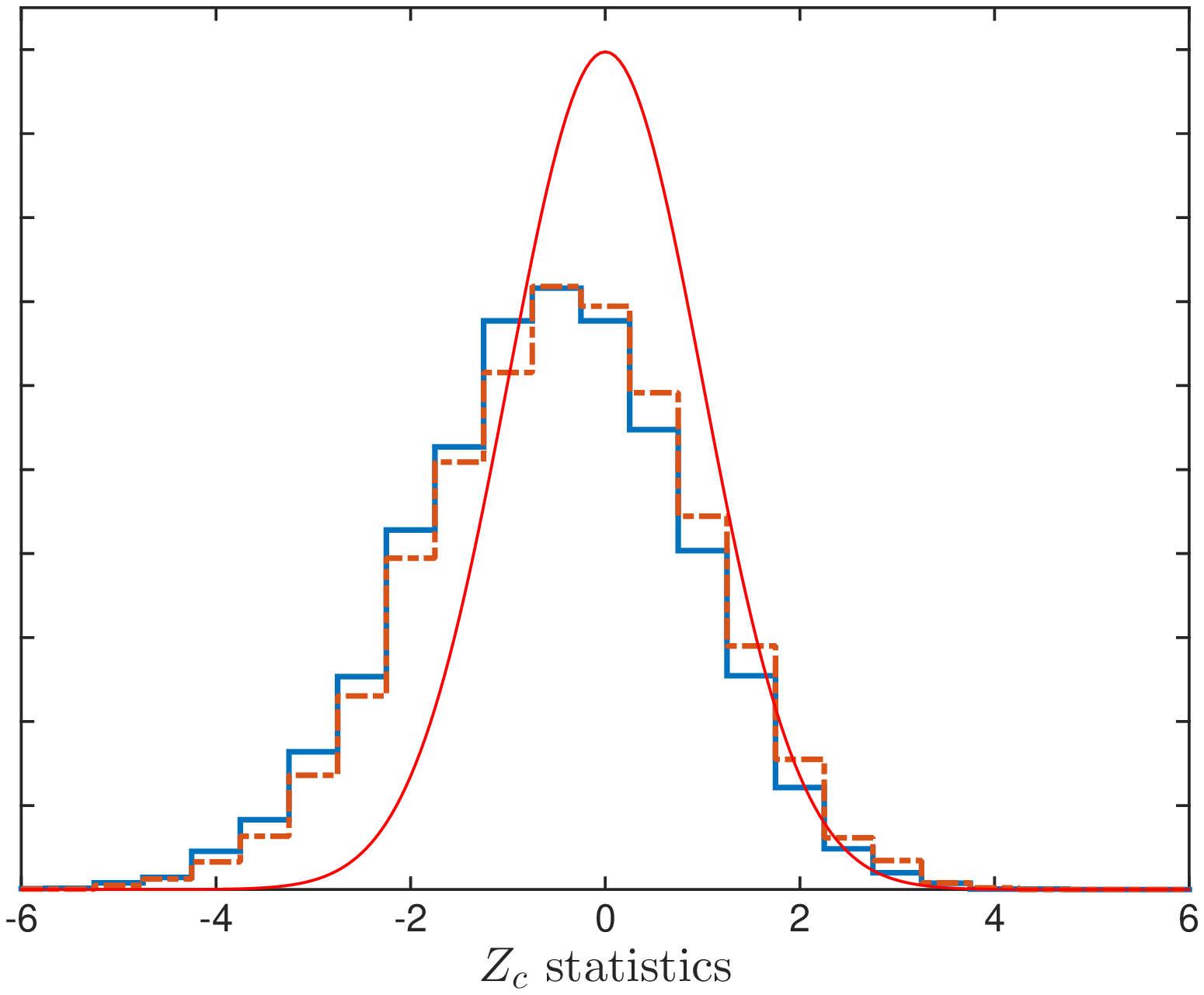}
\caption{\small{The distribution of the $Z_c$ statistics for the two versions of
the Z test. The distribution corresponding to the original test of Andrews \&
Wasserman (\citealt{Bietenholz1986}) is plotted in dot-dashed orange and the
modified version of \citet{Hutsemekers1998} in blue.
The distributions are computed for a sample of 1450 quasars introduced
in Chapter~\ref{Ch:PH-1}, the groups of nearest objects are built in 2D
(see Section~\ref{substat_ParallelTransport}) with $n_v = 80$ and a
number $N_{\rm{sim}}$ of 10\,000 Monte Carlo simulations where the PAs are
randomly generated according to a uniform distribution.
The red over-plotted curve corresponds to a normal distribution that one
would obtain if the $r_i$ were independent such that the central limit theorem
would apply.}}
\label{fig:Zc_statistics}
\end{minipage}
\end{figure}

Apart from this variation, the modified statistics is computed in the same
way as the original test of Andrews \& Wasserman discussed above.
In this work, we use this modified version when we refer to the Z test
as it should be more sensitive to local alignments and thus, more
adapted to the search for such features.
In Fig.~\ref{fig:Zc_statistics}, we compare for the first time the distribution
of the statistics $Z_c$ of the original test with the one of the modified
version. We see that both statistics are approximately normally distributed
and that they are actually only slightly different. The individual values of
$Z_c$ from the two versions, though, are found to be different. A deeper
understanding of the difference between the two version of the Z test
would require the simulation of data set with the introduction of predefined
alignment patterns in it. This is beyond the scope of this section and
should be further studied elsewhere.

\subsection{One-free-parameter statistics and its physical interpretation}
\label{substat_nv-param}
The S and Z test presented above evaluate the mean level at which the
polarization vectors of groups of sources are coherently oriented among
a sample and test its significance against randomness. The size of these
groups is defined through the only free parameter of the statistics which
is the number of nearest neighbours, $n_v$.

To build the groups of nearest-neighbour objects, the relative distances
between sources is required.
The appropriate distance measure in cosmography is the
line-of-sight comoving distance. It is the distance that we would measure
locally between two nearby objects (i.e. close in redshift or distance) today
if they were locked into the Hubble flow, i.e. if their peculiar velocities
were negligible\footnote{This will be assumed throughout this work.}.
The comoving distance between two objects remains the same with the
epoch. To compute this distance, we follow \citet{Peebles1993}.
We introduce the function
\begin{equation}
E(z) \equiv \sqrt{\Omega_M \left(1 + z \right)^3 + \Omega_k \left(1 + z  \right)^2 + \Omega_\Lambda}
\label{eq:Egeneral}
\end{equation}
where $\Omega_M$, $\Omega_k$ and $\Omega_\Lambda$ are the three
dimensionless density parameters of the Universe that reflect its matter
density, the curvature of space and the value of the
cosmological constant, respectively. These three parameters completely
determine the geometry of the Universe and the critical density $\Omega$,
which is the sum of the three, is normalized to unity (see \citealt{Hogg2000}
for a brief summary).
Assuming a flat and isotropic Universe ($\Omega_k=0$), $E(z)$ takes the form
\begin{equation}
E(z) = \sqrt{1 + \Omega_M \left( (1+z)^3 - 1 \right)}\;.
\label{eq:Eflat}
\end{equation}
The total line-of-sight comoving distance $r(z)$ between a source of redshift
$z$ and an observer at $z=0$ is given by the integration of $dz/dE(z)$ along
the photon path (see \citealt{Hogg2000} and \citealt{Peebles1993} for details)
as
\begin{equation}
r(z) = \frac{c}{H_0} \int_0^z \frac{dz'}{E(z')}
\label{eq:D_Comoving}
\end{equation}
where $c$ and $H_0$ are respectively the speed of light and the Hubble
parameter as measured today.

From Eq.~\ref{eq:D_Comoving}, the rectangular coordinates of each source
are evaluated in a flat Universe through
\begin{eqnarray}
x & = & r \cos \delta \cos \alpha	\\ \nonumber
y & = & r \cos \delta \sin \alpha		\\ \nonumber
z & = & r \sin \delta
\label{eq:xyz-coordinates}
\end{eqnarray}
where $\delta$ and $\alpha$ denote the declination and the right
ascension of the object in the equatorial coordinate system.
The relative distances between sources are then simply computed
through the use of their rectangular coordinates.

Note that the Hubble parameter in Eq.~\ref{eq:D_Comoving} only
acts as a global scaling factor. Therefore, the built neighbouring
groups do not depend on its exact value.
This is not true, however, for the value of $\Omega_M$ as it enters
in the argument of the integral. The resulting groups of neighbours
depend of the adopted cosmological parameters.

\medskip

Regarding the use of the S and Z tests, is is clear that a
three-dimensional analysis of the data set is only feasible for samples
for which redshift measurements are available.
Nevertheless, it is worth mentioning that a two dimensional analysis,
where we consider the nearest neighbours on the celestial sphere
rather than in the 3D space, is not devoid of interest. It is applicable
for all samples if we impose $r=1$ while we build the groups of
nearest neighbours.

In both S and Z tests, the number of nearest neighbours $n_v$ is a free
parameter which has to be explored.
Indeed, this parameter is not devoid of physical meaning as it is related
to a characteristic scale of the nearest-neighbour groups, in two or three
dimensions. This would be strictly true for a sample of sources homogeneously
distributed on the whole sky. However, as in general observed samples show
deviations from homogeneity and, more importantly, as only part of the entire
celestial sphere is commonly covered, the parameter $n_v$ does not show a
straight correspondence with a typical physical size of the groups. A dispersion
is naturally expected. Nevertheless, if correlations between polarization
orientations occur for a typical scale or if some sub-samples, well delimited
in space, present such alignments, it is clear that the SL will be smaller for
the corresponding value of $n_v$ than for others.
Therefore, to test the uniformity of the polarization orientations and explore
their characteristics, it is necessary to estimate the SL across a wide range
of values of $n_v$.
Note, however, that the lowest value of SL does not provide an accurate
estimate of the overall significance of the correlations that might be detected
but rather reveals the value of $n_v$ at which the departure from uniformity
is the most significant.

\cite{Jain-Narain-Sarala2004} and \citet{Tiwari-Jain2015} used the S test
where they define the groups not with the parameter $n_v$ but rather with a cut
in the comoving distance between sources\footnote{Note, however, that they
fixed the redshift of all the sources of their sample to be $z=1$. Hence, their
comoving distance cut must be seen roughly as a cut in angular separation.}.
Given this modification, they showed that the S test can be used in the same
way as before. They also explored the dependence of the values of the
significance level on the value of the introduced cut.

This shows explicitly that the number of nearest neighbours has to be
explored and that it contains physical information on the alignment patterns.

\subsection{Coordinate dependence of the S and Z test results}
\label{substat_CoordDependence}
\begin{figure}[h]
\centering
\begin{minipage}{0.8\textwidth}
\centering
\includegraphics[width=\columnwidth]{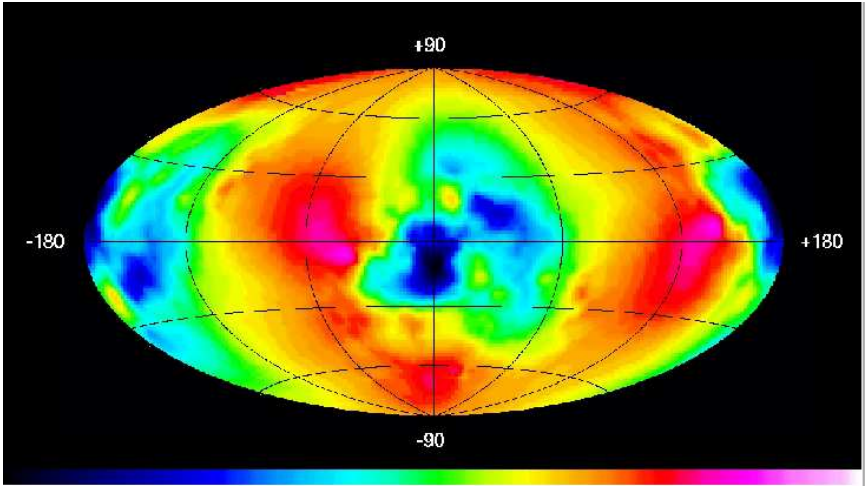}
\caption{\small{Hammer-Aitoff projection of the $S_D$ statistics
averaged over $n_v = 37$ to $43$, as a function of the equatorial
coordinates $\alpha_p$ and $\delta_p$ of the northern pole of the
coordinate system in which the S test is performed.
The less significant the statistics for a given pole, the darker the
corresponding $(\alpha_p,\,\delta_p)$ point on the map.
Figure taken from \citet{Hutsemekers-et-al2005}.}}
\label{fig:SD_CoordDep}
\end{minipage}
\end{figure}
The PAs of quantities which are perpendicular to the line of sight to their
corresponding sources, i.e. projected in the planes orthogonal to
these directions, are dependent on the coordinate system in which the
source positions are reported. Let consider a source having $\alpha$
and $\delta$ as equatorial coordinates to which corresponds a PA expressed
in that coordinate system, $\psi$. Then, if we consider an arbitrary
system of coordinates for which the equatorial coordinates of its northern
pole are $\alpha_p$ and $\delta_p$, the PA $\psi_N$ corresponding to
the source in this new system of coordinates is given by
\begin{equation}
\psi_N = \psi - \arctan \left( \frac{\cos \delta_p \, \sin (\alpha_p - \alpha)}
{\sin \delta_p \, \cos \delta - \sin \delta \, \cos \delta_p \, \cos (\alpha_p - \alpha)} \right) \; .
\label{eq:psi_coord}
\end{equation}
As the changes of PA values depend on the source positions, the
distribution of various PAs from angularly separated sources depends on
the coordinate system. Accordingly, the significance level of statistical tests
is expected to vary.
Fig.~\ref{fig:SD_CoordDep} illustrates this dependence for the application
of the S test to the sample of 355 quasars compiled in
\citet{Hutsemekers-et-al2005}. The color coded SL was computed for a
large set of spherical coordinate systems, spanning the entire sphere, and
reported at the northern pole of these.
Such coordinate-system dependence of statistical results is obviously
unwanted. However, as pointed out by \cite{Hutsemekers-et-al2005},
this dependence can be used to identify the location on the sky where
the alignment are the most significant. Indeed, consider quasars close
the equatorial equator with their polarization vectors perfectly aligned.
Then, if one chooses a coordinate system with a pole located just in
the middle of aligned objects, the polarization angles will range from
$0^\circ$ to $180^\circ$ and no coherent orientation will be detected
by the tests in this new coordinate-system. The regions of minima in
Fig.~\ref{fig:SD_CoordDep} indeed point towards the A1 and A3
regions\footnote{Note that this figure is symmetric by construction.
$(\alpha_p,\,\delta_p)$ is equivalent to
$(\alpha_p+180^\circ,\,-\delta_p)$}.

\subsection{Parallel transport for coordinate invariance}
\label{substat_ParallelTransport}
\citet{Jain-Narain-Sarala2004} developed a workaround to overcome
the coordinate dependence of the results returned by the S and Z
tests. Instead of computing statistics directly from the PAs that are
evaluated with respect to its own meridian, they introduced
corrections to the PAs that involve the relative sky positions of the
sources.
For two sources, $a$ and $b$, separated on the celestial sphere,
they proposed to parallel transport the polarization vector of one
source onto the other before comparing
them\footnote{Note that the parallel transport was already introduced
in the context of the studies of the alignments of the spins of galaxies
(\citealt*{Pen-Lee-Seljak2000}) and is also used to study of the
cosmic microwave background polarization (e.g.,
\citealt{Keegstra-et-al1997} and \citealt{PlanckXIX_int2015}).}.
The parallel transport is performed along the sphere geodesic passing
through the positions of the two sources.
The angle between the polarization vector and the vector tangent to the
geodesic is preserved during the parallel transport. The correction to apply to
the PAs is simply given by the difference between the angles that the
geodesic makes with one of the basis vectors at the locations of the two sources.
At location $P_a$, the plane tangent to the sphere is defined by the vectors of
the local basis $(\mathbf{\hat{e}}_{\theta_a},\,\mathbf{\hat{e}}_{\phi_a})$ in
spherical coordinates, where $\mathbf{\hat{e}}_{\theta_a}$ is pointing toward
the southern pole and $\mathbf{\hat{e}}_{\phi_a}$ towards the East.
The angle $\xi_a$ between the tangent vector of the geodesic
$\mathbf{\hat{t}}_{a}$ and the basis vector $\mathbf{\hat{e}}_{\phi_a}$ is given
by the two--parameter arctangent function
$\arctan ( - \, \mathbf{\hat{t}}_{a} \cdot \mathbf{\hat{e}}_{\theta_a} , \, \mathbf{\hat{t}}_{a} \cdot \mathbf{\hat{e}}_{\phi_a} )$.
This function takes care of the overall addition of $180^\circ$ if the angle lies in
the third or fourth trigonometric quadrants, i.e. if
$(- \, \mathbf{\hat{t}}_{a} \cdot \mathbf{\hat{e}}_{\phi_a})$ is negative or
$(- \, \mathbf{\hat{t}}_{a} \cdot \mathbf{\hat{e}}_{\theta_a})$ and
$(\mathbf{\hat{t}}_{a} \cdot \mathbf{\hat{e}}_{\phi_a})$ are both negative.
The angle $\xi_b$ is obtained in the same manner at the location $P_b$ of the
source $b$.

Assume that the polarization vector of the source $a$ forms an angle $\kappa_a$
with the basis vector $\mathbf{\hat{e}}_{\phi_a}$ at location $P_a$.
At location $P_b$, the transported vector -- from $P_a$ to $P_b$ -- then forms
the angle $\kappa^{(b)}_a = \kappa_a -\xi_a + \xi_b$ with the basis vector
$\mathbf{\hat{e}}_{\phi_b}$.

Now, adopting the IAU convention\footnote{In the International Astronomical
Union convention, the position angles are positively measured East-to-North.
$0^\circ$ corresponds to a vector pointing towards the North pole of the
coordinate system and $90^\circ$ to the East.}, the corrections to apply to
the PAs $\psi_k$'s of a sample of sources in order to compare them in a
coordinate-invariant fashion at the location of the source $i$ are given by
\begin{eqnarray}
\psi^{(i)}_k & = & \psi_k + \Delta_{k \rightarrow i} \nonumber	\\
 				& = & \psi_k + \xi_k - \xi_i
\label{eq:TPcorrection}
\end{eqnarray}
The introduction of these simple corrections to the PAs while evaluating
Eq.~\ref{eq:d_theta} (or Eq.~\ref{eq:d_theta_J}) and Eq.~\ref{eq:Y_i} leads
to coordinate-invariant statistics as was shown by
\citet{Jain-Narain-Sarala2004} and \citet{Hutsemekers-et-al2005}.

In astronomy, when one has to build statistics on the
position angles and to circumvent the dependence of the
statistics on the coordinate system, one rotates locally the PAs
to a coordinate system where the sources are located at the equator
(e.g., \citealt{PlanckXIX_int2015}).
In principle, it could be demonstrated that the parallel transport
corrections mathematically mimic this rotation at least in some cases.
The advantage of the parallel transport, thought, is that it can be
used for widely dispersed sources on the celestial sphere where it
would be impossible to find a coordinate system for which a large
patch of the sky could be at the equator.

\medskip

\begin{figure}[h]
\centering
\begin{minipage}{0.8\textwidth}
\centering
\includegraphics[width=\columnwidth]{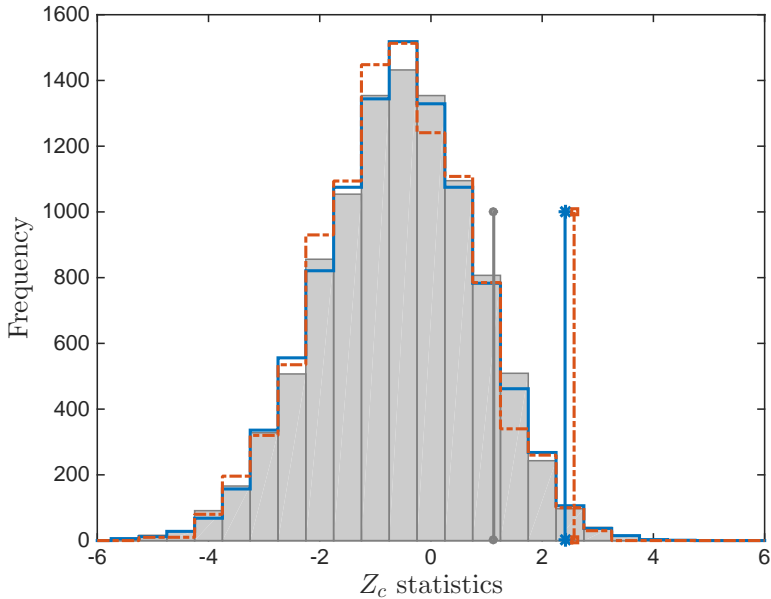}
\caption{\small{The distributions of the $Z_c$ statistics for the two
discussed possibilities for parallel transporting the polarization
vectors along with the distribution obtained without correction.
The Z test which we equipped with the parallel transport is the
modified version of \citet{Hutsemekers1998}. The result of the
first algorithm (following Eq.~\ref{eq:Dij_p1}) is shown by the blue
line and the second (following Eq.~\ref{eq:Dij_p2}) with the
dot-dashed orange line. The gray histogram is without parallel
transport. The vertical lines with corresponding color show the
statistics obtained for the real data sets. The same sample as in
Fig.~\ref{fig:Zc_statistics} is used with $n_v = 80$ and
$N_{\rm{sim}}=10\,000$.}}
\label{fig:Zc_statistics_p}
\end{minipage}
\end{figure}
While there is no ambiguity to equip the S test with the
parallel transport correction, the situation is not so clear for the Z
test, as it was already pointed out by
\citet{Jain-Narain-Sarala2004}. Indeed, for the S test, one should
just parallel transport the polarization vectors of the $n_v$
neighbours of source $i$ at the location of the latter when
evaluating Eq.~\ref{eq:d_theta} (or Eq.~\ref{eq:d_theta_J}).
For the Z test, though, there are a least two different manners to
introduce the parallel transport corrections:
\begin{itemize}
\item One can first evaluate all the mean resultant vectors
$\bmath{{Y}}_j$ by transporting the polarization vectors of the
neighbours to the location $j$ and, afterwards, parallel transport
these $\bmath{{Y}}_j$ to the location of the source $i$ to evaluate
the coordinate-invariant $\tilde{D}_{i,j}$ which is thus expressed as
\begin{eqnarray}
\tilde{D}_{i,j}	& = & \frac{1}{n_v}
								\sum_{k=1}^{n_v} \cos \left[ 2 \left(\psi_i - (\psi_k^{(j)} +\Delta_{j \rightarrow i}) \right) \right]
								\nonumber	\\
					& = & \frac{1}{n_v}
								\sum_{k=1}^{n_v} \cos \left[ 2 \left(\psi_i - (\psi_k + \Delta_{k \rightarrow j} +\Delta_{j \rightarrow i}) \right) \right]
\label{eq:Dij_p1}
\end{eqnarray}
\item Another algorithm is to parallel transport all the polarization
vectors to all the locations of the sources. Then, for each source
$i$, compute the $\bmath{{Y}}_j$ of all $j$ and finally compute
the coordinate-invariant $\tilde{D}_{i,j}$ as
\begin{eqnarray}
\tilde{D}_{i,j} 	& = & \frac{1}{n_v} \sum_{k=1}^{n_v} \cos \left[ 2 \left(\psi_i - \psi_k^{(i)} \right) \right]	 \nonumber	\\
					& = & \frac{1}{n_v} \sum_{k=1}^{n_v} \cos \left[ 2 \left(\psi_i - (\psi_k + \Delta_{k \rightarrow i}) \right) \right]
\label{eq:Dij_p2}
\end{eqnarray}
\end{itemize}
Despite the fact that the two $\tilde{D}_{i,j}$ are both
coordinate-invariant, they are in general different. This comes from
the property of the parallel transport which states that the result
depends on the chosen path.
Therefore, $\Delta_{k \rightarrow j} + \Delta_{j \rightarrow i}$ and
$\Delta_{k \rightarrow i}$ are generally different. The only
configuration for which they are equal are those where the three
locations $i$, $j$, and $k$ belong to the same geodesic of
the sphere.

Whereas the parallel transport is a suitable mathematical trick that
enables one to build statistics that are coordinate-invariant for
circular data tangent to a sphere, there is no real reason to consider
the first or the second way presented above, or even imagine more
complicated ones.
However, the first algorithm (Eq.~\ref{eq:Dij_p1}) is perhaps more
consistent with the fundamental concepts of the Z test (see
\citealt{Bietenholz1986}) and thus, that it seems more appropriate.
Nevertheless, this algorithm has not been used in the literature.
It is the second algorithm (Eq.~\ref{eq:Dij_p2}) that has
been used in (\citealt{Jain-Narain-Sarala2004};
\citealt{Hutsemekers-et-al2005} and
\citealt{Pelgrims-Hutsemekers2015}  (reworked in
Chapter~\ref{Ch:PH-1})).
One reason, perhaps, could be the fact that it takes a simpler form
(Eq.~\ref{eq:Dij_p2}) than that of the first. Beside the credit we just
gave to the first algorithm, it is worth noticing that for computational
purpose the second algorithm should also be disfavoured as it
requires $N$ times more computational steps than the first
one\footnote{The reason why we will not use the first algorithm
(Eq.~\ref{eq:Dij_p1}) in Chapter~\ref{Ch:PH-1} is that we realized the
above only afterwards.}.

\begin{figure}[h]
\centering
\begin{minipage}{0.8\linewidth}
\centering
\includegraphics[width=\columnwidth]{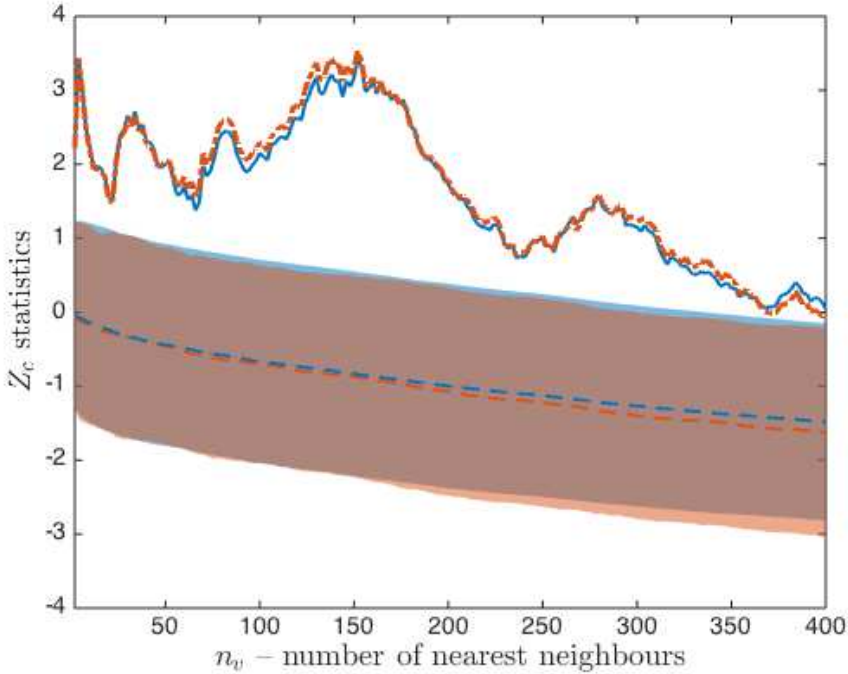}
\caption{\small{The distributions of the $Z_c$ statistics for the two
discussed possibilities for parallel transporting the polarization
vectors along with the statistics obtained for the observed data set.
The Z test which we equipped with the parallel transport is the
modified version of \citet{Hutsemekers1998}. The statistics obtained
with the first algorithm (following Eq.~\ref{eq:Dij_p1}) is shown by the
blue line and the second (following Eq.~\ref{eq:Dij_p2}) with the
dot-dashed orange line. The $1\sigma$ confidence intervals (assuming
normal distributions) of the prior distributions from the two algorithms are
displayed with corresponding color-shaded regions around the means.
The same sample as in Fig.~\ref{fig:Zc_statistics} is used with
$N_{\rm{sim}}=10\,000$ and $n_v \in \left[2,\, 400\right]$.}}
\label{fig:Zc_statistics_p2}
\end{minipage}
\end{figure}
The only disadvantage that we found to the parallel transport trick
is that there is no objective criterion following which one has to
choose the way the parallel transport is performed. As a consequence,
different statistical results could emerge for the same data set,
as shown in Figs.~\ref{fig:Zc_statistics_p}
and~\ref{fig:Zc_statistics_p2}.
However, despite the fact that the distributions obtained through
simulations and that the values of the statistics obtained for the
observed data set differ from the two method, the returned
significance levels agree with each other.
Indeed, we found that the values of the significance levels from
the two implementations of the parallel transport on the Z test
differ by less than a factor of two for the sample that we
considered and for the wide range of $n_v$-values that we
considered. This agreement is enough for our purpose.

\subsection{Generation of simulated data sets}
\label{substat_RsampleGen}
As we stated before, once we use the S or the Z test, simulated data
sets are necessary to assess the significance level with which the
observations differ from randomness. The need of simulations is due
to the mutual dependence of the groups of nearest neighbours which
is impossible to estimate analytically. For both tests, and whatever
the version of the test, the SL of alignment in a sample is defined as
the percentage of generated data sets that give a statistics as
extreme as the one corresponding to the observations.
In order to generate a random catalogue, there are two different
methods that were already proposed and used in the original paper of
\citet{Hutsemekers1998}.

For both methods, the sky positions of the sources are kept fixed.
Then, the polarization position angles have to be generated.
The first method relies on the assumption that at each source location
the PAs have to be uniformly distributed. Hence, for this method of
data set generation, we generate the $N$ polarization PAs according to
a uniform distribution in the interval $0^\circ - 180^\circ$ and reproduce
this step $N_{\rm{sim}}$ times.
The second method uses a re-sampling approach in the sense that the PAs
from the observed sample are randomly shuffled among the positions of
the sources in order to produce the generated data sets. Namely, the
shuffling procedure consists of a bijective mapping of the polarization PAs
onto the source positions. Given a coordinate system, let the position vectors
of the sources be denoted by $\bmath{\hat{e}}_{r}^{(i)}$, for $i = 1,\, ...,\, N$.
To each source corresponds a position angle $\psi$. The observation data
set is thus made of couples $(\bmath{\hat{e}}_{r}^{(i)},\,\psi_i)$.
The shuffling procedure randomly permutes the PAs in such a way that,
for a given randomized sample, the PA at location
$\bmath{\hat{e}}_{r}^{(i)}$ is given by
\begin{equation}
\psi_i^R = \psi_j
\end{equation}
for a given $j$ determined by the random permutation of the list
$\left[1,\,...,\,N \right]$.

\citeauthor{Hutsemekers1998} argued that this procedure is more
appropriated to the detection of correlated PAs with their
corresponding sky locations. He further explained that a possible
systematic bias introduced in the data set would be taken into
account (\citealt{Hutsemekers1998}). The final significance level
computed via reshuffling method is thus thought to be free of
systematic bias. Notice that this is also true if a global physical
alignment is contained in the data set. As
\citet{Jain-Narain-Sarala2004} showed, a comparison between
the SLs obtained with the two methods of data set generations
(through shuffling or uniform randomization) can thus help to
unveil a global coherence in the polarization PAs.

In the previous section, we discussed the problem of the production
of statistical results that are independent of the coordinate system.
We have seen that the introduction of the parallel transport helps to
solve this issue. One may wonder if the procedures of data set
generation that we introduced above lead to distributions that are
independent of the system of coordinates.

The distributions obtained with the data sets that are produced through
the generation of uniformly distributed polarization PAs in their range of
definition are definitely coordinate independent, as long as the number
of simulated data sets is large.

\begin{figure}[h]
\centering
\begin{minipage}{0.8\textwidth}
\centering
\includegraphics[width=\columnwidth]{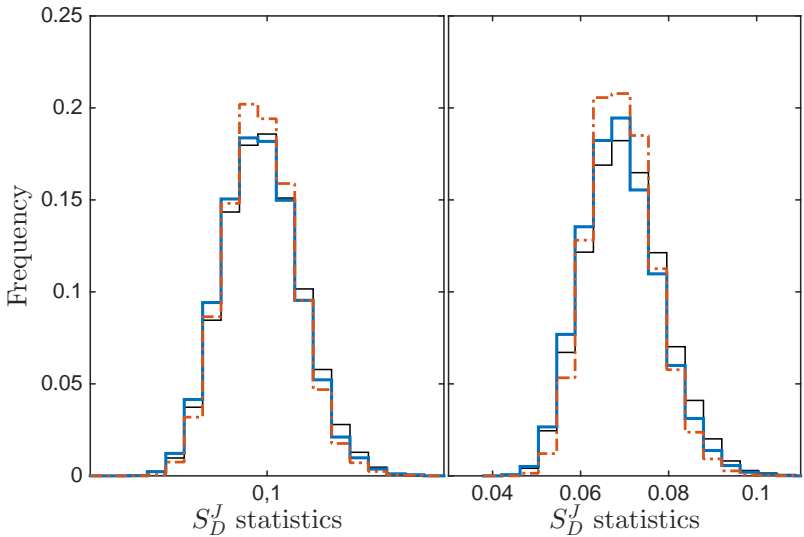}
\caption{\small{The distributions of the $S_D$ statistics for the two
discussed possibilities for generating the randomized data sample
while the shuffling procedure is adopted with parallel transport
corrections, for two values of $n_v$.
The distribution from the simplest mapping
(Eq.~\ref{eq:PTCorr_sh1}) is plotted in blue and the more
sophisticated one (Eq.~\ref{eq:PTCorr_sh2}) in dot-dashed orange.
The thin-black line corresponds to the distribution generated with
PAs drawn from a uniform distribution. The same sample as for
Fig.~\ref{fig:Zc_statistics_p} is used, $N_{\rm{sim}}=10\,000$ and
$n_v = 80$ on the left and $n_v = 160$ on the right.}}
\label{fig:SD_statistics_sh}
\end{minipage}
\end{figure}

The situation is not so clear when the random data sets are generated
via the shuffling of the polarization position angles. What is clear,
though, is that the distribution of the statistics (Eq.~\ref{eq:S_D} and
Eq.~\ref{eq:Z_c}) will depend on the system of coordinates if no
particular care is taken. To overcome this problem, one has to
shuffle the polarization vectors in a coordinate-independent fashion.
Again, the introduction of parallel transport helps but a similar
ambiguity to the one we discussed in the previous section arises.
There is no unique prescription to equip the shuffling with the parallel
transport. We give two such possibilities in the following and compare
the significance levels that they lead to.

\begin{itemize}
\item The simplest coordinate-independent mapping that one can
imagine is that the random position angle at location $j$ from a
generated catalogue $\psi_j^R$ is given by
\begin{eqnarray}
\psi_j^R	& = &\psi_k^{(j)}									\nonumber \\
			& = &\psi_k + \Delta_{k \rightarrow j}\;,
\label{eq:PTCorr_sh1}
\end{eqnarray}
for a given $k$ obtained by the random permutation.
Readily, compared to the real data sample, the position angle of the
source $k$ has migrated to the source location $j$ through parallel
transport on the sphere. This is done for all the PAs of the sample without
repetition, either on the PAs or on the source positions. Starting with the
generated sample, the usual procedure to evaluate the statistics has to be
applied (with the correction for the parallel transport).
\item Another possible algorithm can be imagined as follows. All the
polarization vectors of the sample are parallel transported at each
source location $i$. Then, at each location, the shuffling of the
corrected position angles is done via random permutations. If
necessary, then, the shuffled-polarization vectors have to be parallel
transported back to their ``initial position''.
\begin{eqnarray}
\psi_j^{R\,(i)}	& = & 	\psi_k^{(i)}	 + \Delta_{i \rightarrow j}			\nonumber \\
					& = &	\psi_k + \Delta_{k \rightarrow i} + \Delta_{i \rightarrow j}\;.
\label{eq:PTCorr_sh2}
\end{eqnarray}
In this case, it is important to realize that for the same generated sample,
the value of the PA attached to a given source varies according to the
source location at which it will be used for the evaluation of the statistics.
This justifies the super-script index in the left hand-side of
Eq.~\ref{eq:PTCorr_sh2}. Note that the back parallel transport of the
polarization vectors is only necessary for the computation of the Z
test if the first algorithm (Eq.~\ref{eq:Dij_p1}) is chosen.
\end{itemize}

\begin{figure}[h]
\centering
\begin{minipage}{0.8\textwidth}
\centering
\includegraphics[width=\columnwidth]{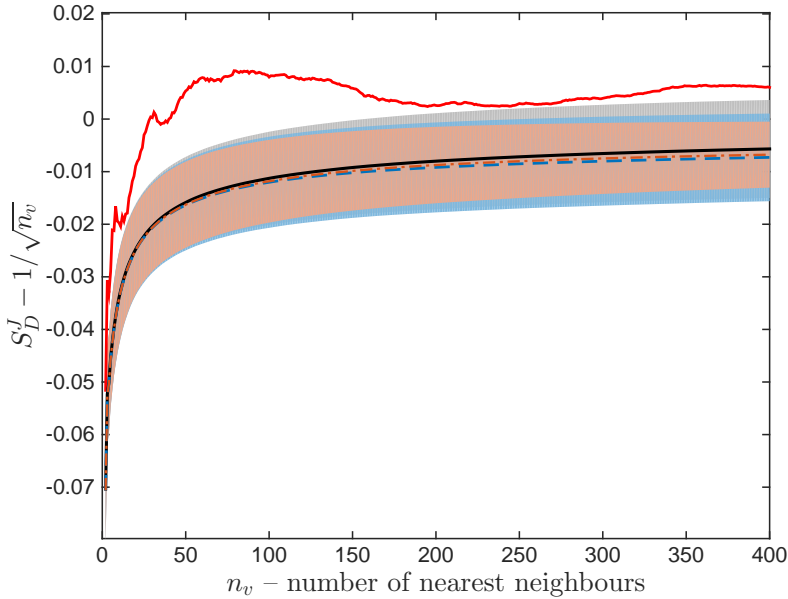}
\caption{\small{The distributions of the $S_D$ statistics for the two
discussed possibilities for generating the randomized data sample
while the shuffling procedure is adopted along with uniform
randomization, against a large range of value of $n_v$. The means
of the distributions from the simplest mapping is plotted in dashed
blue and the more sophisticated one in dot-dashed orange, the
black-thick line is for uniform randomization. The color-shaded regions
with corresponding colors mark the $1\sigma$ confidence intervals of
the distributions. We assume normal distributions to compute these
intervals. The thick-red line shows the statistics for the real data
sample. We use the same sample as for
Figs.~\ref{fig:Zc_statistics_p} and~\ref{fig:SD_statistics_sh} and the
distribution are computed for $N_{\rm{sim}}=10\,000$ and $n_v$
ranging from $2$ to $400$ by step of $2$.}}
\label{fig:SD_statistics_sh-nv}
\end{minipage}
\end{figure}

As we show in Fig.~\ref{fig:SD_statistics_sh} the two possible ways
of generating randomized data sets via parallel-transport-corrected
reshuffling lead to different distributions of the statistics. Indeed,
comparison of the histograms reveals that the second algorithm
leads to narrower distributions than the first one which, in turn, is
narrower than the distribution given by uniform PA generations.
The resultant significance level of the observations are thus expected
to be different given the adopted method and these differences may
become significant as the number of neighbours increases. Indeed,
the difference between the values of the standard deviations of the
distributions obtained with the different possibilities increases with
the number of neighbours $n_v$. This is nicely illustrated in
Fig.~\ref{fig:SD_statistics_sh-nv}. The three methods to generate the
prior distributions are shown for a very large range of $n_v$ values.
As an illustration, we also plot (in red) the statistics obtained for the
real data. A comparison of the red curve with the different distribution
shows that one would obtain SL that varies, for the same sample
and for the same value of the parameter of the statistics.
This would be uncomfortable when thorough analyses are required
and, again, there is no physical criterion suggesting one procedure
to be preferred.
However, we have compared the values of the significance levels
returned by the two parallel-transport-corrected shuffling procedures
and it turned out that they agree by a factor of two, upto $n_v = 400$.
The differences between the two methods are thus irrelevant for our
purpose.

\section{The polarization cap test}
\label{stat_PCTest}
Noting that the large-scale polarization alignments of
Hutsem{\'e}kers et al. (\citeyear{Hutsemekers1998};
\citeyear{Hutsemekers-Lamy2001}; \citeyear{Hutsemekers-et-al2005})
had always been characterized by means of the two same statistical
methods, we found important to independently confirm them.
In this respect, we have developed a new and independent statistical
test. It was originally presented in \citet{Pelgrims-Cudell2014} and
additional details were given in \citet{Pelgrims2014}.
The basics of the new method rely on
the introduction of a new polarization space, a 2-sphere, and on the
treatment of the polarization vectors as points on it. A dedicated study
of the density of these polarization points within the cap of the
polarization sphere allows for the comparison of the observed
distribution of the polarization vector orientations with the distribution
that one would expect to observe by chance. The main features of this
test are that it is intrinsically coordinate invariant, that the likelihood of
the observations against randomness can be analytically derived, that
it directly returns the preferred orientation of the polarization vectors and,
hence, that it should ease the interpretation of the alignment patterns.
We shall refer to this new test as the polarization cap (PC) test.
All details of its construction and its working are explained below.

As we will show, the new method allows us to compare the polarization
vectors of sources located at different angular coordinates and leads to
the characterization of the effect through an unbiased analysis of the
data. The basic idea is to consider the physical polarization vectors as
3-dimensional objects rather than 2-dimensional ones embedded in
their polarization plane. These 3-dimensional objects are the directions
of the electric field oscillations and they are the physical objects which
are measured. As we are dealing with a number of vectors, the
definition of a preferred (3-dimensional) direction naturally follows.
As we shall see in Chapter~\ref{Ch:PC-1Analysis}, the method can be
used to study the dependence of the alignment characteristics with
redshift, position in the sky or degree of linear polarization by
imposing cuts on these variables and repeating the study for the
corresponding sub-sample. Naturally, it can also be used to study the
distribution of the polarization angles inside groups of sources, similar
to the S and Z tests. We use it on this way in Chapter~\ref{Ch:PH-1}.

\subsection{A new coordinate-invariant statistical test for polarization data}
\label{substat:coordInvTest}
When an electromagnetic wave is partially or fully linearly polarized, a
polarization vector is introduced. Its norm reflects the degree of linear
polarization of the radiation while its direction is that of the oscillating
electric field. This vector is embedded into the plane orthogonal to the
radiation direction of propagation, the polarization plane.
Since the electric field is oscillating, the polarization vector is an axial
quantity, rather than a true vector, so that the polarization angle is
determined up to $180^\circ$.

We consider sources as being points on the unit celestial sphere and
we choose a spherical polar coordinate system defined by the orthonormal
3-vectors $(\bmath{\hat{e}}_{r},\,\bmath{\hat{e}}_{\theta},\,\bmath{\hat{e}}_{\phi})$,
with $\bmath{\hat{e}}_{\theta}$ pointing to the South pole. 
Polarization vectors are tangent to this unit sphere. For a given source in the direction
$\bmath{\hat{e}}_{r}$, a polarization vector must lie in the plane
defined by the two unit vectors $\bmath{\hat{e}}_{\phi}$ and $\bmath{\hat{e}}_{\theta}$.
We choose the angle $\kappa$ between the polarization vector $\bmath{p}$
and the basis vector $\bmath{\hat{e}}_{\phi}$, defined in the range
$[-90^\circ,\,90^\circ[$, to be the polarization angle. The normalized
polarization vector can then be written
\begin{equation}
\bmath{\hat{p}}=\cos \kappa \,\bmath{\hat{e}}_{\phi} - \sin \kappa \,\bmath{\hat{e}}_{\theta}\;.\label{eq:1}
\end{equation}
Each measurement $i$ of the data set (\citealt{Hutsemekers-et-al2005})
is equivalent to a position 3-vector $\bmath{\hat{e}}_{r}^{(i)}$ associated
with a normalised polarization direction $\bmath{\hat{p}}^{(i)}$ and
polarization magnitude $|\bmath{p}^{(i)}|$. Contrarily to the various
angles, $\bmath{\hat{e}}_{r}^{(i)}$ and $\bmath{\hat{p}}^{(i)}$
are physical, i.e. they do not depend on the choice of the coordinate system.
As we are interested in polarization alignments, we shall consider
mostly the $\bmath{\hat{p}}^{(i)}$ in the following. 

\begin{figure}
\begin{center}
\begin{minipage}{0.8\textwidth}
\centering
\includegraphics[width = \columnwidth]{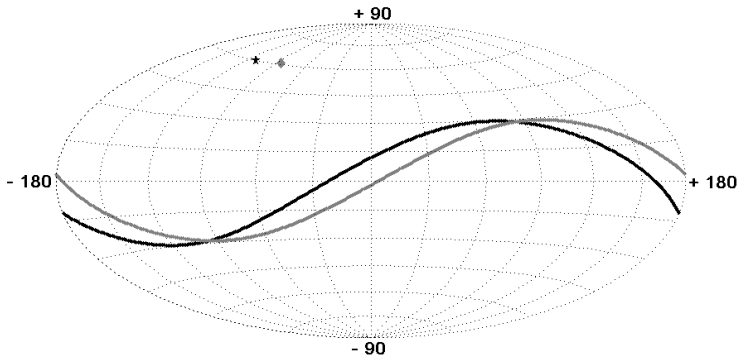}
\caption{\small{Superposition of the Hammer-Aitoff
projections of the celestial sphere and the polarization sphere (in
Galactic coordinates). Two quasars (B1115+080 (in black) and B1157+014
(in grey)) are displayed on the celestial sphere with the corresponding
geometric loci of their polarization point on the polarization sphere.
The source position and the corresponding geometric locus of the polarization
point are printed in same brightness.}}
\label{fig:2Qexample}
\end{minipage}
\end{center}
\end{figure}

\begin{figure}
\begin{center}
\begin{minipage}{0.8\textwidth}
\centering
\includegraphics[width = .8\columnwidth]{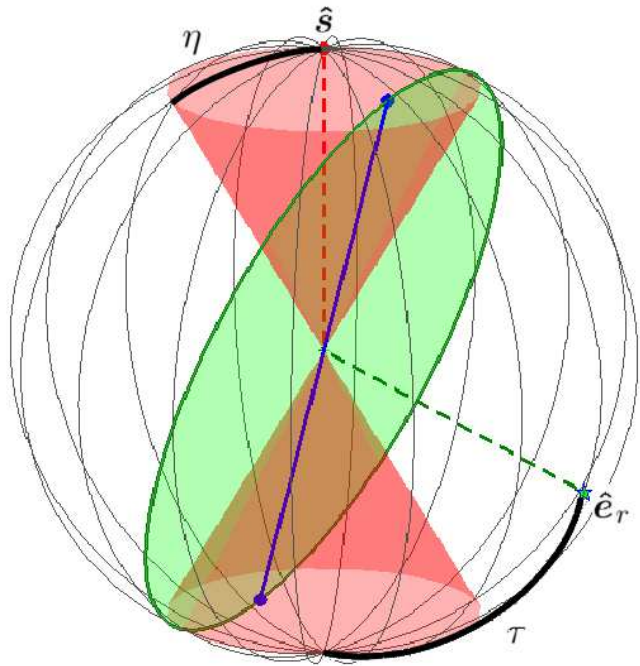}
\caption{\small{Illustration of the building of the PC test. The
considered source is represented by the star with the pointing
vector in dashed green. The green disk is the polarization plane
drawn at the origin of the coordinate system, the dark green circle
is the geometrical locus of all possible polarization points which
correspond to the source. The blue axis corresponds to the
axial-polarization vector attached to the source. The blue points at
its ends, at the intersection with the unit sphere, are the observed
polarization points. The red point (here at the pole of the sphere) is
the centre of a spherical cap of half-aperture angle $\eta$ defined
by the intersection of the cone (in red) and the unit sphere. Under
the hypothesis that the polarization position angle are drawn from
a uniform distribution, the probability that the polarization points fall
inside the spherical cap depends only on the acute angle $\tau$
(Eq.~\ref{eq:3}) between the vectors pointing towards the source
($\bmath{\hat{e}}_r$) and towards the centre of the cap
($\bmath{\hat{s}}$).}}
\label{fig:QandCone2}
\end{minipage}
\end{center}
\end{figure}

The problem is then as follows: we have a number of normalised vectors,
and we want to decide if they are abnormally aligned. We can draw
them from the same origin, and their ends, which we shall call the
polarization points, have to lie on a unit 2-sphere, which we shall
refer to as the polarization sphere. The problem is that, even when
the polarization angles $\psi$ (or $\kappa$)\footnote{If $\psi$ are the
polarization position angles given in the IAU convention, then we have
$\kappa = 90^\circ - \psi$. We developed our formalism in terms of
$\kappa$ for simplicity.} are uniformly distributed, the polarization
sphere is not uniformly covered by the points: they have to lie on
great circles on the 2-sphere. Indeed, for each source, the polarization
vectors are constrained to be in the plane defined by the basis vectors
$(\bmath{\hat{e}}_{\phi}^{(i)},\,\bmath{\hat{e}}_{\theta}^{(i)})$.
The intersection of the plane with the polarization sphere is
a great circle, which is the geometric locus where the polarization
vector attached to the source $i$ may intersect the sphere, as
shown in Fig.~\ref{fig:2Qexample}. 
See Fig.~\ref{fig:QandCone2} for an illustration of the notations and
of the concept of the above.
Note that the Fig.~\ref{fig:2Qexample} is symmetric as polarization vectors
are defined up to a sign. In the following, we choose to show the full sphere,
although a half-sphere could be used to represent the polarization space.

As a result, simple spherical data analysis such as those presented in
\citet{Fisher-Lewis-Embleton1987} are not applicable in the case of
transverse quantities because of the constrained geometrical locus of
the points. One can nevertheless evaluate the density of points at each
location on the unit 2-sphere by adopting Kamb-like methods
(e.g., \citealt{Vollmer1995}). Namely, to compare the observations with
what one would expect if the polarization points were drawn from a
random distribution of polarization angles, we need to select a region on
the polarization sphere, count the number of polarization points within this
region, and compare this number with the prediction.
One could do this by Monte-Carlo techniques but the probabilities turn
out to be rather low so that a detailed study would prove difficult.

However, we found that a particular choice of shape for the region
on the sphere considerably simplifies the evaluation of the probabilities.
We consider cones in which the polarization vectors fall, or equivalently
spherical caps of fixed aperture angle. The probability distribution
of a given number of points in a given spherical cap can be computed
analytically when densities are evaluated through a standard step function,
as explained below. A scan of the whole polarization half-sphere with such
cones leads to a map of expected densities which constitutes the statistical
background. At any location on the half-sphere, the hypothesis of uniformity
can then be tested by calculating the probability of the observed number of
polarization points. An alignment of polarization vectors from different
sources will be detected when an over-density between data points and the
background is significant.

\subsection{Construction of the probability distribution}
\label{substat:PnDistConst}
As mentioned above, the locus of a polarization point is a half-circle in the
plane normal to the source position vector. The probability that a polarization
point lies inside a spherical cap is then given by the length of the arc of circle
intercepted by the cap, divided by the whole length of the half-circle ($\pi$).
Let $\eta$ being the half-aperture angle of the spherical cap, and
$\bmath{\hat{s}}$ the unit vector pointing to its centre. If $\bmath{\hat{p}}^{(i)}$
is a normalised polarization vector attached to the source $i$, with position
vector $\bmath{\hat{e}}_{r}^{(i)}$, the corresponding polarization point lies
inside the spherical cap centred at $\bmath{\hat{s}}$ if and only if
\begin{equation}
|\bmath{\hat{p}}^{(i)}\cdot \bmath{\hat{s}}|\geq\cos \eta
\label{eq:1b}
\end{equation}
is verified. Adopting the decomposition of $\bmath{\hat{p}}^{(i)}$ along two
vectors in the polarization plane
\begin{equation}
\bmath{\hat{p}}^{(i)} = A\, \left(\bmath{\hat{s}} -
								\left(\bmath{\hat{s}} \cdot\bmath{\hat{e}}_{r}^{(i)}\right)
								\bmath{\hat{e}}_{r}^{(i)} \right) + B\, \bmath{\hat{t}}^{(i)}  
\label{eq:2}
\end{equation}
where $\bmath{\hat{t}}^{(i)}=\left(\bmath{\hat{e}}_{r}^{(i)} \times
\bmath{\hat{s}}\right)/|\bmath{\hat{e}}_{r}^{(i)} \times \bmath{\hat{s}}|$,
a straightforward calculation involving the normalisation of $\bmath{\hat{p}}^{(i)}$
and the condition for being inside the spherical cap leads to the arc length
$L^{(i)}$ of the geometric locus lying inside the considered area. The result
takes a simple form in terms of $\tau^{(i)}\in[0^\circ,\,90^\circ[$, the acute
angle between $\bmath{\hat{e}}_{r}^{(i)}$ and $\bmath{\hat{s}}$: the condition
in Eq.~\ref{eq:1b} becomes $\sin \tau^{(i)}\geq \cos \eta$ and, by integration
of the line element, the arc length within the cap is found to be
\begin{equation}
L^{(i)} =
\begin{cases}
	2\,\arccos \left(\frac{\cos\eta}{\sin\tau^{(i)}}\right) & \mbox{if}\:\sin\tau^{(i)}\geq\cos\eta \\
	0 & \mbox{otherwise} \, .
\end{cases}
\label{eq:3}
\end{equation}
\newcommand{\ppi}{\ell}
Therefore, the probability $\ppi^{(i)}$ that the $i$-th source of the sample
leads to a polarization point inside a given spherical cap is
\begin{equation}
\ppi^{(i)}=\frac{L^{(i)}}{\pi}.\label{eq:4}
\end{equation}
Hence, this probability only depends on the chosen aperture angle
of the spherical cap and on the angle between the source position
and the cap centre. This probability is thus completely independent
of the system of coordinates.

\medskip

For each cap, the set of probabilities $\ppi^{(i)}$, corresponding to a
set of polarized sources, leads to the construction of the probability
distribution $P_{n}$ of observing exactly $n$ points of polarization
inside the spherical cap.
If $N$ is the sample size, we have:
\begin{eqnarray}
P_{0} & = & \prod_{i=1}^{N}\left(1-\ppi^{(i)}\right)\label{eq:5}\\
P_{1} & = & \sum_{j=1}^{N}\ppi^{(j)}\,\prod_{i\neq j}\left(1-\ppi^{(i)}\right)\label{eq:6}\\
P_{2} & = & \frac{1}{2}\,\sum_{k=1}^{N}\ppi^{(k)}\,\sum_{j\neq k}\ppi^{(j)}\,\prod_{i\neq j\neq k}\left(1-\ppi^{(i)}\right)\label{eq:7}\\
 & \vdots\nonumber \\
P_{N} & = & \frac{1}{N!}\,\sum_{l=1}^{N}\ppi^{(l)}\,\ldots\,\sum_{j\neq prev.indices}\ppi^{(j)}\,\prod_{i\subset\{\emptyset\}}\left(1-\ppi^{(i)}\right)\nonumber \\
 	 & = & \prod_{l=1}^{N}\ppi^{(l)}\quad.\label{eq:8}
\end{eqnarray}
Note that following the previous definitions, it is possible to write,
for each $n \geq 1$,
\begin{equation}
P_{n}=\frac{1}{n}\,\sum_{j=1}^{N}\ppi^{(j)}\, P_{n-1}{}_{\backslash j}
\label{eq:9}
\end{equation}
where $P_{n-1}{}_{\backslash j}$ is the probability to observe $n-1$ points
of polarization (and only $n-1$) after the $j$-th element is removed from the
original sample, making the new sample size $N-1$.

\subsection{A fast algorithm for generating the $P_{n}$}
\label{substat:PnDistAlgo}
\newcommand{\pp}{\rlap{\large /}{\ell}}

Starting with the entire sample of size $N$, let us consider the probability
$P_{0}$ to observe no polarization point within the cap. We remove the
$k$-th element from this sample. Then, from Eq.~\ref{eq:5}, the probability
to observe no polarization point within this reduced sample, denoted by
$P_{0}{}_{\backslash k}$, is related to $P_{0}$ through
$P_{0}=\pp^{(k)}\, P_{0}{}_{\backslash k}$, where we introduced the
following notation for the probability that the source $k$ does not lead to a
polarization point in the concerned area :
${\pp^{(k)}}\equiv\left(1-\ppi^{(k)}\right)$. 

\begin{figure}[h]
\begin{center}
\begin{minipage}{0.8\textwidth}
\centering
\includegraphics[width = \columnwidth]{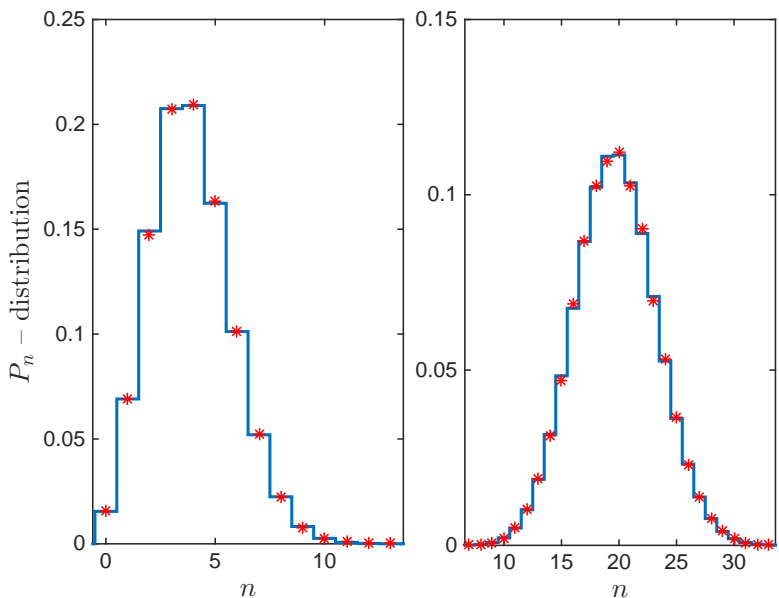}
\caption{\small{Comparison of the $P_n$ distribution built from our
algorithm presented in Section~\ref{substat:PnDistAlgo} (in blue) and
the normalized histogram (red asterisk) obtained by running 100\,000
random simulations where the positions of the sources are kept fixed
and the PAs uniformly distributed. For this simulation, we take the A1
region (see Table~\ref{tab:A123Huts}), choose arbitrarily a spherical
cap and set $\eta = 12^\circ$ for the left panel and $\eta = 35^\circ$
for the right.}}
\label{fig:PnDist_expl}
\end{minipage}
\end{center}
\end{figure}

First consider the probability $P_{1}$ to observe one and only one
polarization point:
\begin{eqnarray}
P_{1} & = & \sum_{j=1}^{N}\ppi^{(j)}\,\prod_{i\neq j}{\pp^{(i)}}\nonumber \\
 & = & \sum_{j=1}^{N}\ppi^{(j)}\, P_{0}{}_{\backslash j}\nonumber \\
 & = & \sum_{j\neq k}\frac{\ppi^{(j)}}{{\pp^{(j)}}}\, P_{0}+\ppi^{(k)}\, P_{0}{}_{\backslash k}\nonumber \\
 & = & \pp^{(k)}\,\left(\sum_{j\neq k}\frac{\ppi^{(j)}}{\pp^{(j)}}\, P_{0}{}_{\backslash k}\right)+\ppi^{(k)}\, P_{0}{}_{\backslash k}\nonumber \\
 & = & \pp^{(k)}\, P_{1}{}_{\backslash k}+\ppi^{(k)}\, P_{0}{}_{\backslash k}\quad.\label{eq:10}
\end{eqnarray}
A similar calculation leads to $P_{2}=\pp^{(k)}\, P_{2}{}_{\backslash k}+\ppi^{(k)}\, P_{1}{}_{\backslash k}$.
One can prove by induction that the following relation holds for every $m \in \mathbb{N}$:
\begin{equation}
P_{m}=\pp^{(k)}\, P_{m}{}_{\backslash k}+\ppi^{(k)}\, P_{m-1}{}_{\backslash k}
\label{eq:11}
\end{equation}
Indeed, assuming the relation is true for $m\leq n-1$, it is easy to show
that it is then true for $m=n$:
\begin{eqnarray}
P_{n} & = & \frac{1}{n}\,\sum_{l=1}^{N}\ppi^{(l)}\, P_{n-1}{}_{\backslash l}\nonumber\\
 & = & \frac{1}{n}\,\sum_{l\neq k}\ppi^{(l)}\, P_{n-1}{}_{\backslash l}+\frac{1}{n}\,\ppi^{(k)}\, P_{n-1}{}_{\backslash k}\nonumber\\
  & = & \pp^{(k)}\,\left(\frac{1}{n}\,\sum_{l\neq k}\ppi^{(l)}\, P_{n-1}{}_{\backslash l\backslash k}\right)
		+\ppi^{(k)}\,\left(\frac{1}{n}\,\sum_{l\neq k}\ppi^{(l)}\, P_{n-2}{}_{\backslash l\backslash k}\right)
		+\frac{1}{n}\,\ppi^{(k)}\, P_{n-1}{}_{\backslash k}\nonumber\\
 & = & \pp^{(k)}\, P_{n}{}_{\backslash k}
		+ \ppi^{(k)}\, \left(\frac{n-1}{n}\, \left[\frac{1}{n-1}\,\sum_{l \neq k}\ppi^{(l)}\, P_{n-2}{}_{\backslash l\backslash k} \right] \right)
		+ \ppi^{(k)}\,\frac{1}{n}\,P_{n-1}{}_{\backslash k}\nonumber\\
 & = & \pp^{(k)}\, P_{n}{}_{\backslash k}+\ppi^{(k)}\, P_{n-1}{}_{\backslash k}\quad.
\end{eqnarray}
Therefore, Eq.~\ref{eq:11} holds\footnote{Afterwards, we realized
that Eq.~\ref{eq:11} was first introduced by \citet{Howard1972} and
its numerical behaviour was extensively discussed by
\citet{Chen-Liu1997} about computational techniques for the
Poisson-binomial probabilities. The algorithm presented here is equivalent
to that given in \citet{Chen-Liu1997}.} for every $m\in\mathbb{N}$.
Assuming that, for a sample of size $N$ and for a given spherical
cap, we have all elementary probabilities $\ppi^{(i)}$, we use
the following algorithm for numerically computing the probability distribution:

\renewcommand{\theenumi}{\roman{enumi}}
\begin{enumerate}
\item We introduce a column vector $V$ of size $N+1$, initialized to
zero, except for $V_{0}$ which is set to 1. The $V_{n}$ are the
$P_{n}$ for an empty data set (with $n=1,...,N$).
\item We add one data point at a time, and update $V$ according to Eq.~\ref{eq:11}.
\item After $N$ iterations, the $V_{n}$ give the $P_{n}$ distribution
for the studied sample.
\end{enumerate}
We test our implementation of the algorithm by comparing its results to those
obtained via a Monte-Carlo treatment, for the A1 region of
\citet{Hutsemekers-et-al2005}. For the generation of the Monte-Carlo
samples, we proceed as follow. We keep the quasar positions fixed and we
generate random polarization PAs according to a uniform distribution. For
various spherical caps, we count the number of polarization points falling
effectively in these caps and build the expected distributions through the
simulated data sets. We then compare these distributions to the $P_{n}$
distributions from our algorithm and Eqs.~\ref{eq:3} and~\ref{eq:4}. As
illustrated in Fig.~\ref{fig:PnDist_expl} for an arbitrary spherical cap, we
obtain very good agreement between theory and simulation. We checked
that the same conclusions are obtained for arbitrary sub-samples of quasar
polarization measurements presented in \citet{Hutsemekers-et-al2005} and
for arbitrary spherical cap.

\subsection{A first example}
\label{substat:StatBackground}
\begin{figure}[h]
\begin{center}
\begin{minipage}{0.8\linewidth}
\centering
\includegraphics[width=\columnwidth]{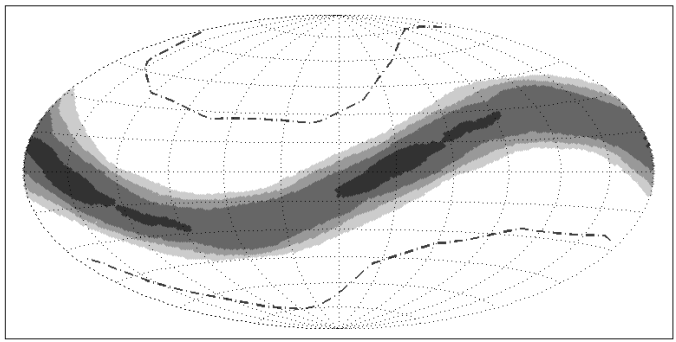}
\caption{\small{Hammer-Aitoff projection (Galactic coordinates) of
the polarization sphere associated to the A1 region. Expected
density regions are displayed following the legend: white:
$\bar{N^{a}}<4$; light grey: $4\leq\bar{N^{a}}<5$; grey:
$5\leq\bar{N^{a}}<6$; dark grey: $6\leq\bar{N^{a}}<7$; black:
$\bar{N^{a}}\geq7$. White regions towards poles which are delimited
by dashed curves are regions where polarization points cannot fall at
all. For this map, we arbitrarily fix $\eta = 17^\circ$.}}
\label{fig:DensityRegionA1}
\end{minipage}
\end{center}
\end{figure}

To illustrate the use of the above, we show in Fig.~\ref{fig:DensityRegionA1}
a map of the expected background for region A1 of
\citet{Hutsemekers-et-al2005}, as defined in Table~\ref{tab:A123Huts}.
\begin{table}[h]
\begin{center}
\begin{minipage}{135mm}
\small{
\begin{tabular*}{135mm}{@{\extracolsep{\stretch{1}}}*{1}{lcccc}}
\hline
\\ [-1.5ex]
region & declination & right ascension & redshift & number of quasars \\ [0.5ex]
\hline
\\ [-1.5ex]
A1 & $\delta\leq50^\circ$ & $168^\circ\leq\alpha\leq217^\circ$ & $1.0\leq z\leq2.3$ & 56\\
A2 & $\delta\leq50^\circ$ & $150^\circ\leq\alpha\leq250^\circ$ & $0.0\leq z <   0.5$ & 53\\
A3 &  							 & $320^\circ\leq\alpha\leq360^\circ$ & $0.7\leq z\leq1.5$ & 29\\ [0.5ex]
\hline 
\end{tabular*}}
\caption{\small{The three regions of alignment of \citet{Hutsemekers-et-al2005}
in equatorial coordinates B1950.}}
\label{tab:A123Huts}
\end{minipage}
\end{center}
\end{table}
At each point $a$ of the polarization sphere we associate a probability
distribution $P_{n}^{a}$ through the use of spherical caps.
The mean values $\bar{N^{a}}=\sum_{n}nP_{n}^{a}$ determine the
expected number of polarization points. From those numbers, we build
iso-density regions on the polarization sphere in order to visualize
the structure that the statistical background takes. We arbitrarily choose
here caps of half aperture $\eta=17^\circ$. The dependence of the results
on $\eta$ will be discussed in Section~\ref{subsubstat:depEta}.

Due to the non uniformity of the source locations, there are regions of
maxima (and minima) in the expected densities of polarization points
as well as regions where polarization points are forbidden. For this
sample, a close look at Fig.~\ref{fig:DensityRegionA1} shows that a
quadrupole is naturally expected in the density structure on the
polarization sphere. This shows that the use of the $P_{n}$ distributions
is mandatory, as the expected density is not flat.

\subsection{Further refinements of the method}

\subsubsection{Optimal set of centres for the spherical caps}
\label{subsubstat:Border effect}
The method presented so far has two problems:
\begin{itemize}
\item Several spherical caps can contain the same polarization points, so
that several probabilities are assigned to the same set
of data points. 
\item Among the caps containing the same data points, the most significant
ones will be those for which several of the $\ppi^{(i)}$ will be
small, i.e. for which the loci of several polarization points are
almost tangent to the caps. This enhanced significance is an artefact
of our method which is due border effects.
\end{itemize}
In order to minimize these problems, we do not allow all caps to be
considered, but rather focus on those that correspond to cones with
an axis along the vectorial sum of the normalised polarization vectors
inside them. Hence the effective polarization vector corresponding to
the centre of the cap is
\begin{eqnarray}
\bmath{s}_{centre}			& = &	\sum_{i\in cap}\bmath{\hat{p}}^{(i)}\;,			\nonumber \\
\bmath{\hat{s}}_{centre}	& = &	\frac{\bmath{s}_{centre}}{|\bmath{s}_{centre}|}\:.
\end{eqnarray}
These centres are first determined by iteration before applying the
algorithm explained above.

\subsubsection{Local p-value of the data}
\label{subsubstat:pvalue}
The study of alignments is performed separately for each cap $a$
on the polarization sphere, for which we derive probability distributions
$P_{n}^{a}$. In each cap, we count the number $o_{a}$ of observed
polarization points, and $P_{o_{a}}^{a}$ gives us the probability
that the presence of $o_{a}$ polarization points in cap $a$ is due
to a background fluctuation. The probability that a generation from
a uniform background has a density greater than the observed one is
given by the p-value $p^{a}=\,\sum_{n\geq o_{a}}\, P_{n}^{a}$. The
latter quantity gives us the significance level of a specific polarization
point concentration in one given direction. As already mentioned,
Eq.~\ref{eq:4} shows that the probabilities are coordinate invariant. It in
fact provides a generalisation of the binomial test used in
Hutseme{\'e}kers et al. (\citeyear{Hutsemekers1998};
\citeyear{Hutsemekers-Lamy2001}; \citeyear{Hutsemekers-et-al2005}).
In other words, the $p^{a}$'s give the cumulative Poisson-Binomial
probabilities that, given the set of polarized sources and under the
assumption that the polarization angles are uniformly distributed,
there are $o_{a}$ or more polarization points in the cap of
half-aperture $\eta$ and centred on $a$.

For each sample, we can consider the cap $a_{\rm{min}}$ that gives
the most significant p-value $p_{\rm{min}}={\min}_{a}(p^{a})$ which we
call the significance level. This defines a direction in polarization space,
and a plane in position space.

\subsubsection{Dependence on the spherical cap aperture}
\label{subsubstat:depEta}
The only free quantity in this method is the aperture half-angle
of the spherical caps.
\begin{figure}[h]
\centering
\begin{minipage}{0.8\linewidth}
\centering
\includegraphics[width=\columnwidth]{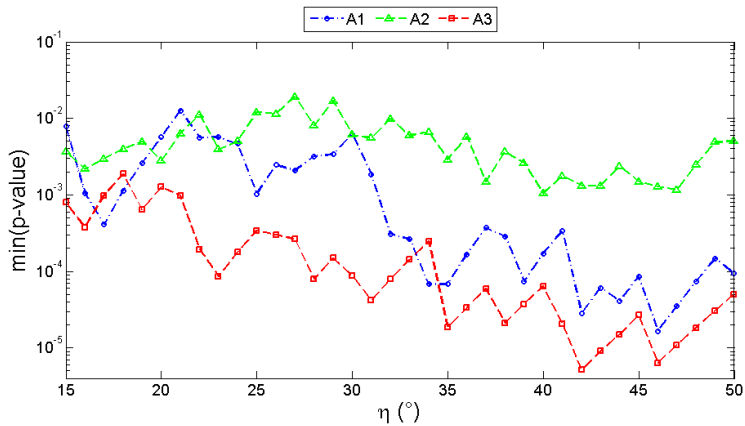}
\caption{\small{Dependence of the significance level with $\eta$, the
half-aperture angle of spherical caps (in degree). The regions are
defined as in Table~\ref{tab:A123Huts}.}}
\label{fig:SLvsCAp}
\end{minipage}
\end{figure}
More than a simple bin width as in histogram-based statistics, the
value of $\eta$ is also somewhat similar to that of the number of nearest
neighbours used in (\citealt{Hutsemekers1998}; \citealt{Jain-Narain-Sarala2004}
and \citealt{Hutsemekers-et-al2005}), and it has a physical meaning.

First of all, each polarization cap corresponds to a band in the sky,
which has an angular width of $2 \, \eta$. Hence, $\eta$ selects part
of the celestial sphere. Secondly, as the sources are angularly separated
and as quasar polarization vectors are always perpendicular to the line of
sight, their projections to the centre of the polarization sphere will always
be spread. $\eta$ takes this spread into account. Finally, $\eta$ is linked
to the strength of the effect (more on this in
Section~\ref{subsecPC-1:NaiveInterp}). A very strong alignment will gather
the polarization points in a small cap, due only to the spread of the sources.
A weaker one will necessitate larger caps, as the effect will be added to a
random one that produces a large spread on the polarization sphere. We
thus see that $\eta$ is determined by physical parameters: the spread of
the sources and the strength of the effect. It thus seems reasonable to
determine its optimal value, which we shall do in the next subsections.

For a given sample of sources we perform the study for a wide range
of half-aperture angle. For each of them we determine the optimum cap
centres, and calculate $p_{\rm{min}}$ as a function of $\eta$.
Fig.~\ref{fig:SLvsCAp} shows $p_{\rm{min}}$ as a function of $\eta$ for
the sub-samples A1, A2 and A3 defined in Table~\ref{tab:A123Huts} and
for $\eta$ taking all integer values between $15^\circ$ and $50^\circ$.
Fig.~\ref{fig:SLvsCAp} shows that the different samples present
significant over-densities of polarization points. We see that $p_{\rm{min}}$
is smaller for $\eta$ between $30^\circ$ and $50^\circ$, depending on the
sample.

\subsubsection{Global significance level of the effect}
\label{subsubstat:GSLofsample}
So far, we have considered the probability that an over-density in
a given cap be due to a background fluctuation. A more relevant probability
maybe that of the occurrence of such an over-density anywhere on the
polarization sphere. To calculate this, we have resorted to a Monte-Carlo
treatment, generating for each data sample $N_{\rm{sim}}$ simulated data
sets, in which we consider only the quasars of that data set, keep their
positions fixed on the sky, and randomly vary their polarization angles
according to a flat distribution. For a given data sample, we introduce a
global significance level $p^{\sigma}$ defined as the percentage of random
sets which produce p-values smaller or equal to $p_{\rm{min}}$ somewhere
on the polarization sphere. 

\begin{figure}[h]
\centering
\begin{minipage}{0.8\linewidth}
\centering
\includegraphics[width=\columnwidth]{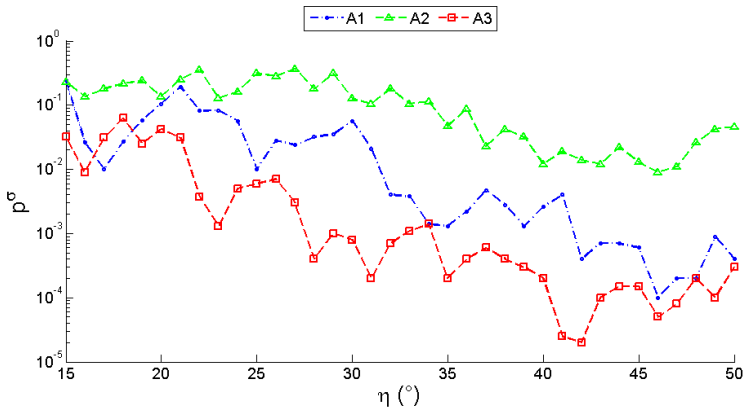}
\caption{\small{Behaviour of the global significance level with the half-aperture
angle for the A1, A2 and A3 regions of Table~\ref{tab:A123Huts}.}}
\label{fig:p^sigma vs eta}
\end{minipage}
\end{figure}

\subsubsection{Optimal angle for the spherical caps}
\label{subsubstat:bestEta}
Fig.~\ref{fig:p^sigma vs eta} shows the behaviour of the global significance
level $p^{\sigma}$ with the aperture angle of the spherical caps for the
sub-samples A1, A2 and A3. Comparing Figs.~\ref{fig:SLvsCAp}
and~\ref{fig:p^sigma vs eta}, we note that $p^{\sigma}$ and $p_{\rm{min}}$
follow the same trend. Clearly, the relation between them must involve
the number of possible caps $N_{c}$ and $p^{\sigma}$ would be equal
to $N_{c}\: p_{\rm{min}}$ if the caps did not overlap and if all simulated
data sets had the same number of caps. Hence we expect $N_{c}$ to
be of the order of the area of the half-sphere divided by the area
of a cap, $p^{\sigma}\,\approx p_{\rm{min}} / \left(1-\cos \eta \right)$.
We found empirically that this relation underestimates $p^{\sigma}$
by a factor smaller than $4$, for all the (sub-)samples we analysed. 

Table~\ref{tab:A123BestParam} shows the significance levels $p_{\rm{min}}$
of over-densities obtained for the different samples of quasars, compared with
the binomial probability $P_{\rm{bin}}$ reported by \citet{Hutsemekers-et-al2005}.
Note that a spherical cap is in general sensitive only to sources along a band of
the celestial sphere so that only part of the entire data sample can contribute to it.
We thus compare the number of polarization points in the cap $o_{a}$ to the
maximum number of points possible in that cap, $o_{a}^{\rm{max}}$.

\begin{figure}[h]
\centering
\begin{minipage}{0.8\linewidth}
\centering
\includegraphics[width=\columnwidth]{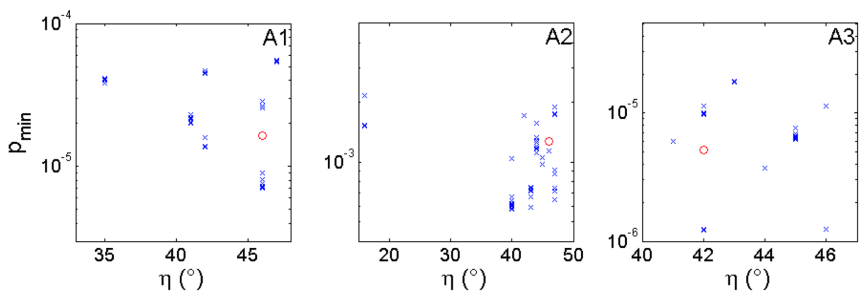}
\caption{\small{Result of the jackknife methods for regions A1, A2 and A3.
The red circles correspond to the results of Table~\ref{tab:A123BestParam}.}}
\label{fig:jackknife}
\end{minipage}
\end{figure}

We see from Table~\ref{tab:A123BestParam} that the best half-aperture angle
depends on the region, and that it is large: 42 or 46 degrees. We also see that
the regions A1 and A3 defined by \citet{Hutsemekers-et-al2005} are the most
significant with our algorithm. However, we need to know whether the difference
between $P_{\rm{bin}}$ and $p_{\rm{min}}$ is important. We shall then study
the errors on the significance level and on $\eta$ and see that the discrepancies
are reasonable.
To do so, we perform a jackknife analysis, removing in turn each quasar from a
given sample, and performing the analysis again. The results are show in
Fig.~\ref{fig:jackknife}.
We see that the errors on $\eta$ are large, and that $p_{\rm{min}}$ can go up
or down by a factor of the order of 3. Hence it seems that our method really
agrees with the estimates of (\citealt{Hutsemekers-et-al2005}). One also clearly
sees that region A2 is less significant than A1 and A3.
\begin{table}[h]
\centering
\begin{minipage}{130mm}
\small{
\begin{tabular*}{130mm}{@{\extracolsep{\stretch{1}}}*{1}{lccccc}}
\hline
\\ [-1.5ex]
Region & $P_{\rm{bin}}$ & $p_{\rm{min}}$ & $\eta$ ($^\circ$) & $o_{a}/o_{a}^{\rm{max}}$ & $p^{\sigma}$\\ [0.5ex]
\hline
\\ [-1.5ex]
A1 & $3.3\,10^{-6}$ & $1.7\,10^{-5}$ & $46$ & $43/56$ & $1.0\:10^{-4}$\\
A2 & $-$ & $1.7\,10^{-3}$ & $46$ & $32/47$ & $0.9\,10^{-2}$\\
A3 & $2.6\,10^{-5}$ & $5.1\,10^{-6}$ & $42$ & $25/29$ & $2.7\:10^{-5}$\\ [0.5ex]
\hline 
\end{tabular*}}
\caption{\small{Significance levels for various data samples. $o_{a}$
is the number of polarization points inside the spherical cap where the
minimum significance level (minimum p-value) $p_{\rm{min}}$ is observed,
$o_{a}^{\rm{max}}$ is the maximum number of polarization points that might
fall inside this cap, $\eta$ is the half-aperture angle of the cap, $P_{\rm{bin}}$
is the binomial probability obtained by \citet{Hutsemekers-et-al2005}
(Table 1) and $p^{\sigma}$ is the global significance level of the region obtained
through the method explained above.}}
\label{tab:A123BestParam}
\end{minipage}
\end{table}

\subsubsection{Mean position angle for local groups}
\label{subsubstat:mPAPC_locGroups}
During our analysis in Section~\ref{subsecPH-1:VisibleWindows}, we consider
groups of neighbouring sources and report the mean position angle
$\bar{\psi}_{\rm{PC}}$ of their corresponding polarization PA distributions.
The latter is computed with respect to the spherical basis vectors
$\left(\bmath{e}_\phi, \, -\bmath{e}_\theta \right)_{\rm{CM}}$ at the location
given by the normalized vectorial sum of the 2-dimensional positions of the sources.
The mean position angle corresponds to the projection of the normalized
sum of the polarization vectors of the sources onto the plane tangent to the
sphere at this mean position.
This quantity has a meaning only when the maximum angular separation between
studied sources is not large and when the position angle distribution is not
uniform.

\subsection{The PC test: conclusion}
\label{substat:Ccl_PCTest}
We have presented a new one-parameter coordinate-invariant method
designed to detect and characterize polarization alignments from sparse
data. The basics of this method are independent from the previous tests.
The new method can thus be used to independently study the orientation
distribution in polarization data catalogues.

This method has the considerable advantage to return significance levels
that are semi-analytically computed. Indeed, our algorithm leads to rapid
evaluation of the probabilities that, by chance, the polarization vectors
point towards given directions, without the need of random catalogue
generations.
Hence, the use of this method allows us to determine unambiguously
the direction of polarization alignments in space and to test for their
significance against randomness.
The remaining drawbacks, however, are that the determination of the global
significance levels relies on very time-consuming Monte Carlo simulations
and that the only parameter of the method might induce edge effects in the
same way as the bin width value does in histogram-based studies. The edge
effects are minimized but not removed by the arbitrary (and time consuming)
selection of spherical caps for which the density studies are performed (see
Section~\ref{subsubstat:Border effect}). These unwanted effects might
eventually be dimmed by adopting another function than the step function for
the evaluation of the densities. However, we did not yet find any function that
still leads to an analytical determination of elementary probabilities
(Eq.~\ref{eq:4}), which is one of the pillar of the method.

All in all, we provide a new statistical test that allows for the study of
the distribution of the polarization PAs from astronomical sources.
Compared to the S and Z tests, our method is intrinsically
coordinate-invariant via physically motivated arguments. However,
and given its current version, this new test still contains the weak points
discussed above. It is worth mentioning that our method can be used
to compare directly the polarization vectors from sources that are
oppositely located on the celestial sphere. In the framework of
cosmological scenarios this constitutes a considerable advantage
compared to the S and Z tests.
Finally, this new statistical test cannot replace the other methods
presented earlier in this chapter. It is complementary to them.
This is especially true for the Z test which, in addition to testing
for local alignments of the polarization vectors, also studies the
correlation of the alignment patterns with their specific locations
on the sky.

\medskip

\begin{figure}
\begin{center}
\begin{minipage}{0.8\linewidth}
\centering
\includegraphics[width=0.8\columnwidth]{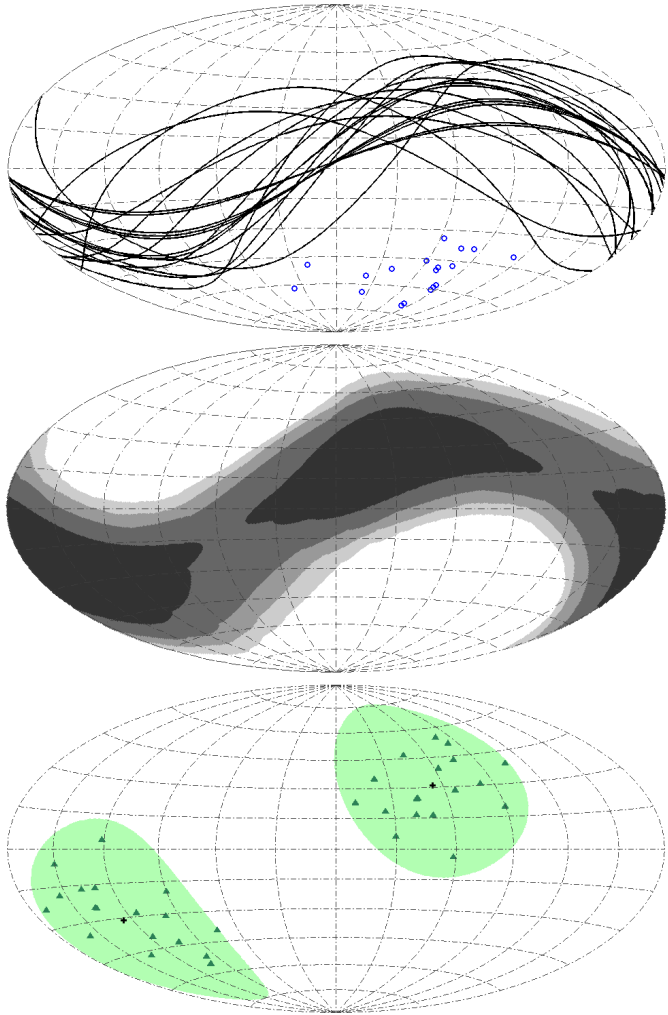}
\caption{\small{The new method at work. Top: Small blue circles are source
positions. Continuous lines are geometrical loci of the corresponding
polarization points. The probability distributions are computed at each location
from the arclength of geometrical loci embedded in the cap.
Middle: Density contours of the mean densities of the distributions evaluated
at each location of the sphere. The darker the shades, the higher the expected
densities.
Bottom: Observed densities $o_{a}$ are evaluated by counting the number of
polarization points (triangles) falling in each cap. We evaluate the cumulative
probabilities from the $o_{a}$ and the $P_n^a$ distributions. The alignment
direction (black crosses) is defined as the centre of the cap (green patches) showing
the most unexpected over-density.
Hammer-Aitoff projected maps are centred on the Galactic centre with positive
Galactic latitude at the top and longitude increasing to the right.}}
\label{fig:IllustrationMethod}
\end{minipage}
\end{center}
\end{figure}

The working of the method is summarized in Fig.~\ref{fig:IllustrationMethod}.
For this illustration, we use a sub-region of the sample of the 355 quasars
with optical polarization measurement compiled in \citet{Hutsemekers-et-al2005}.
This sub-sample, named the S2+ region, is identified in
Section~\ref{subsecPC-1:bestOptRegion} where we performed an unbiased
analysis in order to extract the most aligned regions from the optical sample.
For this region of 18 quasars, the local and global probabilities are found
to be $p_{\rm{min}}= 1.9 \times 10^{-6}$ and $p^\sigma = 1.0 \times 10^{-5}$
when the half-aperture angle of the spherical cap being used is arbitrarily set
to $\eta = 45^\circ$.

\section{The Hawley--Peebles Fourier method}
\label{stat_HPTest}
In Chapter~\ref{Ch:PH-1}, we will make use of an additional statistical test
commonly used to study the alignments of galaxy morphologies.
This test, called the Hawley--Peebles test after \citet{Hawley-Peebles1975},
is based on fitting the observed distribution of PAs by a model of the form

\begin{equation}
n(\psi_i) = \bar{n} \left(1+\Delta_1 \cos 2\psi_i +\Delta_2 \sin 2\psi_i \right)
\label{eq:HPmodel}
\end{equation}
where $\bar{n}$ is the mean of the number of objects per bin and $n(\psi_i)$
is the observed number of objects in the bin centred in $\psi_i$.
The number of bins, and thus their width, is a free parameter.
$\Delta_1$ and $\Delta_2$ are the coefficients of the wave model which
describe the degree of deviation of the distribution from being uniform.
If the PAs are not uniformly distributed, the mean position angle is given by
$\bar{\psi}=({1}/{2})\arctan\left({\Delta_2}/{\Delta_1} \right)$.
A good measure of departure from uniformity is the total amplitude
$\Delta^2={\Delta_1}^2+{\Delta_2}^2$. As easily understood, the larger the
value of $\Delta$, the less uniform the distribution.
The probability that the total amplitude exceeds by chance a given value of
$\Delta$ is computed to be approximately

\begin{equation}
P_{\rm{HP}}=\exp\left( - n \Delta^2 / 4 \right)
\label{eq:HPprob}
\end{equation}
where $N$ is the number of objects in the sample.
However, as far as small samples are considered, random simulations are
required as the distribution of $\Delta$ differs from a normal Gaussian and
as this approximate relation fails far out in the tails
(see \citealt{Godlowski2012} for a detailed discussion). For the simulated
samples, polarization PAs are uniformly generated and distributed among
the sources. The probability $P_{\rm{HP}}$ is then simply given by the
percentage of random realizations having a $\Delta$ value larger than that
of the data.

In our Chapter~\ref{Ch:PH-1}, we tested all reported probabilities using
random simulations and we only found marginal differences compared to
those given by the approximate relation (Eq.~\ref{eq:HPprob}). These are
actually smaller than the variations caused by the choice of the number of
bin. We decided to report only the probabilities computed through the
Gaussian approximation.
It is worth mentioning that this statistical test also depends on the
coordinate system in which PAs are defined. However, as we will use
this test to study the uniformity of the polarization PA distributions
inside relatively small regions of the sky and because the declinations
of these regions are not too high, we expect the changes to be small.

\chapter[Analysis of quasar optical polarization alignments]{A new analysis of quasar optical polarization alignments}
\label{Ch:PC-1Analysis}
This chapter is devoted to the application of our new statistical
method to the catalogue of optical polarization measurements
for the 355 quasars compiled by \citet{Hutsemekers-et-al2005}.
This analysis was originally presented in (\citealt{Pelgrims-Cudell2014}).

Since the PC test is entirely independent of the S and Z tests that have
been applied by Hutsem{\'e}kers et al. (\citeyear{Hutsemekers1998};
\citeyear{Hutsemekers-Lamy2001}; \citeyear{Hutsemekers-et-al2005})
and by \citet{Jain-Narain-Sarala2004}, this new analysis is an
independent search for the characterization of the polarization vector
orientations of this sample.

The results returned by the PC test depend only on the value $\eta$ of
the aperture half-angle of the spherical caps which is used.
From the jackknife treatment in Section~\ref{subsubstat:depEta}, we
have seen that, for a same parent sub-sample, the value of $\eta$ that
maximizes the detection of the alignments strongly fluctuates and that
the local probability $p_{\rm{min}}$ can go up or down by a factor of
the order of 3. This behaviour comes from the fact that $\eta$ takes
into account the spread of the sources on the sky and that the PC test
is subject to edge effects.
To circumvent this problem in this analysis, we choose to fix the angle
$\eta$ at $45^\circ$. Therefore, the local and global significance levels
that we give in the remainder could be slightly improved by choosing
a different value of $\eta$ for each sample.
With this fixed value of $\eta$, we can scan the polarization sphere
with caps, and assign a value of  $p_{\rm{min}}$ to each. The most
significant deviations can be kept and we can numerically evaluate the
global significance $p^{\sigma}$ for the same sample.
This can be done not only on the full data sample, but also on
sub-samples corresponding to regions of redshift, declination or right
ascension, or to cuts on the degree of linear polarization.

\medskip

In the following sections, we thus apply the PC test to the sample first
globally and then to slices in redshift. There, we also consider the
dependence of the alignments on the various parameters, extract the
most significant regions exhibiting an anomalous alignment of
polarization vectors and finally highlight the possibility of a
cosmological alignment that involves sources from both hemispheres.

\section{Full sample}
\label{secPC-1:ResultWNS}
The full sample of quasars is naturally split into Galactic North and Galactic South  
because the observations are away from the Galactic plane, so that besides the whole 
sample, we shall also consider all the northern quasars or all the
southern ones. Each sample has respectively 355, 195 and 160 sources.
\begin{table}
\begin{center}
\small{
\begin{tabular}{lccccc}
\hline
\\ [-1.5ex]
Sample & $p_{\rm{min}}$ & $\left(\delta,\,\alpha\right)_{a_{\rm{min}}}$ ($^\circ$) & $o_{a}/o_{a}^{\rm{max}}$  & $\left(\delta,\,\alpha\right)_{\left\langle \bmath{e}_{r}\right\rangle }$($^\circ$) & $p^{\sigma}$ \\ [0.5ex]
\hline 
\\ [-1.5ex]
Whole & $1.5\,10^{-2}$ & $(48.6,\,283.7)$ & $163/318$ & $(5.5,\,185.0)$ & $0.14$ \\
Northern sky & $9.3\,10^{-2}$ & $(23.1,\,294.0)$ & $\:82/173$ & $(12.2,\,197.2)$ & $0.58$\\
Southern sky & $5.1\,10^{-5}$ & $(39.7,\,270.6)$ & $\:89/142$ & $(-0.7,\,358.5)$ & $6.0\,10^{-4}$\\ [0.5ex]
\hline 
\end{tabular}}
\caption{\small{Parameters of the most significant caps obtained with $\eta = 45^\circ$.}}
\label{tab:WNSBestParam}
\end{center}
\end{table}
We consider all the possible spherical caps, and show the most significant
ones in Table~\ref{tab:WNSBestParam}. The first column gives the most significant p-value,
the equatorial coordinates in the polarization space $\left(\delta,\,\alpha\right)_{a_{\rm{min}}}$ of
the centre of the most significant cap, and the ratio of the number of quasars within the cap to the maximum number, $o_{a}/o_{a}^{\rm{max}}$.
We also give the angular coordinates of the vector $\left\langle \bmath{e}_{r}\right\rangle $ resulting from the normalized sum of the position vectors of the $o_{a}$
sources and the global significance level $p^{\sigma}$ of the alignment.

From this table, one sees that nothing is detected in the whole sample or in the northern 
one. On the other hand, an alignment is detected towards the Galactic South. 
One may wonder then how it was possible to find the significant alignments A1 and A2 
towards the Galactic North, as in Tables~\ref{tab:A123Huts} and~\ref{tab:A123BestParam}. 
The reason for this is that so far we have considered all data points, i.e. all 
redshifts, all declinations, and all right ascensions. The fact that there is an 
alignment to the South and not to the North tells us that the effect depends on the 
physical position of the sources. Hence when we consider all sources, we average the effect, and can simply destroy it.

\begin{figure}[h]
\begin{center}
\begin{minipage}[t]{0.6\textwidth}
\mbox{}\\[-\baselineskip]
\begin{center}
\includegraphics[width=\columnwidth]{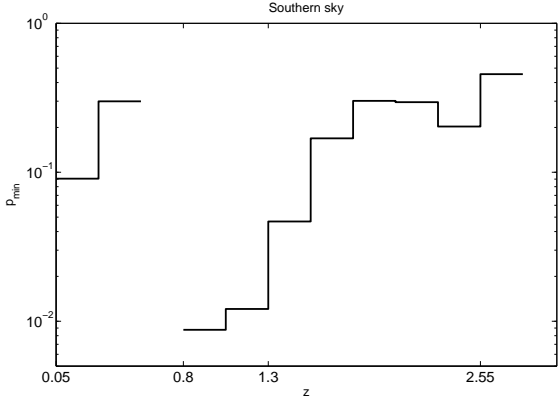}
\end{center}
\end{minipage}\hspace{.02\textwidth}
\begin{minipage}[t]{0.25\textwidth}
\mbox{}\\[-\baselineskip]
\\
\caption{\small{The p-value distribution of the slices of redshift that
show an excess of aligned polarizations towards the direction of the
most significant cap identified in Table~\ref{tab:WNSBestParam}, for
quasars of the southern sample. Redshift bins have a width of
$\delta z = 0.25$.}}
\label{fig:z-S}
\end{minipage}
\end{center}
\end{figure}

To illustrate this, we can consider the redshift distribution of the quasars contributing to the alignment seen towards the Galactic South. Simply counting the aligned quasars in regions of redshift is not enough, though, as the statistics of the sample varies, and as only some quasars have trajectories in polarization space that can intercept the considered cap. However, we have already the required tool: for a fixed cap, we can consider slices of redshift and their p-value. Fig.~\ref{fig:z-S} shows the p-values of slices in which there is an excess of density. We can clearly see in it that the alignment is concentrated in a reduced region of redshift starting at $z=0.8$.

It is indeed known that the directions of large-scale alignments of optical
polarization orientations of quasars show a dependence on the redshift
of the sources (see Hutsem{\'e}kers et al. \citeyear{Hutsemekers1998};
\citeyear{Hutsemekers-Lamy2001}; \citeyear{Hutsemekers-et-al2005};
\citeyear{Hutsemekers-et-al2010} and \citealt{Jain-Narain-Sarala2004}).
Hence studying the effect globally may not make sense, and different
alignments at different redshifts may cancel each other. Also, if only some
regions of redshift have an alignment effect, then it can get washed  out
globally.
Concentrating on the most significant region of Fig.~\ref{fig:z-S} is not
consistent either, as the cap which it is built from is influenced by the
unaligned quasars at high and low redshift. In the next section, we shall
develop a method to determine the regions of redshift where the quasars
are strongly aligned.

\section{Redshift dependence}
\label{secPC-1:Redshift-dependence}
The problem is thus to make a blind analysis of the redshift dependence of the alignment. To do so, we consider a slice of redshift $[z_{\rm{min}}, z_{\rm{max}}]$ and calculate the p-value of the quasars falling in it. We then vary $z_{\rm{min}}$ and $z_{\rm{max}}$ on a grid. The size of the steps $\delta z$ in $z_{\rm{min}}$ and $z_{\rm{max}}$ will of course depend on the statistics of the data.

\begin{figure}
\begin{center}
\begin{minipage}[t]{0.6\textwidth}
\mbox{}\\[-\baselineskip]
\centering
\includegraphics[width=\columnwidth]{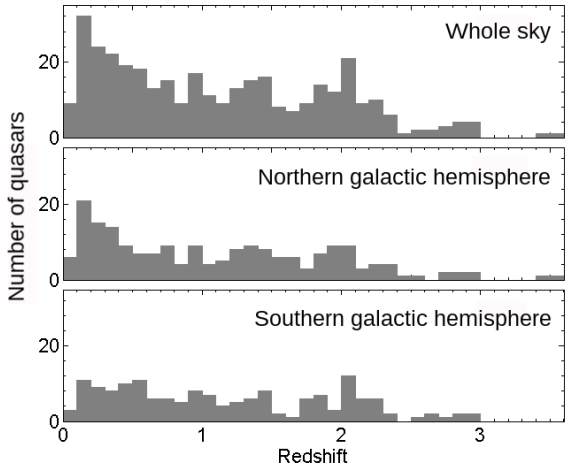}
\end{minipage}\hspace{.02\textwidth}
\begin{minipage}[t]{0.25\textwidth}
\mbox{}\\[-2.\baselineskip]
\caption{\small{The redshift distribution of the sample of 355 quasars
with bin width of $\delta z=0.1$ are shown for the whole sky, the
northern Galactic hemisphere and the southern Galactic hemisphere.
The last bin in the histograms of the whole sky and northern part contains the
quasar at $z=3.94$.}}
\label{fig:z-distrib_WNS}
\end{minipage}
\end{center}
\end{figure}

We show the redshift distribution of the data in Fig.~\ref{fig:z-distrib_WNS}. We see that the high-redshift data points ($z>2.5$) are few, and that there is another deficit in the southern sample in the region $[1.5,\,1.7]$. Also, we see that bins of width $\delta z=0.1$ allow reasonable statistics for most redshifts.

\medskip

\begin{figure}
\begin{center}
\begin{minipage}{0.8\linewidth}
\centering
\includegraphics[width=\columnwidth]{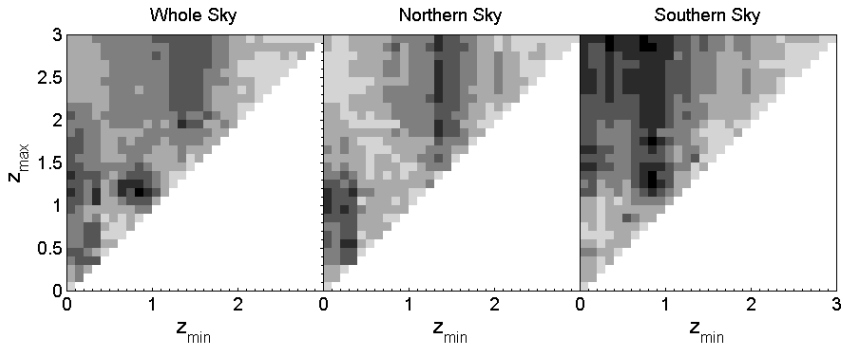}
\caption{\small{Contour plots of $p_{{min}}$ as a function of
the minimum and maximum values of the redshift, for the whole sample,
for the Galactic North and for the Galactic South. Values from $10^{-6}$
to $10^{-5}$ are in black, and the different nuances of grey correspond
to factors of 10, up to the white regions, which are for $p_{\rm{min}}$
between 0.1 and 1.}}
\label{fig:zminzmax}
\end{minipage}
\end{center}
\end{figure}

We can now consider all the values of 
$z_{\rm{min}}$ and $z_{\rm{max}}$ on a grid of spacing $0.1$ (we
also exclude the one quasar with $z>3$). As our test does not use
the quasar position (although it depends on it), we do not need to
introduce further cuts by hand as in (Hutsem{\'e}kers et al.
\citeyear{Hutsemekers1998} and \citeyear{Hutsemekers-et-al2005}).
We nevertheless consider the whole sample, and the northern and
southern regions separately.
We show in Fig.~\ref{fig:zminzmax} the result of this study.
For a given region $[z_{\rm{min}},\,z_{\rm{max}}]$ we show the value of
$p_{\rm{min}}$ as different shades of grey, the darkest regions being the
most significant. Clearly, the dependence on redshift does not seem to
be continuous: the alignment is present for some redshift and not for
others. In particular, all regions present alignments at small $z_{\rm{min}}$,
the northern hemisphere has one further clear alignment starting at $z=1.3$,
whereas the southern hemisphere has a significant alignment starting at
$z=0.8$.
We see that for each sample, the redshift slices that show significant
alignment are  grouped in several islands in the
$(z_{\rm{min}},\,z_{\rm{max}})$ plane. For each island we retain the most
significant sub-sample. The parameters of these nine sub-samples  and of
the  corresponding most significant caps are given in Table~\ref{tab:WNS}.
In this table, sub-samples are quoted by letter which indicates their original
samples; namely, W, N and S indicate if they are extracted from the whole
sky, from the northern sky or from the southern sky (in Galactic coordinates).
Note that the sub-sample named WCo will be introduced and discussed in
Section~\ref{subsecPC-1:WCo}.

\medskip

It may be worth insisting on the fact that cuts in redshift, (or in
declination and  right ascension, see further subsections) amount to the
consideration of data sub-samples  with lower statistics. In that case, our
method leads to higher values of $p_{\rm{min}}$ if an alignment effect is
present, or to a similar value of $p_{\rm{min}}$ if there is no effect. The 
fact that one can markedly increase the significance of the effect by using
such cuts indicates that the effect of alignment is stronger for some regions
of redshift (or for some regions on the celestial sphere).

\begin{figure}
\begin{center}
\begin{minipage}{0.8\linewidth}
\centering
\includegraphics[width=.9\columnwidth]{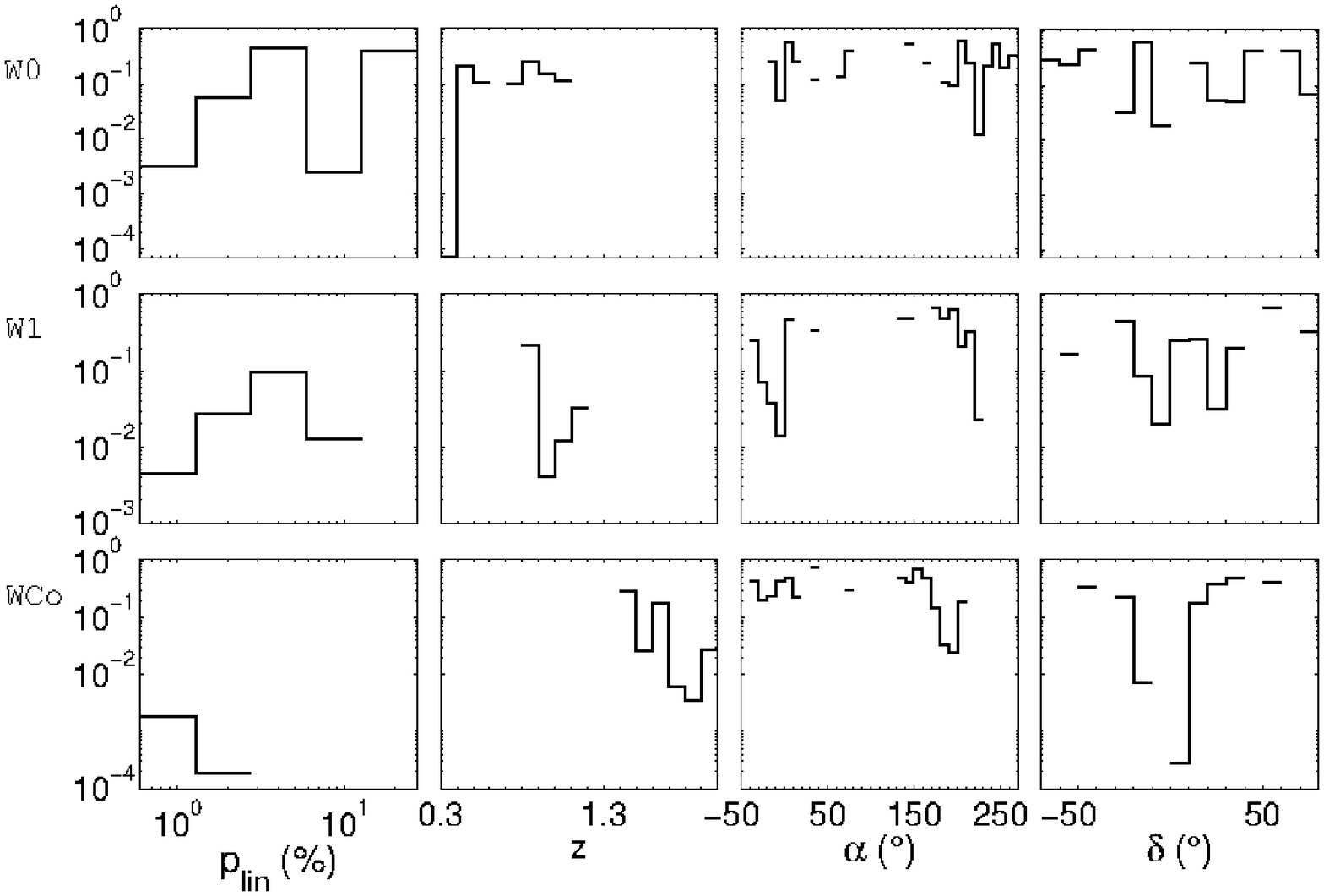}
\includegraphics[width=.9\columnwidth]{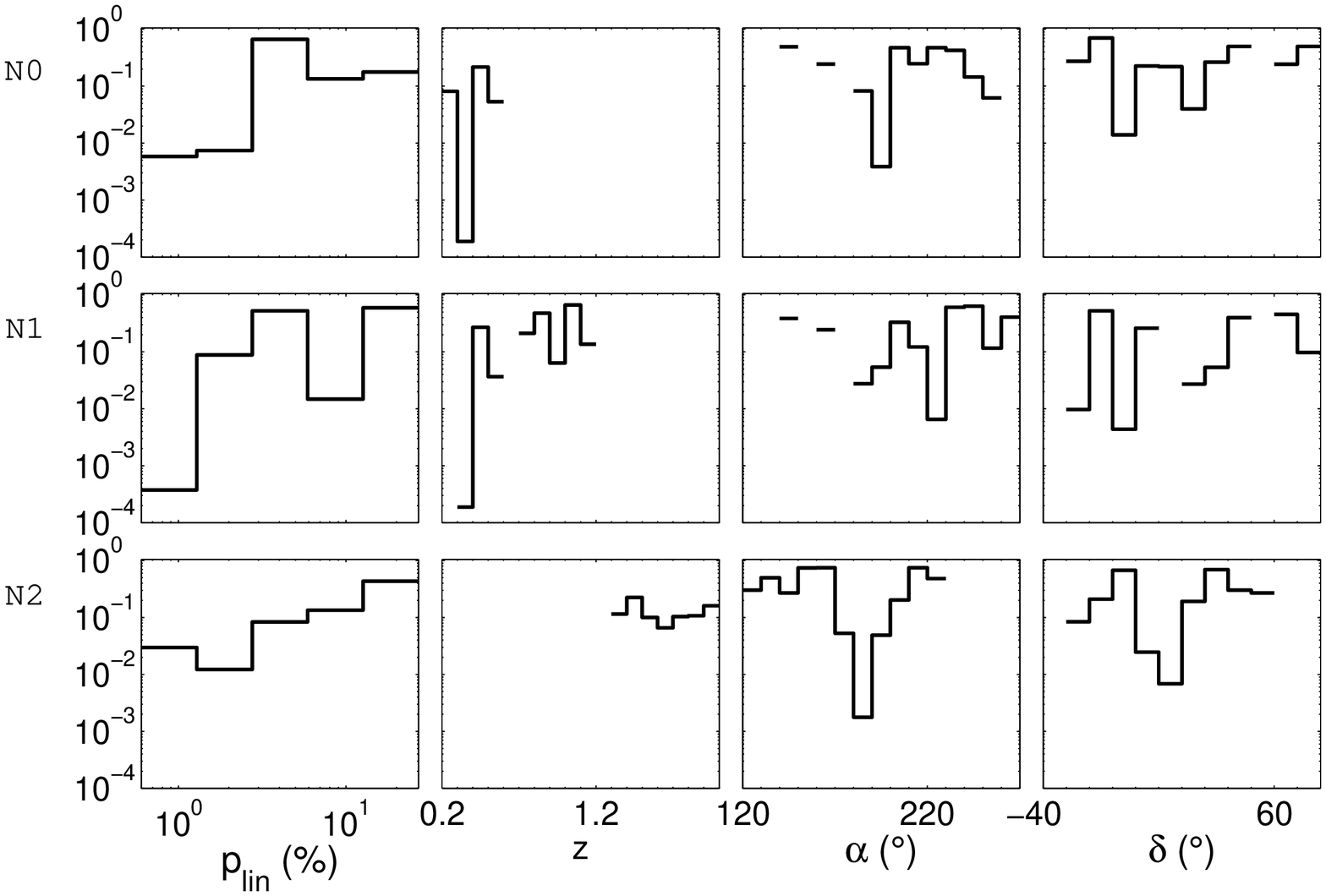}
\includegraphics[width=.9\columnwidth]{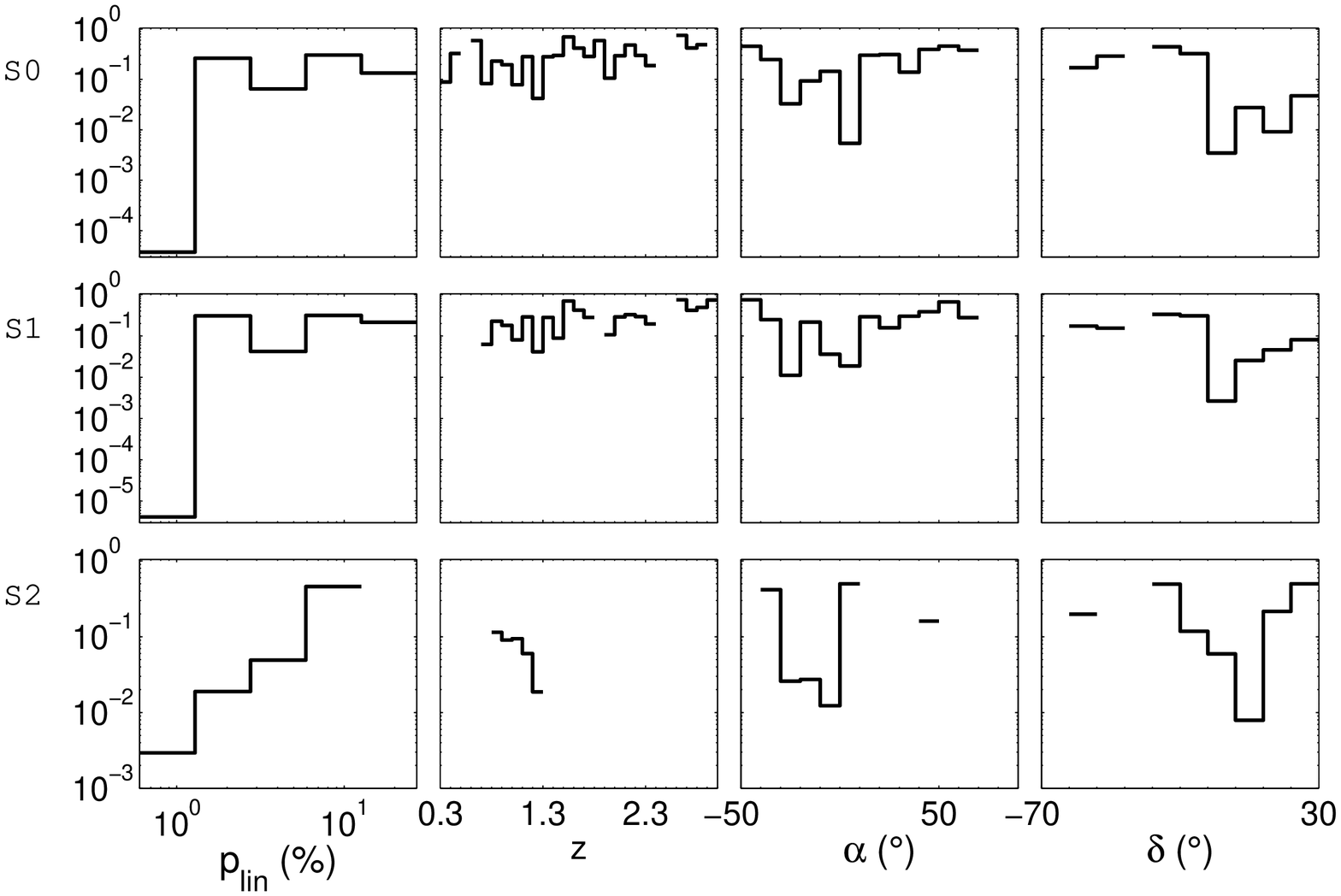}
\caption{\small{Fine structure of the regions of Table~\ref{tab:WNS}.
The ordinates gives the p-values for excess in the sub-regions defined
by the bins in the abscissae. Deficits are not shown.}}
\label{fig:finestruct}
\end{minipage}
\end{center}
\end{figure}

\begin{table}
\centering
\begin{minipage}{137.5mm}
\footnotesize{
\begin{tabular*}{137.5mm}{@{\extracolsep{\stretch{1}}}*{1}{lccccccc}}
\hline
\\ [-1.5ex]
Sample & $z_{\rm{min}}$ & $z_{\rm{max}}$ & $p_{\rm{min}}$ & $\left(\delta,\,\alpha\right)_{a_{\rm{min}}}$ 
($^\circ$) & $o_{a}/o_{a}^{\rm{max}}$ & $\left(\delta,\,\alpha\right)_{\left\langle 
\bmath{e}_{r}\right\rangle }$ ($^\circ$) & $p^{\sigma}$\\ [0.5ex]
\hline
\\ [-1.5ex]
W0 & $0.3$ & $1.1$ & $2.9\,10^{-5}$ & $\left(20.7,\,304.3\right)$ & $60/99$ & $
\left(16.9,\,203.3\right)$ & \\
W1 & $0.8$ & $1.2$ & $7.3\,10^{-6}$ & $\left(25.4,\,278.1\right)$ & $31/40$ & $
\left(12.2,\,181.2\right)$ & \\
W2 & $1.3$ & $2.0$ & $8.3\,10^{-5}$ & $\left(76.0,\,304.7\right)$ & $53/80$ & $
\left(10.2,\,32.4\right)$ & \\
{\bf WCo} &$\mathbf{1.3}$ &$\mathbf{2.0}$ & $\mathbf{4.3\,10^{-6}}$ &$\mathbf{\left(65.9,
\,293.1\right)}$ & $\mathbf{39/50}$ & $\mathbf{\left(-9.3,\,3.5\right)}$ & $\mathbf{2.7\,
10^{-5}}$\\
\\ [-2ex]
{\bf N0} & $\mathbf{0.2}$ & $\mathbf{0.6}$ & $\mathbf{1.4\,10^{-5}}$ & $
\mathbf{\left(15.0,\,308.2\right)}$ & $\mathbf{28/37}$ & $\mathbf{\left(22.3,\,
208.4\right)}$ & $\mathbf{1.8\,10^{-4}}$ \\
N1 & $0.3$ & $1.2$ & $1.5\,10^{-5}$ & $\left(12.7,\,305.0\right)$ &$40/58$ &
$\left(19.3,\,206.4\right)$ &\\ 
{\bf N2} & $\mathbf{1.3}$ & $\mathbf{2.0}$ & $\mathbf{3.5\,10^{-5}}$ & $
\mathbf{\left(78.2,\,298.1\right)}$ & $\mathbf{35/47}$ & $\mathbf{\left(5.8,\,
186.6\right)}$ & $\mathbf{3.4\,10^{-4}}$\\
\\ [-2ex]
S0 & $0.3$ & $2.9$ & $8.1\,10^{-6}$ & $\left(44.7,\,273.8\right)$ & $79/120$ & $
\left(-5.0,\,357.1\right)$ &\\
S1 & $0.7$ & $3.0$ & $3.1\,10^{-6}$ & $\left(43.6,\,272.3\right)$ & $62/89$ & $
\left(-7.0,\,357.6\right)$ &\\
{\bf S2} & $\mathbf{0.8}$ & $\mathbf{1.3}$ & $\mathbf{3.9\,10^{-6}}$ & $
\mathbf{\left(31.8,\,263.9\right)}$ & $\mathbf{19/20}$ & $\mathbf{\left(-7.8,\,
348.3\right)}$ & $\mathbf{3.0\,10^{-5}}$\\ [0.5ex]
\hline 
\end{tabular*}}
\caption{\small{Significant sub-samples from the scan on redshift performed
on the whole sample of 354 quasars and the northern and southern samples
of 194 and 160 sources, respectively. Best cap parameters are shown as in
Table~\ref{tab:WNSBestParam} as well as the lower and  upper limits in
redshift of sub-samples. Note that region WCo is detected for
$p_{\rm{lin}}\leq1.5\%$. Bold characters stress the most significant
independent regions (see the text for a discussion).}}
\label{tab:WNS}
\end{minipage}
\end{table}
The first thing to notice is that we indeed find possible regions of alignment
towards the Galactic South. However, we must decide whether they are all
significant and independent, as a very significant region can always be
somewhat extended by adding to it some noise. 
To decide, we can proceed as in the case of Fig.~\ref{fig:z-S}, and cut this
time each sample in slices of redshift, declination and right ascension.
The results of such a  study are shown in Fig.~\ref{fig:finestruct} for all the
regions of Table~\ref{tab:WNS}.
If for now we concentrate on the last three columns of the southern regions
(last three lines) in Fig.~\ref{fig:finestruct}, we see the structure of S0, S1
and S2. The distributions in right ascension and declination tell us that the
quasars that contribute most are in the same region of the celestial sphere,
which is confirmed by the 7th column of Table~\ref{tab:WNS} that gives the
average position on the sky. Also, the 5th column of Table~\ref{tab:WNS}
shows that the alignment is in the same direction for S0 and S1, and almost
in the same direction for S2 (remember that the caps have an aperture of
$45^\circ$). Hence it seems that there is a very strong alignment (to which
19 quasars out of 20 contribute) for the limited redshift region $0.8<z<1.3$,
and that alignment can be extended to higher or lower redshift, without
changing its significance much. As increasing the number of sources should
markedly decrease the p-value if they also have aligned polarization vectors,
we believe that only S2 is significant.

\medskip

We can perform the same analysis for the northern quasars. Considering
again the last three columns, and this time the fourth, fifth and sixth lines
of Fig.~\ref{fig:finestruct}, we see that N0 and N1 are populated by quasars
in the same region of the celestial sphere, and that N1 is the same as N0,
but extended in redshift. Table~\ref{tab:WNS} confirms that the average
position on the sky is very close, and that the preferred directions of
polarization are almost identical. It thus seems that N0 is the significant
region, as N1 has more statistics, but less significance. On the other hand,
Table~\ref{tab:WNS} clearly shows that N2 is disjoint from N0 and N1 in
redshift and that the preferred directions of alignments are significantly
different. Indeed, the angular change is of the order of $70^\circ$ which is
reminiscent of result already obtained by \citet{Hutsemekers-et-al2005}
which is somehow illustrated in Fig.\ref{fig:alignedRegions_H05}.

\medskip

Finally, we can consider the first two lines of Fig.~\ref{fig:finestruct}.
We see, looking at the plot in right ascension, that the most significant
part of W0 is towards the Galactic North, whereas W1 is more significant
towards the Galactic South. Table~\ref{tab:WNS} shows that the direction
of alignment of W0 (resp. W1) is compatible with that of N0 (resp. S2).
Furthermore, we see that the most significant quasars of W0 fall in the
same redshift bin as those of N0. Hence it seems that W0 is really a
reflection of N0. Similarly, the p-values are smallest in W1 for the same
redshift as for S2, and it seems that W1 is really generated by S2.

We have checked these conclusions by separating W0 and W1 into their
northern and southern parts and by performing the study independently
for these two parts. If p-values of both parts are all higher than the value
of $p_{\rm{min}}$ of the whole, and point towards the same preferred
direction, then it is clear that the observed alignment is produced by
sources from both hemispheres. In the case of W0 (resp. W1) we find
that the northern (resp. southern) alignment is much more significant.

\subsection{Fine structure and best regions}
\label{subsecPC-1:bestOptRegion}
We can study the structure of each region, and check whether it can be
better defined by using further cuts. Consider the first column of
Fig.~\ref{fig:finestruct}, i.e. cuts on linear polarization. We do not find, for
N0, N2 and S2, that cuts in linear polarization increase the effect
significantly (i.e. that $p_{\rm{min}}$ gets reduced by more than a factor
of two). The reduced significance of the bins with large polarization is due
to their lower statistics.

\medskip

On the other hand, the dependence on right ascension and declination
suggests that some regions of the sky are more significantly aligned.
From this observation, we can define even more significant regions, by
placing cuts on right ascension and declination. This does not lead to a
significant difference, except for regions N2 and S2. Following the above
argument, it seems that we have detected three independent regions of
alignment, which are significant. We summarise their parameters in
Table~\ref{tab:best}. Note that N0, N2+ and S2+ are improved versions
of A2, A1+ and A3 defined by \citet{Hutsemekers-et-al2005}.
\begin{table}
\centering
\begin{minipage}{\textwidth}
\footnotesize{
\begin{tabular*}{\textwidth}{@{\extracolsep{\stretch{1}}}*{1}{lcccccccc}}
\hline
\\ [-1.5ex]
Sample & $z_{\rm{min}}$ & $z_{\rm{max}}$ & $p_{\rm{min}}$ & $\left(\delta,\,\alpha\right)_{a_{\rm{min}}}$ ($^\circ$) & $o_{a}/o_{a}^{\rm{max}}$ & $\delta$ interval ($^\circ$) & $\alpha$ interval ($^\circ$) & $p^{\sigma}$ \\ [0.5ex]
\hline
\\ [-1.5ex]
N0 & $0.2$ & $0.6$ & $1.4\,10^{-5}$ & $\left(15.0,\,308.2\right)$ & $28/37$ & $[-25,\:80]^{*}$ & $[135,\;265]^{*}$ & $1.8\, 10^{-4} $ \\
N2+ & $1.3$ & $2.0$ & $4.5\,10^{-6}$ & $\left(79.8,\,289.3\right)$ & $30/35$ & $[-30,\:35]~$ & $[165,\:230]~$ & $5.0 \, 10^{-5}$ \\
S2+ & $0.8$ & $1.3$ & $1.9\,10^{-6}$ & $\left(31.8,\,261.2\right)$ & $18/18$ & $[-55,\;25]^{*}$ & $[-40,\:20]~$ & $1.0\, 10^{-5}$ \\ [0.5ex]
\hline 
\end{tabular*}}
\caption{\small{Best independent regions of alignment.
The regions in $\delta$, $\alpha$ marked by an asterisk describe the
data sample, the others are cuts imposed on the data. N0 is the same
as in Table~\ref{tab:WNS}. S2+ and N2+ are restrictions of S2 and N2
to a smaller region of the celestial sphere.}}
\label{tab:best}
\end{minipage}
\end{table}

\subsection{A possible cosmological alignment}
\label{subsecPC-1:WCo}
Although cutting on polarization does not improve significantly the
previous probabilities, we detected a rather surprising alignment,
as it is very significant only when the North sample is considered
together with the southern one. Indeed, if we consider only small
linear polarizations, with $p_{\rm{lin}}\leq1.5$ per cent, then there
is a North-South alignment with $p_{\rm{min}}<5\:10^{-6}$, as shown
in the sample WCo of Table~\ref{tab:WNS}. This alignment is much
less significant in the North ($p_{\rm{min}}\approx2\,10^{-4}$) or in
the South ($p_{\rm{min}}\approx10^{-3}$), but it becomes significant
once both hemispheres are considered together. It must also be noted
that it is significant only after the cut on linear polarization.

\subsection{A naive interpretation}
\label{subsecPC-1:NaiveInterp}
One can imagine that a systematic oscillating electric field
$\bmath{E}$ is at work in each of the regions we defined. We can
try to determine its norm and take it parallel to the centre of the
polarization cap $\hat{\bmath{s}}_{centre}$, in such a way that the
alignments we found disappear if we subtract that systematic effect
from the samples we defined (in practice we impose that
$p_{\rm{min}}\geq0.1$).
Of course, we have first to project $\bmath{E}$ onto the plane normal
to the direction of propagation, then remove it from the polarization. If
we perform this exercise, the resulting values of $|\bmath{E}|$ are
given in Table~\ref{tab:best-1} for the most significant regions of
Table~\ref{tab:WNS}. It is remarkable that the vectors we have to remove
from the data have roughly the same norm. Due to the projection of the
vector $\bmath{E}$, this naive model could explain why polarization
vectors are not all seen to be aligned.

\begin{table}[h]
\begin{center}
\begin{minipage}{0.4\linewidth}
\small{
\begin{tabular*}{\columnwidth}{@{\extracolsep{\stretch{1}}}*{1}{lc}}
\hline
\\ [-1.5ex]
Sample & $\bmath{|E}|$ (\%)\\ [0.5ex]
\hline
\\ [-1.5ex]
N0 & $0.65-0.70$ \\
N2 & $0.6-0.7$   \\
S2 & $0.8-1.2$	 \\
WCo & $0.5-0.9$  \\ [0.5ex]
\hline
\end{tabular*}}
\caption{\small{Norm of a systematic 3-vector accounting for the effect.}}
\label{tab:best-1}
\end{minipage}
\end{center}
\end{table}

\section{Concluding remarks}
\label{secPC-1:Ccl}
In this chapter, we performed a new analysis of the catalogue of the
355 quasar polarization measurements at optical wavelengths compiled
by \citet{Hutsemekers-et-al2005}. Based on this data set, and from
earlier versions, previous analyses have reported evidence for coherent
orientations of quasar polarization vectors at very large scales.
The quasars involved in these correlations are indeed separated by
gigaparsec-scale comoving distances (see \citealt{Jain-Narain-Sarala2004}
and \citealt{Hutsemekers-et-al2005}).
These analyses, however, rely on the use of the same two statistical
methods, the S and Z tests that we discussed in Sections~\ref{stat_Stest}
and~\ref{stat_Ztest}. Our analysis, though, is performed with the statistical
test that we introduced in Section~\ref{stat_PCTest} which is independent
from the two others. Hence, our analysis is complementary, although the
same data set is under study.
Furthermore, the new method allowed us to extract in an unbiased way
the sub-samples that show significant correlated polarization orientations.

\medskip

The application of the PC test to the optical sample of quasar polarizations
confirms the large-scale polarization alignments. We showed that we
automatically recover regions previously found, and we refined their limits
based on unbiased criteria (see Table~\ref{tab:best}).
We believe that this new analysis puts the alignment effect on stronger
grounds as the global significance level is as low as $3.0 \times 10^{-5}$ for
some regions of space.

\medskip

However, one has to note that the significance levels obtained in this
chapter and those reported by \citet{Hutsemekers-et-al2005} are not
in full agreement. The main reason that accounts for these differences
is that these authors generated random catalogues through the
reshuffling in order to assess the significance levels of the correlations
with the S and Z tests. The advantage is that any systematic effect
vanishes automatically through this method. The disadvantage is that it
washes out global effects, or alignments present for a large number of
quasars. Random generation of polarization angles, as used here or
also by \citet{Jain-Narain-Sarala2004}, has the opposite features: we
can detect global alignments, but we are sensitive to systematic effects.
Hence the two methods do not need to be in full agreement. One should
note, however, that the sample of optical polarization measurements of
quasars considered here comes from many independent observational
campaigns, so that a common bias is unlikely (see Hutsem{\'e}kers et al.
[\citeyear{Hutsemekers1998}; \citeyear{Hutsemekers-et-al2005}] for
discussion).
Furthermore, \citet{Jain-Narain-Sarala2004} have addressed this
question of global systematic effect with the sample of 213 quasar
polarization measurements available at the time by comparing analyses
with the S and Z tests using uniform polarization angle distributions and
distributions made by reshuffling. For some sub-samples, they found
that the significance levels increase by factors of the order three when
the generation of random data set is made through reshuffling.
We expect to obtain the same difference for the global significance level
obtained with our test ($p^\sigma$) if we generate random catalogue
with reshuffling instead of generating the position angles according to a
uniform distribution. Note that this difference is also of the same order
as that we estimated using the jackknife method.
To conclude, our analysis is found to be in good agreement with the
previous ones.

\medskip

Through the PC test, we identified the following main features of the
alignments. The directions of alignments show a dependence on the
redshift of the sources. Although this dependence seems discontinuous,
one should note that we detected significant alignments for redshift
intervals where the distribution of data peaks.
Thus, more data in regions of redshift with poor statistics are required
in order to study this dependence in more details.
Further, as seen in Fig.~\ref{fig:finestruct} for a given redshift interval,
alignments seem to be mainly due to quasars well localized towards
specific directions of the sky, again in agreement with previous studies.
However, no strong evidence has been found for a dependence on the
degree of linear polarization. This result is slightly at odds with previous
claims even though \citet{Hutsemekers-et-al2005} specifically cautioned
that their findings can be due to selection biases in their data set.

\medskip

As a result of the application of our new method to the present sample of
optical polarization measurements of quasars, and in agreement with
\citet{Hutsemekers-et-al2005}, we found several distinct sub-samples of
sources well localized in space that show unexpected alignments of their
polarization vectors. We established two regions towards the North Galactic
pole, one at low and the other at high redshift, and only one towards the
South Galactic pole at intermediate redshift, which possibly dominates
the whole southern sky.
Besides the regions previously detected, or their improved version, we also
showed that there exists the possibility of a cosmological alignment.

\chapter{A search for quasar radio polarization alignments}
\label{Ch:PH-1}
\citet{Jackson-et-al2007} compiled the JVAS/CLASS 8.4-GHz sample of
flat-spectrum radio sources (FSRS), paying particular attention to
instrumental biases. This sample contains polarization position angle (PA)
measurements. As they have
shown, rotation measures of the polarization vectors induced by Faraday
rotation at 8.4 GHz are too small to destroy information about the intrinsic
PAs. Therefore, any observed correlation of PAs among sources can be
thought to be intrinsic to the sources themselves.
\citet{Joshi-et-al2007} extracted from this sample 4290 FSRSs with polarized
flux higher than $1\,\rm{mJy}$, and searched for systematic alignments of
radio polarization vectors of the type reported at optical wavelengths by
Hutsem{\'e}kers and collaborators, i.e.  at cosmological scales.
Their analysis did not reveal such large-scale alignments at radio
wavelengths. From this claim, the recognized wavelength dependence of
the polarization vector alignments has brought the model of axion-like
particle (e.g., \citealt{Das-et-al2005};
\citealt*{Payez-Cudell-Hutsemekers2008}; \citealt*{Agarwal-Kamal-Jain2011})
as the favourite candidate to explain alignments at optical wavelengths.
This model has however been observationally ruled out since it predicts
non-negligible circular polarization which is not detected
(\citealt{Hutsemekers-et-al2010}; \citealt*{Payez-Cudell-Hutsemekers2011}).

\medskip

Beside this analysis, \citet{Tiwari-Jain2013} tested the uniformity of the
polarization PAs  considering roughly the same sample. They found
significant evidence for alignments at distance scales of the order of
$150\,\rm{Mpc}$\footnote{Attention has to be paid regarding this scale
as these authors defined the comoving distances assuming a redshift of
one for all objects. Their analysis is thus a 2-dimensional one rather than
a 3-dimensional one, contrary to what the reported scale might suggest.}.
As the correlations are found at different distance scales, their study does
not contradict the analysis of \citet{Joshi-et-al2007}.
More recently, \citet{Shurtleff2014} studied the correlation of the PAs for
sources grouped in circular regions of $24^{\circ}$ radius. He reported PA
alignments in two regions of the sky, although not at a very significant level.

\medskip

Despite these analyses which involve different statistical tests and
different samples which correspond to different cuts of the original
data set, the status of polarization PA correlation at radio wavelengths
was not clear. Moreover, an analysis taking properly the redshift of the
sources into account was still missing.
The redshift dependence is an important characteristic of the
alignments of quasar polarization vectors at optical wavelengths and it
seemed important to take it into account in the analysis of the radio
sample, especially if one seeks the same signature at radio wavelengths
as at optical wavelengths.
Therefore, we performed a careful analysis of the uniformity of the
polarization PAs of FSRSs belonging to the JVAS/CLASS 8.4-GHz
surveys.
This analysis (published in \citealt{Pelgrims-Hutsemekers2015}) is
presented in this chapter which is structured as follows.

\medskip

The data samples which are studied throughout this analysis are introduced
in Section~\ref{secPH-1:DataSample}. The sample with robust polarization
measurements is made of 4155 objects and spectroscopic redshift
information is collected for 1531 of them.
In Section~\ref{subsecPH-1:VisibleWindows}, taking the redshift of the
sources into account, we shall investigate the polarization PA distributions
of the FSRSs located in regions of the sky where the optical polarization
alignments are the most significant. Stimulated by the detection of alignment
in one of these regions, we perform a complete analysis of the entire data set
in Section~\ref{secPH-1:uniformity_in_JVAS}, with and without accounting for
the redshift.
Having highlighted significant alignment signatures in the sample of quasars,
we search for their characterization in Section~\ref{secPH-1:IdentifRegions}.
We finally summarize our results in Section~\ref{secPH-1:DiscussionFinal},
present arguments against and for the hypothesis of biases in the data set
and discuss a possible interpretation of the data.
We conclude in Section~\ref{secPH-1:Conclusion} either that the data set of
the polarization angle measurements of the JVAS/CLASS 8.4-GHz surveys are
not exploitable due to biases or that they suggest large-scale
alignments at radio wavelengths.
Indeed, as we shall see, two statistical analyses (one in two dimensions and
the other in three dimensions when distance is available), detect significant
large-scale alignments of polarization vectors for samples containing only
quasars among the varieties of FSRSs.
While these correlations prove difficult to explain either by a physical
effect or by biases in the data set, the fact that the quasars which have
significantly aligned polarization vectors are found in regions of the sky
where optical polarization alignments were previously found is striking.

\section{Data sample}
\label{secPH-1:DataSample}
The JVAS/CLASS 8.4-GHz catalogue is made of the JVAS (Jodrell-VLA
Astrometric Survey) and the CLASS (Cosmic Lens All-Sky Survey) surveys
that were gathered by \citet{Jackson-et-al2007} to build the largest catalogue
of polarization measurements of compact radio sources at that time, paying
attention to avoid biases on polarization measurements.
We refer to \citet{Jackson-et-al2007} and references therein for a complete
description of the catalogue and the surveys.
In this catalogue, the total number of object having polarization measurements
is 12\,743 (see the on-line catalogue%
\footnote{http://vizier.u-strasbg.fr/viz-bin/VizieR-3?-source=J/MNRAS/376/371}).
Adopting the prescription given by \citet{Jackson-et-al2007} and
\citet{Joshi-et-al2007}, we keep the sources for which the polarized flux is
higher or equal to $1\,\rm{mJy}$ in order to select significant polarization
detections only and to obtain an unbiased sample. \citet{Jackson-et-al2007}
claim that this cut in polarized flux corresponds to a significant polarization
detections at the level of approximately $4\sigma$. When there is more than
one object in a radius of 1 arcsec on the sky, we select the object with the
highest polarized flux. This selection, which also eliminates multiple
measurements, leaves us with a sample size of 4265 objects. If we only
remove multiple measurements, we recover the source number of 4290
studied by \citet{Joshi-et-al2007}. We nevertheless choose to add the above
constraint for an efficient source separation.

The data table of \citet{Jackson-et-al2007} contains the following
information for the sources: right ascension, declination, Stokes
parameters ($I$, $Q$ and $U$), the corresponding errors
($\sigma_I$, $\sigma_Q$ and $\sigma_U$) and the derived
polarization position angle ($\psi = (1/2) \arctan (U/Q)$) in the
East-of-North convention.
From these data and according to the usual definition for the degree
of linear polarization, we define $p_{\rm{lin}} = (u^2 + q^2)^{1/2}$,
where we make use of the normalized Stokes parameters $u=U/I$ and
$q=Q/I$. We then compute the standard error $\sigma_p$ and
$\sigma_\psi$ for $p_{\rm{lin}}$ and $\psi$ which are given by the
Serkowski's formulae (\citealt{Serkowski1958};
\citealt{NaghizadehKhouei-Clarke1993})
\begin{equation}
\sigma_p = \frac{\left(q^2 \sigma_q^2 + u^2 \sigma_u^2 \right)^{1/2}}{p_{\rm{lin}}}
\label{eq:sigma_plin}
\end{equation}
and
\begin{equation}
\sigma_\psi = \frac{\left(u^2 \sigma_q^2 + q^2 \sigma_u^2 \right)^{1/2}}{2 p_{\rm{lin}}^2}\;.
\label{eq:sigma_PA_real}
\end{equation}
Given that $\sigma_p \simeq \sigma_q \simeq \sigma_u$ is verified in
general (e.g., \citealt{Lamy-Hutsemekers2000}), the usual equation for
the uncertainty on the polarization position angle reads
\begin{equation}
\sigma_\psi = \frac{\sigma_p}{2 p_{\rm{lin}}}\, \rm{rad} = 28.^\circ65 \, \frac{\sigma_p}{p_{\rm{lin}}}\;.
\label{eq:sigma_PA}
\end{equation}
Led by a selection criterion used at optical wavelengths to ensure a
significant polarization detection (e.g., \citealt{Hutsemekers1998};
\citealt{Sluse-et-al2005}), we further constrain the sample asking that
the uncertainty of the position angle, computed as in
Eq.~\ref{eq:sigma_PA}, verify $\sigma_{\psi} \leq 14^{\circ}$.
While the cut at $1\,\rm{mJy}$ ensures significant polarization detection,
our additional cut has to be seen as an additional quality criterion.
Out of the 4265 sources, 4155 satisfied the criterion. This sample, which
we call $All$ in the reminder, constitutes the largest one for which we
have robust polarization PA measurements from the JVAS/CLASS
8.4-GHz surveys.
The sky projection of the sample $All$ is shown in
Fig.~\ref{fig:J4155_GalProj}.

\medskip

Using the NASA Extragalactic Database\footnote{http://ned.ipac.caltech.edu/}
(NED), we identified a total of 3858 sources. We first used the automated mode
"Near-Object/Position List" with a search radius of $0.1$ arcsec. After manual
selection among multiple identifications, 3446 objects were kept.
For the 709 objects left, we used a search radius of $0.5$ arcsec and found
412 additional sources, after having again manually took care of the multiple
identifications.
We stopped the procedure at this value of the search radius in order to ensure
proper identifications.
Out of the 3858 retrieved objects, 1531 have spectroscopic, and thus reliable,
measurements of redshift, $z$.
The use of NED also leads to the classification of the sources.
Table~\ref{tab:NED_classification} reports the number of sources identified for
each class of FSRS as well as the number of these sources for which we have
redshift information.
\begin{table}[h]
\begin{minipage}[t]{0.55\textwidth}
\mbox{}\\[-\baselineskip]
\begin{center}
\small{
\begin{tabular*}{.8\columnwidth}{@{\extracolsep{\stretch{1}}}*{1}{llrc}}
\hline
\\ [-1.5ex]
$z$					& Object Type			& $N$	& Acronym		\\ [0.5ex]
\hline
\\ [-1.5ex]
\multirow{5}{*}{no}	& All 						& 3858	& 	$-$			\\
							& QSOs 				& 1450 	& 	$QSO$		\\
							& Radio Sources 	& 1379 	& 	$RS$ 		\\
							& Galaxies 			& 381 	& 	$G$ 			\\
							& Other	Objects	 	& 648 	& 	$VO$ 		\\
\hline
\multirow{5}{*}{yes}& All						& 1531	&	$All(z)$		\\
							& QSOs 				& 1325 	& 	$QSO(z)$	\\
							& Radio Sources 	& 11 		& 	$-$			\\
							& Galaxies 			& 184 	& 	$-$			\\
							& Other	Objects 		& 11 		& 	$-$			\\ [0.5ex]
\hline
\end{tabular*}}
\end{center}
\end{minipage}\hfill
\begin{minipage}[t]{0.45\textwidth}
\mbox{}\\[-\baselineskip]
\caption{\small{Number ($N$) of the object from different source species
as retrieved from the NED database among the sample of 4155 sources
with reliable polarization PA measurements, with and without redshift
information, $z$. The last column contains the acronyms used for the
samples that we analyse in this work. The category named ``Other Objects''
contains various species with small number of members.}}
\label{tab:NED_classification}
\end{minipage}
\end{table}

As it can be clearly seen, the QSOs represent $86 \%$ of the sample with
redshift measurements. Hence, analyses and results involving samples with
redshift information will mainly concern those objects.
Following \citet{Jackson-et-al2007}, the core-dominated FSRSs are
predominantly quasars or BL Lac objects in which the jet is oriented close
to the line of sight. We nevertheless choose to adopt the notation of the
NED database. Throughout this chapter, we will thus refer to the sub-samples
of QSOs and Radio Sources via the acronyms $QSO$ and $RS$.

\section{Uniformity of radio-polarization PAs}
\subsection{Regions of optical polarization alignments}
\label{subsecPH-1:VisibleWindows}
As we already discussed in the previous chapters, specific regions of the
sky for which polarization PAs of quasars are found to be aligned at optical
wavelengths have been identified.
In the original analysis (\citealt{Hutsemekers1998}), the two most significant
regions were identified by eye and were called A1 and A3.
\citet{Hutsemekers-et-al2005} showed that it is actually from these regions
that comes from most of the alignment signal (see for example
Fig.~\ref{fig:locStat_H05}). In the latter and independent identification
(presented in Chapter~\ref{Ch:PC-1Analysis}), we used an unbiased method
and highlighted the regions N2 and S2 containing significant alignments of the
quasar polarization vectors. While less extended, the latter two regions were
consistently found at similar locations in the 3-dimensional space.
Here, as the sky coverage of the radio surveys and the optical catalogue are
different (see below), we choose to consider the most extended regions
(A1 and A3) to ensure an overlap as big as possible. Furthermore, those
regions have been the subject of various studies in the past. We recall their
boundaries here:
\begin{itemize}
\item A1: $168^\circ \leq \alpha \leq 218^\circ$ ; $\delta \leq 50^\circ$ and $1.0 \leq z \leq 2.3$
\item A3: $320^\circ \leq \alpha \leq 360^\circ$ ; $\delta \leq 50^\circ$ and $0.7 \leq z \leq 1.5$
\end{itemize}
where $\alpha$ and $\delta$ refer to the right ascension and the declination
of the sources, respectively.

\begin{figure}[h]
\centering
\begin{minipage}{0.8\linewidth}
\centering
\includegraphics[width=\columnwidth]{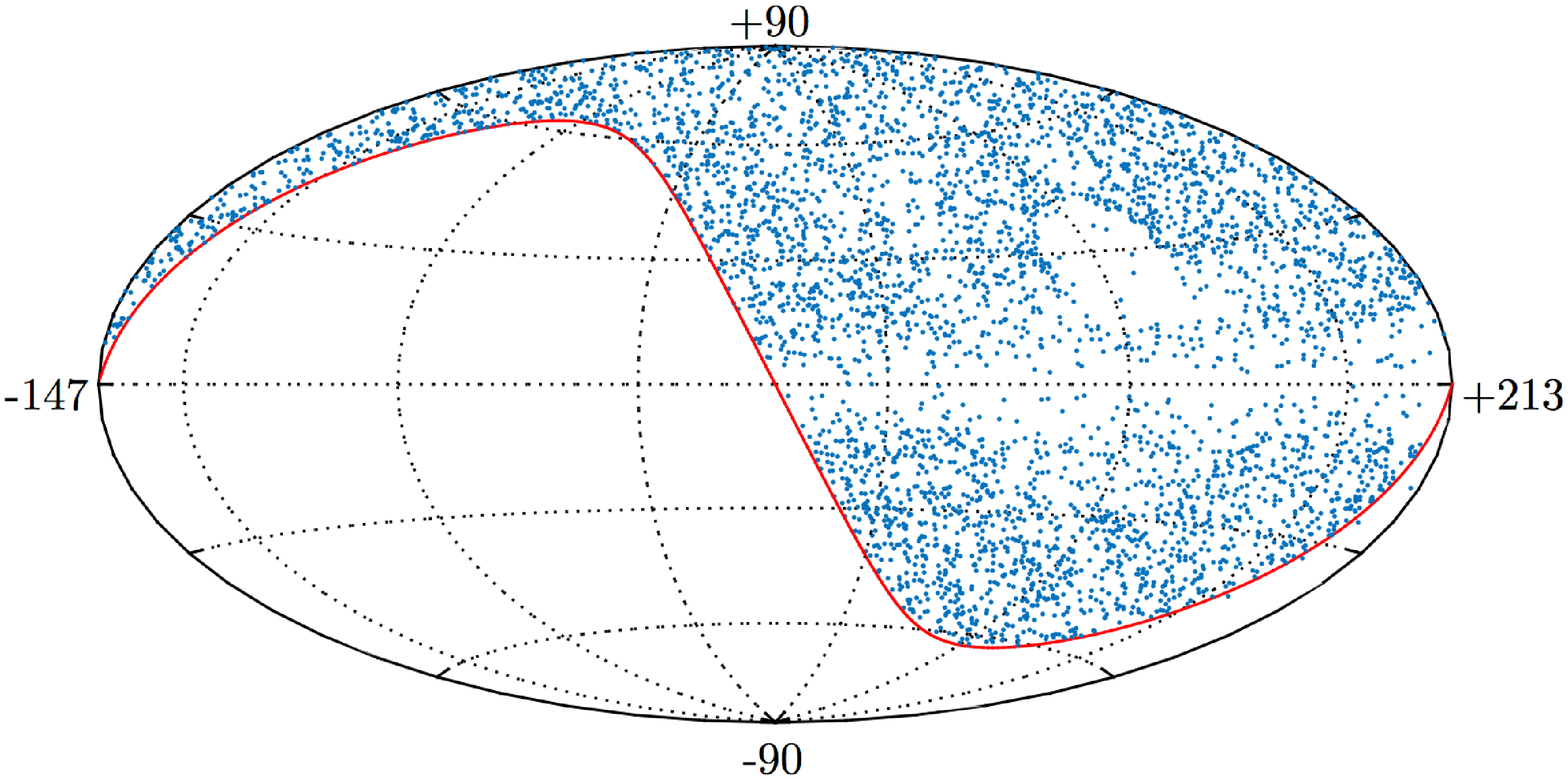}
\caption{\small{Hammer-Aitoff projection of the positions of the 4155
FSRSs in the sample $All$ (blue dots) in Galactic coordinates. The red
line corresponds to the equator of the equatorial coordinates.
The map is centred at Galactic longitude $l_{\rm{gal}} = 33^\circ$ for
comparison of the sky distribution of the optical sample in
Fig.~\ref{fig:locStat_H05}.}}
\label{fig:J4155_GalProj}
\end{minipage}
\end{figure}
\citet{Joshi-et-al2007} addressed the question of uniformity of the polarization
PAs of FSRSs from the JVAS/CLASS 8.4-GHz surveys in these regions and
reported no obvious alignment. However, they did not introduce the cuts in
redshift and thus, only considered the windows towards the A1 and A3 regions,
defined by cuts in right ascension and declination only.
As the redshift of the sources is an important characteristic of the alignments
of optical polarization PAs, we perform a new analysis of these regions.
It is nevertheless important to realize that the sky coverages of the JVAS/CLASS
FSRSs and the studied quasars at optical wavelengths are different. In particular,
the surveys at radio wavelengths do not contain data at $\delta<0^\circ$ and the
optical data set is limited in Galactic latitude as $|b_{\rm{gal}}|\geq 30^\circ$.
The differences of the sky coverages of the two sample can be noted
by comparison of Figs.~\ref{fig:locStat_H05} and~\ref{fig:J4155_GalProj}.

\medskip

To study the polarization PA distributions of the A1 and A3 regions, we
use the Hawley--Peebles test (\citealt{Hawley-Peebles1975} and
\citealt{Godlowski2012}) and the PC test introduced in
Chapter~\ref{Ch-Statistics}. We refer to Sections~\ref{stat_PCTest}
and~\ref{stat_HPTest} for a description of these statistical tests.
Considering samples of sources that are in a small region of the sky,
these tests return the probability that the observed distribution of PAs
is random and define the mean PAs
($\bar{\psi}$ and $\bar{\psi}_{\rm{PC}}$, resp.) which are relevant
only in the case of non-uniformity.

The Hawley--Peebles test analyses PA histograms. The number of bins is a
free parameter. We decide to use 18 bins of $10^\circ$ each, spanning the
range $0^\circ \,-\,180^\circ$. This choice does not maximize the reported
probabilities but is somehow justified by the fact that a bin width of $10^\circ$
corresponds approximately to twice the mean error of the polarization PAs
under study.

The PC test involves spherical caps of equal area. This area is fixed by the
angular aperture of the cap ($2\eta$) which is the only free parameter of the test.
As discussed in the Section~\ref{subsubstat:depEta}, it is useful to investigate
a large range of values for $\eta$. Therefore, we explore here the range
$2^\circ \,-\,90^\circ$ for $\eta$ and we report the probabilities $p_{\rm{min}}$
and $p^\sigma$ corresponding to the value of $\eta$ for which $p^\sigma$ is
the smallest.

Results of the tests applied to the sub-samples extracted from the $All(z)$ and
the $QSO(z)$ samples are shown in Table~\ref{tab:HP-PC_A1A3}.
\begin{table*}
\centering
\begin{minipage}{125mm}
\begin{center}
\small{
\begin{tabular*}{\columnwidth}{@{\extracolsep{\stretch{1}}}*{1}{lcccc}}
\hline
\\ [-1.5ex]
	& \multicolumn{2}{c}{A1 region} 	& \multicolumn{2}{c}{A3 region} 	\\
	& 						$All(z)$	& $QSO(z) $ & $All(z)$	& $QSO(z)$			\\ [0.5ex]
\hline
\\ [-1.5ex]
$N$ 						& $141$		& $139$		& $50$		& $45$				\\
$P_{\rm{HP}}\,(\%)$				& $95.43$	& $93.15$	& $4.19$ 	& $0.96$			\\
$\bar{\psi}\,({}^\circ)$	& $-$		& $-$		& $64$		& $62$ 				\\ [0.5ex]
\hline
\\ [-1.5ex]
$p_{\rm{min}}\,(\%)$ 					& $\geq4$	& $\geq5$	& $0.01$ & $7.2\,10^{-4}$	\\
$\eta\,({}^\circ)$ 			&$20\,-\,90$&$20\,-\,90$& $58$		& $56$				\\
$p^{\sigma}\,(\%)$ [$N_{\rm{sim}}$]	& $\geq13$ [$10^2$] 	& $\geq10$ [$10^2$]
							& $0.07$ [$5\,10^4$] 	& $4.0\,10^{-3}$ [$5\,10^4$]\\
$\bar{\psi}_{\rm{PC}}\,({}^\circ)$	& $-$	& $-$		& $68$ 	& $67$		 	\\ [0.5ex]
\hline
\end{tabular*}}
\caption{\small{Results of the
Hawley--Peebles and the PC tests performed on the sub-samples
corresponding to the A1 and A3 regions of optical polarization
alignments. Sub-samples are obtained from both the $All(z)$ and the
$QSO(z)$ samples. $N$ is the size of the sub-samples, $P_{\rm{HP}}$
is the probability given by the Hawley--Peebles test that the PAs are
drawn from a uniform parent distribution and $\bar{\theta}$ is the mean
polarization PA returned by this method. $p_{\rm{min}}$ is the local
probability obtained with the PC test for the half-aperture angle $\eta$
to which corresponds the minimum global probability $p^{\sigma}$
computed with $N_{\rm{sim}}$ random simulations. The mean angle
$\bar{\theta}_{PC}$ is computed as explained in
Section~\ref{subsubstat:mPAPC_locGroups}.
Probabilities are given in percent.}}
\label{tab:HP-PC_A1A3}
\end{center}
\end{minipage}
\end{table*}
The hypothesis of uniformity of the polarization orientations is rejected
at the level of at least $95\%$ in the A3 region but not in the A1
region\footnote{At first glance, the fact that we found alignment in the A3
region but not in the A1 region could be caused by a bad spatial overlap
between radio and optical data in the A1 region. But the overlapping is
not better in the A3 region.} in both sub-sample $All(z)$ and $QSO(z)$.
The detection of an alignment is very intriguing, especially given the claim
by \citet{Joshi-et-al2007}.
To reconcile the analyses, we extracted the A3 window from the entire
sample of 4155 FSRSs. Applying our tests to this sub-sample of 385
sources, we did not find any evidence for alignment. This test confirms
the negative result of \citet{Joshi-et-al2007} stating that there is not
alignment of the radio polarization vectors inside the A1 and the A3 windows
(see Table~\ref{tab:A3window} and Section~\ref{subsecPH-1:PrelResults}
for further related discussions).

Our analysis of the A1 and A3 regions, however, contradicts the claim of
\citet{Joshi-et-al2007} which states that no alignment is present at radio
wavelengths (8.4 GHz) inside the regions where the optical polarization
vectors are found to be coherently oriented.
This contradiction is likely due to the redshift cuts but could also be due
to the fact that the radio polarizations of QSOs are aligned whereas the
polarization of the other FSRSs are not.

It is therefore of interest to find out whether the alignment tendency
observed in this region is an isolated structure inside the sample of FSRSs
or part of a major trend which was not recognized earlier.
To this end, we shall address the question of uniformity of the polarization
PAs for the complete JVAS/CLASS 8.4-GHz surveys, without restriction on
the sky location, taking redshift into account and considering the subdivision
of the sample into the different species.

\subsection{Full sky coverage}
\label{secPH-1:uniformity_in_JVAS}
To study the uniformity of polarization angle distributions for the sample and
sub-samples of the FSRSs that are sparse and non-uniformly scattered
on the celestial sphere, we shall use the S and Z tests that we introduced in
Chapter~\ref{Ch-Statistics}. As we have seen, the S and Z tests are
appropriate to assess the probability that the distributions of polarization
PAs of local groups are due to statistical fluctuations considering the overall
sample. We use these tests in this section. The intrinsically
coordinate-invariant PC test (Section~\ref{stat_PCTest}) being more useful
for the characterization of correlations is used in
Section~\ref{secPH-1:IdentifRegions}.
We do not use the other statistical tests as they are coordinate-dependent and
that this dependence grows with the angular distance between sources. Hence,
they are thus not adequate to test the uniformity of the PA distribution over large
scales.

\medskip
In Sections~\ref{stat_Stest} and~\ref{stat_Ztest}, we discussed extensively
the coordinate-invariant S and Z tests.
These nearest-neighbour tests compute the probability that the polarization
PAs are uniformly distributed in spatially defined groups of objects, making
use of Monte Carlo simulations.
For each realization, the PAs are reshuffled among the sources of the entire
sample and a statistics is computed for each group of $n_v$ nearest
neighbours. The percentage of Monte Carlo simulations having an average
statistic ($S_D$ or $Z_c$) as extreme as the one of the data defines the
significance level (SL) of the test, i.e., the probability that the observed PA
correlations inside groups can be attributed to statistical fluctuations in the
entire sample.
For the samples of Table~\ref{tab:NED_classification}, we explore a wide
range of values of the parameter $n_v$ (see Section~\ref{substat_nv-param}
for the motivation of doing this) and use a number of random simulations
$N_{\rm{sim}} = 1000$, except contraindication.
We span the range $4 \,-\,400$ (with steps of $2$ and $20$ for ranges
$4\,-\,18$ and $20\,-\,400$, respectively), except for the sample of galaxies
where we stop at $n_v=200$ for obvious reasons.
Note that the lowest value of the SL does not provide an accurate estimate of
the overall significance but gives instead the value of $n_v$ at which the
departure from uniformity is the most significant.

\medskip

We will first consider samples for which reliable redshift measurements are
available. For these samples, both 2- and 3-dimensional analyses are applied,
i.e. defining nearest-neighbour groups on the celestial sphere or in the
3-dimensional comoving space respectively.
We will then turn to the 2-dimensional analysis of the samples of
Table~\ref{tab:NED_classification} which are not constrained by redshift.
For convenience, we give in Table~\ref{tab:Summary_SZtest} a summary of
the results of these two tests applied to all the considered samples.

\subsubsection{Samples with redshift measurements}
\label{SZ_3DAnalysis}
\begin{figure}
\begin{center}
\begin{minipage}{0.8\linewidth}
\centering
\includegraphics[width=\columnwidth]{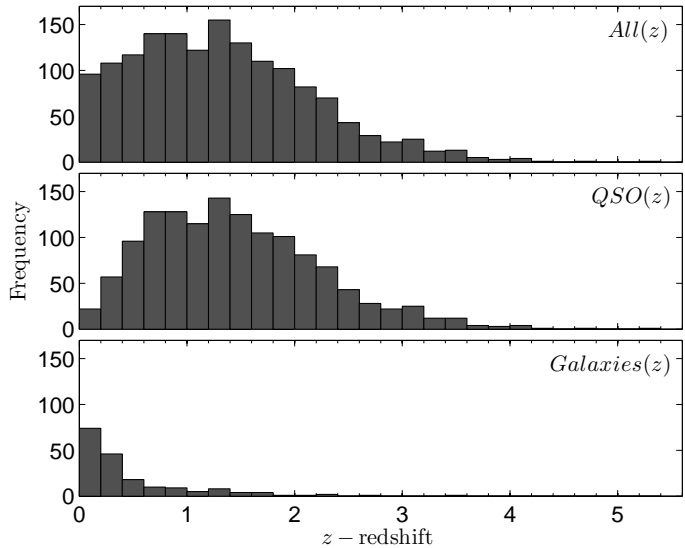}
\caption{\small{Redshift distributions of the sample $All(z)$ and its
sub-samples of QSOs and galaxies.}}
\label{fig:zDistrib}
\end{minipage}
\end{center}
\end{figure}

As already mentioned, the sample of 1531 sources for which reliable redshift
measurement are available is composed at $86\%$ of QSOs. The second
important population of this sample is that of galaxies.
The redshift distributions of these samples are shown in Fig.~\ref{fig:zDistrib}.
Of course, the redshift distributions of the sample $QSO(z)$ and that of
galaxies do not follow the same trend.

\begin{figure}
\begin{center}
\begin{minipage}{0.8\linewidth}
\centering
\begin{tabular}{@{}c}
\includegraphics[width=0.8\columnwidth]{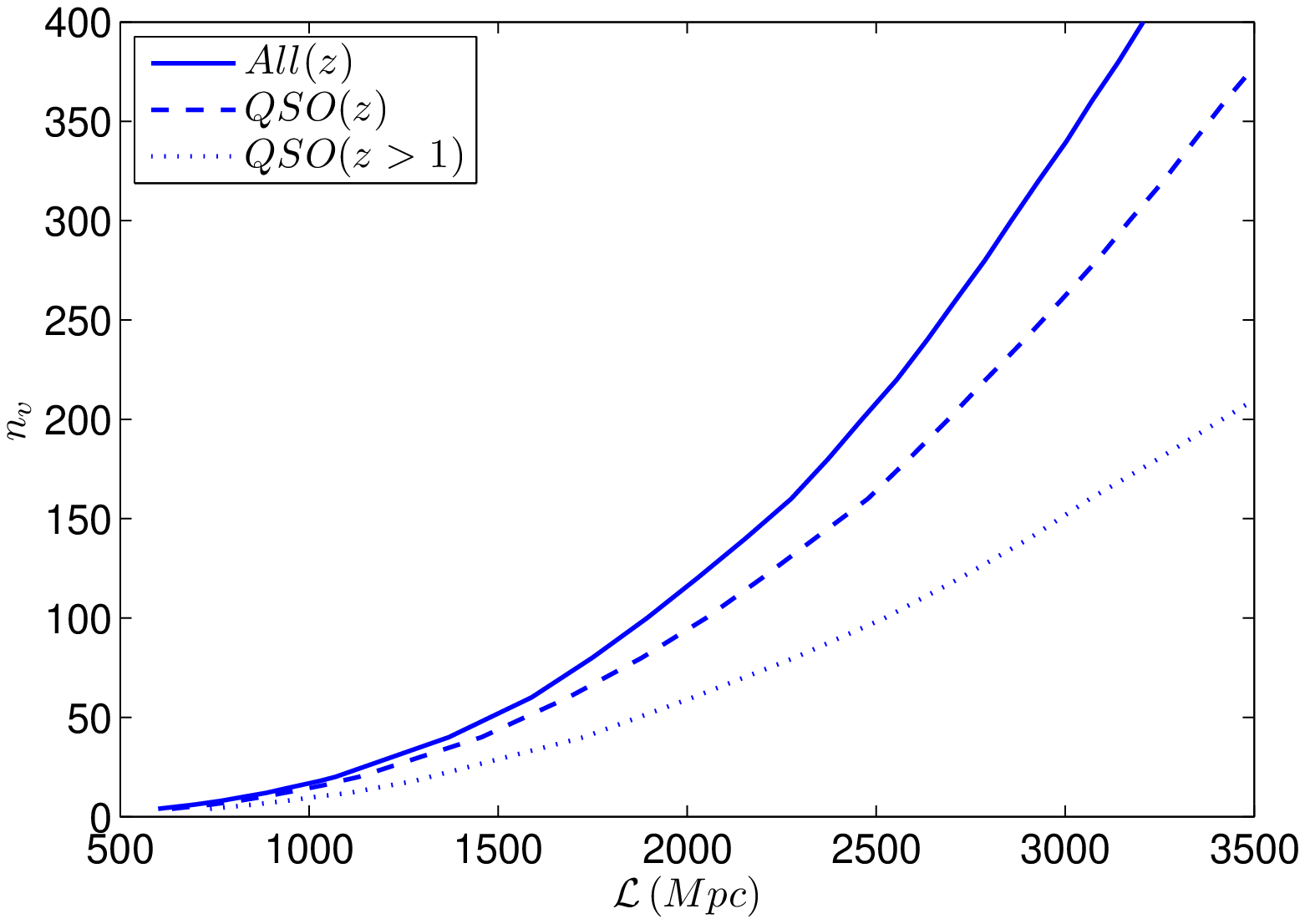}		\\
\includegraphics[width=0.8\columnwidth]{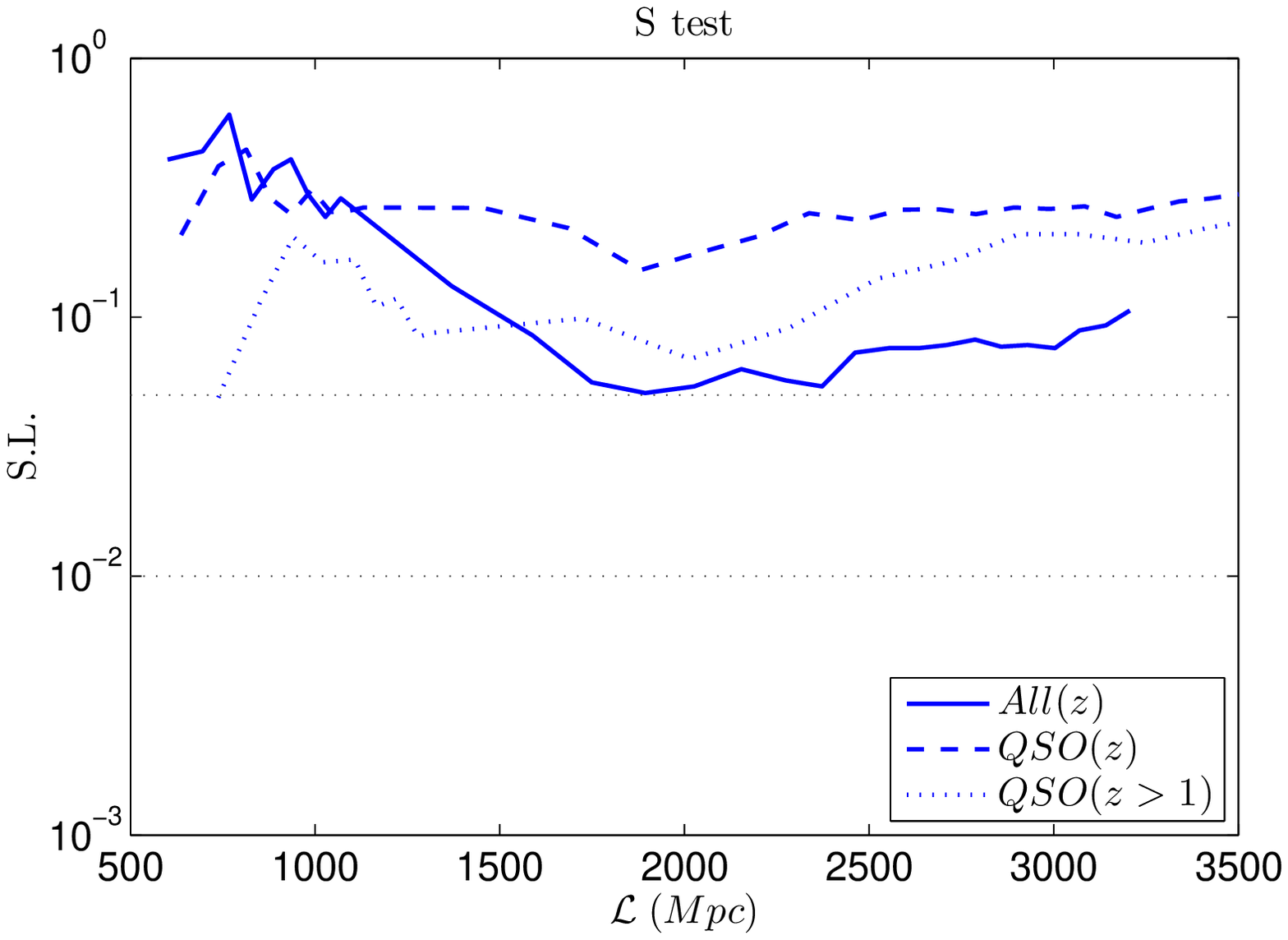}	\\
\includegraphics[width=0.8\columnwidth]{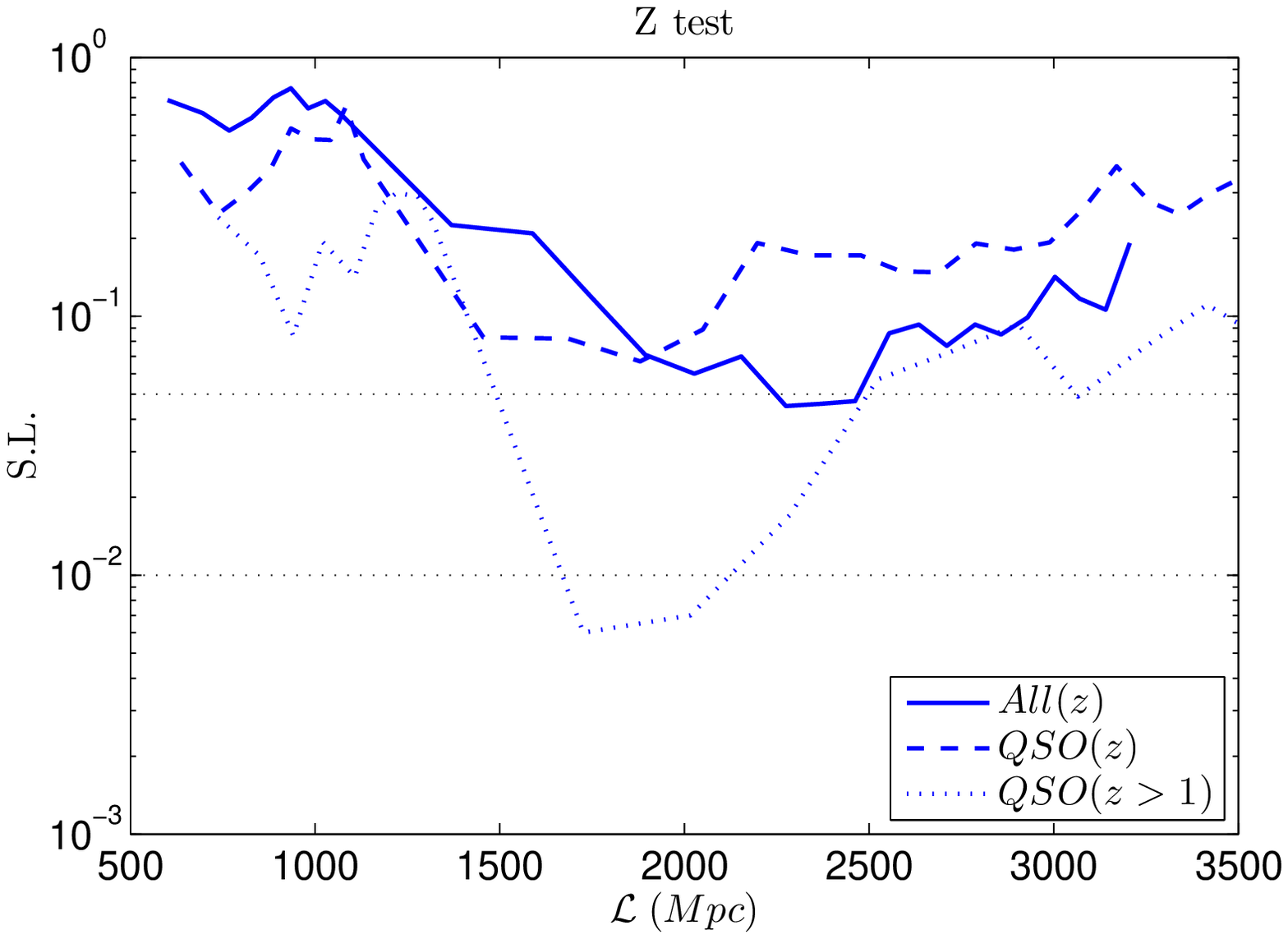}	\\
\end{tabular}
\caption{\small{3D analysis of the samples with redshift measurements.
(\textit{top}): Relation between the parameter $n_v$ and the typical comoving
separation $\mathcal{L}$ in $\rm{Mpc}$ for the samples $All(z)$, $QSO(z)$
and $QSO(z>1)$. (\textit{middle} and \textit{bottom}): Significance level obtained with
the S and Z tests (resp.) as a function of the typical comoving distance
$\mathcal{L}$ for the three samples.
The $5\%$ and $1\%$ SL are indicated with dotted-horizontal lines.}}
\label{fig:3Danalysis-zSample}
\end{minipage}
\end{center}
\end{figure}

\paragraph{3-dimensional analysis.}
We discuss our results in terms of a typical comoving distance $\mathcal{L}$
instead of the free parameter $n_v$ of the statistical tests. This typical scale
is defined as the median of the comoving distances between each central
object and its $n_v$'th nearest neighbour, the median being evaluated over
the full sample under consideration. The line-of-sight comoving distances
(see Section~\ref{substat_nv-param}) are computed assuming a flat
Universe with the cosmological parameters:
$\Omega_{M}=0.31$ and $H_0 = 68 \, \rm{ km \, s}^{-1} \rm{Mpc}^{-1}$,
following \citet{PlanckXVI_2013}.
We show in Fig.~\ref{fig:3Danalysis-zSample} (\textit{top}) the relation between the
parameter value $n_v$ and the typical comoving distance for the samples
we analyse in 3 dimensions.
We applied the statistical tests to the sample of 1531 objects and to the
subcategory of QSO, namely $All(z)$ and $QSO(z)$. We also considered
the high redshift part of the latter, imposing $z>1$. This restricted sub-sample
is populated by 894 sources and is denoted $QSO(z>1)$.

\medskip

Results of the S and Z tests are shown in Fig.~\ref{fig:3Danalysis-zSample}
(\textit{middle} and \textit{bottom}, respectively).
We did not find any significant evidence ($\rm{SL} < 5 \%$) over a wide range
of value of $n_v$ (or $\mathcal{L}$ ) for alignment of the polarization PAs in
the samples $All(z)$ and $QSO(z)$.
However a redshift dependence is possibly detected with the Z test as
suggested in Fig.~\ref{fig:3Danalysis-zSample} (\textit{bottom}).
Indeed, in the high-redshift QSO sample, correlations of polarization PAs
of sources inside groups of typical comoving radius $\sim 2\,\rm{Gpc}$ show
a probability smaller than $1\%$ of being due to statistical fluctuations.

\begin{figure}
\begin{center}
\begin{minipage}{0.8\linewidth}
\centering
\begin{tabular}{@{}c}
\includegraphics[width=.8\columnwidth]{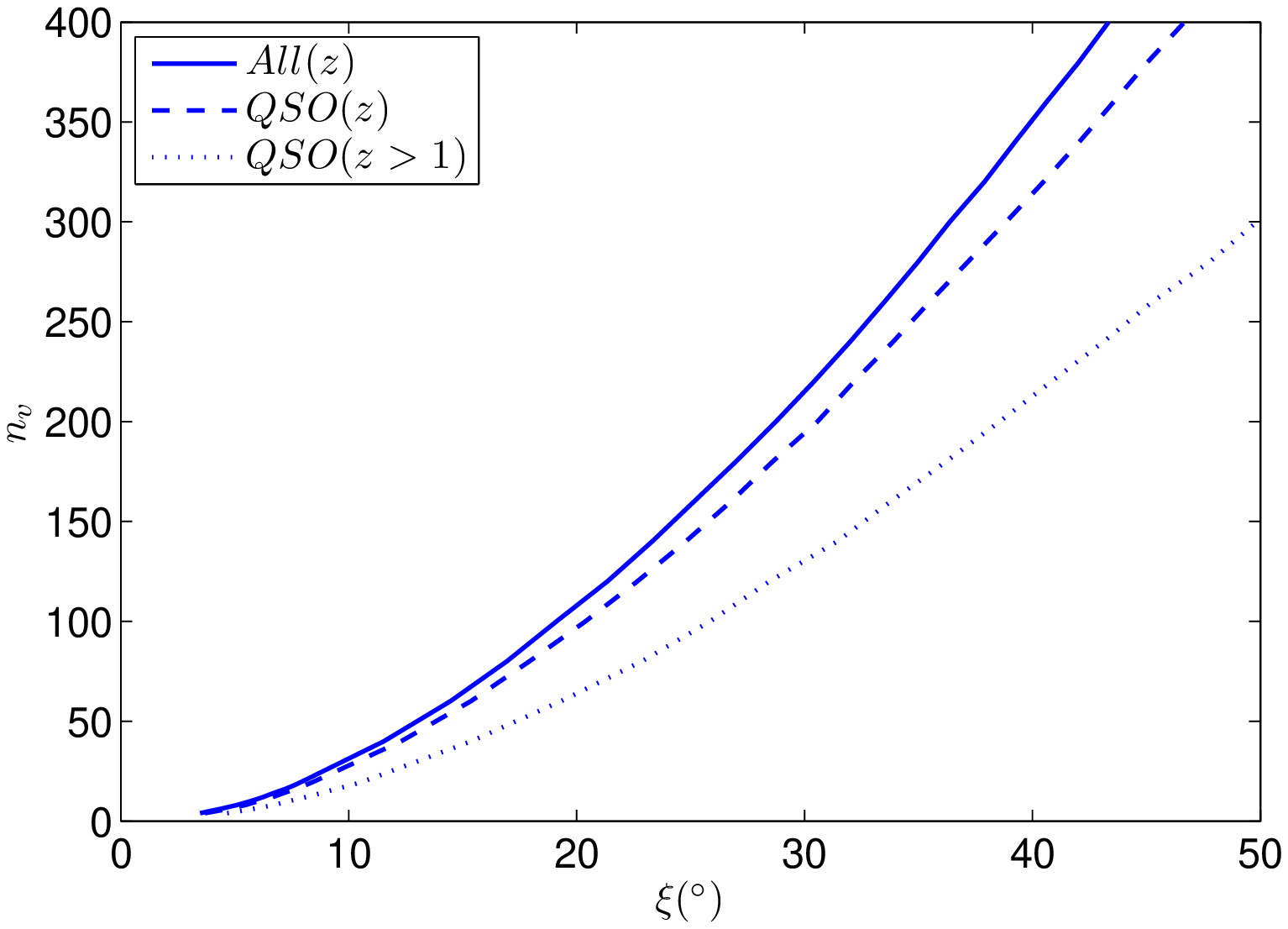}	\\
\includegraphics[width=.8\columnwidth]{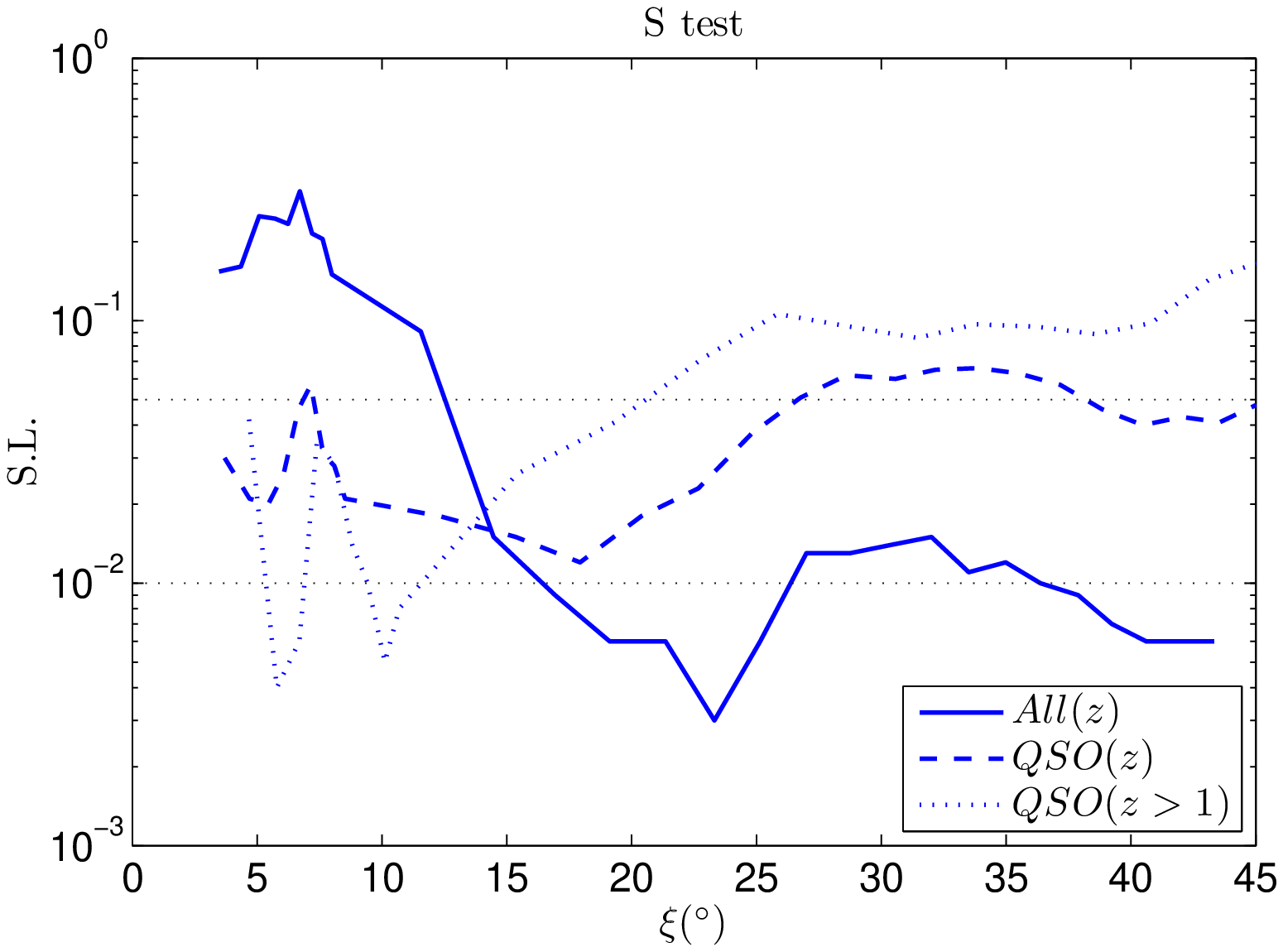}	\\
\includegraphics[width=.8\columnwidth]{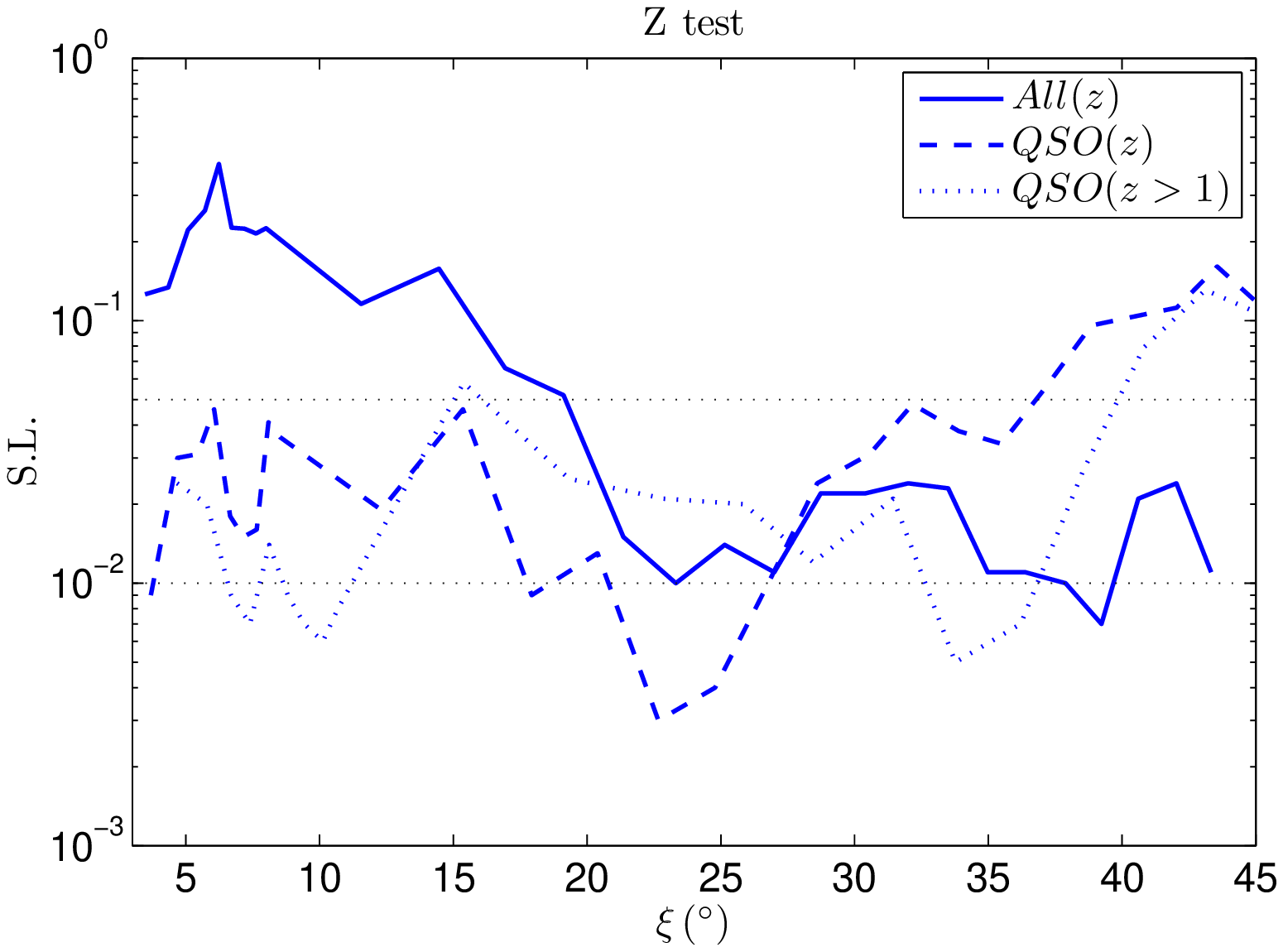}	\\
\end{tabular}
\caption{\small{2D analysis of the samples with redshift measurements.
(\textit{top}): Relation between the parameter $n_v$ and the typical
angular separation $\xi$ in degree for the samples $All(z)$, $QSO(z)$
and $QSO(z>1)$. (\textit{middle} and \textit{bottom}): Significance level
obtained with the S and Z tests (resp.) as a function of the typical angular
separation $\xi$ for the three samples.
The $5\%$ and $1\%$ SL are indicated with dotted-horizontal lines.}}
\label{fig:2Danalysis-zSample}
\end{minipage}
\end{center}
\end{figure}

\paragraph{2-dimensional analysis.}
The radial coordinates of the sources are
fixed to $r=1$ (see Eq.~\ref{eq:xyz-coordinates}) for the 2-dimensional
analysis, even though redshift
measurements are available. We discuss our results in terms of the
typical angular separation $\xi$. The latter is defined as the median of
the angular separation between each object of the sample and its $n_v$'th
nearest neighbour. Fig.~\ref{fig:2Danalysis-zSample} (\textit{top}) shows the
relation between $\xi$ and $n_v$ for the three samples.

We show the dependence of the SL on the typical angular separation
$\xi$ in Fig.~\ref{fig:2Danalysis-zSample}. Significant correlations
($\rm{SL}< 5\%$ over a wide range of $\xi$ value) of the polarization PAs
inside groups is observed for the three samples ($All(z)$, $QSO(z)$ and
$QSO(z>1)$) although the minima occur at different typical angular
separations.
Indeed, for the S test, the sample $All(z)$ shows its minimum SL at
$0.3\,\%$ for $\xi \approx 23^\circ $ and $QSO(z)$ shows a small dip
for the range of $\xi \approx 8^\circ\,-\,26 ^\circ$ with a minimum
$\rm{SL}=1.2\,\%$ for $\xi = 18 ^\circ$. The high-redshift part of the
QSO sub-sample ($QSO(z>1)$) exhibits values of the SL below $1\%$
for smaller angular separation ($\xi \leq 10^\circ$). These features are
confirmed by the Z test as seen from Fig.~\ref{fig:2Danalysis-zSample}
(\textit{bottom}). For this test, the minimum SL value of the sample $QSO(z)$
is found to be as low as $0.3\,\%$ for $\xi \approx 23^\circ$ and the
sample $QSO(z>1)$ shows SL below $1\%$ for $\xi\leq10^\circ$ with an
additional dip around $\xi=34^\circ$.

\subsubsection{Full samples with different object types}
\label{SZ_2DAnalysis}
\begin{figure}
\begin{center}
\begin{minipage}{0.8\linewidth}
\centering
\begin{tabular}{@{}c}
\includegraphics[width=.8\columnwidth]{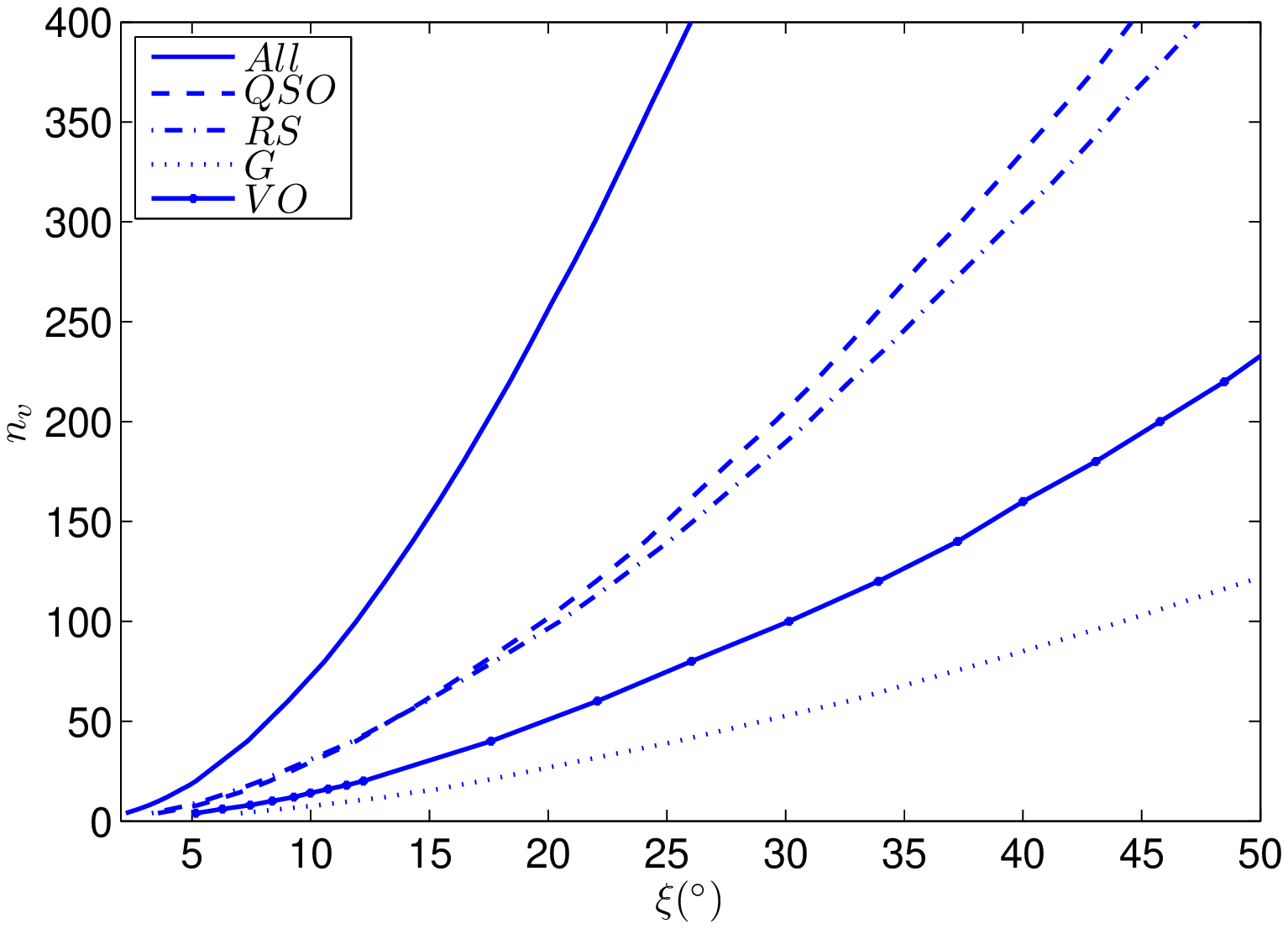}	\\
\includegraphics[width=.8\columnwidth]{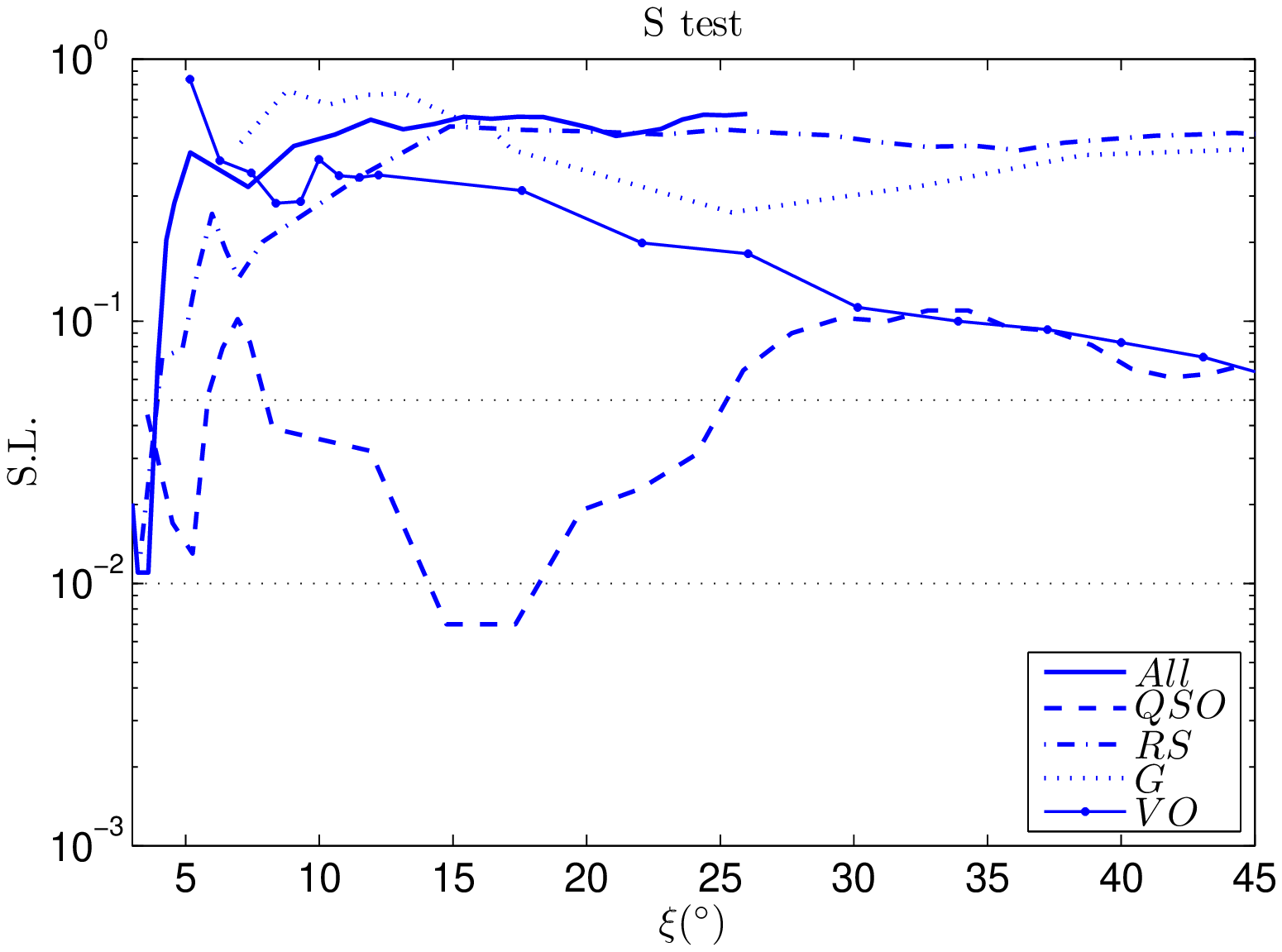}	\\
\includegraphics[width=.8\columnwidth]{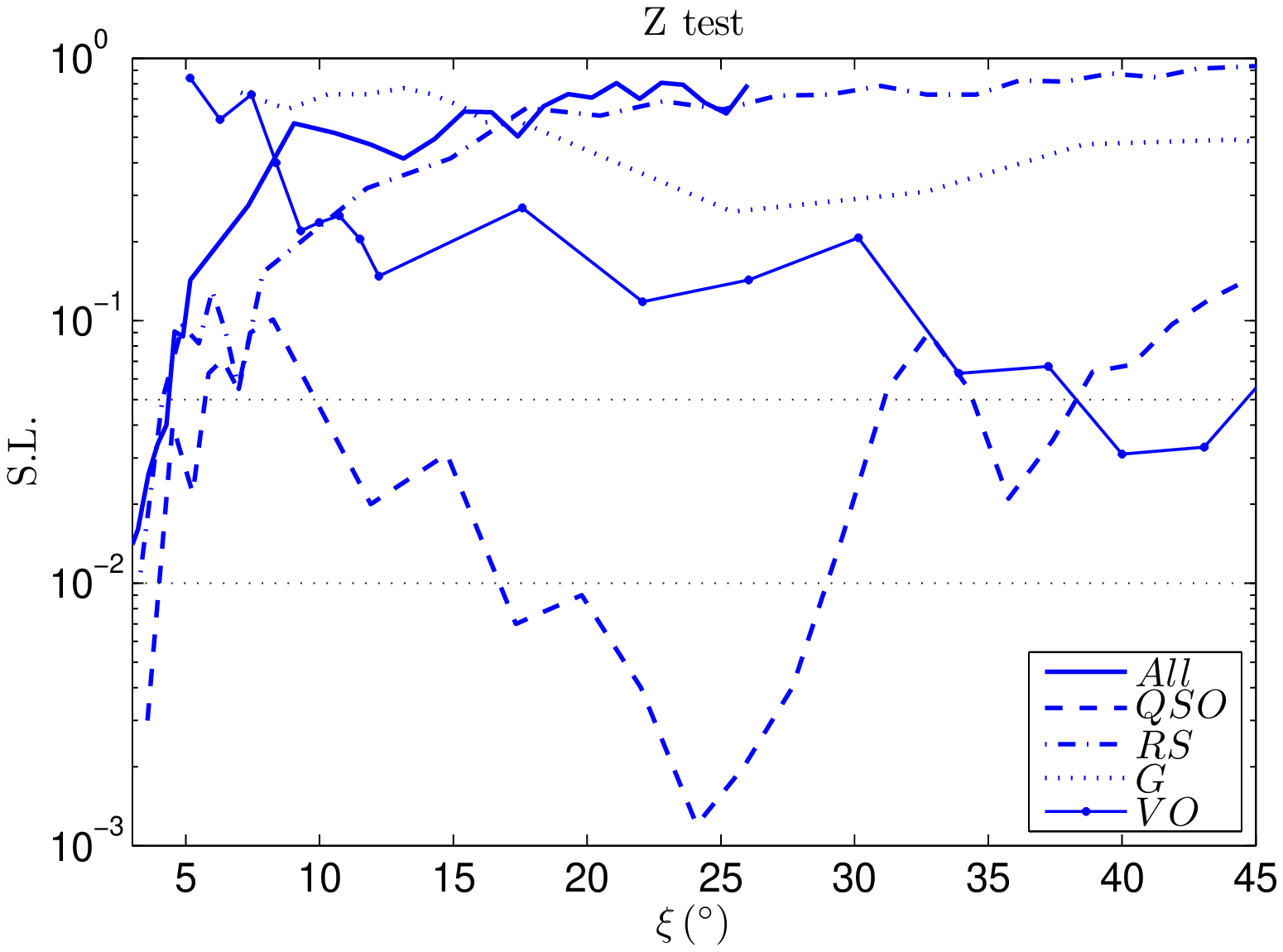}	\\
\end{tabular}
\caption{\small{2D analysis of the full samples. (\textit{top}): Relation between
the parameter $n_v$ and the typical angular separation $\xi$ for the
samples $All$, $QSO$, $RS$, $G$ and $VO$ of
Table~\ref{tab:NED_classification}. (\textit{middle} and \textit{bottom}): Significance
level obtained with the S and Z tests (resp.) as a function of the typical
angular separation $\xi$ for the five samples. Note that the SL value of
the sample $QSO$ for $\xi \approx 24^\circ$ ($n_v=140$) has been
computed with $10^{4}$ random simulations.}}
\label{fig:2Danalysis-fullSample}
\end{minipage}
\end{center}
\end{figure}

Analysing samples with redshift measurements in two dimensions,
we have found significant correlations ($\rm{SL}<5\%$ with minima
$<1\%$). It is therefore interesting to also perform the 2-dimensional
analysis on the other samples of Table~\ref{tab:NED_classification},
i.e. on the samples that are not restricted by the availability of the
redshifts of the sources.
We thus consider the sample of 4155 sources as well as its four
sub-samples with different object types.
We show in Fig.~\ref{fig:2Danalysis-fullSample} (\textit{top}) the relations
between $n_v$ and $\xi$ for these five samples. The results of the
S and Z tests are shown in Fig.~\ref{fig:2Danalysis-fullSample}
(\textit{middle} and \textit{bottom}, respectively).

For small values of $n_v$ (from $6$ to $10$), we found indications
of alignments in the three samples $All$, $RS$ and $QSO$ as the
S and Z tests return SL values at the percent level ($1.1 \%$ and
$1.2 \%$ for the sample $All$).
The indications of alignments in the sample $All$ are reminiscent of
the correlations highlighted by \citet{Tiwari-Jain2013} at the scale of
$\sim 150\,\rm{Mpc}$\footnote{See the footnote 1 of this chapter.}.
The reasons why we found correlations with
lower significance are likely that we consider a different
sample\footnote{When we built the data set from the JVAS/CLASS
8.4--GHz catalogue, we removed duplicate measurements while
\citet{Tiwari-Jain2013} did not (Jain 2015 (private communication)
and \citealt{Tiwari-Jain2015}).}.
We also thought (in \citealt{Pelgrims-Hutsemekers2015}) that the use
of a different definition of the $S_D$ statistics could cause these
differences in SL. However, we have seen in Section~\ref{stat_Stest}
that the two definitions are equivalent.

For large values of $n_v$, alignments are detected with SL below
$5\%$ over a wide range of $\xi$ only for the sample $QSO$.
The SL of the S test applied to $QSO$ exhibits a dip for $n_v$
between $40$ and $140$, reaching the value of $0.7\,\%$ for
$n_v=60$ and $80$. The range of typical angular separations involved
in these correlations is $\xi\approx12^\circ\,-\,24^\circ$, with stronger
correlations for the range $14.5^\circ\,-\,17.5^\circ$.
The Z test exhibits a large dip for the range $n_v = 40\,-\,200$ with the
minimum SL of $ 0.12\,\%$ for $n_v=140$, implying correlations at
$\xi \approx 24^\circ$.
Those correlations of polarization PAs involve QSOs separated by large
distances on the celestial sphere and confirm the detection made in the
sample $QSO(z)$ (with redshift measurements) in the previous sub-section.

For such angular scales, the distributions of the polarization PAs of the
other samples ($All$, $RS$, $G$ and $VO$) are  in good agreement
with the hypothesis of uniformity.
Let us emphasize that these large-scale correlations are not observed
for the category of radio sources ($RS$), even though the sample size
is comparable to that of $QSO$ (see Table~\ref{tab:NED_classification}).

\subsection{Intermediate outcomes}
\label{subsecPH-1:PrelResults}
So far in the chapter, we have studied the distribution of the polarization PAs
of different samples drawn from the JVAS/CLASS 8.4-GHz surveys.
We have found significant
alignments in some of these samples; first, in one of the regions where the
optical polarization vectors were found to be aligned and second, in the
QSO all-sky survey.

Regarding the A3 region, a few reasons might lead to the differences
between our findings and the conclusions of \cite{Joshi-et-al2007}.
As already mentioned, when these authors analysed the so-called A1
and A3 regions, they did not constrain their sample with regard to the
redshift, which is an important characteristic of the optical polarization
alignments. They applied their ``Nearest Neighbour Test'' to the full sample
restricted to the sky window of the A3 region, i.e. introducing cuts in right
ascension and declination only (see Section~3.3 and~5 of
\citealt{Joshi-et-al2007}). The first possible cause of divergent results is a
redshift dependence of the alignments at radio wavelengths as at optical
wavelengths, so that taking all the sources in the window regardless of their
redshift blurs the alignment.
Alternatively, the comparison of the last two columns of
Table~\ref{tab:HP-PC_A1A3} suggests that alignments are more pronounced
for QSOs compared to the other types of objects.

In order to test these two scenarios, we performed an analysis of the samples
obtained by imposing the A3 window cut on the different samples of
Table~\ref{tab:NED_classification}.
Results of the Hawley--Peebles test and the PC test of are shown in
Table~\ref{tab:A3window}.
Correlations between polarization PAs are observed when we consider
the A3 window cut of the samples $QSO$, $QSO(z)$ and $All(z)$, but
no deviation from uniformity is detected for the A3 window cut of $All$,
in agreement with the result of \citet{Joshi-et-al2007}.
Comparison of the last two columns of Table~\ref{tab:A3window} teaches
us that adding the other species to the $QSO$ sample completely blurs
the alignments.
This simple observation argues for the scenario in which the species
selection is at the origin of the detection of the correlations. This scenario
is reinforced when we consider the A3 window cut of the sample $RS$.
For this sub-sample of 138 objects, we found that the distribution of the
polarization PAs is in agreement with the hypothesis of uniformity.
Comparison of Tables~\ref{tab:HP-PC_A1A3} and~\ref{tab:A3window}
does not allow us to conclude on a possible redshift dependence of the
polarization alignments.
This is partially due to the lack of redshift information for non-QSO species.

\begin{table*}[h]
\centering
\begin{minipage}{0.8\linewidth}
\centering
\small{
\begin{tabular*}{\columnwidth}{@{\extracolsep{\stretch{1}}}*{1}{lcccc}}
\hline
\\ [-1.5ex]
							&		\multicolumn{3}{c}{A3 window cut on}		\\
							& $QSO(z)$				& $All(z)$	
							& $QSO$					& $All$		\\ [0.5ex]
\hline
\\ [-1.5ex]
$n$ 						& $100$					& $115$	
							& $114$					& $385$		\\
$P_{\rm{HP}}\,(\%)$				& $0.36$			 	& $1.27$ 
							& $1.19$				& $29.6$	\\
$\bar{\psi}\,({}^\circ)$	& $68$ 					& $72$ 		
							& $68$ 					& $-$		\\ [0.5ex]
\hline
\\ [-1.5ex]
$p_{\rm{min}}\,(\%)$ 					& $7.0\,10^{-4}$		& $1.9\,10^{-3}$ 	
							& $5.9\,10^{-3}$		& $0.14$	\\
$\eta\,({}^\circ)$ 			& $52$					& $52$
					 		& $52$		 			& $68$		\\
$p^{\sigma}\,(\%)$ [$n_{sim}$]	& $0.02$ [$5\,10^4$]	& $0.01$ [$10^4$]
							& $5.4\,10^{-2}$ [$5\,10^4$]	& $1.5$ [$10^3$]	\\
$\bar{\psi}_{\rm{PC}}\,({}^\circ)$	& $69$		& $70$
							& $68$					& $59$		\\ [0.5ex]
\hline
\end{tabular*}}
\caption{\small{Same as Table~\ref{tab:HP-PC_A1A3} but for the
sub-samples obtained by application of the A3 window on the
samples of $1325$ QSOs with redshift ($QSO(z)$), $1531$
sources with redshift ($All(z)$), $1450$ QSOs regardless of the
redshift information ($QSO$), and $4155$ flat-spectrum radio
sources ($All$).}}
\label{tab:A3window}
\end{minipage}
\end{table*}

In the full sample we highlighted alignments involving sources separated
by typical angular scales of about $20^\circ$. For the samples with redshift
measurements, the large-scale correlations are observed to be more
significant in the 2-dimensional analysis than in the 3-dimensional one.
Considering samples that are not limited by the redshift availability, we also
pinpointed that the large-scale correlations mainly concern the category of
QSOs as was already suggested during the study of the A3 window in
Section~\ref{subsecPH-1:VisibleWindows}.

As a conclusion of the analysis of the all-sky (sub-)samples, we find that
the polarization PAs of the JVAS/CLASS 8.4-GHz surveys show correlations
in groups of QSOs with an angular radius of about $20^\circ$.
The significance level at which these correlations can be attributed to
statistical fluctuations in the sample of QSO is found to be as low as
$\sim 0.1\%$ for $\xi \approx 24^\circ$ with the Z test.
\begin{table*}[h]
\centering
\begin{minipage}{137.5mm}
\centering
\footnotesize{
\begin{tabular*}{\columnwidth}{@{\extracolsep{\stretch{1}}}*{1}{lc ccc ccc}}
\hline
\\ [-1.5ex]
	& \multicolumn{3}{c}{S} 	& \multicolumn{3}{c}{Z} 	\\
	3D		&	min(SL) ($\%$)&	$\mathcal{L}$ (Gpc)	&	$n_v$	&	 &
				min(SL) ($\%$)&	$\mathcal{L}$ (Gpc)	&	$n_v$		\\ [0.5ex]
\hline
\\ [-1.5ex]
$All(z)$	&	$-$			&	$-$				&	$-$		&	& 
					$-$			&	$-$					&	$-$			\\
$QSO(z) $ 	&	$-$			&	$-$				&	$-$		&	& 
					$-$			&	$-$					&	$-$			\\
$QSO(z>1)$	&	$-$			&	$-$				&	$-$		&	& 
					$0.6$		&	$\sim 1.7$			&	$40$		\\ [0.5ex]
\hline
\\ [-1.5ex]
	2D		&	min(SL) ($\%$)&$\xi \,({}^\circ)$	&	$n_v$	&	 &
				min(SL) ($\%$)&	$\xi \,({}^\circ)$	&	$n_v$		\\ [0.5ex]
\hline
\\ [-1.5ex]
$All(z)$	&	$0.3$	&	$\sim 23$		&	$140$	&	& 
					$1.0$		&	$\sim 23$			&	$140$		\\
$QSO(z) $ 	&	$1.2$	&	$\sim 18$		&	$80$	&	& 
					$0.3$	&	$\sim 23$			&	$120$		\\
$QSO(z>1)$	&	$0.5$	&	$\sim 10$		&	$18$	&	& 
			$0.6$ / $0.5$&	$10$ / $34$	&	$18$ / $160$		\\ [0.5ex]
\hline
\\ [-1.5ex]
$All$		&	$1.1$	&	$3\,-\,4$		&	$8\,-\,10$	&	& 
					$1.2$	&	$\sim 3$			&	$6\,-\,8$	\\
$QSO$		&	$0.7$	&	$14.5\,-\,17.5$	&	$60\,-\,80$	&	& 
					$0.12\,{}^*$	&	$\sim 24$			&	$140$		\\
$RS$		&	$1.3$	&	$\sim 3$		&	$4$			&	& 
					$1.1$	&	$\sim 3$			&	$4$			\\
$G$			&	$-$			&	$-$				&	$-$			&	 &
					$-$			&	$-$					&	$-$			\\
$VO$		&	$-$			&	$-$				&	$-$			&	 &
					$3.0$ 	&	$\sim 40$			&	$160$		\\ [0.5ex]
\hline
\end{tabular*}}
\caption{\small{Summary of the application of the S and Z statistical
tests to all samples
of Table~\ref{tab:NED_classification}. For each test, we report the value
of the minimum SL with the corresponding $n_v$ parameter and its
attached typical scale ($\mathcal{L}$ or $\xi$ for the 3- or 2-dimensional
analysis, resp.). We only show results when the SL of the sample is found
to be below the threshold of $5\%$ for a wide range of $n_v$. All SL have
been evaluated with 1000 Monte Carlo simulations except the smallest one
(marked by an asterisk) for which we had to use $10\,000$ simulations.}}
\label{tab:Summary_SZtest}
\end{minipage}
\end{table*}

\section{Identification of regions of aligned polarizations}
\label{secPH-1:IdentifRegions}
For the correlations highlighted in the previous section, it would be of
interest to figure out if the alignments detected at typical scales of
$\xi \approx 15^\circ\,-\,25^\circ$ are due to a global trend across the
whole sky coverage of the survey or if they are prominent in some
regions of the sky, as seems to be the case at optical wavelengths.

\medskip

To this end, we proceed to the identification of the groups of sources
with distributions of polarization PAs that show significant departure
from uniformity. The fact that these groups are clustered in space or
not tells us whether the correlations of polarization orientations are
due to well localized objects or to a general trend.
This identification can, inter alia, be achieved with the help of the S
and Z tests. For clarity, we give the details for the S test.
Also, note that we limit our search to the 2-dimensional analysis of the
sample $QSO$ since it revealed the most convincing evidence for
departure from uniformity with a confidence level higher than $99\%$
for the range of $\xi \approx 14.5^\circ \, - \, 17.5^\circ$.

\medskip

In Section~\ref{secPH-1:uniformity_in_JVAS}, local statistics $S_i$
were computed for each nearest-neighbour group and these statistics
have been computed for each simulated data set. We attribute to each
central source $i$ the quantity $s_i$ which tells us how much the
corresponding group of nearest neighbours contributes to the global
statistics $S_D$. This quantity is defined as
$s_i = \left( {\left\langle S_i \right\rangle - S_i^\star} \right)/{2\sigma_i}$,
where $S_i^\star$ is the statistics obtained for the observed data set
(see Eqs.~\ref{eq:d_theta} and~\ref{eq:S_D}) and where
$\left\langle S_i \right\rangle$ and $\sigma_i$ are the mean and the
standard deviation of this statistics evaluated over the whole set of
simulations assuming that the local statistics are normally distributed.
The larger the value of $s_i$, the more the group contributes to $S_D$.
If the local statistics were normally distributed, then $s_i$ would be the
number of sigma with which the observations differ from randomness,
divided by 2. However, the local statistics are not normally distributed
and $s_i$ is just an empiric measure of the significance of the local
alignment. A group of nearest-neighbour objects is considered as
contributing significantly to $S_D$ if $s_i \geq s_c$, for an arbitrary
threshold $s_c$.

\medskip

Of course we shall search for the identification of the most significant
groups, i.e. consider the $s_i$ quantities computed with the parameter
$n_v$ chosen such that $S_D$ is the smallest
(see Table~\ref{tab:Summary_SZtest}).
To visualize the sky location of the most significant groups we produce
maps which highlight their central sources. Note that these maps do not
critically depend on the choice of $n_v$. These maps are equal-area
Schmidt projection (e.g., \citealt{Fisher-Lewis-Embleton1987}) of the
northern hemisphere (in equatorial coordinates). This choice is suitable
for the considered data set as it covers only positive declinations.
We also plot (in grey bold lines) the limits of the A1 and A3 windows
defined in Section~\ref{subsecPH-1:VisibleWindows}.
Let us insist on the fact that only the northern part of these limits are
shown: the A1 and A3 windows defined from the analysis at optical
wavelengths extend to the South equatorial hemisphere which is not
displayed here.

\medskip

\begin{figure}
\begin{center}
\begin{minipage}{0.8\linewidth}
\centering
\includegraphics[width=\columnwidth]{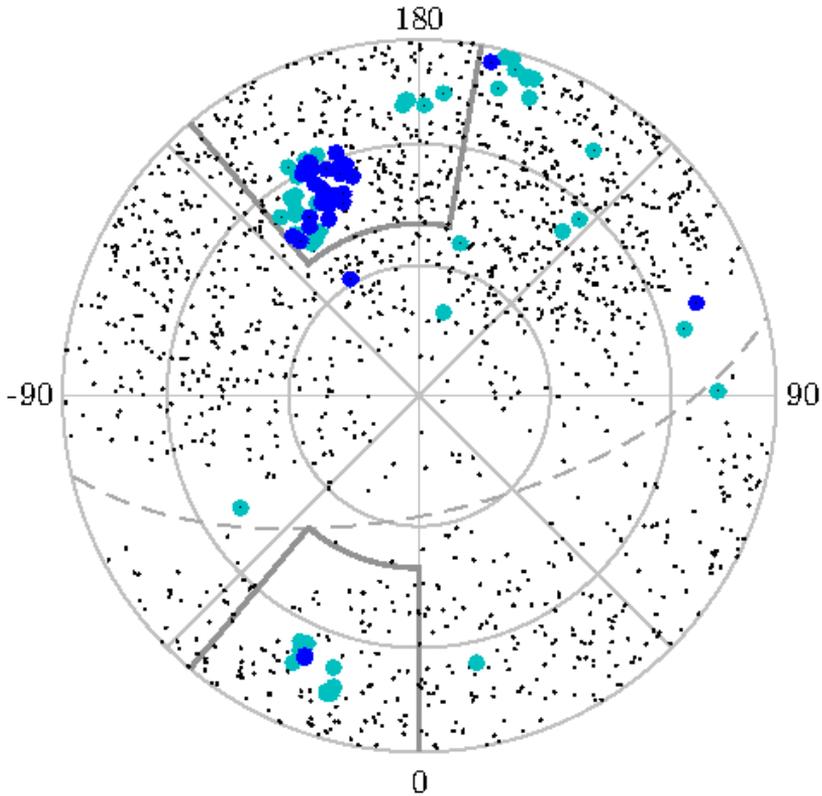}
\caption{\small{Identification map for the sample $QSO$ using the S
test in 2D. The parameter is fixed to $n_v=80$
(see Fig.~\ref{fig:2Danalysis-fullSample} (\textit{middle})) and we
adopt two threshold values: $s_c=2.5$ (lighter points) and $s_c=3.0$
(darker points),
respectively cyan and light blue. Identification maps are equal area
Schmidt projection of equatorial coordinates. Only the equatorial north
hemisphere is displayed with the north pole at the centre of the map.
Grey circles are parallels of declinations $0^\circ$, $30^\circ$ and
$60^\circ$ and grey diagonals are meridians of right ascensions being
multiple of $45^\circ$. The curved dashed line is the Galactic equator,
the North and the South Galactic caps being respectively above and
below the line. Grey bold lines are northern boundaries of the A1 and
A3 regions of optical polarization alignments (see text). Small black
dots are the locations of the 1450 sources of the sample. Highlighted
sources are objects for which corresponding neighbours show a
polarization PA distribution that is unlikely due to chance.}}
\label{fig:S2D1450_identif}
\end{minipage}
\end{center}
\end{figure}

\begin{figure}[h]
\begin{minipage}[t]{0.66\textwidth}
\mbox{}\\[-\baselineskip]
\centering
\includegraphics[width=.9\columnwidth]{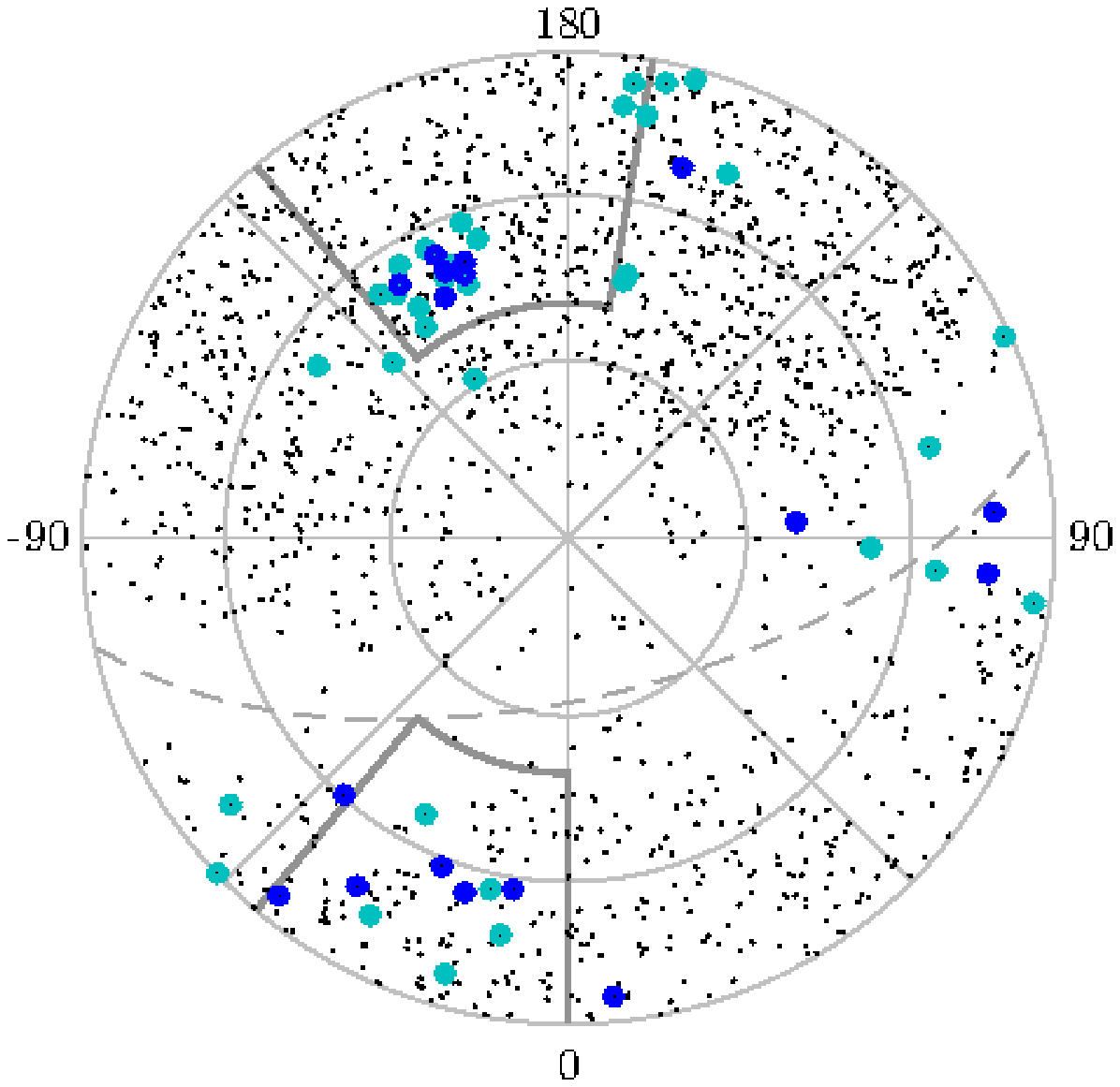}
\end{minipage}
\begin{minipage}[t]{0.3\textwidth}
\mbox{}\\[-1.5\baselineskip]
\caption{\small{Identification map for the sample $QSO$ using the
Z test in 2D. The parameter is fixed to $n_v=140$ (see
Fig.~\ref{fig:2Danalysis-fullSample} (\textit{bottom})) and we adopt two
threshold values: $s_c=1.65$ (lighter points) and $s_c=1.75$ (darker
points), respectively cyan and light blue. Please note that the
thresholds $s_c$ for the S and Z tests do not refer to the same
quantities and have thus different values (compare Eqs.~\ref{eq:d_iMAX}
and~\ref{eq:Z_i} for instance).}}
\label{fig:Z2D1450_identif}
\end{minipage}
\end{figure}

The identification map corresponding to the S test (in 2D) for the sample
$QSO$ with the parameters $n_v=80$ and $s_c=2.5$ is shown in
Fig.~\ref{fig:S2D1450_identif}. As one can see, the highlighted central
sources cluster in three or four groups along with other more sparse
and/or isolated locations.
When pushing $s_c$ up to $3$ (darker points), only the cluster with
right ascension $\alpha \sim 206^\circ$ and declination $\delta \sim 38^\circ$
remains. Following this analysis, it is likely that the significant departure
from uniformity in this sample is due to polarization alignments in a few
groups of QSOs. It is intriguing that two of them are found in the A1 and
A3 windows.
In order to put the latter identification of aligned regions on stronger
grounds, we may use other tests. The Z test also reveals significant
non-uniformity. For the $QSO$ sample and the parameter value
$n_v=140$, it leads to the map shown in Fig.~\ref{fig:Z2D1450_identif}
which is in relatively good agreement with Fig.~\ref{fig:S2D1450_identif}
although it shows more scattered clusters.
\begin{figure}[h]
\begin{minipage}[t]{0.66\textwidth}
\mbox{}\\[-\baselineskip]
\centering
\includegraphics[width=.9\columnwidth]{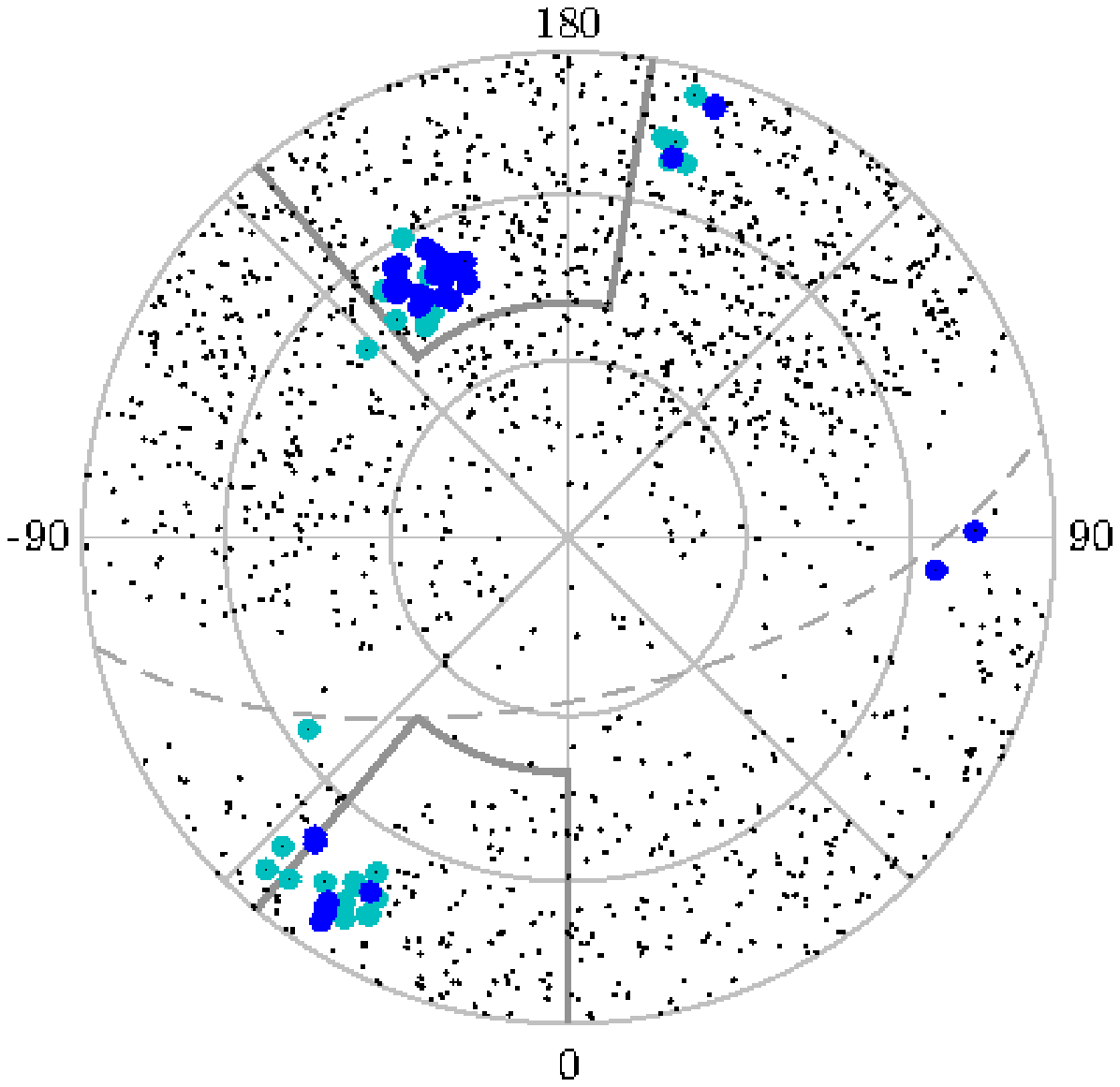}
\end{minipage}
\begin{minipage}[t]{0.3\textwidth}
\mbox{}\\[-1.5\baselineskip]
\caption{\small{Identification map for the sample $QSO$ with the PC test.
Parameters are fixed as $\Omega=20^\circ$ and $\eta=40^\circ$ for the
North Galactic Cap and $\Omega=30^\circ$ and $\eta=50^\circ$ for the
South Galactic Cap. Highlighted sources correspond to groups showing
$p_{\rm{min}} \leq 10^{-3}$ (lighter points) and $p_{\rm{min}} \leq 3\,10^{-4}$
(darker points), respectively cyan and light blue.}}
\label{fig:PC-1450_identif}
\end{minipage}
\end{figure}

\medskip

Also, we find relevant to proceed to a complementary identification using
the PC test. As this test does not depend explicitly on the number of nearest
neighbours (while it is encapsulated within the statistics), local groups can
be defined by a physical angular scale, denoted $\Omega$.
In order to carry out an identification as close as possible to those produced
with the S and the Z tests, we found necessary to split each sample in its
two Galactic hemispheres to determine the physical scale at which local
groups have to be defined. Indeed, the density of the data points in the
North Galactic Cap and the South Galactic Cap are different.
In the sample $QSO$, the typical angular separation corresponding to
$n_v=80$ is $\xi \approx 17^\circ$ and corresponding to $n_v=140$ is
$\xi \approx 24^\circ$ (cf. Fig.~\ref{fig:2Danalysis-fullSample} (Top)).
However, by splitting the sample in its northern and southern Galactic parts,
we obtain $\xi_N \approx 16^\circ$ and $\xi_S \approx 23^\circ$ for
$n_v=80$ and $\xi_N \approx 21^\circ$ and $\xi_S \approx 33^\circ$ for
$n_v=140$, respectively. Given these values, we decided to define local
groups in 2 dimensions with angular scales $\Omega_N = 20^\circ$ for the
North and $\Omega_S = 30^\circ$ for the South.

As we search for the characterization of the polarization PA distribution of
each group taken as a whole, we shall not investigate values of $\eta$ (the
free parameter of the method) below the angular separation of the group,
i.e. below the imposed angular scales.
We arbitrarily chose $\eta=40^\circ$ and $\eta=50^\circ$ for the North and
the South, respectively. The identification map computed with these
parameters is shown in Fig.~\ref{fig:PC-1450_identif}.
We checked the robustness of the map with other pairs of values such as
$(\Omega_N,\,\Omega_S) = (15^\circ,\,25^\circ)$ and $(25^\circ,\,35^\circ)$.
We also checked the stability of our results using other values of $\eta$.
Note that we did not search for the optimal value of $\eta$, i.e. the one
which would give the lowest probabilities, as the method undergoes edge
effects. We rather spanned the range of $20^\circ$ to $60^\circ$ with step
of $5^\circ$ and found consistent maps.

\bigskip

Although a close examination shows discrepancies in the precise locations
of central sources of neighbouring groups, a comparison of the maps
presented in Figs.~\ref{fig:S2D1450_identif}{,}~\ref{fig:Z2D1450_identif}
and~\ref{fig:PC-1450_identif} shows a relatively good agreement, especially
for the cluster at $\left( \alpha,\,\delta\right) \sim \left(206^\circ,\,38^\circ \right)$.

\medskip

In order to define more precisely the limits of regions of polarization
alignment, we proceed as follows. To each central source corresponds a
group of nearest objects (defined via the parameter $n_v$ or $\Omega$).
A highlighted central source is said to form a cluster along with (an)other
highlighted source(s) if it belongs to the group of nearest objects of the
latter. A central source is discarded from a cluster if it is not in the
neighbourhood of a sufficient percentage of central sources forming this
cluster (e.g. $\sim 60\%$).
Reproducing this test for all highlighted objects, we end up with
identification of independent clusters. We finally add to the cluster the
nearest neighbouring objects of each central sources, paying attention
to duplication. Although this procedure is rudimentary, it is sufficient for
our goal.
We thus end up with three regions for each of the three tests. We decide
to define our final regions as the intersection of the regions from the
different tests. We report the final regions of alignments in
Table~\ref{tab:1450_identif-2D} which also gives some of their characteristics
and the result of the application of the Hawley--Peebles test on their
polarization PA distributions.

\medskip

As a result, we identified three well-defined regions of the sky in which
QSOs show coherently oriented polarization vectors.
Two of these regions are located in the North Galactic hemisphere of
the sky and one towards the South. Considering the southern cap, it is
worth remarking that more than $85\%$ of the sources of the sub-sample
identified here belong to the A3 window defined from the region of
optical alignment discovered by Hutsem{\'e}kers et al.
(\citeyear{Hutsemekers1998}; \citeyear{Hutsemekers-Lamy2001};
\citeyear{Hutsemekers-et-al2005}).
To the North, the identified regions are located at the edges of the A1
window of optical alignment, one outside at low declination and the
other inside at high declination. We call them RN1 and RN2, respectively.
It is again worth mentioning that more than $70\%$ of the sources of the
RN2 sub-sample identified here belong to the A1 window.
It is remarkable that our region RN2 coincides with the main aligned
cluster resulting from the independent analysis of \citet{Shurtleff2014}.
Consistently with our previous results, we report a stronger alignment
than he did as we only consider the species of QSO.

\medskip

\begin{figure}
\begin{center}
\begin{minipage}{\textwidth}
\centering
\begin{tabular}{@{}cc}
\includegraphics[width=0.5\columnwidth]{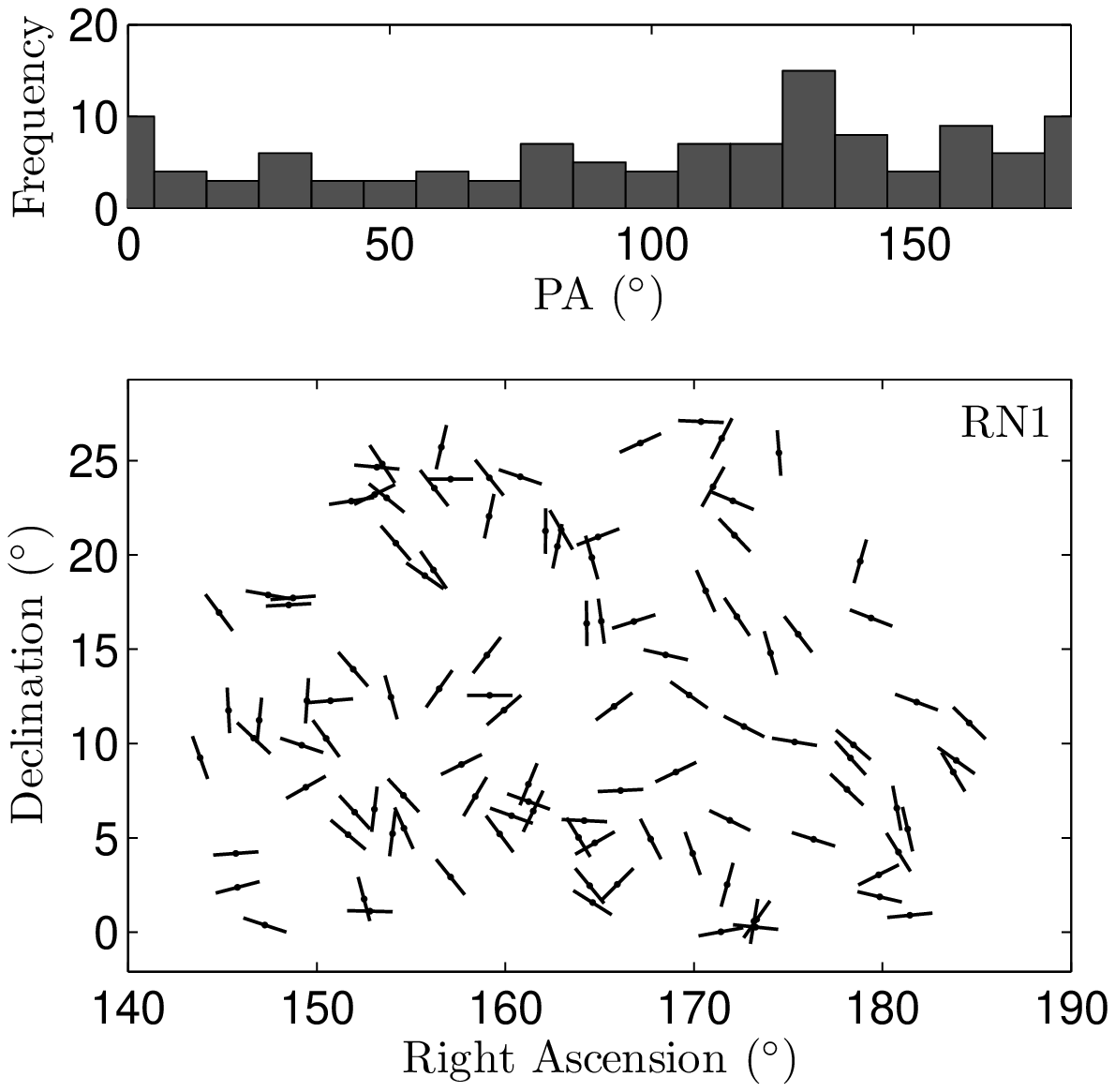} &
\includegraphics[width=0.5\columnwidth]{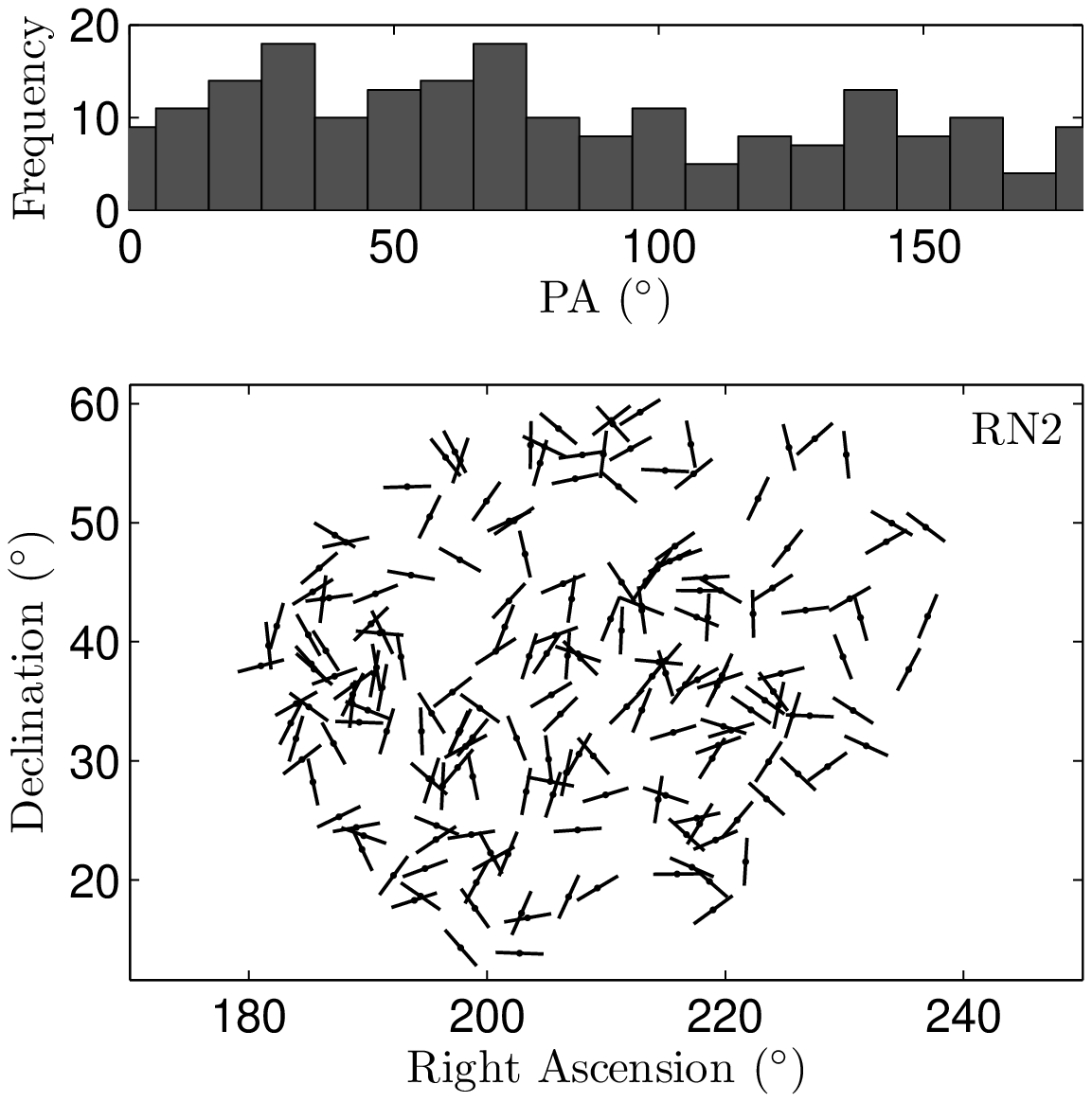} \\
[1.5ex]
\multicolumn{2}{c}{
\includegraphics[width=0.5\columnwidth]{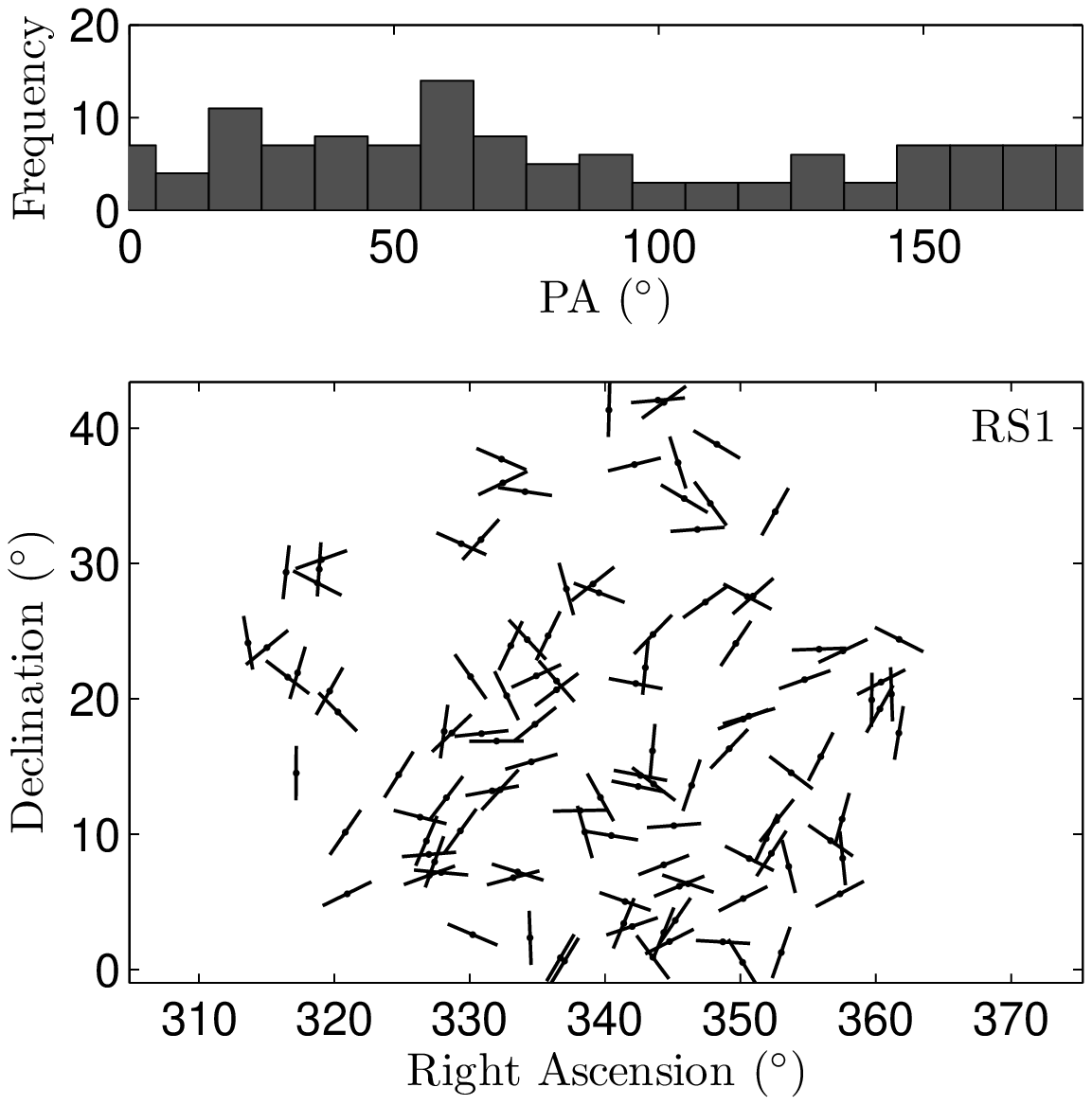}} \\
\end{tabular}
\caption{\small{Maps of polarization vectors in the identified regions
along with their corresponding polarization PA histograms. Polarization
vectors are normalized to the same length in each map. (\textit{top-left} and
\textit{top-right}): the two regions in the North Galactic hemisphere at low
and high (equatorial) declination, respectively. (\textit{bottom}): the region
identified in the South Galactic hemisphere. Properties of these regions
are given in Table~\ref{tab:1450_identif-2D}.}}
\label{fig:IdentifRegionsMap}
\end{minipage}
\end{center}
\end{figure}

In order to visualize the alignment patterns, we show in
Fig.~\ref{fig:IdentifRegionsMap} the equatorial-coordinate maps of the
normalized polarization vectors of the identified regions along with their
corresponding polarization PA histogram. Some structures can be spotted
out by eye. This is better seen in the region RN1 (see
Fig.~\ref{fig:IdentifRegionsMap} (Top)). The statistical tests used
throughout this analysis do not allow us to search and characterized better
such structures. This task is far beyond the scope of this chapter and would
request dedicated algorithms to compute the likelihood of structures of
aligned polarization vectors in a random sample.
\begin{table}[h]
\centering
\begin{minipage}{0.8\linewidth}
\centering
\small{
\begin{tabular*}{\columnwidth}{@{\extracolsep{\stretch{1}}}*{1}{lrrrrrr}}
\hline
\\ [-1.5ex]
& $N$	&	$(\alpha,\,\delta)_{\rm{CM}}\,({}^\circ)$	&	$\bar{\xi}\,({}^\circ)$	&
$\xi_{\rm{max}}\,({}^\circ)$	& $P_{\rm{HP}}\,(\%)$	&	$\bar{\psi}\,({}^\circ)$	\\ [0.5ex]
\hline
\\ [-1.5ex]
RN1 &	$108$		&	$(163,\,12)$ 	&	$12$ 	& $21$	&	$0.45$	&	$131$	\\
RN2 &	$191$		&	$(206,\,38)$	&	$14$ 	& $25$	&	$1.17$	&	$42$ 	\\
RS1 &	$116$		&	$(340,\,18)$	&	$15$ 	& $25.2$&	$1.45$	&	$57$	\\ [0.5ex]
\hline
\end{tabular*}}
\caption{\small{Identified regions from the 2-dimensional analysis of the
sample $QSO$ with the S, Z and PC tests. Regions are intersections of
those given by each test (see text). The two first lines are for the regions
located in the North Galactic cap and the third is for the region of the
South part. They are named RN1, RN2 and RS1, respectively.
$N$ is the number of members belonging to the region,
$(\alpha,\,\delta)_{\rm{CM}}$ refers to the position of the normalized
vectorial sum of the sky location of the sources of the region, $\bar{\xi}$ and
$\xi_{\rm{max}}$ are the mean and the maximum value of the angular
separations of sources to $(\alpha,\,\delta)_{\rm{CM}}$.
$P_{\rm{HP}}$ and $\bar{\psi}$ are the results of the Hawley--Peebles test.}}
\label{tab:1450_identif-2D}
\end{minipage}
\end{table}

\section{Interpreting the results}
\label{secPH-1:DiscussionFinal}
So far we have found that the polarization vectors of QSOs which are
in groups that have angular radii of about $20^\circ$ have correlated
orientations. We showed that these groups cluster in three independent
regions of the sky and that to each of these corresponds a different
preferred polarization PA.

\subsection{Are the data contaminated?}
The preferred angles for the two northern regions are found to have
values close to $45^\circ$ and $135^\circ$. These values, are very
particular (see \citealt*{Battye-Browne-Jackson2008}) and lead us to
consider the possibility that the correlations we found are due to biases
in the data set; despite our careful selection of polarized sources in
Section~\ref{secPH-1:DataSample}, following the prescriptions of
\citet{Jackson-et-al2007}. This hypothesis is a priori difficult to reconcile
with the local character of the alignment features but could potentially
explain that they are better detected with the 2-dimensional analysis
than with the 3-dimensional one.
In this sense, and contrarily to what \citet{Jackson-et-al2007} and
\citet{Joshi-et-al2007} claimed, we also find evidence for a global
non-uniformity inside the polarization data set. Using the Hawley--Peebles
test, the probability that the distribution of the 4155 objects is uniform is
found to be $P_{\rm{HP}}= 2.7\%$ (with $\bar{\psi}\sim 51^\circ$).
This non-uniformity of the overall polarization distribution of the sample
$All$ argues for the hypothesis of a biased data set.
However, consistently with our previous results, this non-uniformity is
found to come from the sub-category of QSO as we find
$P_{\rm{HP}}= 1.1\%$ (with $\bar{\psi}\sim 57^\circ$) for this sample
and that removing the QSOs from the sample $All$ leads to
$P_{\rm{HP}}= 44.5\%$\footnote{The polarization PA distribution of the
sample $RS$ is also in good agreement with uniformity
($P_{\rm{HP}}= 79.1\%$).}.
This result together with the previous evidence for alignment of QSOs and
not for the other species is awkward to reconcile with an observational bias,
as there is no reason for a contamination of the polarization data for the
species of QSO and not for the others, as we shall see.

\medskip

\begin{figure}[h]
\begin{minipage}[t]{0.6\textwidth}
\mbox{}\\[-\baselineskip]
\centering
\includegraphics[width=\columnwidth]{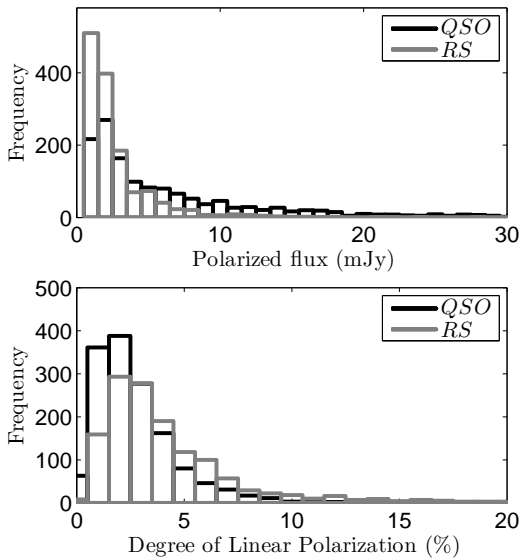}
\end{minipage}\hspace{0.03\textwidth}
\begin{minipage}[t]{0.26\textwidth}
\mbox{}\\[-\baselineskip]
\\[-2.5ex]
\caption{\small{Polarized flux (\textit{top}) and degree of linear polarization
(\textit{bottom}) of the samples $QSO$ and $RS$. A two-sample
Kolmogorov--Smirnov test reveals that the polarized flux, as well as the
degree of linear polarization, of the two samples have a probability much
below $1\%$ to be drawn from the same underlying parent distribution.}}
\label{fig:QSO-vs-RS-lightPropeties}
\end{minipage}
\end{figure}

Comparing the properties of the samples $QSO$ and $RS$ (which have a
comparable number of objects), we note some differences. As illustrated by
the Fig.~\ref{fig:QSO-vs-RS-lightPropeties}, their polarization characteristics
at radio wavelengths do not follow the same parent distribution. However, we
do not find obvious reasons why the QSOs would be more affected by
observational biases than the sample $RS$ as the QSO sample shows higher
total and polarized flux.

\begin{figure}[t]
\begin{center}
\begin{minipage}{.6\textwidth}
\mbox{}
\centering
\\[1.ex]
\includegraphics[width=\columnwidth]{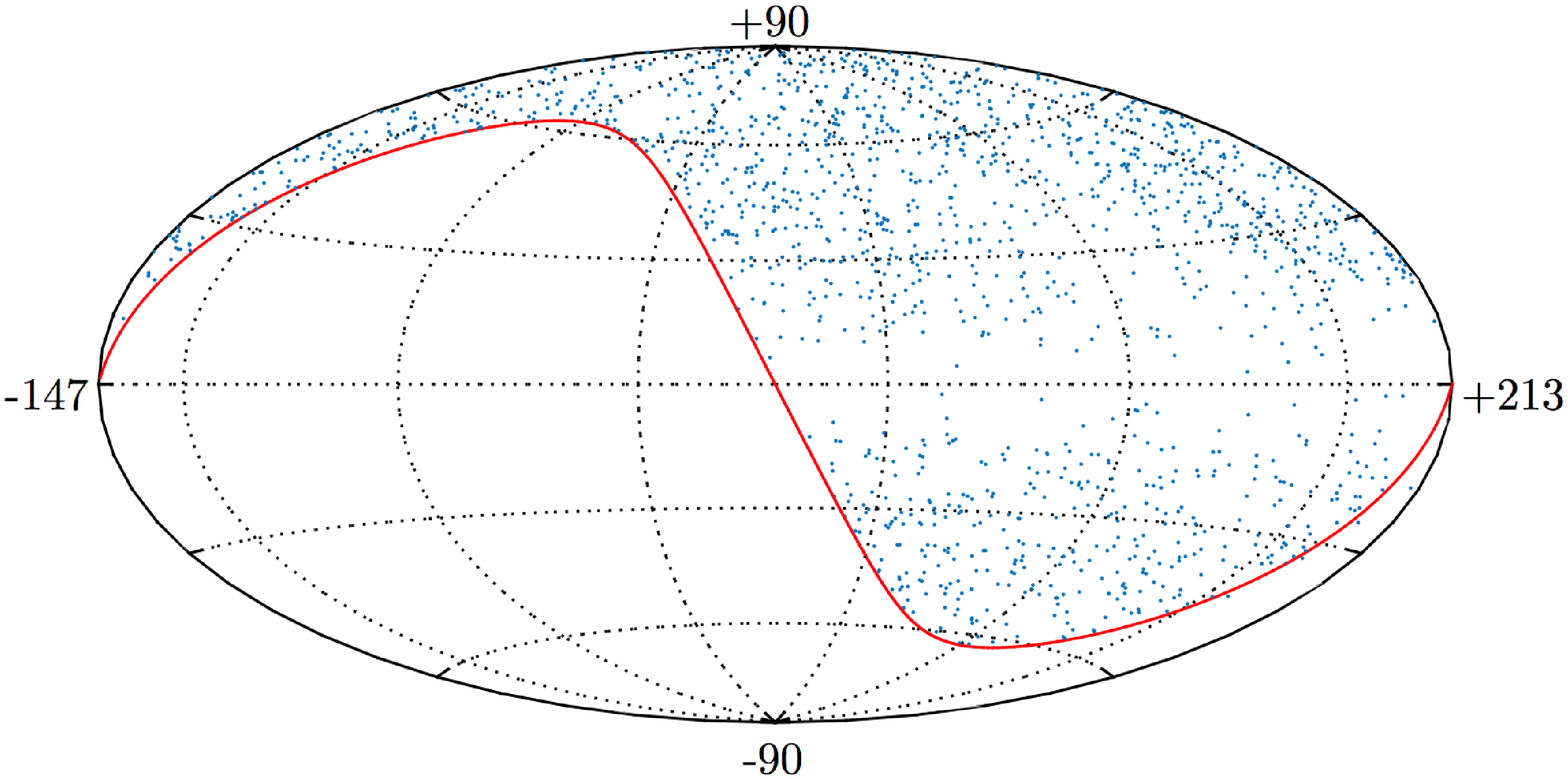}\\
\includegraphics[width=\columnwidth]{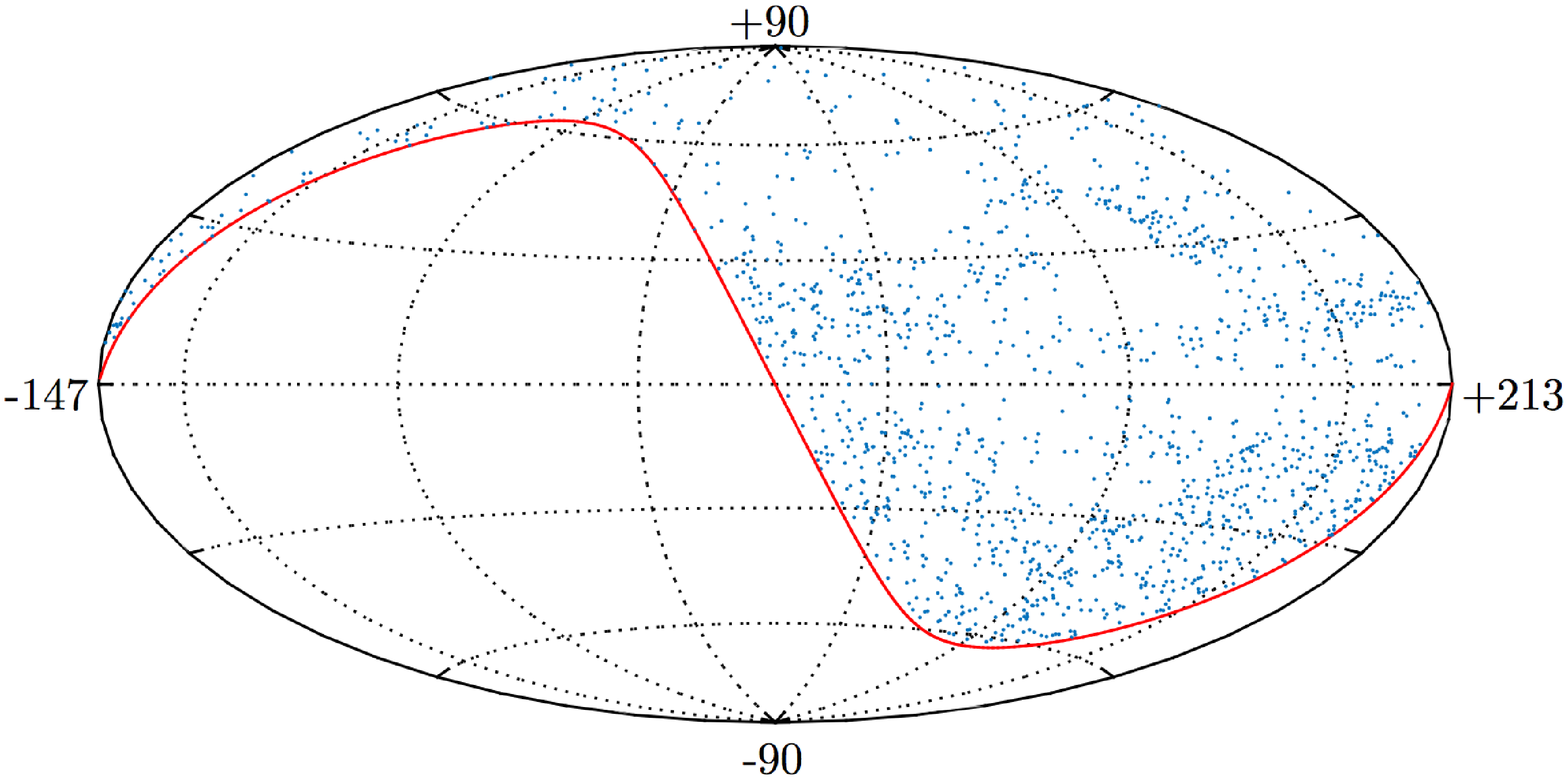}
\end{minipage}\hspace{0.03\textwidth}
\begin{minipage}{0.26\textwidth}
\mbox{}\\[-\baselineskip]
\\[-1.5ex]
\caption{\small{Hammer-Aitoff projection of the positions in Galactic
coordinates of the sub-samples $QSO$ (\textit{top}) and $RS$
(\textit{bottom}) that contain 1450 and 1379 FSRSs respectively.
The red line corresponds to the equator of the equatorial coordinates.
The map is centred at Galactic longitude $l_{\rm{gal}} = 33^\circ$ for
comparison of the sky distribution of the optical sample in
Fig.~\ref{fig:locStat_H05} and of the entire sample of the 4155 FSRSs
in Fig.~\ref{fig:J4155_GalProj}. The difference of the sky coverages
of the sub-sample $QSO$ and $RS$ can be spotted by eye.}}
\label{fig:J1450-1379_GalProj}
\end{minipage}
\end{center}
\end{figure}

The distributions on the sky of the two samples are also different as
shown in Fig.~\ref{fig:J1450-1379_GalProj}. While the sample $QSO$ is
almost homogeneously distributed over the sky, the sample $RS$ is far
from being so.

In order to test the possibility that it is the difference of the distributions on the
sky that is responsible for the detection of alignments for $QSO$ and not for
$RS$, we test the uniformity of the polarization PA of the RS's belonging to the
regions of alignments of Table~\ref{tab:1450_identif-2D}.
The overlap is very poor for the regions in the North Galactic hemisphere: only
32 RS's are found in each of the RN1 and RN2 regions.
To the South, however, there are 165 RS's in the RS1 region.
The Hawley--Peebles test does not reveal departure from uniformity, neither
taking RS's alone or mixing them with the QSOs of this region\footnote{This
was expected from the analysis of the A3 window (see
Section~\ref{subsecPH-1:PrelResults}).}.
Therefore, while the bad overlap between the $QSO$ and $RS$ samples in
the northern regions could explain the difference in the alignment detection,
this is not the case to the South. We thus conclude that the difference of the
distributions on the sky is unlikely responsible for this difference.

Similarly to an instrumental bias, contamination by foreground polarization
would affect more strongly the sample of RS's than that of QSOs as the
polarized flux is globally smaller for $RS$ (see
Fig.~\ref{fig:QSO-vs-RS-lightPropeties} (\textit{top})). This is again in contradiction
with what is observed. The contamination by foreground polarization is thus
unlikely to be responsible for the observed correlations of the polarization PAs
of QSOs.

\subsection{Are the polarization alignments real?}
As polarization is usually correlated to the morphological axis of the object
(e.g., \citealt{Saikia-Salter1988}; \citealt{Lister2001};
\citealt*{Pollack-Taylor-Zavala2003}; \citealt{Smith-et-al2004};
\citealt{Marin2014}), there might be real differences between the classes of
QSO and RS. Indeed, the core dominated FSRSs are predominantly quasars
or BL Lac objects in which the jet is oriented close to the line of sight
(see \citealt{Jackson-et-al2007}). The majority of the sources belonging to the
sample $RS$ is thus expected to be BL Lac objects which are thought to be
viewed at very small angles to the line of sight. Consequently, they are
expected to show rapid variations ($<2$ years) of their polarization PA and to
be more strongly polarized than quasars, which is what we observe in
Fig.~\ref{fig:QSO-vs-RS-lightPropeties} (Bottom). Also, for this class of objects,
no net correlation between the jet orientation and the polarization PA has been
reported (e.g., \citealt{Pollack-Taylor-Zavala2003} and references therein).
These observational facts could explain the absence of alignment signatures
for the sample of RS's within the hypothesis that polarization alignments reflect
morphological-axis alignments of the sources, as supported by the recent
discovery at optical wavelengths of such correlation for one of the most largest
known quasar group at $z\sim1.3$ (see Chapter~\ref{Ch:HBPS2014}).
The latter hypothesis is also reinforced by the discovery of large-scale alignments
of the jet position angles of active galactic nuclei in the ELAIS N1 field by
\citet{Jagannathan-Taylor2014} and \citet{Jagannathan2014}.
The fact that radio and optical alignments are found in the same parts of the sky
also supports a real effect.
In this framework, one would have to compare the alignment patterns observed
at optical wavelengths with these at radio wavelengths.
However, given the bad overlap between the sky coverage of the radio and
optical catalogues, a detailed comparison is not straightforward.
\begin{figure}[h]
\centering
\begin{minipage}{.8\linewidth}
\centering
\includegraphics[width=\columnwidth]{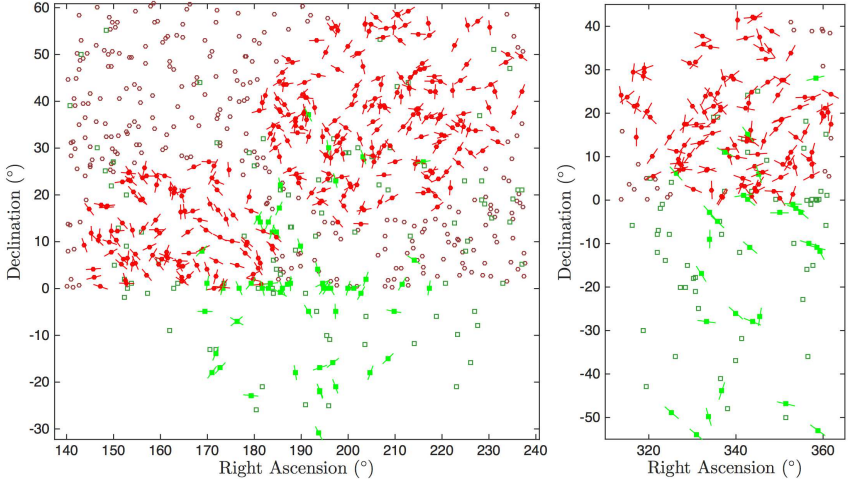}
\caption{\small{Comparison of the optical sample and radio sample
in the sky neighbourhood of the align regions. (\textit{left}) towards the North Galactic Cap
and (\textit{right}) toward the South Galactic Cap.
For each panel we show all the sources from both samples that have
(equatorial) right ascension and declination in the range
$\alpha \in \left[140^\circ, \, 238^\circ \right]$ and
$\delta \in \left[-31^\circ, \, 61^\circ \right]$ for the North and
$\alpha \in \left[313^\circ, \, 362^\circ \right]$ and
$\delta \in \left[-55^\circ, \, 43^\circ \right]$ for the South.
Sources from the sub-sample $QSO$ are displayed by brown circles and
those from the optical sample by dark green squares. Sources contained in
aligned regions (accounting for redshift limits at optical wavelengths) are
highlighted with filled symbols in light colors (red and green, respectively)
and their polarization vectors are shown with fixed but arbitrary length.
}}
\label{fig:NGC-SGC_comp_JV}
\end{minipage}
\end{figure}
To illustrate the weak overlaps between the radio and optical samples,
we show in Fig.~\ref{fig:NGC-SGC_comp_JV} equatorial maps of the
sources from both samples that are found in the neighbourhood of the
regions of alignments. Sources that we defined as being part of the
RN1, RN2 and RS1 regions are highlighted in light red and sources
belonging to the A1 and A3 regions of \citet{Hutsemekers-et-al2005}
in light green. Again, the alignment regions in the optical that we
defined in Chapter~\ref{Ch:PC-1Analysis} are smaller than A1 and A3,
the overlaps being even worth. We also show the polarization vector
orientations of sources that belong to regions of alignments.

\section{Concluding remarks}
\label{secPH-1:Conclusion}
We tested the hypothesis that the polarization position angles are randomly
distributed among the FSRSs contained in the JVAS/CLASS 8.4-GHz surveys
presented in \citet{Jackson-et-al2007}.
We performed the analysis in two and three dimensions, accounting for the
distribution of the sources on the sky (2D) and additionally for their line-of-sight
comoving distances (3D).
The polarization orientations of quasars (the $QSO$ sample) show low
probabilities to be consistent with the hypothesis of randomness. This
departure from uniformity is likely to be due to correlations of polarization
vectors of QSOs in groups of angular radius of about $20^\circ$. A basic
identification procedure has shown that these groups cluster in three distinct
regions of the sky.
Two of them fall in the A1 and A3 windows of the sky where optical
polarization alignments were found in Hutsem\'ekers et al.
(\citeyear{Hutsemekers1998}; \citeyear{Hutsemekers-Lamy2001};
\citeyear{Hutsemekers-et-al2005}).
Among sources in the JVAS/CLASS sample, only the sub-sample of QSO
exhibits such large-scale correlations.
If real, such alignments at radio wavelengths would support the
interpretation of alignments at optical wavelengths by spin-axis alignments
(\citealt{Hutsemekers-et-al2014} and Chapter~\ref{Ch:HBPS2014}).
However, our findings prove to be difficult to interpret either as resulting of
biases in the data set or as being the signature of a physical effect.
Indeed, one can find arguments for and against each scenario. Among them,
the fact that the alignments are more pronounced in 2D than in 3D and that
the mean PAs are multiple of $45^\circ$ in some regions would suggest a
biased data set whereas the detection of alignments for one class of object
but not for the others and the clustering of aligned sources in a few regions
of the sky consistent with those found at optical wavelengths might be seen
as the signature of a physical effect.

\medskip

To conclude, we highlighted correlations between the quasar radio-polarization
vectors which could demonstrate the presence of the same
kind of alignment effect as seen at optical wavelengths, or alternatively,
which could demonstrate that the radio polarization catalogue is affected
by observational biases and thus cannot be used to study the polarization
orientations of flat-spectrum radio sources.
Therefore, the claim by \citet{Joshi-et-al2007} stating that, at radio
wavelengths, there is no alignment signature of polarization vectors on
cosmological scales of the type found at optical wavelengths should be
taken with caution. The consequences regarding the optical polarization
alignments of quasars and the conclusions on possible mechanisms
that produce them should be revisited accordingly.

\medskip

More data are clearly needed to assess either the reality of polarization
alignments at radio wavelengths or the presence of residual biases in the
JVAS/CLASS 8.4-GHz radio polarization samples.
Furthermore, more optical polarization measurements of quasars in the
regions of alignments of the radio sample, and vice versa, would help to
disentangle the two possibilities and, in the mean time, allow us to
perform a careful comparison of the alignments from both spectral
bands.

\chapter[Alignment of quasar polarizations with LSS]{Alignment of quasar polarizations with large-scale structures}
\label{Ch:HBPS2014}
The reported alignments of the optical polarizations of quasars involve
sources that are located in regions of the comoving space that extend
over $\rm{Gpc}$ scales at redshift $z \sim 1.0$ (Hutsem{\'e}kers et al.
\citeyear{Hutsemekers1998}; \citeyear{Hutsemekers-Lamy2001};
\citeyear{Hutsemekers-et-al2005}; \citealt{Jain-Narain-Sarala2004};
\citealt{Shurtleff2013}; \citealt{Pelgrims-Cudell2014}).
Possible effects modifying the polarization of light along the line of sight,
in particular mixing with axion-like particles, have been investigated in
detail (e.g., \citealt{Das-et-al2005}; \citealt{Agarwal-et-al2012}). 
However, because of the absence of comparable circular polarization,
these mechanisms have essentially been ruled out
(\citealt{Hutsemekers-et-al2010}; \citealt{Payez-Cudell-Hutsemekers2011}).

Since quasar polarization is often related to the geometry of the
object, another interpretation would be that quasar morphological
axes themselves are aligned. This hypothesis is supported by the
detection of quasars polarization alignments at radio wavelengths
(Chapter~\ref{Ch:PH-1}) as this implies a mechanism of polarization
alignment independent of the wavelength. Note that this is true only if
the direction of the polarization vectors at the considered wavelength
is correlated to the morphologies of the objects.
One has then to search for a mechanism that would align the quasar
structural axes over huge distances.

In the framework of the tidal torque theory, the angular momenta of
the galaxies and of the massive black holes that they harbour is
transferred from the surrounding matter density field during collapse
and accretion into proto-galaxies and proto-clusters (e.g.,
\citealt{White1984}; \citealt{Heavens-Peacock1988};
\citealt{Lee-Pen2001}). If the quasar structural axis orientations
themselves are correlated due to such a mechanism of angular
momentum transfer, they are then expected to correlate to the shape
of the structure they belong to, which can be the filaments or the
sheets that form the cosmic web.

\medskip

In the concordance model of cosmology, the matter density field is not
expected to contain superstructures exceeding the homogeneity scale
of the Universe. Based on fractal analyses of the matter distribution,
this scale is known to be at most of the order of $\sim 370\,\rm{Mpc}$
(\citealt*{Yadav-Bagla-Khandai2010}).
Rationally, the angular momenta of galaxies are thus not expected to be
correlated over scales larger than that. As this scale is almost one order
of magnitude smaller than the typical scale of the optical polarization
alignments, it seems unlikely that the aforementioned
hypothesis is responsible for the quasar-structural-axis alignments.
However, \citet{Clowes-et-al2013} studied the space distribution of
quasars in the high-redshift sample from the Sloan Digital Sky Survey
Data Release 7 (SDSS DR7) (\citealt{Schneider-et-al2010})
and claimed the discovery of an extreme-scale structure at redshift
$z \sim 1.3$. This elongated large quasar group (LQG), named the
Huge-LQG by the authors, has its largest axis running over more than
$1 \, \rm{Gpc}$. While the physical origin of this structure, which is the
largest structure in the Universe following the words of
\citet{Clowes-et-al2013}, and its potential to challenge the cosmological
principle is still controversial (\citealt{Nadathur2013};
\citealt{Einasto-et-al2014}; \citealt{Park-et-al2015} and
\citealt*{EneaR-Cornejo-Campusano2015}), this LQG offered for the
first time the opportunity to test the hypothesis following which the
quasar polarization alignments are correlated to the large-scale
structures of the Universe and thus could reflect the alignment of the
quasar structural axes over cosmological scales.

\begin{figure}[t]
\centering
\begin{minipage}{0.8\textwidth}
\centering
\includegraphics[width=\columnwidth]{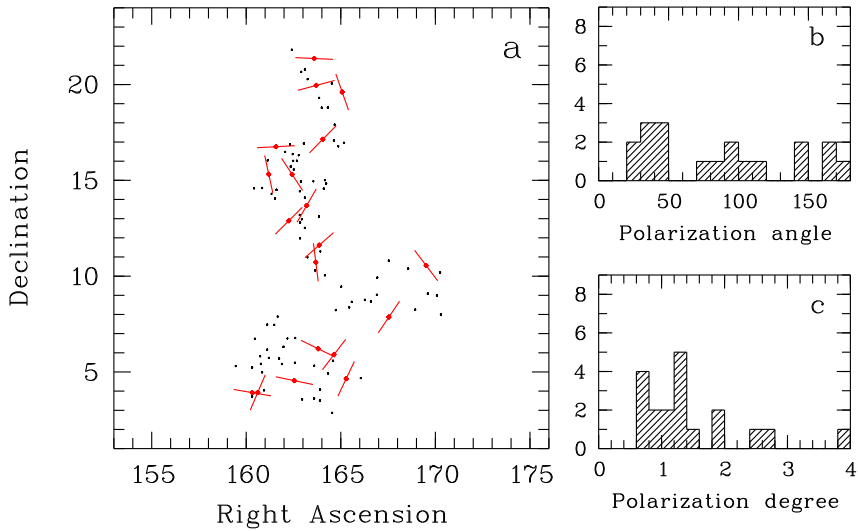}
\caption{\small{The polarization of the 19 quasars with
$p_{\rm{lin}} \geq 0.6 \%$. (a) Map of polarization vectors over the
large-scale structure; right ascensions and declinations are in degree;
the length of the polarization vectors is arbitrary. (b) Distribution of
polarization angles (in degree). (c) Distribution of polarization degrees
(in $\%$).}}
\label{map6}
\end{minipage}
\end{figure}
To test this hypothesis, we have measured the polarization of quasars
belonging to two large quasar groups at redshift $z \sim 1.3$ described
by \citet{Clowes-et-al2013}. These groups are the U1.27 and the U1.28
called the Huge-LQG and the CCLQG, respectively.
It is interesting to note that the Huge-LQG as well as the smaller
one located in its cosmic neighbourhood (named the CCLQG
after it discovery by \citealt{Clowes-Campusano1991}) are located on
the outskirts of the A1 region (or equivalently the N2+ region) of the
quasar-optical-polarization alignments. Moreover, these LQGs are both
located inside the RN1 window of the sky in which we found that the
radio-polarization of quasars are aligned (see
Section~\ref{secPH-1:IdentifRegions}). Whether these sky locations is
coincidental or physically real is still an open question.

\medskip

Out of the 93 observed quasars, 19 are significantly polarized
($p_{\rm{lin}} \geq 0.6 \%$). As we shall see in the next sections, we
found from this sample that quasar polarization vectors are either
parallel or perpendicular to the directions of the large-scale structures
to which they belong. Statistical tests indicate that the probability that
this effect is compatible with randomly oriented polarization vectors is
of the order of $1\%$. We also found that quasars with polarization
preferentially perpendicular to the host structure have large emission
line widths while objects with polarization preferentially parallel to the
host structure have small emission line widths.
Considering that quasar polarization is usually either parallel or
perpendicular to the accretion disk axis depending on the inclination with
respect to the line of sight, and that broader emission lines originate from
quasars seen at higher inclinations, we conclude that quasar spin axes
are likely parallel to their host large-scale structures.
In the next sections of this chapter, we present the polarization data and
their analysis as published in \citet{Hutsemekers-et-al2014}.

\section{Observations and polarization  measurements}
\label{sec:obs}
\begin{figure}[h]
\begin{center}
\begin{minipage}[t]{0.55\linewidth}
\mbox{}\\[-\baselineskip]
\begin{center}
\includegraphics[width=.9\columnwidth]{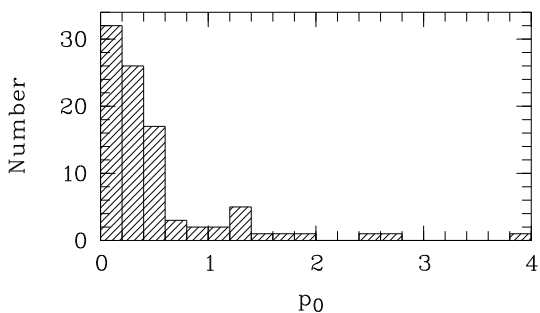}
\end{center}
\end{minipage}\hspace{3mm}
\begin{minipage}[t]{0.25\linewidth}
\mbox{}\\[-\baselineskip]
\caption{\small{The distribution of the debiased polarization
degree $p_0$ (in \%) measured for the sample of 93 quasars.}}
\label{fig1}
\end{minipage}
\end{center}
\end{figure}
Observations were carried out at the European Southern
Observatory\footnote{http://www.eso.org/public/unitedkingdom/},
Paranal, on March 22--26, 2014, using the Very Large Telescope
equipped with the FORS2 intrument in the standard imaging polarimetry
mode IPOL\footnote{FORS User Manual, VLT-MAN-ESO-13100-1543,
Issue 92.0}. Linear polarimetry is performed by inserting in the parallel
beam a Wollaston prism which splits the incoming light rays into two
orthogonally polarized beams separated by $22 ''$.
Image overlapping is avoided by inserting a special mask in the focal
plane. To measure the normalized Stokes parameters $q$ and $u$, four
frames are obtained with the half-wave plate rotated at four position
angles, $0^\circ$, $22.5^\circ$, $45^\circ$ and $67.5^\circ$.
This procedure allows us to remove most of the instrumental polarization.
The linear polarization degree $p_{\rm{lin}}$ and the polarization position
angle $\psi$ are derived using $p_{\rm{lin}} = (q^2 + u^2)^{1/2}$ and
$\psi = (1/2) \arctan \, (u/q)$ so that $q= p_{\rm{lin}} \cos 2\psi$ and
$u = p_{\rm{lin}} \sin 2\psi$.
Since orthogonally polarized images of the object are simultaneously
recorded, the measured polarization does not depend on variable
transparency or seeing.
\begin{figure}[h]
\centering
\begin{minipage}{0.8\textwidth}
\centering
\includegraphics[width=.8\columnwidth]{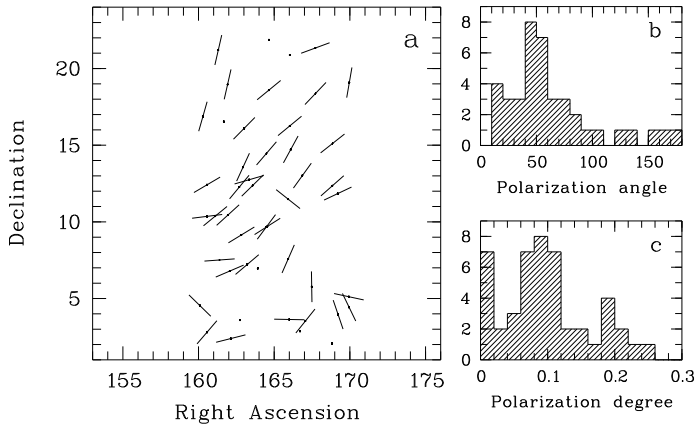}
\caption{\small{The interstellar polarization in the region of the sky
corresponding to the quasar large-scale structure under study (data
from \citealt*{Berdyugin-Piirola-Teerikorpi2014}).
(a) Map of the polarization vectors; right ascensions and declinations
are in degrees; the length of the polarization vectors is arbitrary.
(b) Distribution of the polarization angles (in degrees).
(c) Distribution of the polarization degrees (in $\%$).}}
\label{maps}
\end{minipage}
\end{figure}

All observations were obtained using the FORS2 V$\_$high filter
($\lambda_0 = 555 \, \rm{nm}$, FWHM~$= 123\,\rm{nm}$).
Data reduction and measurements were performed as detailed in
\citet{Sluse-et-al2005}. The instrumental polarization was checked
using the unpolarized stars WD0752--676 and WD1615--154
(\citealt{Fossati-et-al2007}) and found to be
$p_{\rm{lin}} = 0.05 \pm 0.06 \%$, which is consistent with
zero\footnote{We also observed HD~64299 which turned out to be
polarized with $p_{\rm{lin}} = 0.17 \pm 0.04 \%$, in agreement with
\citet{Masiero-et-al2007}.}.
We did not use field stars to estimate the instrumental polarization
because of spurious off-axis polarization in FORS1/2
(\citealt{Patat-Romaniello2006}). To fix the zero-point of the polarization
position angle, polarized standard stars have been observed:
NGC~2024-1, Ve~6-23, CD-28$^\circ$ 13479, HD~316232,
BD-14$^\circ$ 922 (\citealt{Fossati-et-al2007}). The offset --to subtract
from the raw polarization angle-- was determined to be
$2.5^\circ \pm 0.5^\circ$ in the V$\_$high filter.

\medskip

The linear polarization of all 73 quasars of the Huge-LQG and of 20
out of the 34 quasars of the CCLQG has been obtained, i.e. for a total
of 93 quasars. These measurements\footnote{Also publicly available in
electronic form at http://www.aanda.org} were summarized in Table~1 of
\citet{Hutsemekers-et-al2014}.
The error on the polarization degree is between $0.06\%$ and $0.23\%$,
with a mean value of $0.12\%$. The distribution of the debiased
polarization degree is illustrated in Fig.~\ref{fig1}.
It shows a peak near the null value, in agreement with other polarization
measurements of radio-quiet non-BAL quasars (\citealt{Berriman-et-al1990};
\citealt*{Hutsemekers-Lamy-Remy1998}). All objects are at Galactic
latitudes higher than $50^\circ$ which minimizes contamination by
interstellar polarization. In this region of the sky, the interstellar polarization
is around $p_{\rm is} \simeq 0.1\%$ with a peak of the polarization position
angles near $50^\circ$ (Fig.~\ref{maps}).

As in Hutsem{\'e}kers et al. (\citeyear{Hutsemekers1998};
\citeyear{Hutsemekers-Lamy2001}; \citeyear{Hutsemekers-et-al2005}),
we consider that polarization is essentially intrinsic to the quasar when
$p_{\rm{lin}} \geq 0.6 \%$ (\citealt{Berriman-et-al1990};
\citealt{Hutsemekers-Lamy-Remy1998}; \citealt{Sluse-et-al2005}).
Out of 93 quasars, 19 have $p_{\rm{lin}} \geq 0.6 \%$. Their properties
are given in Table~\ref{poltab2} at the end of this chapter. For these 19
polarized quasars, the uncertainties on the polarization PAs (computed
as in Eq.~\ref{eq:sigma_PA}) are found to be
$\sigma_{\psi} \leq 10^\circ$ with an average value around $3^\circ$.
\begin{figure}[h]
\begin{center}
\begin{minipage}[t]{0.6\linewidth}
\mbox{}\\[-\baselineskip]
\begin{center}
\includegraphics[width=.92\columnwidth]{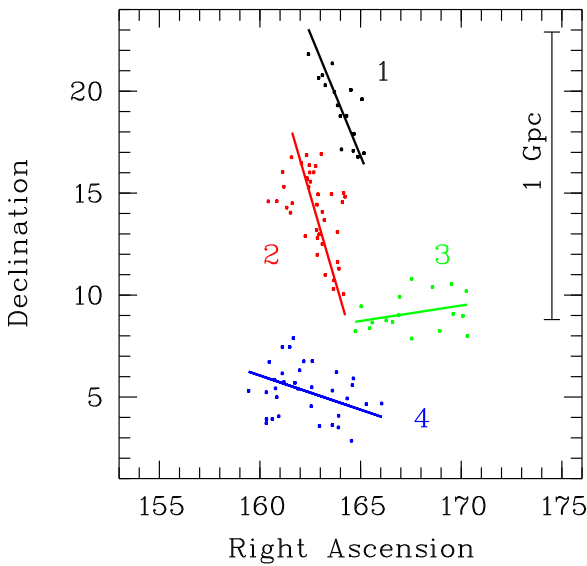}
\end{center}
\end{minipage}\hspace{5mm}
\begin{minipage}[t]{0.25\textwidth}
\mbox{}\\[-\baselineskip]
\caption{\small{The quasar groups and their orientations on the sky. Right
ascensions and declinations are in degrees. The superimposed lines
illustrate the orientations of the four groups labelled 1, 2, 3,
4. The comoving distance scale at redshift $z = 1.3$ is indicated
assuming a flat Universe with $H_0 = 70\, \rm{km}\, \rm{s}^{-1}\, \rm{Mpc}^{-1}$ and
$\Omega_M = 0.27$.}}
\label{stru}
\end{minipage}
\end{center}
\end{figure}

\section{Analysis of polarization alignments}
\label{sec:ana}
In Fig.~\ref{map6} we show a map of the quasar polarization vectors
over the LQG structures.
The map does not show any evidence of coherent orientations or
alignments. The distribution of the polarization angles is flat, compatible
with random orientations and with no contamination by interstellar
polarization.

\medskip

In order to compare the quasar polarization angles to the direction of
the local structures, we consider four structures for which we determine
a mean orientation, as illustrated in Fig.~\ref{stru}.
\begin{figure}[h]
\centering
\begin{minipage}[t]{0.55\linewidth}
\mbox{}\\[-\baselineskip]
\begin{center}
\includegraphics[width=.68\columnwidth]{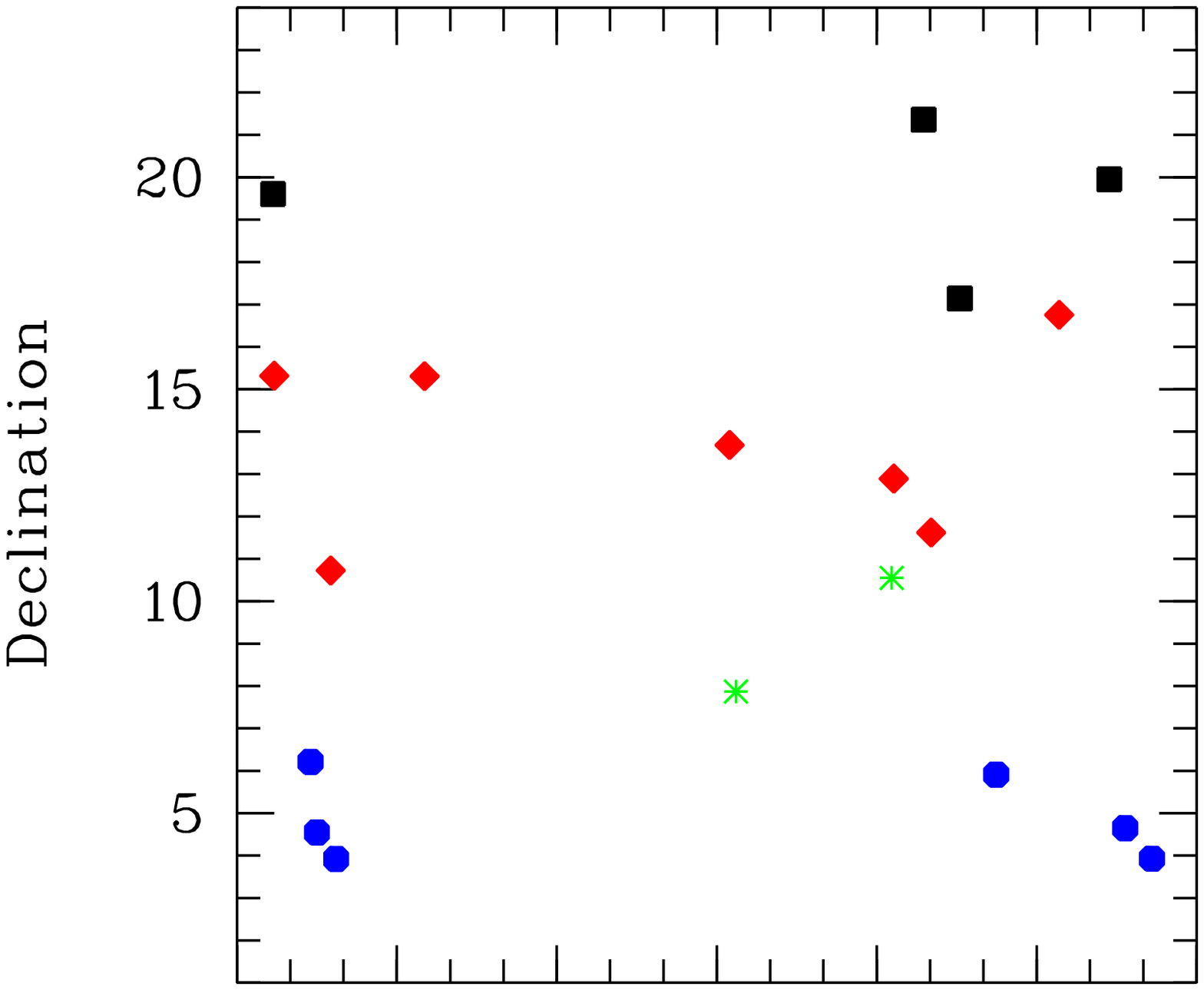}\\
\includegraphics[width=.68\columnwidth]{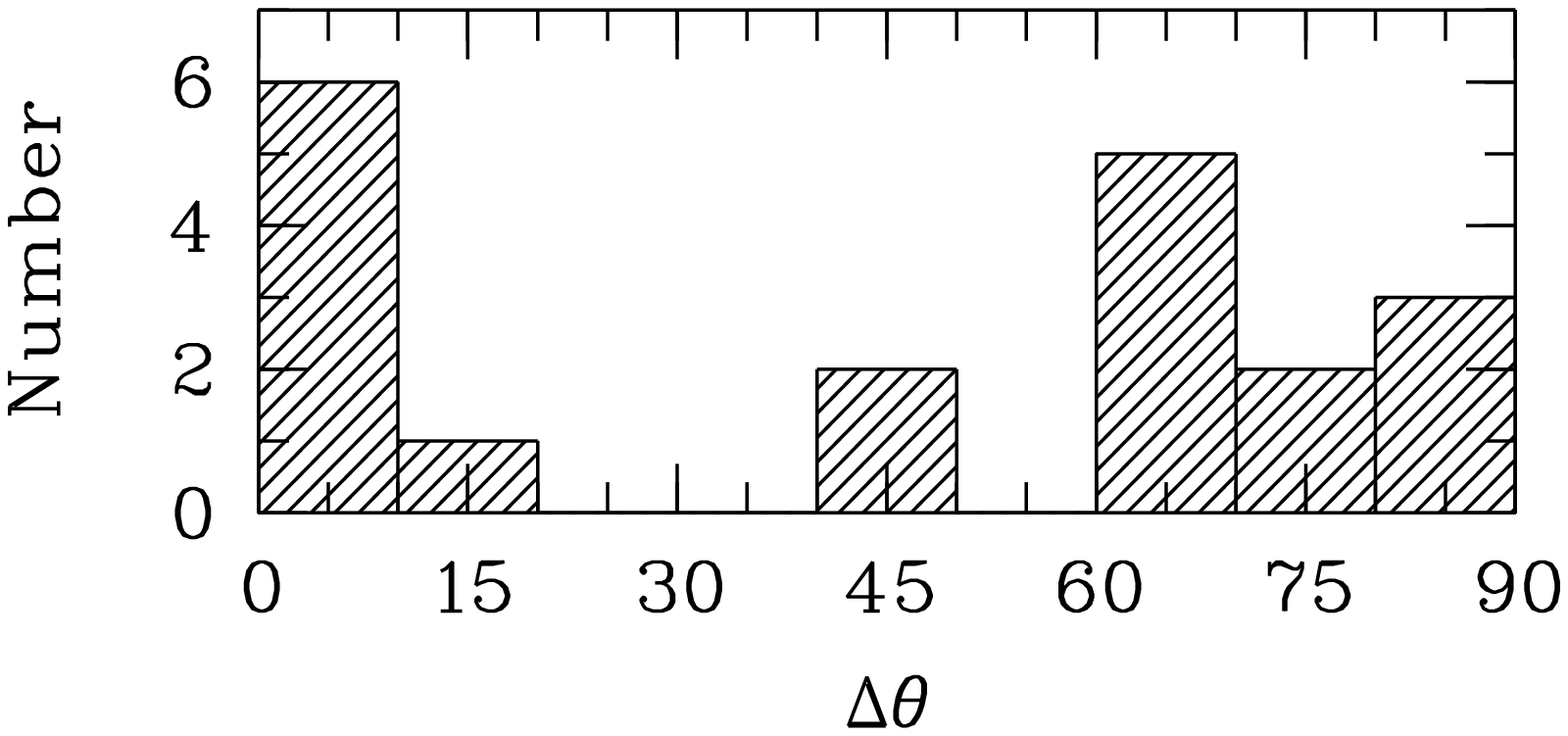}
\end{center}
\end{minipage}\hspace{3mm}
\begin{minipage}[t]{0.25\linewidth}
\mbox{}\\[-\baselineskip]
\caption{\small{\textit{bottom}: The distribution of the acute angle
$\Delta_{\psi \chi}$ (in degree) between quasar polarizations and the
orientation of their host large-scale structure. \textit{top}: $\Delta_{\psi \chi}$
is plotted against the object's declination (in degrees) to illustrate the
behaviour of the different quasar groups (1: squares, 2: lozenges, 3:
asterisks, 4: hexagons; colors as in Fig.~\ref{stru}).}}
\label{dteta6}
\end{minipage}
\end{figure}
Group 4 is the CCLQG defined in (\citealt{Clowes-et-al2012}).
The Huge-LQG is divided in groups denoted 1, 2 and 3. Group 3
corresponds to the branch set of 17 quasars identified by
\citet{Clowes-et-al2013}. The large vertical part of the Huge-LQG is
then separated into groups 1 and 2. The mean projected direction of
the structures is determined by an orthogonal regression in right
ascension, declination (\citealt{Isobe-et-al1990}). For groups 1, 2,
3, 4, we measure the position angles $\chi$ = $157^\circ$, $164^\circ$,
$81^\circ$, and $109^\circ$, respectively. We estimate the acute angle
between the quasar polarization vectors and the PA of the structures to
which they belong using $\Delta_{\psi \chi} = \min \left( |\chi - \psi\,|\, , \,
180^\circ - |\chi - \psi\,| \right) $.

The distribution of $\Delta_{\psi \chi}$ is illustrated in Fig.~\ref{dteta6}
(\textit{bottom}). It shows a bimodal distribution, with both
alignments ($\Delta_{\psi \chi} \simeq 0^\circ$) and anti-alignments
($\Delta_{\psi \chi} \simeq 90^\circ$) in each quasar group (except group
3). The cumulative binomial probability of having nine or more quasars in
the first and the last bins is $P_{\rm bin} = 1.4\%$. The Kuiper test
(\citealt{Arsham1988}; \citealt{Fisher1993}) gives a probability $P_{\rm{K}} =
1.6\%$ that the observed distribution is drawn from an uniform
distribution. These results are robust if we consider the 28 quasars
with $p_{\rm{lin}} \geq 0.5\%$ (in this case $P_{\rm bin} = 1.2\%$
and $P_{\rm{K}} = 1.0\%$).

\begin{figure}[h]
\centering
\begin{minipage}[t]{0.6\linewidth}
\mbox{}\\[-\baselineskip]
\begin{center}
\includegraphics[width=\columnwidth]{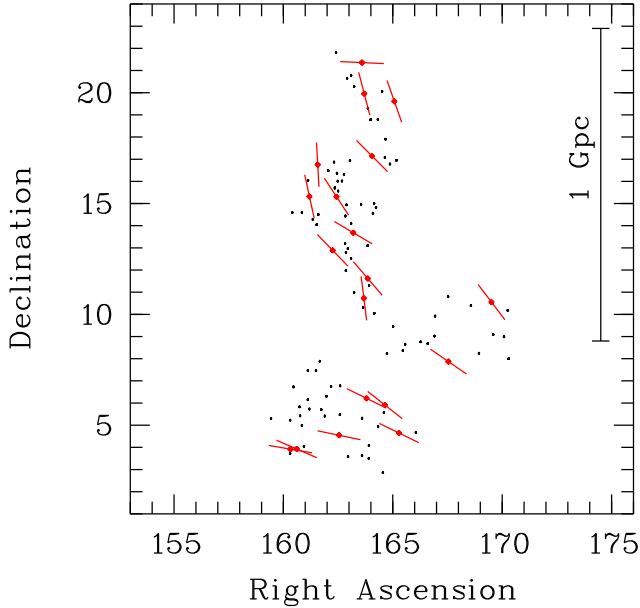}
\end{center}
\end{minipage}\hspace{4mm}
\begin{minipage}[t]{0.28\linewidth}
\mbox{}\\[-\baselineskip]
\caption{\small{The polarization vectors of the 19 quasars with
$p_{\rm{lin}}\geq 0.6\%$ are superimposed on the large-scale
structure after rotation of the polarization angles according to
$\tilde{\psi} = \rm{mod}(\psi,90^\circ) +90^\circ$.
A clear correlation is seen but we nevertheless caution against
exaggerated visual impression since polarization angles are now
in the range $90^\circ - 180^\circ$.
Right ascensions and declinations are in degree. The comoving
distance scale is indicated as in Fig.~\ref{stru}.}}
\label{maprot6}
\end{minipage}
\end{figure}
A bimodal distribution of $\Delta_{\psi \chi}$ is exactly what we expect
if the quasar morphological axes are related to the orientation of the
host large-scale structures. Indeed, the polarization of type 1 active
galactic nuclei (AGN) is usually either parallel or perpendicular to the
AGN accretion disk axis depending on the inclination with respect to
the line of sight (e.g., \citealt{Smith-et-al2004}). We may assume that
higher luminosity AGN (quasars) behave similarly. In Fig.~\ref{maprot6},
the quasar polarization angles modified according to $\tilde{\psi} =
\rm{mod}(\psi,90^\circ) +90^\circ$ are plotted over the LQG structure,
unveiling a remarkable correlation. We stress that such a behaviour
cannot be due to contamination by interstellar polarization which
would align all polarizations similarly.

\medskip

\begin{figure}[h]
\begin{center}
\begin{minipage}{0.8\linewidth}
\centering
\includegraphics[width=0.85\columnwidth]{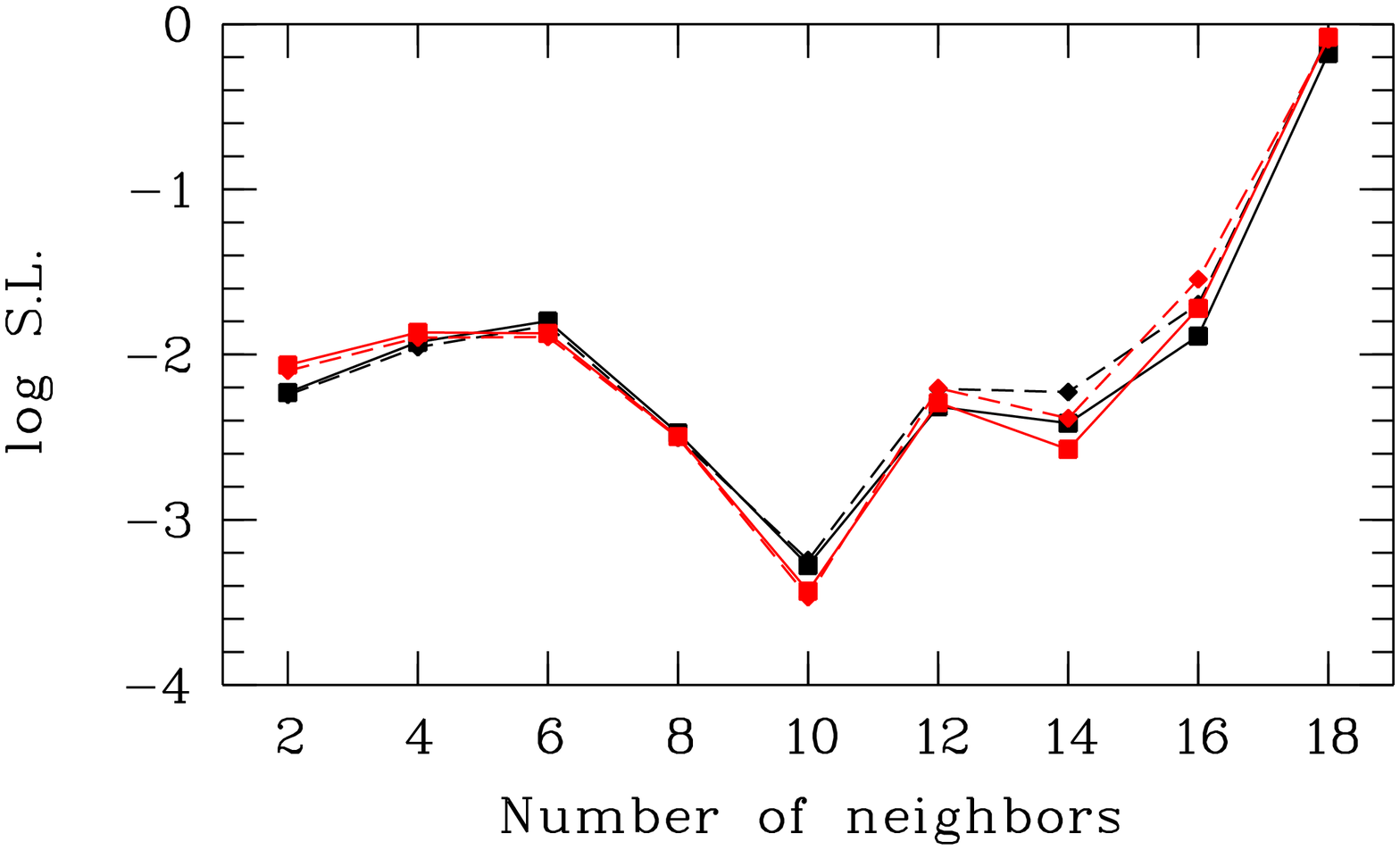}
\caption{\small{The logarithm of the significance level (SL) of the Z
test applied to the sample of 19 polarized quasars, as a function of
the number of nearest neighbours. The solid line refers to simulations
obtained by shuffling angles over positions while the dashed line
refers to simulations obtained by randomly generating angles. The
statistics are computed with (in red) and without (in black) parallel
transport of the polarization vectors.}}
\label{sl6}
\end{minipage}
\end{center}
\end{figure}
To quantify the significance of this correlation independently of the
shape of the host structure, we use the Andrews and Wasserman $Z$
statistical test (\citealt{Bietenholz1986}; \citealt{Hutsemekers1998} and
Section~\ref{stat_Ztest} of this thesis).
This test is best suited to small samples since it does not involve angle
dispersion. The idea of the Andrews \& Wasserman test is to compute
for each object $i$, the mean direction $\bar{\psi}_{i}$ of its
$n_{v}$ nearest neighbours, and to compare this local average to the
polarization angle of the object $i$, $\psi_{i}$. If angles are
correlated to positions, one expects, on the average, $\psi_{i}$ to
be closer to $\bar{\psi}_{j=i}$ than to $\bar{\psi}_{j \neq i}$.
As a measure of the closeness of $\psi_{i}$ and $\bar \psi_{j}$,
one uses $D_{i,j}$ = ${\bf y}_{i} \cdot {\bf \bar Y}_{j}$, where
${\bf y}_{i}$ is the normalized polarization vector of object $i$ and
${\bf \bar Y}_{j}$ the normalized resultant polarization vector of the
$n_{v}$ neighbours of object $j$, excluding $j$. Then $Z_i$ is
computed by ranking $D_{i,j=i}$ among the $D_{i,j=1,N}$ and the final
statistics $Z_c$ is obtained by averaging the $Z_i$ over the sample of
$N$ objects. $Z_{c}$ is expected to be significantly larger than zero
when the polarization angles are not randomly distributed over
positions. To make the test independent of the coordinate system,
polarization vectors can be parallel transported before computing the
resultant polarization vectors (\citealt{Jain-Narain-Sarala2004}).

Here, the polarization vectors are computed using ${\bf y} = (\cos\Psi,
\sin\Psi)$ with the angle $\Psi = 4 \, \rm{mod} (\psi,90^\circ)$ instead of
$\Psi = 2 \psi = 2 \, \rm{mod}(\psi,180^\circ)$ to test for
either alignments or anti-alignments (i.e., dealing with 4-axial data
instead of 2-axial data, see \citet{Fisher1993}). To estimate the
statistical significance, $10^5$ samples of 19 angles were created
through Monte-Carlo simulations either by shuffling the measured
angles over positions, or by randomly generating them in the
[$0^\circ,\,180^\circ$] range (\citealt{Press-et-al1992}). The significance
level (SL) of the test is finally computed as the percentage of
simulated configurations for which $Z_c \geq Z_c^{\star}$ where
$Z_c^{\star}$ is the measured sample statistics. Since all quasars
are in a limited redshift range, we only consider their angular
positions on the sphere to build the groups of nearest neighbours.

The significance level of the $Z$ test is illustrated in Fig.~\ref{sl6}.
It shows that the probability that the polarization angles are randomly
distributed over positions is smaller than $1\%$.
The effect is stronger (SL $< 0.1\%$) when the mean orientation
is computed with 10 nearest neighbours, i.e. roughly half of the
sample. This number corresponds to a mean comoving distance of $\sim
550 \,\rm{Mpc}$, in agreement with the trend seen in Fig.~\ref{maprot6}. 
Parallel transport has little effect since all quasars lie close to
each other and to the equator. We emphasize that a deviation from
uniformity is only detected when using $4\, \rm{mod}(\psi,90^\circ)$
in the $Z$ test and not when using $2\, \rm{mod}(\psi,180^\circ)$,
which means that purely parallel or perpendicular alignments are not
seen (Fig.~\ref{map6}). If we consider the 28 quasars with
$p_{\rm{lin}} \geq 0.5\%$, a similar curve is obtained with the minimum
shifted to $n_v = 20$, which corresponds to a mean comoving distance
of $\sim 650\, \rm{Mpc}$.

\medskip

\begin{figure}[h]
\begin{minipage}[t]{0.68\textwidth}
\mbox{}\\[-\baselineskip]
\begin{center}
\includegraphics[width=0.9\columnwidth]{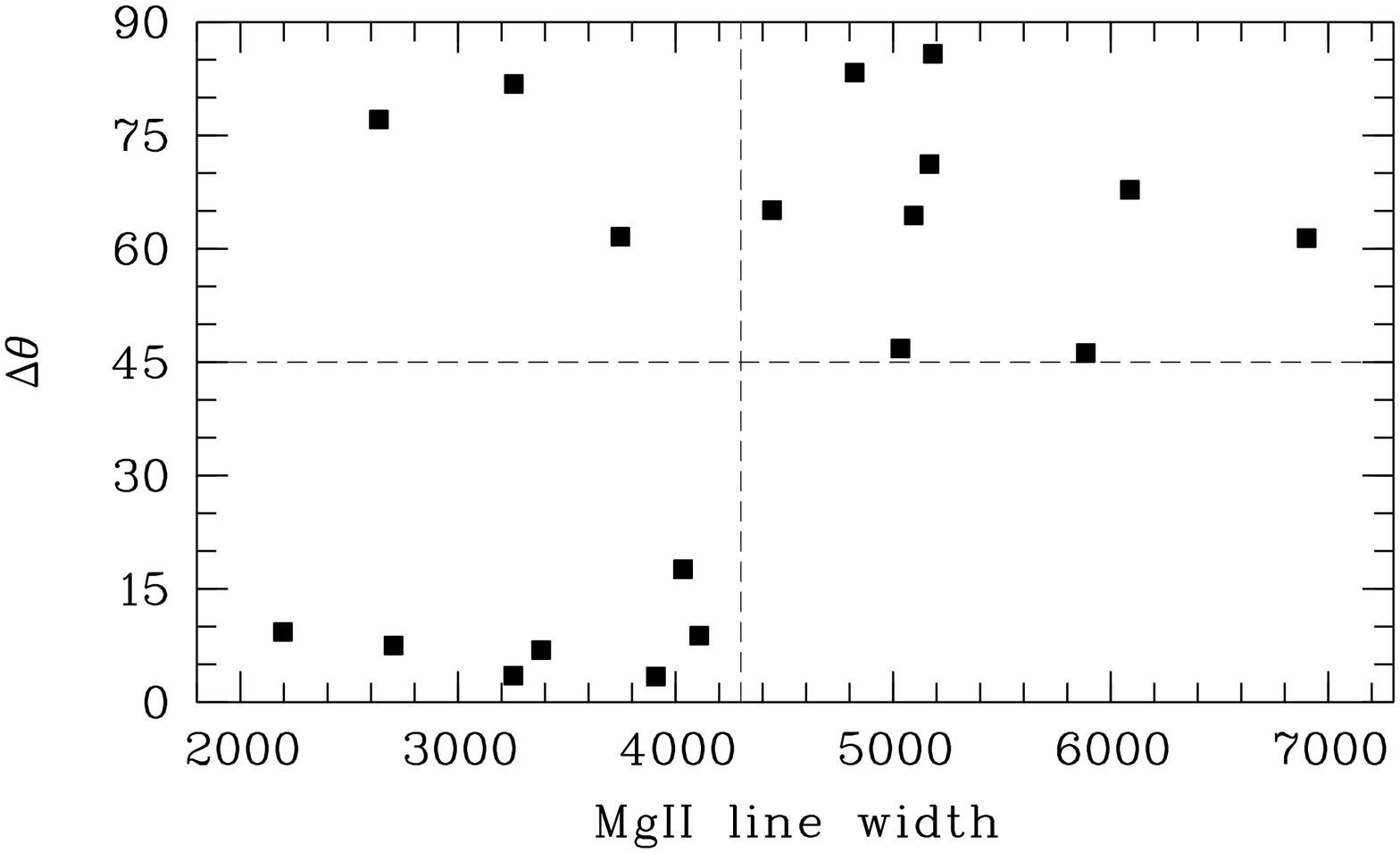}
\end{center}
\end{minipage}
\begin{minipage}[t]{0.25\textwidth}
\mbox{}\\[-\baselineskip]
\caption{\small{The angle $\Delta \theta$ (in degree) between quasar
polarizations and the orientation of their host large-scale structures
as a function of the MgII emission line width (FWHM in
$\rm{km}\, \rm{s}^{-1}$).}}
\label{tewi6}
\end{minipage}
\end{figure}
Since the width of low-ionization emission lines (H$\beta$, MgII)
observed in quasar spectra correlates with the object's inclination
with respect to the line of sight (\citealt{Wills-Browne1986};
\citealt{Brotherton1996}; \citealt{Jarvis-McClure2006};
\citealt{Decarli-et-al2008}), we plot in Fig.~\ref{tewi6} the angle
$\Delta_{\psi \chi}$ as a function of the quasar MgII emission line
width (FWHM from \citet{Shen-et-al2011}). 
We see that most objects with polarization perpendicular to the host
structure ($\Delta_{\psi \chi} > 45^\circ$) have large emission line
widths while all objects with polarization parallel to the host structure
($\Delta_{\psi \chi} < 45^\circ$) have small emission line widths.

\begin{figure}[h]
\begin{minipage}[t]{0.68\textwidth}
\mbox{}\\[-\baselineskip]
\begin{center}
\includegraphics[width=0.95\columnwidth]{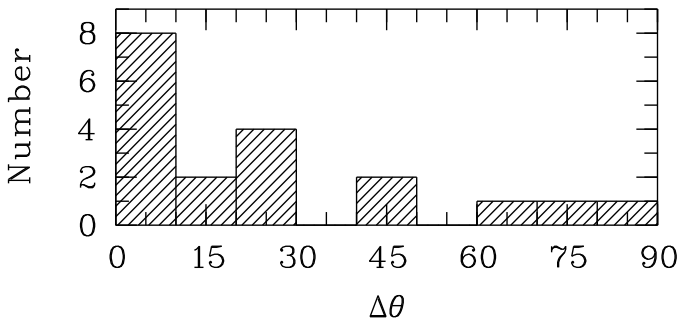}
\end{center}
\end{minipage}
\begin{minipage}[t]{0.28\textwidth}
\mbox{}\\[-1.2\baselineskip]
\caption{\small{The distribution of the acute angle $\Delta_{\psi \chi}$
(in degrees) between quasar polarizations and the orientation of their
host large-scale structure after rotating by $90^\circ$ the polarization
angles of objects with MgII emission line widths larger than
$4300\, \rm{km}\, \rm{s}^{-1}$.}}
\label{histrot6}
\end{minipage}
\end{figure}

A two-sample Kolmogorov--Smirnov test indicates that there is a
probability of only $1.4\%$ that quasars with either perpendicular or
parallel polarizations have emission line widths drawn from the same
parent population.
Quasars seen at higher inclinations\footnote{Face-on: $i = 0^\circ$.
Edge-on: $i = 90^\circ$.} generally show broader low-ionization
emission lines, in agreement with line formation in a rotating disk
(\citealt{Wills-Browne1986}; \citealt{Jarvis-McClure2006};
\citealt{Decarli-et-al2008}).
The relation seen in Fig.~\ref{tewi6} thus supports our hypothesis that
the polarization of quasars is either parallel or perpendicular to the
host structure depending on their inclination. When rotating by
$90^\circ$ the polarization angles of objects with MgII emission line
widths larger than $4300\, \rm{km}\, \rm{s}^{-1}$, a stronger alignment
is seen (Fig.~\ref{histrot6}).
The Kuiper test gives a probability $P_{\rm{K}} = 0.5\%$ that the observed
distribution is drawn from an uniform distribution, but this value should
be seen with caution since the cut at $4300\, \rm{km}\, \rm{s}^{-1}$ is
arbitrary. On the other hand, it should be emphasized that the emission
line width does not only depend on inclination but also on the mass of
the central black hole if the rotating disk is virialized. Quasars with lower
black hole mass will have narrower emission lines whatever their
inclination so that some of them may still appear anti-aligned in
Fig.~\ref{histrot6}.

Since objects seen at higher inclinations preferentially show
polarization perpendicular to their axes (\citealt{Smith-et-al2004}),
we finally infer that quasar spin axes should be predominantly
parallel to the orientation of the structures to which they belong.

\section[Conclusion on optical polarization alignments in LQGs]{Conclusion on the alignment of optical quasar polarizations with their large-scale structure}
\label{sec:conclu}
We have measured the polarization of 93 quasars belonging to
large-scale quasar groups. 19 quasars out of 93 are significantly
polarized with $p_{\rm{lin}}\geq 0.6 \%$.

We found that quasar polarization vectors are either parallel or
perpendicular to the large-scale structures to which they belong,
and correlated to the polarization vectors of their neighbours. The
probability that these results can be attributed to a random
distribution of polarization angles is on the order of $1\%$. Such a
behaviour cannot be due to contamination by interstellar polarization.
Our results are robust if we consider $p_{\rm{lin}} \geq 0.5 \%$ instead
of $p_{\rm{lin}} \geq 0.6 \%$, or if we subtract a systematic
$p_{\rm is} = 0.1\% $ at $\psi_{\rm is} = 50^\circ$ to simulate the
correction of a possible contamination by interstellar polarization
(Fig.~\ref{maps}).

Assuming that quasar polarization is either parallel or perpendicular
to the accretion disk axis as a function of inclination, as observed in
lower luminosity active galactic nuclei, and considering that broader
emission lines originate from quasars seen at higher inclinations, we
inferred that quasar spin axes are likely parallel to their host large-scale
structures.

Galaxy spin axes are known to align with large-scale structures such
as cosmic filaments (e.g., \citealt{Tempel-Libeskind2013};
\citealt{Zhang-et-al2013}; and references therein). Till now, such
alignments are detected up to redshift $z \sim 0.6$ at scales
$\lesssim 100 \, \rm{Mpc}$ (\citealt{Li-et-al2013}).
Detailed interpretations remain complex because the link between
galaxy and halo spin axes is not straightforward, and because the
strength and orientation of the alignments depend on several factors,
in particular the mass of the halo and the cosmic history (e.g.,
\citealt{Hahn-Tessier-Carollo2010};
\citealt{Trowland-Lewis-Bland_H2013}; \citealt{Dubois-et-al2014}).
We have found that quasar accretion disk axes are likely parallel to the
large-scale structures to which they belong over Gpc scales at redshift
$z \sim 1.3$, i.e. one order of magnitude larger than currently known
galaxy alignments. Although the scales involved are much larger, we may
assume that similar mechanisms can explain alignments of quasar and
galaxy axes with their host large-scale structure, keeping in mind that
polarization-related quasar regions (accretion disk, jet, scattering region)
are not necessarily well aligned with the stellar component of the host
galaxy (\citealt{Borguet-et-al2008}; \citealt{Hopkins-et-al2012}), and that
quasars, more prone to strong feedback mechanisms, can have a different
cosmic history (\citealt{Dubois-et-al2014}).

Since coherent orientations of quasar polarization vectors, and then
quasar axes, are found on scales larger than $500 \, \rm{Mpc}$, our
results might also provide an explanation to the very large-scale
polarization alignments reported by Hutsem{\'e}kers et al.
(\citeyear{Hutsemekers1998}; \citeyear{Hutsemekers-Lamy2001};
\citeyear{Hutsemekers-et-al2005}). In this case those alignments
would be intrinsic and not due to a modification of the polarization
along the line of sight. The existence of correlations in quasar axes
over such extreme scales would constitute a serious anomaly for the
cosmological principle.

\begin{table}[h]
\centering
\begin{minipage}{137.5mm}
\centering
\footnotesize{
\begin{tabular*}{137.5mm}{@{\extracolsep{\stretch{1}}}*{1}{lcccccccc}}
\hline
\\ [-1.5ex]
Object & $z$ & LQG & $p_{\rm{lin}}$ & $\sigma_{p}$ & $\psi$ \ & $\sigma_{\psi}$ & FWHM & $\sigma_{\rm FWHM}$\\ 
&  &  & (\%) & (\%) & ($^\circ$) & ($^\circ$) & km s$^{-1}$ & km s$^{-1}$\\ [0.5ex]
\hline
\\ [-1.5ex]
SDSSJ105421.90+212131.2 & 1.2573 & 1 & 1.04 & 0.08 &  92.6 & 2.2 & 5094 & 214 \\    
SDSSJ105446.73+195710.5 & 1.2195 & 1 & 1.89 & 0.23 &  75.2 & 3.5 & 3256 & 363 \\    
SDSSJ105611.27+170827.5 & 1.3316 & 1 & 1.29 & 0.08 &  44.8 & 1.8 & 6088 & 158 \\    
SDSSJ110016.88+193624.7 & 1.2399 & 1 & 1.14 & 0.23 & 160.4 & 5.9 & 3909 & 348 \\    
SDSSJ104445.03+151901.6 & 1.2336 & 2 & 1.25 & 0.11 & 167.5 & 2.5 & 3254 & 196 \\  
SDSSJ104616.31+164512.6 & 1.2815 & 2 & 1.25 & 0.11 &  86.9 & 2.5 & 2635 & 222 \\   
SDSSJ104859.74+125322.3 & 1.3597 & 2 & 0.72 & 0.13 &  45.6 & 5.3 & 3746 & 397 \\    
SDSSJ104941.67+151824.6 & 1.3390 & 2 & 1.31 & 0.13 & 146.4 & 2.9 & 4034 & 633 \\    
SDSSJ105245.80+134057.4 & 1.3544 & 2 & 1.32 & 0.11 &  30.2 & 2.4 & 5885 & 174 \\    
SDSSJ105442.71+104320.6 & 1.3348 & 2 & 0.73 & 0.11 & 172.8 & 4.4 & 4108 & 269 \\    
SDSSJ105525.68+113703.0 & 1.2893 & 2 & 2.55 & 0.10 &  49.1 & 1.1 & 4443 & 399 \\    
SDSSJ111009.58+075206.8 & 1.2123 & 3 & 1.81 & 0.17 &  34.2 & 2.7 & 5032 & 626 \\    
SDSSJ111802.11+103302.4 & 1.2151 & 3 & 3.97 & 0.10 & 142.4 & 0.7 & 6900 &1256 \\   
SDSSJ104116.79+035511.4 & 1.2444 & 4 & 1.55 & 0.11 &  99.7 & 2.0 & 2195 & 296 \\   
SDSSJ104225.63+035539.1 & 1.2293 & 4 & 0.69 & 0.08 &  23.2 & 3.3 & 5182 & 380 \\    
SDSSJ105010.05+043249.1 & 1.2158 & 4 & 2.67 & 0.08 & 101.5 & 0.9 & 2703 & 190 \\   
SDSSJ105512.23+061243.9 & 1.3018 & 4 & 0.98 & 0.12 & 115.9 & 3.5 & 3381 & 299 \\   
SDSSJ105833.86+055440.2 & 1.3222 & 4 & 0.62 & 0.21 &  37.8 &10.3 & 5167 & 410 \\    
SDSSJ110108.00+043849.6 & 1.2516 & 4 & 0.84 & 0.10 &  25.7 & 3.4 & 4823 & 269 \\ [0.5ex]
\hline
\end{tabular*}}
\caption{\small{The sample of 19 quasars with $p_{\rm{lin}} \geq 0.6\%$.
Column~1 gives the quasar SDSS name, column~2 the redshift
$z$, column~3 the quasar group (Fig.~\ref{stru}), columns~4 and 5 the
polarization degree $p_{\rm{lin}}$ and its error $\sigma_p$, columns~6 and 7 the
polarization position angle $\psi$ and its error $\sigma_{\psi}$,
columns~8 and 9 the MgII emission line FWHM and its error from \citet{Shen-et-al2011}.}}
\label{poltab2}
\end{minipage}
\end{table}

\chapter[Quasar radio polarizations align with LQGs]{Quasar radio polarizations align with LQG major axes}
\label{Ch:PH-2}
The co-evolution of the spins of galaxies with their surrounding
cosmic web has been theoretically established for some time
(e.g., \citealt{White1984}; \citealt{Heavens-Peacock1988};
\citealt{Catelan-Theuns1996}; \citealt{Lee-Pen2000} and
\citealt{Hirata-Seljak2004}; see \citealt{Joachimi-et-al2015} for a
recent review). It is predicted that the spin of the dark-matter halo
as well as the spin of the central supermassive black hole (SMBH) of
a galaxy do not point in random directions of space, but instead
point towards particular directions that are determined by the
geometry of the neighbouring cosmic web
(e.g., \citealt{Aragon-Calvo-et-al2007};
\citealt{Codis-et-al2012}; see \citealt{Kiessling-et-al2015} for a
recent review).
These predictions have been supported by numerous observations
(e.g., \citealt{West1994}; \citealt{Pen-Lee-Seljak2000},
\citealt{Lee-Pen2001}, \citealt{Faltenbacher-et-al2009},
\citealt{Jones-et-al2010}; \citealt{Li-et-al2013},
\citealt{Tempel-Libeskind2013}, \citealt{Zhang-et-al2013}; see
\citealt{Kirk-et-al2015} for a recent review). Unfortunately, relying on
the apparent shapes of the galaxies that are used as a proxy of their
spin axes, these studies are limited to the low redshift ($z < 1$)
Universe because the sources need to be resolved to assess their
orientations with respect to their environment.

However, we have shown in the previous chapter that the orientation
of the optical polarization vectors of quasars are correlated to the
orientations of the large quasar groups (LQG) to which they belong,
at redshift $\sim 1.3$. This analysis was carried out within the two
large quasar groups called the CCLQG (with 34 members) and the
Huge-LQG (with 73 members) identified by \citet{Clowes-et-al2013}
and references therein.
Given the correlation at optical wavelengths between the electric vector
position angle and the morphological axis of quasar (e.g.,
\citealt{Smith-et-al2004}), we interpreted our observations as resulting
from the alignment of the spin axes of the quasars with the orientation
of the large-scale structure to which they belong, which is assumed to
be traced by the large quasar groups.

While these alignments take place at very large scales ($\geq 100\,
{h}^{-1}\rm{Mpc}$), they may reflect the recognized co-evolution of the
orientations of the spins of galaxies with the properties of their
surrounding large-scale structure. The study of the polarization of
quasars could then constitute an additional probe of the co-evolution
discussed above because it does not suffer from the observational
constraints inherent to studies relying on the apparent morphologies
of galaxies (\citealt{Kirk-et-al2015}).
Moreover, studies involving quasars can be made at high redshift.
Therefore, it is important to confirm the correlations that involve the
polarization position angles of quasars and the characteristics of
their large-scale environments, traced here by the large quasar groups.
To this end, instead of measuring the polarization of all quasars
belonging to a given LQG, we collect polarization measurements
of quasars that belong to a sample of LQGs and compare their
polarization vectors to the orientations of the groups to which the
quasars belong.

\medskip

In this chapter, we thus consider a large sample of LQGs built from the
Sloan Digital Sky Survey (SDSS) Data Release (DR) 7 quasar catalogue
in the redshift range $1.0 - 1.8$.
For 86 quasars that are embedded in sufficiently rich LQGs, we collect
radio polarization measurements with the goal to study possible
correlations between quasar polarization vectors and the major axes of
their host LQGs.
Assuming the radio polarization vector perpendicular to the quasar spin
axis (or the central SMBH spin axis), we found that the quasar spin is
preferentially parallel to the LQG major axis inside LQGs that have at
least 20 members.
This result independently supports our observations at optical
wavelengths (Chapter~\ref{Ch:HBPS2014}). We additionally found that
when the richness of an LQG decreases, the quasar spin axis becomes
preferentially perpendicular to the LQG major axis and that no correlation
is detected for quasar groups with fewer than 10 members.
This chapter contains the details of our
analysis also presented in \citet{Pelgrims-Hutsemekers2016}.

\section{Data samples and premises}
\label{sec:DataSample-premise}
The CCLQG and the Huge-LQG have first been identified with a
hierarchical clustering method in the quasar catalogue of the SDSS
DR7. Their detection is supported by spatial
coincidence with Mg II absorbers (\citealt{Clowes-et-al2013}) and
with a temperature anomaly in the cosmic microwave background
(\citealt*{EneaR-Cornejo-Campusano2015}).
These large quasar groups have been independently confirmed
(\citealt{Nadathur2013}; \citealt{Einasto-et-al2014};
\citealt{Park-et-al2015}) using other friends-of-friends algorithms
(e.g., \citealt{Huchra-Geller1982}).
However, it is worth mentioning that these studies have revisited
the original claim which stated that these quasar groups challenge
the cosmological principle. It turned out that, following these studies,
the cosmological principle is still safe.

In particular, \citet{Einasto-et-al2014} used a reliable subset of the
SDSS DR7 quasar catalogue to perform their analysis. Their sample
of quasars is defined in the redshift range of $z=1.0 - 1.8$, in the
window of the sky determined by
$\lambda_{SDSS} \in [-55^\circ,\, 55^\circ]$ and
$\eta_{SDSS} \in [-32^\circ,\, 33^\circ],$ where $\lambda_{SDSS}$
and $\eta_{SDSS}$ are the SDSS latitude and longitude,
respectively\footnote{https://www.sdss3.org/dr8/algorithms/surveycoords.php},
and with an additional cut in $i$--magnitude, $i \leq 19.1$.
For this sample of 22\,381 quasars, which we call the Einasto sample,
they produced publicly
available\footnote{http://cdsarc.u-strasbg.fr/viz-bin/qcat?J/A+A/568/A46}
catalogues of LQGs that are found with a friends-of-friends
algorithm using different values of the linking length ($LL$).

\begin{figure}
\centering
\begin{minipage}{.8\linewidth}
\centering
\includegraphics[width=\columnwidth]{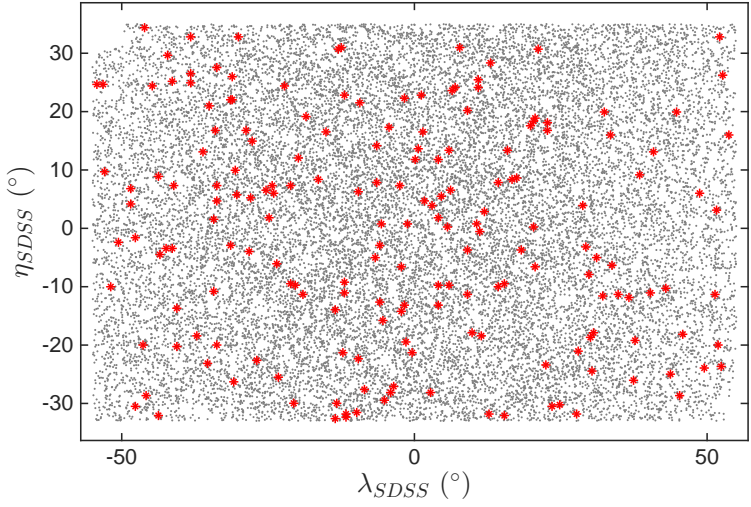}
\caption{\small{Sky map in SDSS coordinates of the 22\,381 quasars
contained in the Einasto sample. The 185 quasars for which we retrieved
radio polarization measurement in the JVAS/CLASS 8.4-GHz surveys
are highlighted in red.}}
\label{fig:MapEinasto-JVAS}
\end{minipage}
\end{figure}
We used the sample of large quasar groups built by
\citet{Einasto-et-al2014}, focusing on those groups defined by
$LL = 70\, {h}^{-1}\rm{Mpc}$. This choice is motivated by two different
reasons. First, in \citet{Hutsemekers-et-al2014}, the alignment of
quasar morphological axes with the large-scale structures was found
in the Huge-LQG and the CCLQG. These two groups are retrieved in
the Einasto sample by using a friends-of-friends algorithm with that
value of the linking length.
Second, the richness (the number of members) of the LQGs has to be
sufficiently high to allow reliable determination of their geometrical
properties. For values of the linking length below
$70\, {h}^{-1}\rm{Mpc}$, there are at most a few LQGs with
a richness above $10$ and none above $20$. For $LL = 70\,
{h}^{-1}\rm{Mpc}$ there are several tens of rich LQGs. Above that
linking length, the percolation process occurs (\citealt{Nadathur2013},
\citealt{Einasto-et-al2014}). The large quasar groups stop to grow
by including neighbouring sources and instead merge among themselves.
The number of independent rich large quasar groups thus starts to decrease
rapidly for $LL \gtrsim 75\, {h}^{-1}\rm{Mpc}$.

We searched for polarization measurements of quasars that belong to
the Einasto sample to compare the polarization position angles to the
position angles of the groups. At optical wavelengths, there are
unfortunately too few LQG members with polarization
measurements in the compilation of \citet{Hutsemekers-et-al2005}.
Since there is a correlation between the orientation of the radio
polarization vector and the axis of the system similar to what occurs
at optical wavelengths \citep{Rusk-Seaquist1985}, we decided to consider
quasar polarization measurements from the JVAS/CLASS 8.4-GHz surveys
compiled by \citet{Jackson-et-al2007}, adopting their quality criterion on the
polarized flux ($\geq 1\, \rm{mJy}$). The choice of this sample is further
motivated below.

\medskip

For the Einasto sample, we therefore searched for JVAS/CLASS polarization
measurements of quasars with a search radius of $0.5$ arcsec.
As in Chapter~\ref{Ch:PH-1}, we constrained our sample even more by
only retaining polarization measurements if the condition
$\sigma_\psi \leq 14^\circ$ was satisfied, where $\sigma_\psi$ is the
error on the position angle of the polarization vector
(Eq.~\ref{eq:sigma_PA}).
After verifying the reliability of the identifications, 185 objects were found.
We show them in Fig.~\ref{fig:MapEinasto-JVAS} along with all the quasars
from the Einasto sample.
For these 185 sources, the median of $\sigma_\psi$ is $1.7^\circ$.
With $LL = 70\, {h}^{-1} \rm{Mpc}$, 30 of the 185 quasars are found to
be isolated sources and 155 belong to quasar groups with richness
$m\geq2$. To determine meaningful morphological position angles for the
LQGs, we considered at least five members as necessary.
The 86 quasars belonging to 83 LQGs with richness $m\geq 5$ constitute
our core sample in which we investigate the possible correlation between
the quasar polarization vectors and the LQG orientations.
According to \citet{Shen-et-al2011}, these sources are all radio-loud
quasars and have SMBH masses in the ranges
$\log_{10} \left( \frac{M_{\rm{BH}}}{M_\odot} \right) \in [8.24, \, 10.12]$
and bolometric luminosities in the range
$\log_{10} L_{\rm{bol}} \in [45.9,\, 47.7],$ where $L_{\rm{bol}}$ is in
$\rm{erg} \, \rm{s}^{-1}$.

\medskip

The principal contamination source of the polarization position angle
measurements at radio wavelengths is the Faraday rotation, which takes
place in our Galaxy, but also at the source. The Faraday rotation is
undesired in our study because it smears out any intrinsic correlation of
the polarization vectors with other axes.
\citet{Jackson-et-al2007} and \citet{Joshi-et-al2007} proved the reliability
of the JVAS/CLASS 8.4-GHz surveys against any sort of biases and also
showed that the Faraday rotation at this wavelength is negligible along
the entire path of the light, from the source to us. Given our conclusion in
Chapter~\ref{Ch:PH-1}, we verify that this is also the case for the quasar
sub-sample that we consider in this chapter.

\medskip

The radio polarization vector from the core of a quasar is expected to
be essentially perpendicular to the (projected) spin axis of its central
engine (e.g.,
\citealt{Wardle2013}; \citealt{McKinney-Tchekhovskoy-Blandford2013}).
The latter can also be traced by the radio-jet axis when it is observed
in the sub-arcsecond core of the quasar. The fact that radio polarization
and jet axis are preferentially perpendicular supports the view that
the radio polarization vectors can be used to trace the quasar spin axes.
Radio polarization vectors and radio jets are known to be essentially
perpendicular (\citealt{Rusk-Seaquist1985}; \citealt{Saikia-Salter1988};
\citealt{Pollack-Taylor-Zavala2003}; \citealt{Helmboldt-et-al2007}).
This holds for the sources contained in the JVAS/CLASS 8.4-GHz
surveys, as shown by \citet{Joshi-et-al2007}.
We verified that this is also true for the sub-sample that we use here.

\begin{figure}[h]
\centering
\begin{minipage}{0.6\textwidth}
\mbox{}\\[-\baselineskip]
\begin{center}
\includegraphics[width=\columnwidth]{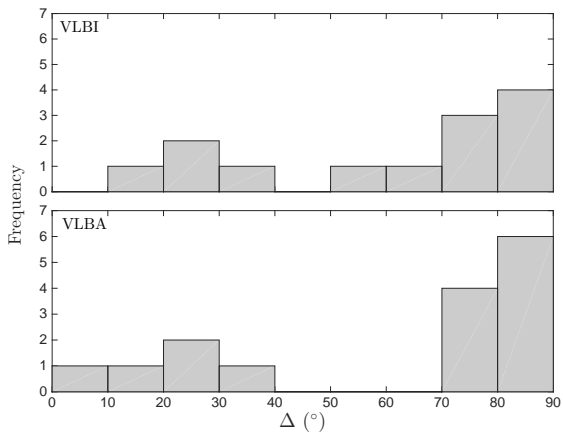}
\end{center}
\end{minipage}\hspace{.01\textwidth}
\begin{minipage}{.25\textwidth}
\mbox{}\\[-1.3\baselineskip]
\caption{\small{Distributions of the acute angles between the radio
polarization vectors and the jet axes of the 13 quasars from the VLBI
compilation of \citet{Joshi-et-al2007} (\textit{top}) and of the 15
quasars from the VLBA sample of \citet{Helmboldt-et-al2007}
(\textit{bottom}), with correction for SDSS J122127.04+441129.7..}}
\label{fig:DPA-vlbi-jet}
\end{minipage}
\end{figure}
For our sample of $41$ quasars\footnote{The cut at $m=10$ is justified
below.} in LQGs with $m\geq10$, we searched for jet axis information in
the VLBI compilation of \citet{Joshi-et-al2007} and in the VLBA sample of
\citet{Helmboldt-et-al2007}. In these catalogues, we found 13 and 15
sources with jet position angle measurements, respectively\footnote{These
two sub-samples are not independent. For the $9$ objects in common, the
jet position angles agree within $\sim 20^\circ$, except for one source that
shows an offset of about $72^\circ$ (SDSS J122127.04+441129.7). After
inspecting the VLBA maps \citep{Helmboldt-et-al2007}, we realized that the
sign of the position angle of the VLBA jet needs to be changed for this
object.}.
For these objects, we computed the acute angle between the polarization
vector and the jet axis.
The distribution of these angles, shown in Fig.~\ref{fig:DPA-vlbi-jet},
demonstrates that even within our small sample the radio polarizations
show a strong tendency to be perpendicular to the radio jets.
Therefore, we safely conclude that in our sample the radio polarization
vectors of the quasars trace the spin axes of the quasars and thus of
their central supermassive black holes (SMBH).
Any correlation found with the polarization vectors could then be
interpreted in terms of the quasar spin axes.

\subsection{Faraday contamination}
\label{subsec:FaradayRotation}
Although \citet{Jackson-et-al2007} and \citet{Joshi-et-al2007} stated
that the Faraday rotation that take place either in our Galaxy or within
the source can be neglected for quasars observed at $8.4\,\rm{GHz}$,
it is important to verify that these contaminations are indeed negligible
for our sample.

\subsubsection{Faraday rotation at the source level}
It was stated in \citet{Jackson-et-al2007} and \citet{Joshi-et-al2007}
that the Faraday rotation of the polarization vectors that takes
place within the source can be neglected for quasars observed at
$8.4\,\rm{GHz}$.
However, according to \citet{Pollack-Taylor-Zavala2003},
\citet{Zavala-Taylor2004} and \citet{Helmboldt-et-al2007}, some
active galactic nuclei can harbour very strong magnetic fields that
imply very high rotation measures of the order of $\sim 500$ to few
thousand $\rm{rad \, m}^{-2}$.
At $8.4\,\rm{GHz}$, this would lead to high Faraday rotation, up to
$\sim 30^\circ$, which would be dramatic for our purpose.
Nevertheless, Fig.~\ref{fig:DPA-vlbi-jet} shows that the polarization
vectors at $8.4\,\rm{GHz}$ have a significant tendency to be
perpendicular to the jet axes. The Faraday rotation taking place at
the source might explain the observed dispersion in
Fig.~\ref{fig:DPA-vlbi-jet} but is not high enough to smear out the
expected correlation between the radio-jet axes and the polarization
vectors of the quasars.
Moreover, an additional scatter in the polarization-vector--jet-axis
correlation might arise due to curved jets
(\citealt{McKinney-Tchekhovskoy-Blandford2013}).
\begin{figure}[h]
\begin{center}
\begin{minipage}{.8\linewidth}
\includegraphics[width=\columnwidth]{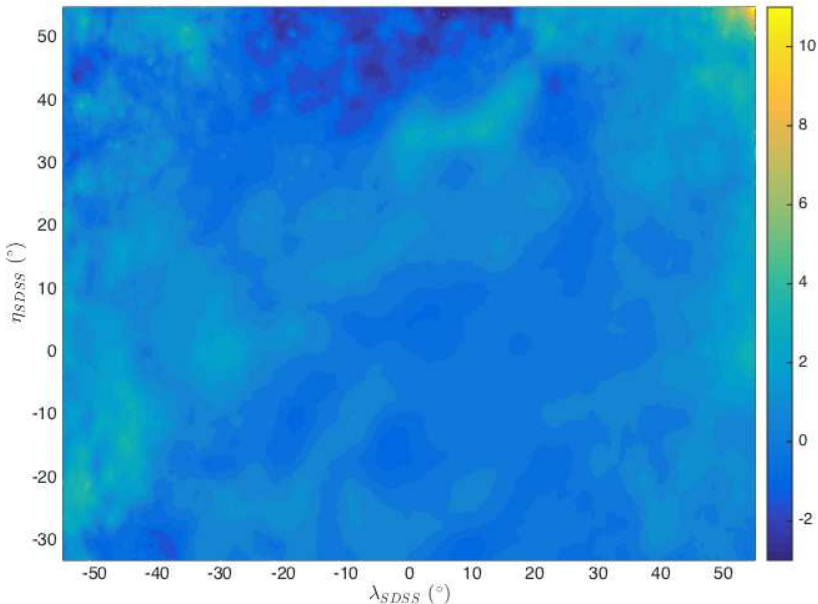}
\caption{\small{Contour plot in SDSS coordinates of the expected
Faraday rotation angles at $8.4\,\rm{GHz}$ within the sky window
of the Einasto sample, extracted from the whole sky rotation measure
map of \citet{Oppermann-et-al2015}.
Colour-codded rotation angles are in degrees.}}
\label{fig:FaradRot_EinastoWindow}
\end{minipage}
\end{center}
\end{figure}

\subsubsection{Galactic Faraday rotation}
We used the all-sky Galactic Faraday map produced by
\citet{Oppermann-et-al2015} to verify that the Galactic Faraday rotation
can be neglected in our analysis. From their map of rotation measures, we
extracted the whole sky window covered by the Einasto sample. We show
in Fig.~\ref{fig:FaradRot_EinastoWindow} the map of the Faraday rotation
angles expected at $8.4\,\rm{GHz}$. For the entire window, the distribution
of the Faraday rotation angles that is due to the Galactic magnetic field has
a mean of $0.6^\circ$ and a standard deviation of about $1^\circ$.

For the source locations of our sample with JVAS/CLASS polarization
measurements (the 185 sources), the contamination is even lower with
a distribution having a mean of $0.5^\circ$ and a standard deviation of
$0.6^\circ$.
We conclude that the Galactic Faraday rotation can be neglected
because the rotation angles are within the error bars of the polarization
data.

\section{Position angles of LQGs}
\label{sec:MPA_det}
To define the position angle of an LQG, we can proceed in two ways. We
can consider a group of quasars as a cloud of points on the celestial sphere,
or we can take the three-dimensional comoving positions of the sources into
account.
For either approach we determine the morphological position angle (MPA) of
an LQG through the eigenvector of the inertia tensor corresponding to the
major axis of the set of points.

\begin{figure}[h]
\centering
\begin{minipage}[t]{.69\textwidth}
\mbox{}\\[-\baselineskip]
\begin{center}
\includegraphics[width=.95\columnwidth]{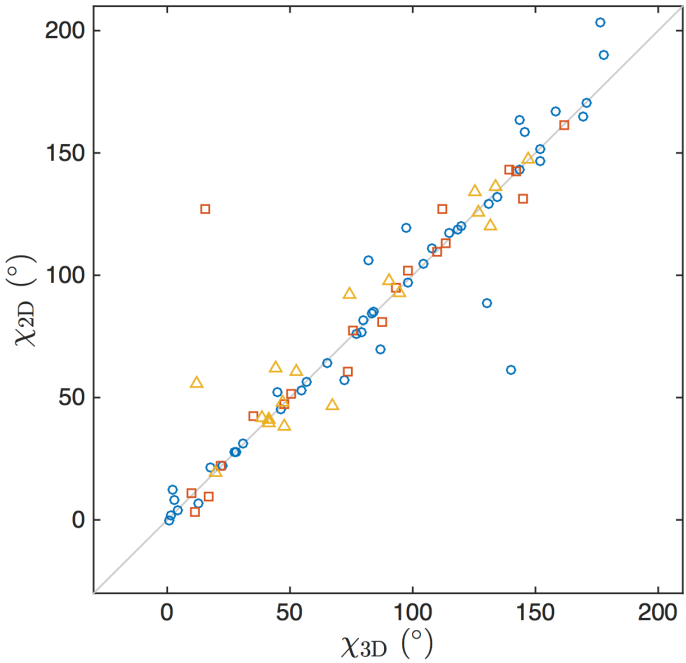}
\end{center}
\end{minipage}
\begin{minipage}[t]{.25\textwidth}
\mbox{}\\[-\baselineskip]
\\[-1.9ex]
\caption{\small{Morphological position angles (in degrees) of the
large quasar groups determined with the two-dimensional
method ($\chi_{2\rm{D}}$) as a function of those determined with
the three-dimensional one ($\chi_{3\rm{D}}$) for the 83 LQGs
with $m \geq 5$. Circles, squares and triangles show LQGs with
richness $m < 10$, $10 \leq m < 20$ and $m \geq 20$,
respectively. Some values of $\chi_{2\rm{D}}$ have been adjusted
by $180^\circ$ for clarity.}}
\label{fig:mPAa-vs-mMPA3}
\end{minipage}\hspace{.06\textwidth}
\end{figure}

\medskip

For the two-dimensional approach, the quasar positions are projected
onto the plane tangent to the celestial sphere. The orientation of the
two-dimensional cloud of points is determined by computing its inertia
tensor, assuming quasars to be unit point-like masses. The position
angle of the eigenvector corresponding to the most elongated axis
defines the morphological position angle of the large quasar group.
We checked that this method returns position angles that are in excellent
agreement (within 1 degree) with those obtained with an orthogonal
regression \citep{Isobe-et-al1990}.
The latter method was used in the previous chapter to define the position
angles of the quasar groups.

For the second approach, the three-dimensional comoving positions (see
Section~\ref{substat_nv-param}) of the quasars are used to determine the
geometrical shape of an LQG by considering its tensor of inertia, assuming
quasars to be unit point-like masses. The decomposition of the tensor in
terms of its eigenvalues and eigenvectors allows us to define the three
principal axes of the group. A simple projection on the plane orthogonal to
the line of sight of the major axis defines the position angle of the large
quasar group.

As a result of the inclination of the system with respect to the line of sight,
the MPAs determined by the two methods may differ. In our case, we found
that they generally agree well. Indeed, we show in
Fig.~\ref{fig:mPAa-vs-mMPA3} a comparison of the position angles of
the LQGs that we obtained by the two- and three-dimensional procedures.
While the two methods often return MPAs that agree well, these quantities
can be largely different owing to the apparent shape and to the inclination
of the system w.r.t. the line of sight.
Because the two methods return similar results, we base our discussion
on the three-dimensional approach, which is more physically motivated.
In our calculation, we assume the same cosmological model
as in \citet{Einasto-et-al2014}, that is, a flat $\Lambda$CDM Universe with
$\Omega_M = 0.27$. The reduced Hubble parameter ${h}$ acts only as a
global scaling factor which is irrelevant for our purpose.

\medskip

For either approach, the morphological position angle of each large
quasar group is derived at the centre of mass of the group. In general,
a quasar for which we retrieved radio polarization measurement is
angularly separated from the centre of mass of its hosting group. Hence,
the acute angle between the two orientations (the polarization vector and
the projected major axis) depends on the system of coordinates that is
used.
To overcome this coordinate dependence, we used parallel transport on
the celestial sphere to move the projected eigenvector from the centre of
mass of the group to the location of the quasar with polarization data.
By introducing $\psi$ for the polarization position angle of a quasar and
$\chi$ for the (parallel-transported) position angle of the LQG to which it
belongs, we compute the acute angle between the two orientations as
\begin{equation}
\Delta_{\psi\chi}=90^\circ - | 90^\circ - |\psi - \chi| | \, \rm{.}
\label{eq:DeltaPsiChi}
\end{equation}
The use of the parallel transport before evaluating the acute angle leads
to coordinate-independent statistics.
Both $\psi$ and $\chi$ are defined in the range $0^\circ - 180^\circ$
and are computed in the east-of-north convention.

\section[Correlation between polarization and LQG PAs]{Correlation between polarization and LQG position angles}
\label{sec:EVPA-MPAcorrelation}
\begin{figure}[h]
\centering
\begin{minipage}[t]{0.57\textwidth}
\mbox{}\\[-\baselineskip]
\begin{center}
\includegraphics[width=\columnwidth]{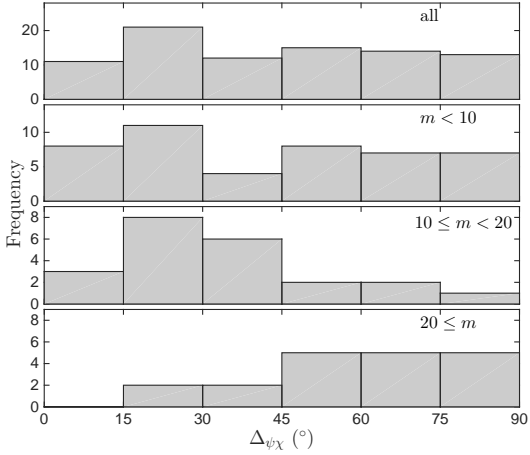}
\end{center}
\end{minipage}
\begin{minipage}[t]{0.22\textwidth}
\mbox{}\\[-\baselineskip]
\\[-.5ex]
\caption{\small{Histogram of the distribution of $\Delta_{\psi\chi}$
(in degrees) for the 86 quasars with polarization--LQG position-angle
measurements (\textit{top}) and for the three sub-samples having
richness $m < 10$, $10 \leq m < 20$ and $m \geq 20$.}}
\label{fig:histDPA_3Dm}
\end{minipage}
\end{figure}
In Fig.~\ref{fig:histDPA_3Dm} (top) we show the distribution of
$\Delta_{\psi\chi}$ for the 86 quasars with polarization and LQG
position-angle measurements. The distribution of the full sample (top)
does not show any departure from uniformity. The probability given by
a one-sample Kolmogorov-Smirnov (KS) test that the distribution is
drawn from a uniform parent distribution is $P_{\rm{KS}} = 88\,\%$.
Since the alignment of optical polarization vectors with LQG orientations
was found in very rich groups, and as the accuracy of the position angle
of an LQG most likely depends on its richness, we divided our sample
into three sub-samples with $m < 10$ (45 objects), $10 \leq m < 20$
(22 objects) and $m \geq 20$ (19 objects).
For the smallest LQGs ($m < 10$), the distribution of $\Delta_{\psi\chi}$
does not show any departure from uniformity. The probability given by a
one-sample KS test that the distribution is drawn from a uniform parent
distribution is $P_{\rm{KS}}= 54 \,\%$.
However, for the larger groups, a dichotomy is observed between the two
sub-samples.
The polarization vector of a quasar belonging to a very rich LQG
($m\geq20$) appears preferentially perpendicular to the projected major
axis of the group ($\Delta_{\psi\chi}>45^\circ$), whereas the polarization
vector of a quasar belonging to an LQG with medium richness
($10 \leq m < 20$) seems to be preferentially parallel
($\Delta_{\psi\chi} < 45^\circ$).
A two-sample KS test tells us that the probability that the two parts of the
sample with $10 \leq m < 20$ and $m \geq 20$ have their distributions of
$\Delta_{\psi\chi}$ drawn from the same parent distribution is $0.05\,\%$.

For the 19 data points of the sub-sample of LQGs with $m \geq 20$,
15 show $\Delta_{\psi\chi} > 45^\circ$. The cumulative binomial probability
of obtaining 15 or more data points with
$\Delta_{\psi\chi} > 45^\circ$ by chance is $P_{\rm{bin}} = 0.96\,\%$.
Of 22 data points of the sub-sample of LQGs with $10 \leq m < 20$,
17 show $\Delta_{\psi\chi} < 45^\circ$, which gives the cumulative binomial
probability $P_{\rm{bin}} = 0.85\,\%$.
These results indicate a correlation between the position angle of the major
axis of an LQG and the radio polarization position angle of its members in
rich ($m\geq10$) quasar groups.

\begin{figure}[h]
\centering
\begin{minipage}[t]{0.57\textwidth}
\mbox{}\\[-\baselineskip]
\begin{center}
\includegraphics[width=\columnwidth]{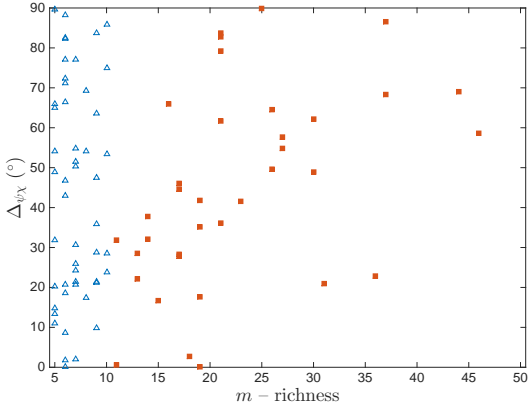}
\end{center}
\end{minipage}
\begin{minipage}[t]{0.22\textwidth}
\mbox{}\\[-\baselineskip]
\\[-.5ex]
\caption{\small{$\Delta_{\psi\chi}$ versus the richness $m$ of the LQGs.
For $m\leq10$ and $m\geq 11$, symbols are triangles and filled squares,
respectively.}}
\label{fig:DPA3D-vs-m}
\end{minipage}
\end{figure}

The dichotomy between the two sub-samples of LQGs with $10\leq m < 20$
and $m\geq20$ is also illustrated in Fig.~\ref{fig:DPA3D-vs-m}, where we
plot the $\Delta_{\psi\chi}$ of each quasar against the richness of its
corresponding LQG.
Surprisingly, for $m \geq 11$, we even see a possible linear correlation
of $\Delta_{\psi\chi}$ with the richness of the large quasar groups.
A Spearman correlation test on the pairs $\Delta_{\psi\chi} - m$ gives
a rank-order correlation coefficient of $0.54$ with a probability of obtaining
this result by chance of $0.08\,\%$.  
For $m < 10$, there is no specific trend of $\Delta_{\psi\chi}$ with the
richness in agreement with the distribution seen in
Fig.~\ref{fig:histDPA_3Dm}.
To understand whether this lack of correlation for $m < 10$ is due to
the uncertainty on the determination of the major axis position angle
of the LQG, as might naively be expected for the smallest groups, we
estimated the confidence interval of the morphological position angle
using the bootstrap method described in the sub-section below.
Keeping only MPAs for which the half-width confidence interval is
lower than $20^\circ$ (27 objects out of 45), the distribution of
$\Delta_{\psi\chi}$ remains uniform with $P_{\rm{KS}} = 72\,\%$.
The absence of alignments in quasar groups with richness $m < 10$
is therefore likely to be real.

On the other hand, the uncertainties, both on the position angles of
the major axes of the LQGs and on the polarization position angles,
cannot account for the correlations that we report.
The introduction of poorly defined orientations in our analysis can
only scramble an existing correlation.
The same argument applies to the contamination of the polarization
position angles by Faraday rotation, which was found to be negligible
in the previous section.

\medskip

In summary, our analysis shows that the quasars that belong to very
rich ($m\geq20$) large quasar groups have polarization vectors
preferentially perpendicular to the projected major axis of their hosting
LQGs.
The polarization vectors then become more often parallel to the LQG
axes when $m$ decreases before no correlation is observed for the
smallest ones ($m<10$).

\medskip

In Table~\ref{tab:dataTable}, we summarize the data for the 41 quasars
hosted in LQGs with $m \geq 10$. For each quasar for which we collected
polarization measurements, we list its SDSS name, its redshift, the
position angle of its polarization vector, the identification index of
the LQG to which it belongs (following the numbering of
\citet{Einasto-et-al2014}), the richness of the group and the position angle
of the projected major axis (MPA).
In Fig.~\ref{fig:skyPlot} we show the projection on the sky of the low and
high richness parts of the LQG sample defined with
$LL = 70 \,{h}^{-1}\rm{Mpc}$ that show correlations, i.e. with
$10 \leq m<20$ and $m\geq 20$, respectively.
We highlighted those for which we retrieved the radio polarization for at
least one member and show the orientations of the projected major axes
along with the polarization vectors of the quasars.

\subsection{Morphological position angles and their uncertainties}
\label{subsec:MPAs-and-err}
To quantify the uncertainties of the morphological position angles that
characterize the large quasar groups, we use the bootstrap method.
This procedure allows properly accounting for the circular nature of the
data (\citealt{Fisher1993}). For a given LQG of richness $m$, we produce
$N_{\rm{sim}}$ bootstrap LQGs with the same number of members,
allowing replicates.
The position angle of each bootstrap LQG is determined through the
inertia tensor procedure used for real groups.
As this procedure is not properly defined for groups resulting from only
one point, we take care in the generation of LQGs to avoid bootstrap
samples that consist of $m$ replications of the same source.
The probability that such a configuration occurs is $m^{1-m}$. Hence,
the rejection procedure can only affect poor large quasar groups.
We note that  even for those poor LQGs, the effect of theses
configurations on the evaluation of the confidence interval is negligible
($\ll 1^\circ$).

For a given group of quasars, we therefore collect a corresponding
distribution of $N_{\rm{sim}}$ estimates of the morphological position
angle. From this distribution, we evaluate the mean and its corresponding
confidence interval. For a distribution of axial-circular quantities $\chi_k$
such as the position angle of the LQG major axes, the mean is computed
as (\citealt{Fisher1993})
\begin{equation}
\bar{\chi} = \frac{1}{2}\, \arctan \left(\frac{\sum_{k=1}^{m} \sin 2\chi_k}{\sum_{k=1}^{m} \cos 2\chi_k} \right)\,.
\label{eq:meanPA}
\end{equation}
Since there is no proper definition of the standard deviation for axial-circular
data, we evaluate the confidence interval of the unknown mean at the
$100\,(1-\alpha)\%$ level as follows\footnote{This procedure
is based on the method described in (\citealt{Fisher1993}).}. We define
\begin{equation}
\gamma_k = \frac{1}{2} \arctan \left( \frac{\sin \left( 2(\chi_k - \bar{\chi}) \right)}{\cos \left( 2(\chi_k - \bar{\chi}) \right)} \right)
\label{eq:gamma_b}
\end{equation}
where $k=1,...,N_{\rm{sim}}$. The $\gamma_k$'s are defined in the
range $\left[-90^\circ,\, 90^\circ \right]$. Then we sort the $\gamma_k$ in
increasing order to obtain
$\gamma_{(1)} \leq ... \leq \gamma_{(N_{\rm{sim}})}$. If $l$ is the integer
part of $\left(N_{\rm{sim}} \alpha +1 \right)/2$ and 
$u=N_{\rm{sim}} - l$, the confidence interval for $\bar{\chi}$ is
$\left[ \bar{\chi} + \gamma_{(l+1)}, \, \bar{\chi} + \gamma_{(u)} \right]$.
We choose to compute the confidence interval of $\bar{\chi}$ at the $68\%$
confidence level. We define the half-width of the confidence interval as
$\rm{HWCI} = \left( \gamma_{(u)} - \gamma_{(l+1)} \right)/2$.
\begin{figure}[h]
\centering
\begin{minipage}[t]{.59\textwidth}
\mbox{}\\[-\baselineskip]
\begin{center}
\includegraphics[width=\columnwidth]{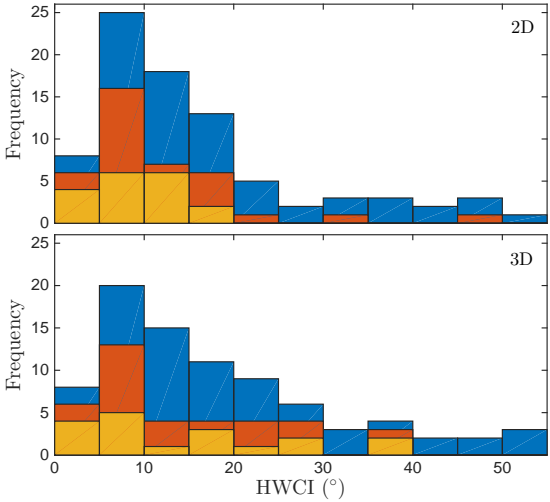}
\end{center}
\end{minipage}
\begin{minipage}[t]{0.25\textwidth}
\mbox{}\\[-\baselineskip]
\\
\caption{\small{Histograms of the $\rm{HWCI}$'s for the MPA
values obtained for the 83 independent LQGs with the two-
(\textit{top}) and three-dimensional (\textit{bottom}) approaches
(see Section~\ref{sec:MPA_det}).
The $\rm{HWCI}$'s are evaluated using the bootstrap method as
explained in the text. Histograms are for $m\geq5$ (blue), $m\geq10$
(orange) and $m\geq20$ (yellow).}}
\label{fig:HWCI-chi2D3D}
\end{minipage}
\end{figure}

Using the bootstrap method with 10\,000 simulations, we evaluated the
half-width confidence intervals for the 83 LQGs of our sample.
The distribution of $\rm{HWCI}$ corresponding to the three-dimensional
evaluation of the morphological position angles is shown in
Fig.~\ref{fig:HWCI-chi2D3D} for different cuts in richness and is compared
to those obtained from the two-dimensional procedure.
In general, the confidence intervals are lower when the two-dimensional
procedure is chosen to estimate the LQG position angles.
The highest values of the $\rm{HWCI}$'s ($\geq 20^\circ$) can in general
mostly be attributed to poor LQGs with $m<10$, as naively expected.

\section{Discussion}
\label{sec:InterpretationRLQG}
As discussed in Section~\ref{sec:DataSample-premise}, the radio
polarization vector of a quasar is expected to be essentially perpendicular
to the spin axis of its central engine.
The correlation that we found between the polarization vectors and the
major axes of the host large quasar groups might thus reflect an existing
link between the spin axes of the quasars and the major axes of the host
LQGs.

Our analysis independently supports the view that the spin axis of
quasars that belong to very rich LQGs are preferentially parallel to the
major axes of their hosting LQGs, as found in Chapter~\ref{Ch:HBPS2014}.
In addition, we found that the quasar-spin axes become preferentially
perpendicular to the LQG's major axes as the richness of the LQGs
decreases, down to $m\geq10$.

Regardless of the richness dependence that we discuss below, our
observation also suggests that the quasar spin axes have an intrinsic
tendency to align themselves within their host large quasar groups. The
observations of \citet{Jagannathan-Taylor2014} that radio jets in the GRMT
ELAIS N1 Deep Field align with each other over scales of
$50 - 75 \, {h}^{-1}\,\rm{Mpc}$ at redshift $z \gtrsim 1$ support our interpretation
\citep{Taylor-Jagannathan2016}.

\medskip

As discussed in the introduction, the fact that the spin axes of black
holes are found to align with their surrounding large-scale structures,
which are assumed to be traced by LQGs, could be understood in the
framework of the tidal torque theory if we accept that these predictions
can be extrapolated to larger scales.
However, and as far as we know, a richness dependence of the relative
orientation is not a predicted feature of this theory. We therefore explore
our data set to determine whether the richness dependence hides another
dependence that could be more physically motivated.

For this exploratory analysis, we relied on our core sample of LQGs with
$m \geq 10$ and for which we have radio polarization measurements for
at least one of their quasar member (Table~\ref{tab:dataTable}).
For LQGs with $m\geq 10$, the median of the diameters (the largest
separation between two members in groups) is about
$310\,{h}^{-1}\,\rm{Mpc}$ and the median separation of the closest
pairs is about $45\,{h}^{-1}\,\rm{Mpc}$.

\subsubsection{Richness dependence and quasar intrinsic properties}
N-body simulations have shown that the direction towards which the spin
axes of dark-matter haloes preferentially point relative to the large-scale
structure depends on their mass (\citealt{Codis-et-al2015}).
As the masses of the SMBH and the host dark-matter halo are thought to be
linked and as their spin axes might be aligned at high redshift
(\citealt{Dubois-Volonteri-Silk2014}), we searched for a possible dependence
of $\Delta_{\psi \chi}$ with the black hole mass.
Using the Spearman correlation test, we found that the SMBH masses (from
\citealt{Shen-et-al2011}) do not show any correlation with $\Delta_{\psi \chi}$
or with the richness of the host LQG.
The same conclusion is reached if we consider other quasar properties
reported in \citet{Shen-et-al2011}, such as the radio-loudness, the width of the
emission lines, or the redshift. There is thus apparently no hidden relation of
$\Delta_{\psi \chi}$ with quasar properties that could explain the richness
dependence of $\Delta_{\psi \chi}$.

\subsubsection{Richness dependence and LQG characteristics}
In Section~\ref{sec:MPA_det} we used the inertia tensors of the large
quasar groups to assign their orientations in the three-dimensional
comoving space. This resulted in fitting ellipsoids to the quasar systems.
In this subsection, we use the relative lengths of the principal axes of the
ellipsoids to characterize their shapes.

The richness of LQGs is of course correlated with their size, both in
comoving space and in the projection on the sky. The larger the comoving
volume of the ellipsoid or the larger the projected area on the celestial
sphere, the better the black hole spins align with the major axis of the
system. However, the correlations of the acute angles between the
polarization vectors and the projected LQG major axes with these
quantities are not stronger than with the richness of the groups that we
observed in Fig.~\ref{fig:DPA3D-vs-m}.

\medskip

Analyses studying the alignments of galaxy axes in the low-redshift
Universe found evidence for a dependence on the geometrical
properties of their surroundings. Namely, given that the studied
galaxies are in the neighbourhood of filaments or sheets (walls) of the
cosmic web, their spin axes tend to point preferentially in different
directions with respect to those defined by their cosmic environment
(e.g., \citealt{Tempel-Libeskind2013}; \citealt{Zhang-et-al2013};
\citealt{Li-et-al2013}; see \citealt{Kirk-et-al2015}). For instance, these
studies show that the spin axes of spiral galaxies in sheets are not
correlated to the vectors normal to the sheets but that they are
significantly correlated with the axes of the neighbouring filaments.
Furthermore, the importance of this correlation is likely to be dependent
on the distances of the galaxies to the filaments. These behaviours are
predicted from theory and are observed in simulated universes
(e.g., \citealt{Codis-et-al2012}).

Let $a$, $b$ and $c$ be the comoving lengths of the major, intermediate
and minor axes of the ellipsoids fitted to the LQGs, respectively.
By definition: $a \geq b \geq c$.
The geometries of the LQGs can be characterized through parameters
derived from the ratio of these lengths. We define $\tilde{b}=b/a$ and
$\tilde{c}=c/a$, which implies $1 \geq \tilde{b} \geq \tilde{c}$. We further
introduce the parameter $\mathcal{O}$ related to the oblateness of the
system. This parameter is defined as
\begin{equation}
\mathcal{O} = \frac{2 \tilde{b} - \left(1 + \tilde{c} \right)}{1 - \tilde{c}}
\label{eq:Oblateness}
\end{equation}
and takes values in the range $\left[ -1,\, 1\right]$.
$\mathcal{O}=0$ corresponds to the cases where the length of the
intermediate axis of the ellipsoid is precisely at the middle of the
lengths of the minor and major axes. This value corresponds to the
transition between prolate ($\mathcal{O} < 0$) and oblate
($\mathcal{O} > 0$) systems. The asymptotic configurations
$\mathcal{O} = -1$ and $\mathcal{O} =1$ correspond respectively to
(thick) filaments and to (thick) sheets, i.e. to cylindrical and disk-like
systems.

\begin{figure}[h]
\centering
\begin{minipage}[t]{.69\textwidth}
\mbox{}\\[-\baselineskip]
\begin{center}
\includegraphics[width=\columnwidth]{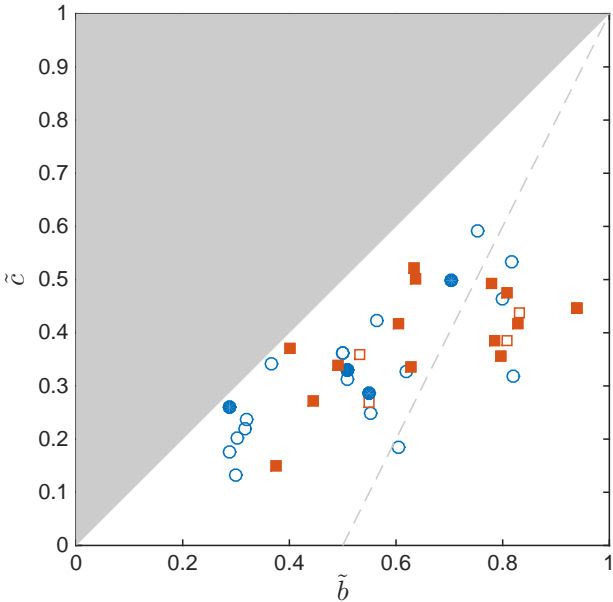}
\end{center}
\end{minipage}
\begin{minipage}[t]{.28\textwidth}
\mbox{}\\[-\baselineskip]
\\
\caption{\small{Morphological plane summarizing the geometrical
properties of the LQGs containing at least one polarized quasars
and having $m\geq10$. The grey region is forbidden by definition of
the parameters $\tilde{b}$ and $\tilde{c}$. The dashed-grey line is for
$\mathcal{O}=0$ and marks the transition between prolate and oblate
systems (see text). LQGs with $m$ below and
above $20$ are displayed by circles and squares, respectively.
Filled symbols are for those where $\Delta_{\psi \chi} > 45^\circ$.}}
\label{fig:MorphPlane_lqg}
\end{minipage}\hspace{.035\textwidth}
\end{figure}
We show in Fig.~\ref{fig:MorphPlane_lqg} the distribution in the plane
($\tilde{b},\, \tilde{c}$) of the 38 independent LQGs that contain at least
one quasar for which we retrieved radio polarization measurements and
that have richness $m\geq 10$.
As seen from this figure, there is no significant difference in the
geometrical properties between LQGs that have a richness
$10\leq m< 20$ and those that have $m\geq20$. Similarly, the large
quasar groups with $\Delta_{\psi \chi}> 45^\circ$ or
$\Delta_{\psi \chi} \leq 45^\circ$ do not distribute differently in the
morphological plane.

Inside this sample of 41 LQGs, we also searched for correlations
between the acute angles $\Delta_{\psi \chi}$ and the geometrical
characteristics of the LQGs such as oblateness or departure from
spherical or cylindrical symmetry and also considered the position
of quasars with radio polarization measurements relative to their host
large quasar groups. We did not find any convincing correlation with
these physical properties that could be hidden in the richness
dependence.

\medskip

A quantity that is related to the richness of an LQG and that could
have a better physical meaning is its density. We can naively define
the density of an LQG as $\rho = m/V$, where $V$ is the comoving
volume of its fitted ellipsoid. In our core sample of LQGs, there is
a relation between the richness and the density: the richer a large quasar
group, the lower its density. We then applied a Spearman correlation
test to the pairs $\Delta_{\psi \chi} - \rho$, which resulted in
rank-order correlation coefficient of $-0.50$ with a probability of
obtaining this result by chance of $0.19 \%$, if we consider all the
LQGs with $m \geq 11$ and at least one $\Delta_{\psi \chi}$
measurement (the sub-sample studied in
Section~\ref{sec:EVPA-MPAcorrelation}). SMBHs spin axes are thus
parallel to the host LQG axis when the density of the latter is low
and perpendicular when the density is high. The strength of this
correlation is similar to the strength of the $\Delta_{\psi \chi} - m$
correlation. The dependence between $\Delta_{\psi \chi}$ and the
richness $m$ could then reflect a dependence between $\Delta_{\psi
\chi}$ and $\rho,$ which might be easier to interpret.
\citet{Codis-et-al2015} have shown that the spin axes of the dark-matter
haloes of galaxies are preferentially parallel to their neighbouring
filaments when the halo masses are low or, equivalently, when the
density of their cosmic environment is low.
When the density of the environment increases, the halo spin axis starts
to avoid alignment with the filaments to finally point preferentially
perpendicular to them.
Our observations might thus support these predictions if we assume that,
at least at redshift $z\geq 1.0$, (\textit{i}) a similar behaviour can be
expected for the spin axes of the central SMBH of quasars, (\textit{ii}) the
density of an LQG reflects the density of the surrounding cosmic web,
(\textit{iii}) the LQGs can be used to trace the large-scale structures, and
(\textit{iv}) correlations that occur between the galaxy spin axes and
filaments also occur at larger scales between quasar-spin axes and LQG
major axes.

\section{Concluding remarks}
In this chapter, we have further studied the correlation between the
spin-axis orientations of the supermassive black holes at the centres
of quasars with the orientations of the large quasar groups to which
they belong. To do this, we considered a large sample of LQGs drawn
from a reliable sub-sample of the SDSS DR7 quasar catalogue in the
redshift interval $\left[1.0,\,1.8\right]$ to which a friends-of-friends
algorithm has been applied with a linking length of
$70\, {h}^{-1}\rm{Mpc}$ (\citealt{Einasto-et-al2014}).
We chose this value of the linking length because it allows to recover
the LQGs in which we found the correlation at optical wavelengths.
Because too few optical polarization data are available for the quasars
that belong to the Einasto sample, we used radio polarization
measurements from the JVAS/CLASS 8.4-GHz surveys
(\citealt{Jackson-et-al2007}). This sample is claimed to be free of biases
that might affect the polarization angles. Furthermore, the polarization
vectors at 8.4 $\rm{GHz}$ have a strong tendency to be perpendicular
to the spin axes of the SMBHs in the quasar cores.

We retrieved reliable polarization measurement for 185 quasars belonging
to the Einasto sample.
Among these, 86 were found in LQGs populated by at least five members,
a minimum value to determine reliable position angles.
To compare the position angles of the quasar polarization vectors with the
position angles of the system to which the quasars belong, we studied the
LQGs thought their inertia tensors in the three-dimensional comoving
space.

For rich quasar groups ($m \geq 20$), we found that the spin axes of
the supermassive black holes are preferentially parallel to the major
axes of their host large quasar groups.
This result adds weight to the previous finding at optical wavelengths
that in two large quasar groups the quasar spin axes (inferred from
polarizations) align with the group axes.
Combined with the initial discovery, our analysis indicates that the
alignments of the SMBH spins axes with the LQG major axes do not
depend on the quasar radio loudness.

Additionally, the use of a large sample of LQGs allowed us to probe
the alignments for a wide variety of quasar systems.  We unveiled a
surprising correlation: the relative orientations of the spin axes of
quasars with respect to the major axes of their host LQGs appear to
depend on the richness of the latter, or equivalently on the density
of objects. The spin axes of SMBHs appear preferentially parallel to
the major axes of their host LQGs when the latter are very rich (or
have a very low density), while the spin axes become preferentially
perpendicular to the LQG major axes when the richness decreases to
$m \geq 10$ or, equivalently, when the quasar density
increases to $1.5\, 10^{-5}\,(h^{-1}\, \rm{Mpc})^{-3}$. No correlation
is observed below this richness or above this density. Possible
interpretations were discussed in Section~\ref{sec:InterpretationRLQG},
but this intriguing feature needs to be confirmed.

Numerical simulations show that the spin axes of the dark-matter
haloes align with preferred directions of the cosmic web (e.g.,
\citealt{Codis-et-al2012}). At high redshift, the axes of the supermassive
black hole spin and of the dark-matter halo spin are predicted to align
(e.g., \citealt{Dubois-Volonteri-Silk2014}).
At high redshift, thus, the SMBH spin axes might preferentially align
with specific directions of the neighbouring large-scale structures as
suggested by \citet{Lagos-Padilla-Cora2009}. While
happening at much larger scales, our observations are in qualitative
agreement with this expectation which could, in principle, be tested from
simulation. Note that, if simulations catch properly the physics of structure
formation, no alignment of the black hole spins with their large-scale
structures are expected at low redshift due to consecutive mergers and
black hole coalescence (e.g., \citealt{Dubois-Volonteri-Silk2014}).

In agreement with the view that the richest large quasar groups at high
redshift most likely represent the progenitors of complexes or chains of
superclusters \citep{Einasto-et-al2014}, the correlation that we found
might be the high-redshift counterpart of the alignments at $z \sim 0$
of clusters of galaxies with the superclusters in which they are
embedded (e.g., \citealt{Einasto-Joeveer-Saar1980}; \citealt{West1999}).

The study of quasar polarization appears to be a promising tool to probe
the correlation of the spins of extragalactic objects across a very broad
range of redshift. It is mandatory, however, to understand better the
connection between the cosmic web and the large quasar groups that are
defined via friends-of-friends algorithms.


\begin{table*}
\begin{center}
\begin{minipage}{137.5mm}
\centering
\footnotesize{
\begin{tabular*}{137.5mm}{@{\extracolsep{\stretch{1}}}*{1}{lcrrrr}}
\hline
\\ [-1.5ex]

SDSS name & $z$ & PPA ($^\circ$) & LQG ID & $m$ & MPA ($^\circ$) \\
\hline
\\ [-1.5ex]

074809.46+300630.4	&	1.6942	&	     13.3		&	       7	&		18		&		15.9		\\
083740.24+245423.1 &	1.1254	&       58.0		&	   184	&		15	 	&		74.8		\\
090910.09+012135.6	&	1.0255	&     130.4		&	   410	&		10	 	&		36.2		\\
091204.62+083748.2	&	1.5388	&     103.3		&	   356	&		30	 	&		41.1		\\
091439.42+351204.5	&	1.0738	&     127.6		&	   267	&		14	 	&		95.5		\\
091641.76+024252.8	&	1.1019	&     162.0		&	   410	&		10	 	&		35.5		\\
091648.90+385428.1	&	1.2656	&     163.7		&	   396	&		17	 	&		12.1		\\
093105.33+141416.4	&	1.0997	&       56.2		&	   548	&		17	 	&		10.2		\\
094148.11+272838.8	&	1.3063	&     161.5 		&	   652	&		19		&		143.9	\\
095956.04+244952.4	&	1.4803	&       91.1		&	   844	&		19	 	&		49.3		\\
104552.72+062436.4	&	1.5091	&       25.7		&	 1199		&		21		&		126.6	\\
104831.29+211552.2	&	1.4810	&         0.2		& 	1215		&		14		&		142.3	\\
105431.89+385521.6	&	1.3662	&       82.3		&	1266		&		23		&		123.8	\\
112229.70+180526.4	&	1.0414	&     156.1		&	1437		&		21	 	&		94.3		\\
112814.74+225148.9	&	1.0809	&     112.7		&	1453		&		26	 	&		48.2		\\
113053.28+381518.6	&	1.7413	&       44.7		&	1507		&		13	 	&		22.7		\\
114658.29+395834.2	&	1.0882	&       89.3		&	1501		&		11	 	&		89.8		\\
115232.86+493938.6	&	1.0931	&     139.0		&	1643		&		10		&		162.7	\\
115518.29+193942.2	&	1.0188	&       16.0		&	1716		&		11	 	&		47.8		\\
120518.69+052748.4	&	1.2956	&	   166.9		&	1857		&		13	 	&		15.5		\\
121106.69+182034.2	&	1.5150	&     163.0		&	1925		&		21	 	&		19.1		\\
122127.04+441129.7	&	1.3444	&       57.9		&	2002		&		36	 	&		35.0		\\
122847.42+370612.0	&	1.5167	&       70.7		&	1996		&		17	 	&		98.5		\\
123505.80+362119.3	&	1.5983	&       54.2		&	1996		&		17	 	&		98.8		\\
123736.42+192440.5	&	1.5334	&     178.2		&	2011		&		37	 	&		66.5		\\
123757.94+223430.1	&	1.4175	&     155.0		&	2011		&		37	 	&		68.3		\\
123954.13+341528.8	&	1.1698	&      111.6		&	2151		&		19		&		111.7		\\
130020.91+141718.5	&	1.1060	&     125.5		&	2136		&		21	 	&		41.7		\\
133915.90+562348.1	&	1.4254	&     115.3		&	2287		&		31	 	&		94.3		\\
134208.36+270930.5	&	1.1898	&       18.0		&	2650		&		44		&		129.0	\\
134821.89+433517.1	&	1.1140	&         8.4		&	2563		&		10		&		113.5		\\
135116.91+083039.8	&	1.4398	&       97.5		&	2580		&		26	 	&		47.9		\\
135351.58+015153.9	&	1.6089	&     111.9		&	2714		&		21	 	&		14.7		\\
140214.81+581746.9	&	1.2673	&     137.9		&	2809		&		16	 	&		71.9		\\
142251.89+070025.9	&	1.4505	&     145.0		&	2813		&		19		&		109.9	\\
142330.09+115951.2	&	1.6127	&       15.3		&	2961		&		30		&		146.4	\\
145420.85+162424.3	&	1.2763	&      110.5		&	3198		&		10		&		139.0	\\
150124.63+561949.7	&	1.4670	&      166.2		&	3126		&		27	 	&		43.8		\\
150910.11+161127.7	&	1.1474	&        75.3		&	3219		&		46		&		134.0	\\
152037.06+160126.6	&	1.4669	&      103.1		&	3408		&		27	 	&		48.4		\\
152523.55+420117.0	&	1.1946	&      164.8		&	3375		&		25	 	&		74.8		\\

\hline
\end{tabular*}}
\caption{\small{Data for the $41$ quasars with $\Delta_{\psi \chi}$
measurements and belonging to rich ($m \geq 10$) LQGs. Column~1
gives the SDSS quasar name, Col.~2 the redshift $z$, Col.~3 the
polarization position angle (PPA) in degrees (East-of-North), Col.~4 the
identification index from the catalogue of \citet{Einasto-et-al2014} of the
LQG to which the quasar belong, Col.~5 the number of member in that
LQG $m$ and Col.~6 the position angle (in degrees, East-of-North) of the
major axis of the LQG when projected on the sky and parallel transported
at the location of the quasar for which we collected polarization
measurements.}}
\label{tab:dataTable}
\end{minipage}
\end{center}
\end{table*}

\begin{figure}
\begin{center}
\begin{minipage}{.8\linewidth}
\centering
\includegraphics[width=\columnwidth]{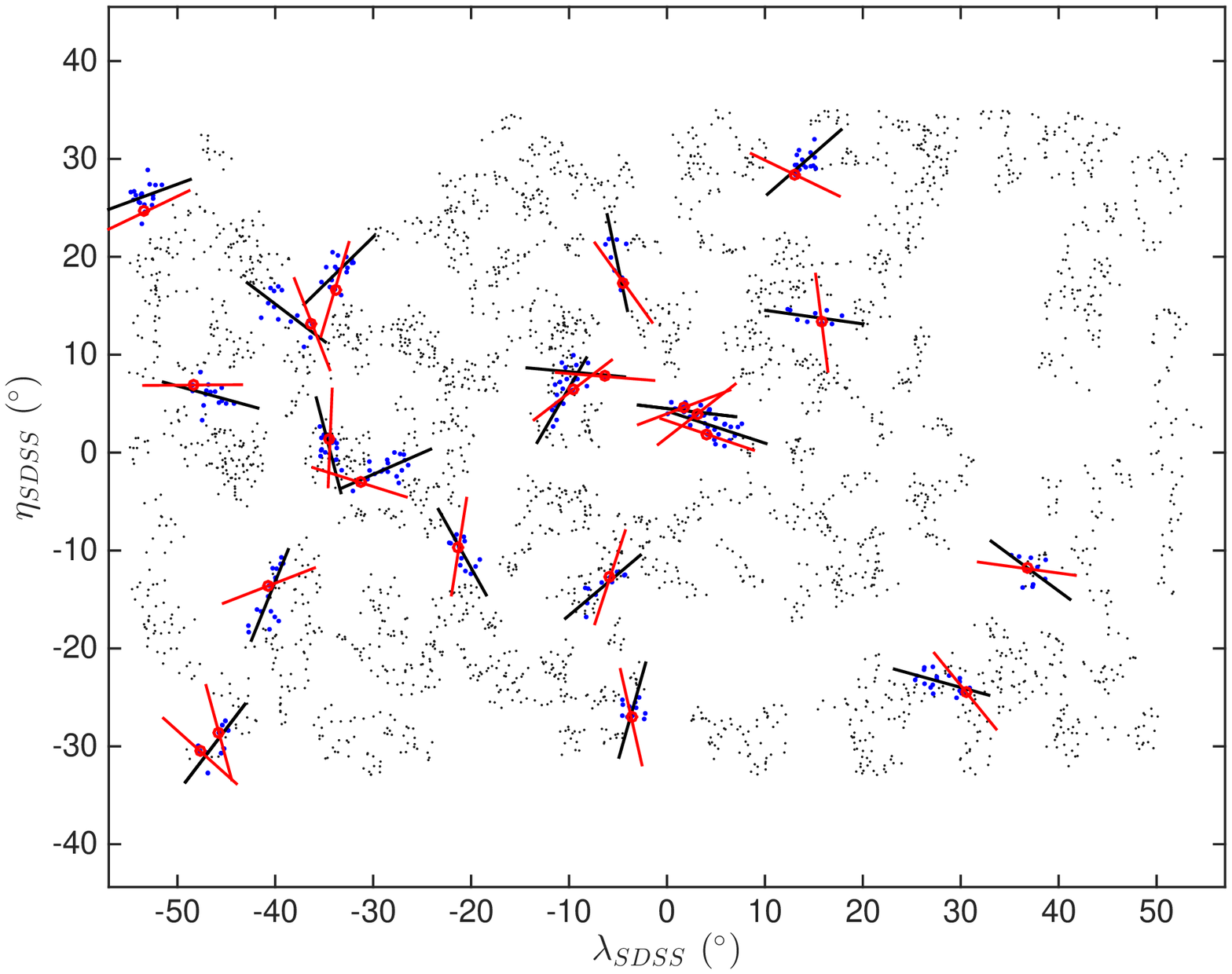} \\[-2.ex]
\includegraphics[width=\columnwidth]{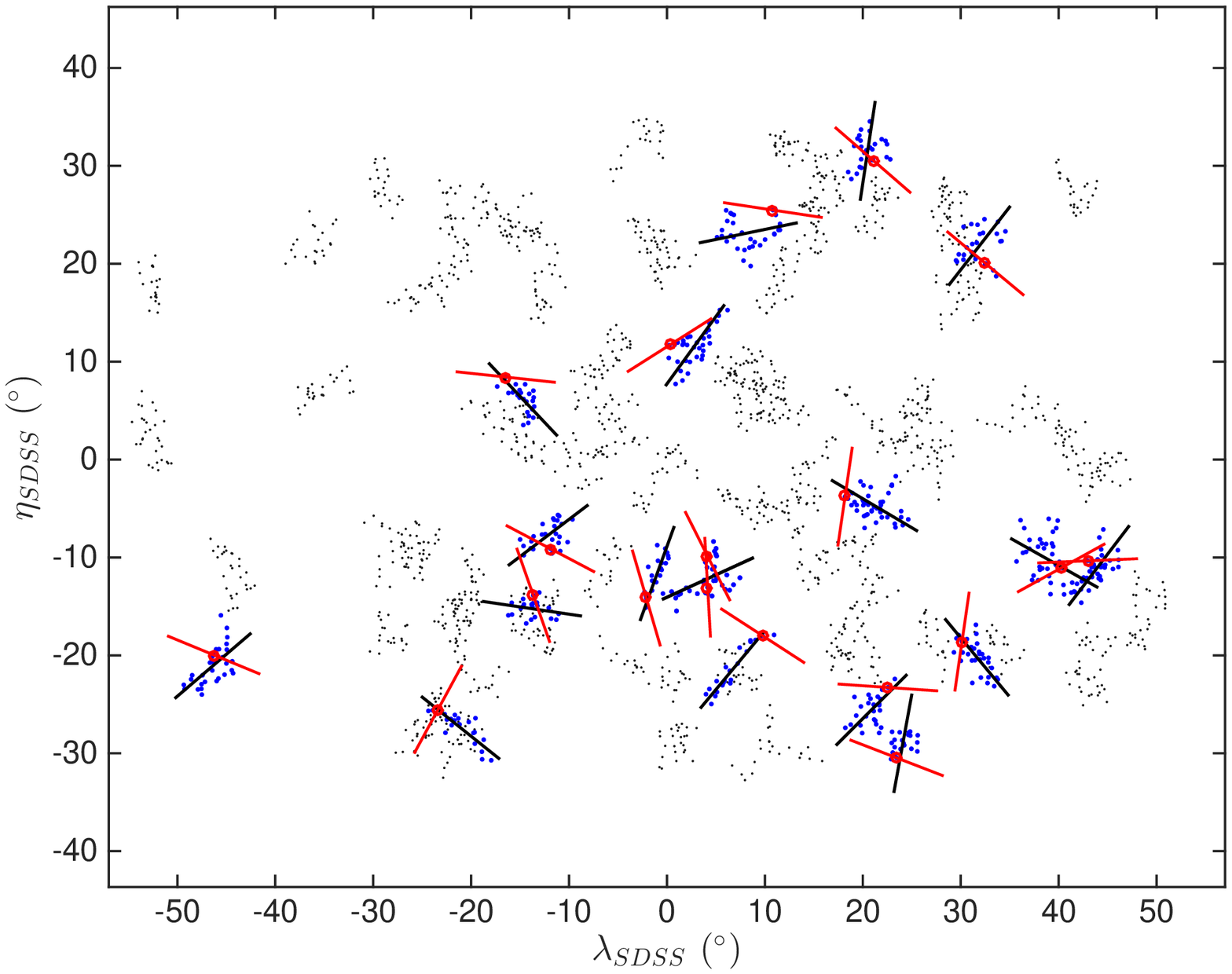}
\caption{\small{Projection on the sky, in SDSS coordinates, of the LQGs
having richness in the range $10 - 19$ (\textit{top}) and larger than $20$
(\textit{bottom}). The LQGs (clouds of grey dots) containing at least one
quasar with polarization measurements (circled in red) are highlighted in
blue. The black lines trace the orientation of the projected major axes of
the groups (here at the centres of masses). The red lines give the
orientations of the polarization vectors. All lines are of equal lengths for
visualization. The polarization vectors are preferentially parallel to the
MPAs of the groups (\textit{top}) and preferentially perpendicular
(\textit{bottom}).}}
\label{fig:skyPlot}
\end{minipage}
\end{center}
\end{figure}

\chaptertoc{Conclusion and outlook}
Quasars, as well as less luminous active galactic nuclei, have the
particularity to emit polarized light due to the anisotropic geometrical
configuration of their components or to their inherent magnetic fields.
The study of polarization thus provides a unique tool to probe the
characteristics of the sources of electromagnetic radiation in quasars
and of the medium through which this radiation propagates. Along the
years, the study of the polarization of quasars has proven its power to
infer their morphology and their orientation with respect to the line of
sight of the observer while they are too far to be resolved by
telescopes.

\medskip

The comparison of the orientations of the polarizations from quasars
that are separated by billions of light-years has led to the striking
discovery that they are aligned instead of being randomly distributed
as expected (\citealt{Hutsemekers1998}).
Since then, this discovery has been confirmed and the significance of
the correlations enhanced (\citealt{Hutsemekers-Lamy2001};
\citealt{Jain-Narain-Sarala2004} and \citealt{Hutsemekers-et-al2005}).
These alignments may challenge our current understanding of the
Universe as they imply correlations over cosmological scales
($\gtrsim 1\,\rm{Gpc}$), much beyond the hypothesized homogeneity
scales of the Universe.

\medskip

These alignments of quasar-polarization vectors did not find
obvious explanation as the variety of scenarios that have been
proposed among the years shows. A non-exhaustive list of these
includes:
fundamental constant variation, cosmic strings, cosmic birefringence,
cosmological magnetic fields, new dark-matter particle candidates,
anisotropic or rotational cosmological models, bad sampling of the
data set,...
(see e.g., \citealt{Chang-Wang-Li2012}; \citealt{Poltis-Stojkovic2010};
\citealt{diSeregoAlighieri2015}; \citealt{Urban-Zhitnitsky2010};
\citealt{Tiwari-Jain2015}; \citealt{Das-et-al2005};
\citealt*{Payez-Cudell-Hutsemekers2008};
\citealt{Hutsemekers-et-al2008}; \citealt*{Agarwal-Kamal-Jain2011};
\citealt{Hutsemekers-et-al2005}; \citealt{Ciarcelluti2012};
\citealt{Kuvschinova-Panov2014}; \citealt{Joshi-et-al2007}).
Nevertheless, none have been successful to account properly for the
observational specificities of the alignments
(e.g., \citealt{Hutsemekers-et-al2010}; \citealt{diSeregoAlighieri2015}).

\medskip

We devoted
this work to a careful analysis of this intriguing and unsolved
anomaly. In Chapter~\ref{Ch-Statistics}, we reviewed statistical tests
that have been extensively used to characterize the polarization
alignments. We then introduced a new and independent statistical method.
The latter allowed us to re-analyse independently the current data set
of optical-polarization measurements of 355 quasars in
Chapter~\ref{Ch:PC-1Analysis}. It further allowed us to proceed to an
unbiased identification of the regions of the space where the
alignments of the quasar polarization vectors are the more significant.
As a result of our analysis, the very-large-scale alignments of the
quasar-optical-polarization vectors are put on stronger grounds.

Given that the status of such alignments at radio
wavelengths was not clear, and especially that the redshift of the
sources had never been properly taken into account, we proceeded in
Chapter~\ref{Ch:PH-1} to a new analysis of the polarization vectors of
a large sample of flat-spectrum radio sources drawn from the
JVAS/CLASS 8.4-GHz surveys which is (claimed) free of biases
(\citealt{Jackson-et-al2007}). Contrary to previous studies, we found
significant large-scale alignment patterns within this sample.
Interestingly, these alignments involve only quasars among the
variety of flat-spectrum radio sources contained in the sample. While
our findings prove difficult to explain either by biases in the data set or
by physical effects, the fact that regions of quasar-radio-polarization
alignments are found either nearby or within the regions of the sky
where optical-polarization alignments are the most significant is
striking and suggests a common explanation.
Unfortunately, as we discussed in Chapter~\ref{Ch:PH-1} a combination
of the two data sets and a comparison of the alignments at optical and
radio wavelengths turned out to be difficult because of poor overlap of
the full samples.
A systematic comparison of the alignments from both radio and optical
samples is still lacking and should be investigated in the future when
more data are available.
Meanwhile, the fact that the alignments features from both spectral
bands may be due to the same physical mechanism is worth keeping
in mind and would imply strong phenomenological constraints on
possible scenarios.

More polarization data are thus clearly needed to provide a better
assessment.
A more homogeneous sample is required at optical wavelengths and
an observational confirmation of the reliability of the radio data set is
mandatory. In this respect, we are looking forward to the upcoming
compilation of quasar optical polarization measurements which will
increase the sample size and provide a better sampling in redshift
(Sluse 2015, private communication).

\medskip

As we were searching for the mechanism responsible for these very
large-scale alignments of the polarization vectors, we
measured the optical polarizations of the quasars belonging to two
very large quasar groups (LQG) at a cosmological redshift
$z\sim 1.3$, i.e. when the Universe was about one third of its age
(\citealt{Clowes-et-al2013}).
As we showed in Chapter~\ref{Ch:HBPS2014}, quasar polarization
orientations were found to be either parallel or perpendicular to the
orientations of the LQGs. Given the established relation between the
orientations of the optical polarization vectors of quasars and the
directions of their structural axes, we inferred that the spin axes of
the supermassive black hole at the centre of quasars tend to align
themselves with their neighbouring large-scale structures assumed
to be traced by the LQGs. This interpretation is supported by spectral
properties of the targeted quasars. Indeed, those that show a broader
Mg II emission spectral line, and thus that are viewed
with higher inclinations, are those that have polarization vectors
preferentially perpendicular to the LQG axes, so that the black-hole
spins are parallel.
To confirm our observations and interpretations we submitted a
proposal for new optical polarization measurements of quasars that
belong to another very rich large quasar group that has an average
redshift of $\sim 1.6$ from the LQG sample of
\citet{Einasto-et-al2014}. Unfortunately, the survey has not been
completed and only three new reliable data points have been
obtained. It is worth mentioning that the three new data points
confirm our expectations.

\medskip

In order to confirm independently our previous results and to question
further the correlations that the black hole spin axes have with the
major axes of the LQGs they belong to, we used quasar radio
polarization measurements from the JVAS/CLASS 8.4-GHz surveys
(\citealt{Jackson-et-al2007}) and a large sample of LQGs drawn from
the high-redshift part of the SDSS DR7 quasar catalogue
(\citealt{Einasto-et-al2014}).
Polarization in this radio sample is thought to be due to synchrotron
emission in the core of the active galactic nucleus where the magnetic
field is expected to be parallel to the central black hole spin axis
(\citealt{Jackson-et-al2007}; \citealt{Wardle2013}).
Hence, irrespective of the inclination of the sources with respect to
the line of sight of the observer, the polarization vectors are expected
to be perpendicular to the spin axes of the black holes and therefore
perpendicular to the LQG major axes if the black hole spin axes
actually align with them. Our analysis, presented in
Chapter~\ref{Ch:PH-2}, reinforces our previous study at optical
wavelengths. We indeed found that black hole spin axes of quasars
that belong to very rich groups align with the major axes of the
large quasar groups.
We additionally found that the preferred spin axis orientations
significantly depend on the richness of the quasar groups.

\medskip

As discussed in Chapters~\ref{Ch:HBPS2014} and~\ref{Ch:PH-2},
what could cause the spin axes of the black holes at the centre of
quasars to be aligned with the large quasar group axes is the angular
momentum transfer from matter during its gravitational collapse to
form galaxies in the young Universe.
Supported by N-body Monte Carlo simulations, the tidal torque theory
tells how galaxies acquire their spin during their formation and how it
evolves while they drift through the cosmic web which is in constant
evolution (e.g., \citealt{White1984}; \citealt{Heavens-Peacock1988};
\citealt{Codis-et-al2012}; \citealt{Laigle-et-al2015}).
Given this scenario the spin of the galaxies and so their apparent
morphologies are expected to be correlated to specific directions in
space. These are defined by the geometrical characteristics of the
gravitational potential of the surrounding matter. Such an
arrangement of matter and angular momentum is not only observed in
simulations but is also inferred from real observations in the low
redshift Universe (e.g., \citealt{Tempel-Libeskind2013};
\citealt{Zhang-et-al2013}; \citealt{Li-et-al2013}).
Our analyses provide the first high-redshift observational confirmation
of this effect, but on larger scales than predicted.
Indeed, according to theory, the intrinsic alignments between the spins
of galaxies are observed to drop rapidly as their separation increases
(see \citealt{Kirk-et-al2015} for a review). Our observations show that
the story might be different and that these correlations may happen at
much larger scales. This would be true at least in the young Universe
and for galaxies that harbour active nuclei.
Our results constitute a hint that there might be a missing ingredient in
our understanding of the large-scale structure formation and more
generally in the current cosmological paradigm.

Moreover, our analyses imply that the intrinsic alignments could have
non-vanishing effects for the upcoming experiments that intend to
constrain the cosmological parameters. This is especially relevant for
those that will investigate the nature of the dark energy through the
study of the deformation of the galaxy images induced by massive
objects located along the path of the light, deformation known as the
weak gravitational lensing effect (\citealt{Laureijs-et-al2011};
\citealt{Amendola-et-al2013}; \citealt{Kirk-et-al2015}).
It is therefore of prime importance to understand these large-scale
correlations. In particular, this requires a better understanding of the
relation between the large-scale structures of the Universe and the
large quasar groups defined through friends-of-friends algorithms
from the quasar distribution.
Indeed, the link between the quasars and the large-scale structures is
not well established.
While quasars are expected to be located in sufficiently dense regions
of the matter distribution for the supermassive black holes to form, it is
not guaranteed that they trace the highest density regions and hence,
the large-scale structures of the Universe (\citealt{Fanidakis-et-al2013}).
This assumption actually underlaid the interpretation given in
Chapters~\ref{Ch:HBPS2014} and~\ref{Ch:PH-2}.

As shown by analyses in the local Universe, quasars are likely to be
located at the peripheries of the clusters and filaments. Nearby
quasars, typically found in small groups and poor clusters of galaxies,
are indeed located in relatively low-density and large-scale
environments that surround the supercluster of galaxies (e.g.,
\citealt*{Sochting-Clowes-Campusano2002},
\citeyear{Sochting-Clowes-Campusano2004};
\citealt{Coldwell-Lambas2006}; \citealt{Lietzen-et-al2009},
\citeyear{Lietzen-et-al2011}).
It has been suggested that, at high redshift, the large quasar groups
mark the precursors of the superclusters that we observe in the local
Universe (e.g., \citealt*{Einasto-Joeveer-Saar1980};
West [\citeyear{West1994}; \citeyear{West1999}]) and that they are
perhaps the seeds of supercluster complexes such as the famous Sloan
Great Wall (\citealt{Einasto-et-al2011}). This still needs to be confirmed.
Forthcoming deep and wide galaxy surveys will definitely help to
understand better the interplay between quasars and their surroundings
as well as to question the physical nature of the large quasar groups.
In this respect, cross-correlations of the quasar distributions with maps
of the cosmic microwave background may help.

Indeed, the photons from the cosmic microwave background (CMB)
have travelled from the surface of last scattering to Earth, have been
deflected, redshifted and blue shifted by the gravitational potentials
of massive structures and have been heated up by energetic free
electrons filling the haloes of these structures (e.g.,
\citealt*{Challinor-Ford-Lasenby2000}; \citealt{Gawiser-Silk2000};
\citealt{Wands-Piattella-Casarini2015}).
These phenomena, encoded in the CMB and known as the weak
lensing, the integrated Sachs-Wolfe and the thermal Sunyaev-Zel'dovich
effects, have been successfully mapped by the Planck satellite and
contain information about the large-scale structures. If the large
quasar groups somehow trace the cosmic web, then characteristic
anisotropies in the CMB should correspond to them.
These imprints could be detected by stacking methods such as those
that allowed the discovery of the integrated Sachs-Wolfe effect
induced by low-redshift superstructures embedded in our
accelerated-expanding Universe (\citealt{Padmanabhan-et-al2005};
\citealt{Granett-Neyrinck-Szapudi2008}; \citealt{PlanckXXI2015}).
Recent pioneering work have been done in that direction (e.g.,
\citealt{DiPompeo-et-al2015};
\citealt*{EneaR-Cornejo-Campusano2015}).
They demonstrate that such an analysis can be conducted to
understand better what is the role of quasars in the evolution of the
large-scale structures and, more importantly for our purpose, to
investigate the physical reality of the large quasar groups.

\medskip

With respect to this, our observation of the correlation of quasar
polarization orientations with the axes of the large quasar groups can
be read as an argument in favour of the reality of these large-scale
quasar structures. Still, many questions need to be carefully addressed.
An important one is the choice of the linking length used to define the
groups. Furthermore, at optical wavelengths, we have considered the
LQGs as clouds of points on the celestial sphere and we have split the
Huge-LQG into three different branches. This suggests that
the Huge-LQG might actually be a complex of large quasar groups
defined with a smaller linking length as noted by \citet{Park-et-al2015}.
When we compared the axes of the large quasar groups to the radio
polarizations, though, we did not consider such subdivision of the
largest quasar groups.
It is clear that we have to address the implication on our results of such
procedures and, more generally, of the impact of the linking length which
is chosen to build the large quasar groups. Such detailed analysis will
required dedicated algorithms and statistical methods.

\medskip

In Chapters~\ref{Ch:PH-1} and~\ref{Ch:HBPS2014}, we mentioned that
the alignments between the supermassive black hole spins and the LQG
axes could provide an explanation for the very
large-scale-quasar-polarization-vector alignments (discussed in the first
chapters of this work). This would constitute an important result. However,
the scales at which the correlations are observed do not match. Indeed,
the very large-scale alignments involve distances that are two to four
times larger than the diameters of the large quasar groups in which we
found correlations.
Beside, one may wonder why the optical polarization vectors inside the
very large-scale regions of alignments have been reported to be parallel
to each other rather than parallel and perpendicular as it is the case
inside LQGs. One possibility to reconcile these two kinds of alignments
could be selection biases inside the sample of the 355
quasar-optical-polarization measurements (Hutsem{\'e}kers 2015, private
communication).
If this hypothesis turns out to be correct, it would imply coherence of
the orientations of the LQGs over cosmological scales, at least for some
regions of the comoving space.

\bigskip

To conclude,
we started this work with an intriguing anisotropy in the distribution
of the quasar polarization vectors which challenges the current
cosmological paradigm. We confirmed this anomaly and
provided evidence for similar alignments at radio
wavelengths. We then found correlations linking the quasar
polarization vectors and the orientations of large quasar groups.
The possibility that the very-large-scale alignments of
quasar-polarization vectors and their correlations with the large
quasar groups are due to the alignment of the supermassive black
hole spin axis with the cosmic web may open a new chapter in the
detection and the characterization of the large-scale structures and
lead to new questions about the evolution of the Universe.

\cleardoublepage

\addcontentsline{toc}{chapter}{Bibliography}
\bibliographystyle{mn2e}
\bibliography{mn-jour,myReferences}

\end{document}